\documentclass[11pt,a4paper]{article}
\pdfoutput=1
\usepackage{jheppub}

\usepackage[english]{babel}
\usepackage{amsfonts,amsmath,amssymb,amsthm,bm,mathtools,slashed,empheq}
\usepackage{bbold}
\usepackage{comment,enumerate,footnote,graphicx,subfloat,relsize,footnote}
\usepackage{array,tabularx,tabu,multirow,framed,capt-of,colortbl}
\usepackage[usenames,dvipsnames]{xcolor}
\usepackage{booktabs,hhline}
\numberwithin{equation}{section}
\makeatletter
\g@addto@macro\bfseries{\boldmath}
\makeatother
\usepackage{cleveref}
\crefformat{pluralequation}{#2\black{eqs.~(}#1\black{)}#3}
\Crefformat{pluralequation}{#2\black{Equations~(}#1\black{)}#3}
\crefformat{pluralfigure}{#2\black{figs.~}#1#3}
\Crefformat{pluralfigure}{#2\black{Figures~}#1#3}
\usepackage{tikz}
\usetikzlibrary{arrows,shapes}
\usetikzlibrary{trees,patterns}
\usetikzlibrary{matrix,arrows} 				% For commutative diagram
\usetikzlibrary{positioning}				  % For "above of=" commands
\usetikzlibrary{calc,through}				  % For coordinates
\usetikzlibrary{decorations.pathreplacing}  % For curly braces
\usepackage{pgffor}							% For repeating patterns

\usetikzlibrary{decorations.pathmorphing}	% For Feynman Diagrams
\usetikzlibrary{decorations.markings}
\tikzset{
>=stealth', %%  (Un)comment for (more) less conventional arrows
vector/.style={decorate, decoration={snake}, draw},
provector/.style={decorate, decoration={snake,amplitude=2.5pt}, draw},
antivector/.style={decorate, decoration={snake,amplitude=-2.5pt}, draw},
bigvector/.style={decorate, decoration={snake,amplitude=4pt}, draw},
fermion/.style={draw=black, postaction={decorate}, 
	decoration={markings,mark=at position .55 with {\arrow[draw=black]{>}}}},
fermionbar/.style={draw=black, postaction={decorate},
    decoration={markings,mark=at position .55 with {\arrow[draw=black]{<}}}},
fermionnoarrow/.style={draw=black},
doublefermion/.style={draw=black,double, postaction={decorate},
	decoration={markings,mark=at position .57 with {\arrow[draw=black]{>}}}},
doublefermionbar/.style={draw=black,double, postaction={decorate},
	decoration={markings,mark=at position .57 with {\arrow[draw=black]{<}}}},
doublefermionnoarrow/.style={draw=black,double},
gluon/.style={decorate, draw=black,
    decoration={coil,amplitude=4pt, segment length=5pt}},
scalar/.style={dashed,draw=black, postaction={decorate},
	decoration={markings,mark=at position .55 with {\arrow[draw=black]{>}}}},
scalarbar/.style={dashed,draw=black, postaction={decorate},
    decoration={markings,mark=at position .55 with {\arrow[draw=black]{<}}}},
scalarnoarrow/.style={dashed,draw=black},
momentum/.style={draw=black, postaction={decorate},
    decoration={markings,mark=at position 1 with {\arrow[draw=black]{>}}}},
antimomentum/.style={draw=black, postaction={decorate},
    decoration={markings,mark=at position 0.1 with {\arrow[draw=black]{<}}}}
}

\tikzstyle{block} = [draw, rectangle, minimum height=3em, minimum width=6em]
    
\newcommand{\nc}{\newcommand}
\nc{\pd}{\partial}
\nc{\bea}{\begin{eqnarray}}
\nc{\eea}{\end{eqnarray}}
\nc{\bal}{\begin{alignedat}}
\nc{\eal}{\end{alignedat}}
\nc{\beq}{\begin{equation}}
\nc{\eeq}{\end{equation}}
\nc{\bit}{\begin{itemize}}
\nc{\eit}{\end{itemize}}
\nc{\benu}{\begin{enumerate}}
\nc{\eenu}{\end{enumerate}}
\nc{\bdes}{\begin{description}}
\nc{\edes}{\end{description}}
\nc{\bma}{\begin{pmatrix}}
\nc{\ema}{\end{pmatrix}}

\newcommand{\black}[1]	{{\color{black} 	#1}}

\nc{\nn}{\nonumber}
\nc{\hc}{\text{h.c.}}
\nc{\cc}{\text{c.c.}}

\nc{\abs}[1]{\left| #1 \right|}
\def\[{\left[}
\def\]{\right]}
\def\({\left(}
\def\){\right)}
\def\<{\langle}
\def\>{\rangle}

\def\GeV{{\rm GeV}}
\def\TeV{{\rm TeV}}

\def\A			{\mathsmaller{A}}
\def\B			{\mathsmaller{B}}
\def\Bstar		{\mathsmaller{B^*}}
\def\Bcal		{\mathsmaller{\cal B}}
\def\D			{\mathsmaller{D}}
\def\Dbar		{\mathsmaller{\bar{D}}}
\def\F			{\mathsmaller{F}}
\def\H			{\mathsmaller{H}}
\def\Hstar		{\mathsmaller{H^*}}
\def\Hdagger	{\mathsmaller{H^\dagger}}
\def\L			{\mathsmaller{L}}
\def\R			{\mathsmaller{R}}
\def\S			{\mathsmaller{S}}
\def\Scal		{\mathsmaller{\cal S}}
\def\T			{\mathsmaller{T}}
\def\V			{\mathsmaller{V}}
\def\Vstar		{\mathsmaller{V^*}}
\def\W			{\mathsmaller{W}}
\def\Wstar		{\mathsmaller{W^*}}

\def\aA			{\alpha_{\A}}
\def\aB			{\alpha_{\Bcal}}
\def\aR			{\alpha_{\R}}
\def\aS			{\alpha_{\Scal}}

\def\aH			{\alpha_{\H}}
\def\ann		{{\rm ann}}
\def\BBSF		{{\B\text{-}\BSF}}
\def\BoffBSF	{{\Bstar\text{-}\BSF}}

\def\BSF		{{\rm \mathsmaller{BSF}}}

\def\dec		{{\rm dec}}

\def\DDbar		{\D\Dbar}

\def\DS			{\D\S}

\def\DP			{\mathsmaller{\rm DP}}

\def\ellS		{\ell_{\Scal}}

\def\EH			{E_{\!\H}}

\def\hH			{h_{\H}}

\def\HBSF		{{\H\text{-}\BSF}}
\def\HoffBSF	{{\Hstar\text{-}\BSF}}
\def\im			{\mathbb{i}}
\def\IP			{\mathsmaller{\rm IP}}

\def\kappaB		{\kappa_{\Bcal}}
\def\kappaS		{\kappa_{\Scal}}

\def\mD			{m_{\D}}
\def\mH			{m_{\H}}

\def\mS			{m_{\S}}
\def\mZ			{m_{\mathsmaller{Z}}}

\def\nD			{n_{\D}}
\def\nDbar		{n_{\Dbar}}
\def\nS			{n_{\S}}
\def\NA			{N_{\A}}
\def\NR			{N_{\R}}
\def\phikell	{\varphi_{|{\bf k}|,\ell}^{}}
\def\PB			{P_{\!\B}}
\def\PBvec		{{\bf P}_{\!\B}}
\def\PH			{P_{\!\H}}
\def\PHvec		{{\bf P}_{\!\H}}
\def\PHdagger	{P_{\!\Hdagger}}
\def\PHdaggervec{{\bf P}_{\!\Hdagger}}

\def\PVvec		{{\bf P}_{_{\!\!\V}}}
\def\PW			{P_{_{\!\!\W}}}
\def\PWvec		{{\bf P}_{_{\!\!\W}}}

\def\Rabelian	{R_{\mathsmaller{U(1)}}}
\def\rad		{{\rm rad}}
\def\RB			{R_{\B}}
\def\RW			{R_{\W}}
\def\RH			{R_{\H}}
\def\RHF		{R_{\H}^{\mathsmaller{F}}}
\def\RHBH		{R_{\H}^{\B\H}}
\def\RHWH		{R_{\H}^{\W\H}}
\def\rA 		{r_{\A}}
\def\rB 		{r_{\B}}
\def\sA 		{s_{\A}}
\def\sB 		{s_{\B}}

\def\SUL		{SU_\mathsmaller{L}(2)}
\def\Svec		{S_{\rm vec}}
\def\Ssc		{S_{\rm scl}}

\def\SS			{\S\S}
\def\SSDDbar	{\S\S\!\mathsmaller{/}\!\D\Dbar}

\def\UY			{U_\mathsmaller{Y}(1)}

\def\vrel		{v_{\rm rel}}
\def\VBSF		{{\V\text{-}\BSF}}
\def\VoffBSF	{{\Vstar\text{-}\BSF}}
\def\WBSF		{{\W\text{-}\BSF}}
\def\WoffBSF	{{\Wstar\text{-}\BSF}}
\def\xB			{x_{\Bcal}}
\def\xS			{x_{\Scal}}
\def\YD			{Y_{\D}}

\def\YH			{Y_{\H}}

\def\yF         {y_{\F}}
\def\zetaA		{\zeta_{\A}}
\def\zetaB		{\zeta_{\Bcal}}
\def\zetaH		{\zeta_{\H}}
\def\zetaR		{\zeta_{\R}}
\def\zetaS		{\zeta_{\Scal}}

\newcommand{\var}[6]{
\arraycolsep=1pt\def\arraystretch{0.7}
\( \begin{array}{ccc} {#1,}&{#2,}&{#3} \\ {#4,}&{#5,}&{#6} \end{array} \)
}
\newcommand*\widefbox[1]{\fbox{\hspace{1ex}#1\hspace{1ex}}}

\begin{document}

\preprint{Nikhef-2021-003}
\title{
Bound states of WIMP dark matter in Higgs-portal models. Part I. 
Cross-sections and transition rates
}

\author{Ruben Oncala and Kalliopi Petraki}

\affiliation{
\href{http://www.lpthe.jussieu.fr/spip/index.php}{\color{black}
Sorbonne Universit\'e, CNRS, Laboratoire de Physique Th\'eorique et Hautes \'Energies, LPTHE, F-75005 Paris, France}}

\affiliation{
\href{https://www.nikhef.nl/en/}{\color{black} Nikhef}, 
Science Park 105, 1098 XG Amsterdam, The Netherlands}

\emailAdd{roncala@nikhef.nl}
\emailAdd{kpetraki@lpthe.jussieu.fr}

\abstract{
We investigate the role of the Higgs \emph{doublet} in the thermal decoupling of multi-TeV dark matter coupled to the Weak interactions of the Standard Model and the Higgs. The Higgs doublet can mediate a long-range force that affects the annihilation processes and binds dark matter into bound states. More importantly, the emission of a Higgs doublet by a pair of dark matter particles can give rise to extremely rapid monopole bound-state formation processes and bound-to-bound transitions. We compute these effects in the unbroken electroweak phase. To this end, we consider the simplest renormalisable fermionic model, consisting of a singlet and a doublet under $SU_{L}(2)$ that are stabilised by a $\mathbb{Z}_2$ symmetry, in the regime where the two multiplets coannihilate. In a companion paper, we use the results to show that the formation of metastable bound states via Higgs-doublet emission and their decay decrease the relic density very significantly.}
\arxivnumber{2101.08666} %%For JHEP

\maketitle

%%%%%%%%%%%%%%%%%%%%%%%%%%%%%%%%%%%%%%%%%%%%%%%%%%%%%%%%%%%%%%%%%%%%
%%%%%%%%%%%%%%%%%%%%%%%%%%%%%%%%%%%%%%%%%%%%%%%%%%%%%%%%%%%%%%%%%%%%
%%%%%%%%%%%%%%%%%%%%%%%%%%%%%%%%%%%%%%%%%%%%%%%%%%%%%%%%%%%%%%%%%%%%
\clearpage
\section{Introduction \label{Sec:Intro}}

Particles coupled to the Weak interactions of the Standard Model (SM), known as WIMPs, have been arguably the most widely considered candidates for dark matter (DM) in the past decades. 
Among the archetypical WIMP models are scenarios where DM is a linear combination of the neutral components of electroweak multiplets that couple to the Higgs doublet. The discovery of the Higgs boson and the measurement of its properties impel the investigation of its implications for such scenarios.

While most related research in the past focused on electroweak-scale WIMP masses, $m \sim 100~\GeV$, the current experimental constraints motivate considering the multi-TeV mass regime more carefully. This is particularly important in view of the numerous existing and upcoming observatories probing high-energy cosmic rays, such as H.E.S.S., IceCube, CTA and KM3Net. The experimental exploration of the multi-TeV scale urges the comprehensive theoretical understanding of the dynamics and possibilities in this regime.

The hierarchy between the multi-TeV and electroweak scales implies the emergence of new effects. In particular, the Weak interactions between particles with multi-TeV mass manifest as \emph{long-range}, since the interaction range $l\sim (100~\GeV)^{-1}$ may be comparable or exceed the de Broglie wavelength $(\mu \vrel)^{-1}$ and/or the  Bohr radius $(\mu \alpha)^{-1}$ of the interacting particles, where $\mu = m/2 \gtrsim\TeV$, $\vrel$ and $\alpha$ are the reduced mass, relative velocity and coupling to the force mediators. The long-range nature of the interactions gives rise to non-perturbative phenomena, the Sommerfeld effect and the existence of bound states.

Bound states form invariably with dissipation of energy. It has been recently shown that the emission of a scalar boson charged under a symmetry alters the effective Hamiltonian between the interacting particles and gives rise to monopole transitions; this renders bound-state formation (BSF) extremely rapid even for small couplings~\cite{Oncala:2019yvj}. For WIMPs, this implies that BSF via emission of a Higgs doublet may be a very significant inelastic process. Moreover, it has been shown in simplified models, that the 125~GeV Higgs boson can mediate a sizeable long-range force between TeV-scale particles, despite being heavier than all SM gauge bosons~\cite{Harz:2017dlj,Harz:2019rro}.

The phenomenological importance of the above is rather large. It has been long known that the Sommerfeld effect~\cite{Sommerfeld:1931,Sakharov:1948yq} -- the distortion of the wavefunction of interacting particles due to a long-range force -- affects the DM annihilation cross-sections at low relative velocities~\cite{Hisano:2002fk}. This, in turn, alters the DM chemical decoupling in the early universe, and consequently the predicted DM mass and couplings to other particles~\cite{Hisano:2003ec}. It also affects the radiative signals expected from the DM annihilations during CMB and inside galaxies today~\cite{Hisano:2004ds}. More recently, it has been realised that the formation and decay of metastable (e.g.~particle-antiparticle) bound states in the early universe can decrease the DM density~\cite{vonHarling:2014kha}, and contribute to the DM indirect detection signals~\cite{Pospelov:2008jd,MarchRussell:2008tu,An:2016kie,Cirelli:2016rnw,Asadi:2016ybp,Baldes:2017gzw,Baldes:2017gzu,Cirelli:2018iax}. Importantly, BSF can be faster than annihilation in a variety of models~\cite{Petraki:2015hla,Petraki:2016cnz,Harz:2018csl,Harz:2019rro,Oncala:2019yvj}, and it can also produce novel indirect signals even in models where annihilation is absent or suppressed~\cite{Pearce:2013ola,Cline:2014eaa,Petraki:2014uza,Pearce:2015zca,Baldes:2020hwx,Detmold:2014qqa,Mahbubani:2019pij}.

Collecting the above considerations, in this work we consider bound states in models of WIMP DM coupled to the Higgs,  and in a companion paper~\cite{Oncala:2021swy} we compute their effect on the DM relic density. We are interested in scenarios that feature a trilinear coupling between the DM and Higgs multiplets, i.e.~$\delta {\cal L} \supset \bar{\chi}_{n} H \chi_{n+1} +$~h.c., where $\chi_n$ is a fermionic or bosonic $n$-plet under $\SUL$, and DM is the lightest linear combination of the neutral $\chi_n$ and $\chi_{n+1}$ components after electroweak symmetry breaking. Such scenarios have been considered extensively in the literature, and appear also in supersymmetry, e.g.~Bino-Higgsino or Higgsino-Wino DM~\cite{Mahbubani:2005pt,DEramo:2007anh,Cohen:2011ec,Cheung:2013dua,Calibbi:2015nha,Banerjee:2016hsk,Fraser:2020dpy,Freitas:2015hsa,Lopez-Honorez:2017ora,Wang:2018lhk,Abe:2019wku,Dedes:2014hga,Yaguna:2015mva,Tait:2016qbg,Beneke:2016jpw,Bharucha:2018pfu,Filimonova:2018qdc,Betancur:2018xtj}. For concreteness, we shall focus on the most minimal such model with a Majorana singlet and a Dirac doublet~\cite{Mahbubani:2005pt,DEramo:2007anh,Cohen:2011ec,Cheung:2013dua,Calibbi:2015nha,Banerjee:2016hsk,Fraser:2020dpy,Freitas:2015hsa,Lopez-Honorez:2017ora,Wang:2018lhk,Abe:2019wku}.

Even neglecting bound state effects during the DM freeze-out, many of these models are predicted to reside in the TeV scale (see~\cite{Lopez-Honorez:2017ora} for a summary). BSF in the early universe increases the DM destruction rate and, for a given set of couplings, pushes the predicted DM mass to even higher values~\cite{vonHarling:2014kha}. If DM is heavier than 5~TeV, freeze-out begins before the electroweak phase transition (EWPT), even though it may be completed much later. Here we will assume electroweak symmetry throughout; this considerably simplifies the anyhow lengthy calculations, and is a practice that has been followed in past literature~\cite{Lopez-Honorez:2017ora}. 
We discuss in detail the extent of validity of this approximation for the DM thermal decoupling in ref.~\cite{Oncala:2021swy}. Here we simply note that, by virtue of the Goldstone boson equivalence theorem, BSF via Higgs-doublet emission remains important in the broken electroweak phase, where it occurs via emission of the Higgs boson and the longitudinal components of the Weak gauge bosons~\cite{Ko:2019wxq}.

In the epoch prior to the EWPT, if the two DM multiplets are nearly mass-degenerate (perhaps as a result of a larger symmetry group), the DM freeze-out is determined by all their self-annihilation and co-annihilation processes~\cite{Edsjo:1997bg}. Our computations focus on this regime. Then, in the class of models considered, the entire Higgs doublet contributes to the (long-range) potential between the DM multiplets. Moreover, bound states can form via emission of a Higgs doublet. As we shall see, the cross-sections for BSF via Higgs-doublet emission can exceed those for annihilation or BSF via vector emission by orders of magnitude, and result in a late period of DM chemical recoupling~\cite{Oncala:2019yvj}. This, in turn, opens the possibility that thermal-relic WIMP DM may be much heavier than anticipated, potentially approaching the unitarity limit~\cite{Oncala:2021swy}, which in the non-relativistic indeed implies (or presupposes) the presence of long-range interactions~\cite{Baldes:2017gzw}.

This work is organized as follows. In \cref{Sec:LongRangeDynamics}, we introduce the DM model and derive the long-range potentials between the DM multiplets in the unbroken electroweak phase. We identify the scattering and bound eigenstates of these potentials by appropriate spin and gauge projections, before computing the DM annihilation processes and bound-state decay rates. In \cref{Sec:BSF}, we calculate all the radiative BSF cross-sections, while in \cref{Sec:BSF_OffShell} we consider BSF via scattering on a relativistic thermal bath. These results are employed in the companion paper~\cite{Oncala:2021swy} to compute the DM thermal decoupling in the early universe. We conclude in \cref{Sec:Conclusion}. Several technical aspects are addressed in the appendices, including a proof of the (anti)symmetrisation of the two-particle-irreducible kernels and wavefunctions in the case of identical particles (\cref{App:IdenticalParticles}), and the analytical computation of monopole bound-to-bound transitions in the Coulomb limit (\cref{App:OverlapIntegral_RBoundToBound}.)

%%%%%%%%%%%%%%%%%%%%%%%%%%%%%%%%%%%%%%%%%%%%%%%%%%%%%%%%%%%%%%%%%%%%
%%%%%%%%%%%%%%%%%%%%%%%%%%%%%%%%%%%%%%%%%%%%%%%%%%%%%%%%%%%%%%%%%%%%
%%%%%%%%%%%%%%%%%%%%%%%%%%%%%%%%%%%%%%%%%%%%%%%%%%%%%%%%%%%%%%%%%%%%
\clearpage
\section{Long-range dynamics in the unbroken electroweak phase \label{Sec:LongRangeDynamics}}

\subsection{The model \label{sec:Model_Lagrangian}}

We introduce a gauge-singlet Majorana fermion
$S = (\psi_\alpha, \psi^{\dagger \dot{\alpha}})^{\T}$ 
of mass $\mS$, as well as a Dirac fermion  
$D=(\xi_{\alpha}, \chi^{\dagger \dot{\alpha}})^{\T}$ 
of mass $\mD$ with SM gauge charges $\SUL \times \UY=(\mathbb{2}, 1/2)$. We assume that $S$ and $D$ are odd under a $\mathbb{Z}_2$ symmetry that leaves all the SM particles unaffected. Under these assignments, the new degrees of freedom (dof) allow to extend the SM Lagrangian by the following interactions
\begin{align}
{\delta \cal L} &= 
 \frac{1}{2} \bar{S} (\im \slashed{\partial} -\mS )S
+\overline{D} (\im \slashed{\cal D} -\mD )D
-(y_{\L} \bar{D}_{\L} H S + y_{\R} \bar{D}_{\R} H S + \hc),
\label{eq:Lagrangian}
\end{align}
where $H$ is the SM Higgs doublet of mass $\mH$ and hypercharge $\YH = 1/2$, and 
$D_{\L} \equiv P_{\L} D = (\xi_{\alpha},0)^{\T}$ and $D_{\R} \equiv P_{\R} D = (0,\chi^{\dagger \dot{\alpha}})^{\T}$, 
with $P_{\R,\L} = (1 \pm \gamma_5)/2$ being the right-handed and left-handed projection operators. In the above, ${\cal D}_\mu \equiv \partial_\mu - \im g_1 Y B_\mu - \im g_2 W_\mu^a t^a$ is the covariant derivative, with $t^a = \frac{1}{2} (\sigma^1, \sigma^2, \sigma^3)$ and $\sigma$ being the Pauli matrices. The particle content of \cref{eq:Lagrangian} is summarised in \cref{tab:particles}.
\begin{table}[h!]
\centering
\begin{equation*}
\begin{tabular}{|c|c|c|c|}
\hline
field & $\SUL$ & $\UY$ & $\mathbb{Z}_2$ \\
\hline
$S$ & 1 & 0   & $-1$\\
$D$ & 2 & 1/2 & $-1$\\
$H$ & 2 & 1/2 & $+1$\\
\hline
\end{tabular}
\end{equation*}
\caption{\label{tab:particles} Particle content and charge assignments}
\end{table}

Here, we have taken the mass parameters $\mS$ and $\mD$ to be real. This can always be achieved by rephasing $\psi$ and either $\xi$ or $\chi$. Rephasing the remaining spinor eliminates the phase of one of the Yukawa couplings. Thus the free parameters of the present model are 4 real couplings (two masses and two dimensionlesss Yukawa couplings), and a phase that allows for CP violation.

We are interested in the regime in which $S$ and $D$ can co-annihilate efficiently before the EWPT of the universe, which occurs if their masses are similar, within about $10\%$. This is because the number density of the heavier species in the non-relativistic regime is suppressed with respect to that of the lighter species by a factor $\exp[-(\delta m/m) x]$. Since $x\equiv m/T \gtrsim 25$ during DM freeze-out, if $\delta m/ m \gtrsim 10\%$ the co-annihilations and the self-annihilations of the heavier species are typically subdominant to the self-annihilations of the lightest species, although their relative importance depends also on the corresponding cross-sections. Such a small discrepancy in the masses does not significantly affect (most of) the cross-sections that we compute in the following; we shall thus take the masses to be equal, which greatly simplifies the computations and allows to obtain analytical results,
\begin{align}
\mD = \mS \equiv m .
\label{eq:MassEquality}
\end{align}
We will very often use the reduced mass of a pair of DM particles,
\begin{align}
\mu \equiv m/2 .
\label{eq:ReducedMass}
\end{align}

Moreover, in order to reduce the number of free parameters, we set
\begin{align}
y_{\L} = y_{\R} \equiv y ,
\label{eq:YukawaEquality}
\end{align}
which we take to be real. (The CP violation is anyway not important for our purposes.) Our computations can of course be extended to more general Yukawa couplings. As is standard, we define the couplings
\begin{align}
\alpha_1 \equiv \frac{g_1^2}{4\pi}, \qquad
\alpha_2 \equiv \frac{g_2^2}{4\pi}, \qquad
\aH \equiv \frac{y^2}{4\pi}.
\label{eq:alphas_def}
\end{align}

Various aspects of its phenomenology and experimental constraints of the model \eqref{eq:Lagrangian} have been considered extensively in the past~\cite{Mahbubani:2005pt,DEramo:2007anh,Cohen:2011ec,Cheung:2013dua,Calibbi:2015nha,Banerjee:2016hsk,Fraser:2020dpy,Freitas:2015hsa,Lopez-Honorez:2017ora,Wang:2018lhk}, and we shall not review them here. We only note that after the EWPT, $S$ and $D$ mix to produce three neutral mass eigenstates, the lightest of which is stable and can play the role of DM. In the following, we focus on computing the long-range effects in the unbroken electroweak phase. In the companion paper \cite{Oncala:2021swy}, we briefly review the mass eigenstates and their interactions after electroweak symmetry breaking for the choice of parameters denoted in \cref{eq:MassEquality,eq:YukawaEquality}, before computing the DM decoupling in the early universe, which alters the predicted mass-coupling relation, thereby affecting all experimental constraints.

\subsection{Static potentials \label{sec:LongRangeDynamics_Potential}}

The $D$, $\bar{D}$ and $S$ fermions interact with each other via the $B$, $W$ and $H$ boson exchanges that give rise to long-range potentials. The kernels generating these potentials are shown in \cref{fig:2PI}.  To compute them, we decompose the incoming and outgoing momenta as
\begin{subequations}
\label{eq:MomentaDecomposition}
\label[pluralequation]{eqs:MomentaDecomposition}
\begin{align}
p_1 = P/2 + p, \qquad p_1' = P/2 +p' , \\ 
p_2 = P/2 - p, \qquad p_2' = P/2 -p' ,
\end{align}
\end{subequations}
where $P$ is the total momentum and $\pm p$ ($\pm p'$) are the momenta of incoming (outgoing) particles in the center-of-momentum (CM) frame. 
\begin{figure}
\centering	
\begin{tikzpicture}[line width=1pt, scale=1]
%%%%%%%%%%%%%%%%%%%%%%%%%%%%%%%%%%%%%%%%%%%%%%%%%%%%%%%%%%%%%%%%%%%%%%%%%%%%%%%%%%%%%%%%%%%%%
%%%%%%%%%%%%%%%%%%%%%%%%%%%%%%%%%%%%%%%%%%%%%%%%%%%%%%%%%%%%%%%%%%%%%%%%%%%%%%%%%%%%%%%%%%%%%
%%%%%%%%%%%%%%%%%%%%%%%%%%%%%%%%%%%%%%%%%%%%%%%%%%%%%%%%%%%%%%%%%%%%%%%%%%%%%%%%%%%%%%%%%%%%%
\begin{scope}[shift={(0,0)}]
\begin{scope}[shift={(-3,0)}]
\node at (-2,1){$D_i^{s_1}$};
\node at (-2,0){$\bar{D}_j^{s_2}$};
\draw[->] (-1.5, 1.3) -- (-1.1, 1.3);\node at(-1.3, 1.6){$p_1$};
\draw[->] (-1.5,-0.3) -- (-1.1,-0.3);\node at(-1.3,-0.6){$p_2$};
\draw[doublefermion] 	(-1.6,1) -- (-0.7,1);\draw 	(+1.6,1) -- (+0.7,1);
\draw[doublefermionbar] (-1.6,0) -- (-0.7,0);\draw	(+1.6,0) -- (+0.7,0);
\draw[->] (+1.1, 1.3) -- (+1.5, 1.3);\node at(+1.3, 1.6){$p_1'$};
\draw[->] (+1.1,-0.3) -- (+1.5,-0.3);\node at(+1.3,-0.6){$p_2'$};
\node at (2,1){$S^{s_1'}$};
\node at (2,0){$S^{s_2'}$};
\draw[fill=white,shift={(0,0.5)}] (-0.9,-0.7) rectangle (0.9,0.7);
\node at (0,0.5){$\im {\cal K}_{\mathsmaller{D\bar{D}\leftrightarrow SS}}$};
\end{scope}
\node at (0  ,0.5){$=$};
\node at (0.8,0.5){$\dfrac{1}{2} \(\vphantom{
\begin{array}{c} a \\ a \\ a \\ a \end{array}
}\right.$};
\begin{scope}[shift={(2.8,0)}]
\node at (-1.3,1){$D_{i}^{s_1}$};
\node at (-1.3,0){$\bar{D}_{j}^{s_2}$};
\draw[->] (-0.9, 1.3) -- (-0.4, 1.3);\node at(-0.6, 1.6){$p_1$};
\draw[->] (-0.9,-0.3) -- (-0.4,-0.3);\node at(-0.6,-0.6){$p_2$};
\draw[doublefermion]	(-1,1) -- (0,1);
\draw[doublefermionbar]	(-1,0) -- (0,0);	
\draw[scalar]	(0,1) -- (0,0);
\node at (0.5,0.5){$H$};
\draw 	(0,1) -- (1,1);
\draw 	(0,0) -- (1,0);	
\draw[->] (+0.4, 1.3) -- (+0.9, 1.3);\node at(+0.6, 1.6){$p_1'$};
\draw[->] (+0.4,-0.3) -- (+0.9,-0.3);\node at(+0.6,-0.6){$p_2'$};
\node at (1.3,1){$S^{s_1'}$};
\node at (1.3,0){$S^{s_2'}$};
\end{scope}
\node at (5,0.5){$+$};
\begin{scope}[shift={(7.2,0)}]
\node at (-1.3,1){$D_{i}^{s_1}$};
\node at (-1.3,0){$\bar{D}_{j}^{s_2}$};
\draw[->] (-0.9, 1.3) -- (-0.4, 1.3);\node at(-0.6, 1.6){$p_1$};
\draw[->] (-0.9,-0.3) -- (-0.4,-0.3);\node at(-0.6,-0.6){$p_2$};
\draw[doublefermion]	(-1,1) -- (0,1);
\draw[doublefermionbar]	(-1,0) -- (0,0);	
\draw[scalar]	(0,1) -- (0,0);
\node at (0.5,0.5){$H$};
\draw 	(0,1) -- (1,1);
\draw 	(0,0) -- (1,0);	
\draw[->] (+0.4, 1.3) -- (+0.9, 1.3);\node at(+0.6, 1.6){$p_2'$};
\draw[->] (+0.4,-0.3) -- (+0.9,-0.3);\node at(+0.6,-0.6){$p_1'$};
\node at (1.3,1){$S^{s_2'}$};
\node at (1.3,0){$S^{s_1'}$};
\end{scope}
\node at (9,0.5){$\left.\vphantom{
\begin{array}{c} a \\ a \\ a \\ a \end{array}
}\)$};
\end{scope}
%%%%%%%%%%%%%%%%%%%%%%%%%%%%%%%%%%%%%%%%%%%%%%%%%%%%%%%%%%%%%%%%%%%%%%%%%%%%%%%%%%%%%%%%%%%%%
%%%%%%%%%%%%%%%%%%%%%%%%%%%%%%%%%%%%%%%%%%%%%%%%%%%%%%%%%%%%%%%%%%%%%%%%%%%%%%%%%%%%%%%%%%%%%
%%%%%%%%%%%%%%%%%%%%%%%%%%%%%%%%%%%%%%%%%%%%%%%%%%%%%%%%%%%%%%%%%%%%%%%%%%%%%%%%%%%%%%%%%%%%%
\begin{scope}[shift={(0,-3.5)}]
\begin{scope}[shift={(-3,0)}]
\node at (-2,1){$D_i^{s_1}$};
\node at (-2,0){$\bar{D}_j^{s_2}$};
\draw[->] (-1.5, 1.3) -- (-1.1, 1.3);\node at(-1.3, 1.6){$p_1$};
\draw[->] (-1.5,-0.3) -- (-1.1,-0.3);\node at(-1.3,-0.6){$p_2$};
\draw[doublefermion] 	(-1.6,1) -- (-0.7,1);\draw[doublefermionbar] 	(+1.6,1) -- (+0.7,1);
\draw[doublefermionbar] (-1.6,0) -- (-0.7,0);\draw[doublefermion] 		(+1.6,0) -- (+0.7,0);
\draw[->] (+1.1, 1.3) -- (+1.5, 1.3);\node at(+1.3, 1.6){$p_1'$};
\draw[->] (+1.1,-0.3) -- (+1.5,-0.3);\node at(+1.3,-0.6){$p_2'$};
\node at (2,1){$D_{i'}^{s_1'}$};
\node at (2,0){$\bar{D}_{j'}^{s_2'}$};
\draw[fill=white,shift={(0,0.5)}] (-0.9,-0.7) rectangle (0.9,0.7);
\node at (0,0.5){$\im {\cal K}_{\mathsmaller{D\bar{D}\leftrightarrow D\bar{D}}}$};
\end{scope}
\node at (0,0.5){$=$};
\begin{scope}[shift={(2.8,0)}]
\node at (-1.3,1){$D_i^{s_1}$};
\node at (-1.3,0){$\bar{D}_j^{s_2}$};
\draw[->] (-0.9, 1.3) -- (-0.4, 1.3);\node at(-0.6, 1.6){$p_1$};
\draw[->] (-0.9,-0.3) -- (-0.4,-0.3);\node at(-0.6,-0.6){$p_2$};
\draw[doublefermion] 		(-1,1) -- (0,1);
\draw[doublefermionbar] 	(-1,0) -- (0,0);	
\draw[vector]	(0,0) -- (0,1);
\node at (0.7,0.5){$B, W^a$};
\draw[doublefermion] 		(0,1) -- (1,1);
\draw[doublefermionbar] 	(0,0) -- (1,0);	
\draw[->] (+0.4, 1.3) -- (+0.9, 1.3);\node at(+0.6, 1.6){$p_1'$};
\draw[->] (+0.4,-0.3) -- (+0.9,-0.3);\node at(+0.6,-0.6){$p_2'$};
\node at (1.3,1){$D_{i'}^{s_1'}$};
\node at (1.3,0){$\bar{D}_{j'}^{s_2'}$};
\end{scope}
\end{scope}
%%%%%%%%%%%%%%%%%%%%%%%%%%%%%%%%%%%%%%%%%%%%%%%%%%%%%%%%%%%%%%%%%%%%%%%%%%%%%%%%%%%%%%%%%%%%%
%%%%%%%%%%%%%%%%%%%%%%%%%%%%%%%%%%%%%%%%%%%%%%%%%%%%%%%%%%%%%%%%%%%%%%%%%%%%%%%%%%%%%%%%%%%%%
%%%%%%%%%%%%%%%%%%%%%%%%%%%%%%%%%%%%%%%%%%%%%%%%%%%%%%%%%%%%%%%%%%%%%%%%%%%%%%%%%%%%%%%%%%%%%
\begin{scope}[shift={(0,-7)}]
\begin{scope}[shift={(-3,0)}]
\node at (-2,1){$D_i^{s_1}$};
\node at (-2,0){$D_j^{s_2}$};
\draw[->] (-1.5, 1.3) -- (-1.1, 1.3);\node at(-1.3, 1.6){$p_1$};
\draw[->] (-1.5,-0.3) -- (-1.1,-0.3);\node at(-1.3,-0.6){$p_2$};
\draw[doublefermion] (-1.6,1) -- (-0.7,1);\draw[doublefermionbar] 	(+1.6,1) -- (+0.7,1);
\draw[doublefermion] (-1.6,0) -- (-0.7,0);\draw[doublefermionbar]	(+1.6,0) -- (+0.7,0);
\draw[->] (+1.1, 1.3) -- (+1.5, 1.3);\node at(+1.3, 1.6){$p_1'$};
\draw[->] (+1.1,-0.3) -- (+1.5,-0.3);\node at(+1.3,-0.6){$p_2'$};
\node at (2,1){$D_{i'}^{s_1'}$};
\node at (2,0){$D_{j'}^{s_2'}$};
\draw[fill=white,shift={(0,0.5)}] (-0.9,-0.7) rectangle (0.9,0.7);
\node at (0,0.5){$\im {\cal K}_{\mathsmaller{DD\leftrightarrow DD}}$};
\end{scope}
\node at (0,0.5){$=$};
\node at (0.8,0.5){$\dfrac{1}{2} \(\vphantom{
\begin{array}{c} a \\ a \\ a \\ a \end{array}
}\right.$};
\begin{scope}[shift={(2.8,0)}]
\node at (-1.3,1){$D_i^{s_1}$};
\node at (-1.3,0){$D_j^{s_2}$};
\draw[->] (-0.9, 1.3) -- (-0.4, 1.3);\node at(-0.6, 1.6){$p_1$};
\draw[->] (-0.9,-0.3) -- (-0.4,-0.3);\node at(-0.6,-0.6){$p_2$};
\draw[doublefermion] (-1,1) -- (0,1);
\draw[doublefermion] (-1,0) -- (0,0);	
\draw[vector]	(0,0) -- (0,1);
\node at (0.7,0.5){$B, W^a$};
\draw[doublefermion] (0,1) -- (1,1);
\draw[doublefermion] (0,0) -- (1,0);	
\draw[->] (+0.4, 1.3) -- (+0.9, 1.3);\node at(+0.6, 1.6){$p_1'$};
\draw[->] (+0.4,-0.3) -- (+0.9,-0.3);\node at(+0.6,-0.6){$p_2'$};
\node at (1.3,1){$D_{i'}^{s_1'}$};
\node at (1.3,0){$D_{j'}^{s_2'}$};
\end{scope}
\node at (5,0.5){$+$};
\begin{scope}[shift={(7.2,0)}]
\node at (-1.3,1){$D_i^{s_1}$};
\node at (-1.3,0){$D_j^{s_2}$};
\draw[->] (-0.9, 1.3) -- (-0.4, 1.3);\node at(-0.6, 1.6){$p_1$};
\draw[->] (-0.9,-0.3) -- (-0.4,-0.3);\node at(-0.6,-0.6){$p_2$};
\draw[doublefermion] (-1,1) -- (0,1);
\draw[doublefermion] (-1,0) -- (0,0);	
\draw[vector]	(0,0) -- (0,1);
\node at (0.7,0.5){$B, W^a$};
\draw[doublefermion] (0,1) -- (1,1);
\draw[doublefermion] (0,0) -- (1,0);	
\draw[->] (+0.4, 1.3) -- (+0.9, 1.3);\node at(+0.6, 1.6){$p_2'$};
\draw[->] (+0.4,-0.3) -- (+0.9,-0.3);\node at(+0.6,-0.6){$p_1'$};
\node at (1.3,1){$D_{j'}^{s_2'}$};
\node at (1.3,0){$D_{i'}^{s_1'}$};
\end{scope}
\node at (9,0.5){$\left.\vphantom{
\begin{array}{c} a \\ a \\ a \\ a \end{array}
}\)$};
\end{scope}
%%%%%%%%%%%%%%%%%%%%%%%%%%%%%%%%%%%%%%%%%%%%%%%%%%%%%%%%%%%%%%%%%%%%%%%%%%%%%%%%%%%%%%%%%%%%%
%%%%%%%%%%%%%%%%%%%%%%%%%%%%%%%%%%%%%%%%%%%%%%%%%%%%%%%%%%%%%%%%%%%%%%%%%%%%%%%%%%%%%%%%%%%%%
%%%%%%%%%%%%%%%%%%%%%%%%%%%%%%%%%%%%%%%%%%%%%%%%%%%%%%%%%%%%%%%%%%%%%%%%%%%%%%%%%%%%%%%%%%%%%
\begin{scope}[shift={(0,-10.5)}]
\begin{scope}[shift={(-3,0)}]
\node at (-2,1){$D_i^{s_1}$};
\node at (-2,0){$S^{s_2}$};
\draw[->] (-1.5, 1.3) -- (-1.1, 1.3);\node at(-1.3, 1.6){$p_1$};
\draw[->] (-1.5,-0.3) -- (-1.1,-0.3);\node at(-1.3,-0.6){$p_2$};
\draw[doublefermion] (-1.6,1) -- (-0.7,1);\draw[doublefermionbar] (+1.6,1) -- (+0.7,1);
\draw (-1.6,0) -- (-0.7,0);\draw	(+1.6,0) -- (+0.7,0);
\draw[->] (+1.1, 1.3) -- (+1.5, 1.3);\node at(+1.3, 1.6){$p_1'$};
\draw[->] (+1.1,-0.3) -- (+1.5,-0.3);\node at(+1.3,-0.6){$p_2'$};
\node at (2,1){$D_{i'}^{s_1'}$};
\node at (2,0){$S^{s_2'}$};
\draw[fill=white,shift={(0,0.5)}] (-0.9,-0.7) rectangle (0.9,0.7);
\node at (0,0.5){$\im {\cal K}_{\mathsmaller{DS\leftrightarrow DS}}$};
\end{scope}
\node at (0  ,0.5){$=$};
\begin{scope}[shift={(2.8,0)}]
\node at (-1.3,1){$D_{i}^{s_1}$};
\node at (-1.3,0){$S^{s_2}$};
\draw[->] (-0.9, 1.3) -- (-0.4, 1.3);\node at(-0.6, 1.6){$p_1$};
\draw[->] (-0.9,-0.3) -- (-0.4,-0.3);\node at(-0.6,-0.6){$p_2$};
\draw[doublefermion]	(-1,1) -- (0,1);
\draw	(-1,0) -- (0,0);	
\draw[scalar]	(0,1) -- (0,0);
\node at (0.5,0.5){$H$};
\draw 	(0,1) -- (1,1);
\draw[doublefermion] 	(0,0) -- (1,0);	
\draw[->] (+0.4, 1.3) -- (+0.9, 1.3);\node at(+0.6, 1.6){$p_2'$};
\draw[->] (+0.4,-0.3) -- (+0.9,-0.3);\node at(+0.6,-0.6){$p_1'$};
\node at (1.3,1){$S^{s_2'}$};
\node at (1.3,0){$D_{i'}^{s_1'}$};
\end{scope}
\end{scope}
%%%%%%%%%%%%%%%%%%%%%%%%%%%%%%%%%%%%%%%%%%%%%%%%%%%%%%%%%%%%%%%%%%%%%%%%%%%%%%%%%%%%%%%%%%%%%
%%%%%%%%%%%%%%%%%%%%%%%%%%%%%%%%%%%%%%%%%%%%%%%%%%%%%%%%%%%%%%%%%%%%%%%%%%%%%%%%%%%%%%%%%%%%%
%%%%%%%%%%%%%%%%%%%%%%%%%%%%%%%%%%%%%%%%%%%%%%%%%%%%%%%%%%%%%%%%%%%%%%%%%%%%%%%%%%%%%%%%%%%%%
\end{tikzpicture}
\caption[]{\label{fig:2PI}
The kernels generating the long-range potentials between pairs of $D$, $\bar{D}$ and $S$ fermions. The double lines represent the $\SUL \times \UY = (\mathbb{2},1/2)$ fermion $D$ while the single lines stand for the gauge singlet $S$. The arrows on the fermion lines denote the flow of Hypercharge. The indices $i,j,i',j'$ and $s_1, s_2, s_1', s_2'$ are $\SUL$ and spin indices, respectively. The factors $1/2$ in the interactions involving identical particles, $D\bar{D} \leftrightarrow SS$ and $DD \leftrightarrow DD$, ensure that the resummation of the kernels does not result in double-counting of loops (cf.~\cref{App:IdenticalParticles}.)}
\end{figure}
For low momentum transfers, we find (see e.g.~\cite{Petraki:2015hla,Petraki:2016cnz,Harz:2018csl,Oncala:2018bvl,Oncala:2019yvj})
\begin{subequations}
\label{eq:Kernels}	
\label[pluralequation]{eqs:Kernels}	
\begin{align}
%%%%%%%%%%%%%%%%%%%%%%%%%
\im [{\cal K}_{\mathsmaller{D\bar{D} \leftrightarrow SS}}]_{ij}^{s_1s_2,s_1's_2'} &\simeq
+\im 4 m^2(y^2 \delta_{ij}) \ \frac{1}{2} 
\[
 \frac{ \delta^{s_1 s_1'} \delta^{s_2 s_2'} }{({\bf p'-p})^2 +\mH^2} 
-\frac{ \delta^{s_1 s_2'} \delta^{s_2 s_1'} }{({\bf p'+p})^2 +\mH^2} 
\], 
\label{eq:K2PI_DDbartoSS} \\
%%%%%%%%%%%%%%%%%%%%%%%%%
\im [{\cal K}_{\mathsmaller{D\bar{D} \leftrightarrow D\bar{D}}}]_{ij,i'j'}^{s_1s_2,s_1's_2'} &\simeq
+\frac{\im 4 m^2}{({\bf p'-p})^2} 
\(g_1^2 \YD^2 \delta_{ii'} \delta_{jj'} + g_2^2 t^a_{i'i} t^a_{jj'} \) 
\delta^{s_1 s_1'} \delta^{s_2 s_2'} ,
\label{eq:K2PI_DDbartoDDbar} \\
%%%%%%%%%%%%%%%%%%%%%%%%%
\im [{\cal K}_{\mathsmaller{DD \leftrightarrow DD}}]_{ij,i'j'}^{s_1s_2,s_1's_2'} &\simeq
-\im 4 m^2 \ \frac{1}{2}
\[
\frac{\(g_1^2 \YD^2 \delta_{ii'} \delta_{jj'} + g_2^2 t^a_{i'i} t^a_{j'j} \) \delta^{s_1 s_1'} \delta^{s_2 s_2'} }
{({\bf p'-p})^2} 
\right. 
\nn \\
&\hphantom{-\im 4 m^2 \ \frac{1}{2} \[\right. ~~}
\left. 
-\frac{\(g_1^2 \YD^2 \delta_{ij'} \delta_{ji'} + g_2^2 t^a_{j'i} t^a_{i'j} \) \delta^{s_1 s_2'} \delta^{s_2 s_1'} }
{({\bf p'+p})^2} 
\],
\label{eq:K2PI_DDtoDD} \\
%%%%%%%%%%%%%%%%%%%%%%%%%
\im [{\cal K}_{\mathsmaller{DS \leftrightarrow DS}}]_{i,i'}^{s_1s_2,s_1's_2'} &\simeq
-\frac{\im 4 m^2}{({\bf p'+p})^2 +\mH^2} 
(y^2 \delta_{ii'})
\delta^{s_1 s_2'} \delta^{s_2 s_1'} .
\label{eq:K2PI_DStoDS}
%%%%%%%%%%%%%%%%%%%%%%%%%
\end{align}
\end{subequations}
In determining the sign of each contribution in the above, we have taken into account the number of fermion permutations needed to perform the Wick contractions. This is the origin of the relative minus sign between the $t$- and $u$-channels of the $D\bar{D} \leftrightarrow SS$ and $DD \leftrightarrow DD$ interactions. The factors $1/2$ appearing in \cref{eq:K2PI_DDbartoSS,eq:K2PI_DDtoDD} ensure that the resummation of the kernels does not double-count the loop diagrams by exchanging identical particles in the loops; this is shown in \cref{App:IdenticalParticles}. 

To obtain the non-relativistic potentials, we must diagonalise the interactions \eqref{eq:Kernels}	in momentum, spin, and gauge space.

\paragraph{Momentum space.} 
The kernels of both the $t$- and $u$-channel diagrams depend only on the momentum transfer, which is however different in the two cases, ${\cal K}_t ({\bf p-p'})$ and ${\cal K}_u ({\bf p+p'})$. This implies that in position space, the $u$-channel potential depends on the orbital angular momentum mode $\ell$ of the state under consideration~\cite[appendix~A]{Oncala:2019yvj}. Specifically, the static potentials generated by $t$- and $u$-channel diagrams are~\cite{Petraki:2015hla,Oncala:2019yvj}
\begin{subequations}
\label{eq:Potential_PositionSpace}
\label[pluralequation]{eqs:Potential_PositionSpace}
\begin{align}
V_t (r) &=  -\frac{1}{\im 4 m^2} \int \frac{d^3 {\bf q}}{(2\pi)^3} 
\, \im {\cal K}_t ({\bf q})
\, e^{\im {\bf q \cdot r}} ,
\label{eq:Potential_t}
\\
V_u (r) &=  -\frac{(-1)^\ell}{\im 4 m^2} \int \frac{d^3 {\bf q}}{(2\pi)^3} 
\, \im {\cal K}_u ({\bf q})
\, e^{\im {\bf q \cdot r}} .
\label{eq:Potential_u}
\end{align}
\end{subequations}
Inserting \cref{eqs:Kernels} into \eqref{eqs:Potential_PositionSpace} yields the well-known Coulomb and Yukawa potentials.

\paragraph{Spin diagonalisation.}  
The factors $\delta^{s_1 s_1'} \delta^{s_2 s_2'}$ and $\delta^{s_1 s_2'} \delta^{s_2 s_1'}$ arising from $t$- and $u$-channel exchanges respectively, can be written in matrix form in the basis 
$\{\uparrow \uparrow,~\uparrow \downarrow,~\downarrow \uparrow,~\downarrow \downarrow \}$ as
\begin{align}
t\text{-channel:}~~
\(
\begin{array}{cccc}
1&0&0&0 \\
0&1&0&0 \\
0&0&1&0 \\
0&0&0&1
\end{array}
\),
\qquad 
u\text{-channel:}~~
\(
\begin{array}{cccc}
1&0&0&0 \\
0&0&1&0 \\
0&1&0&0 \\
0&0&0&1
\end{array}
\) .
\label{eq:SpinMatrices}
\end{align}
Clearly, the $t$-channel interactions conserve spin along each leg of the ladder, and the corresponding spin factor is simply the unity operator. On the other hand, the $u$-channel interactions conserve the total spin only. Indeed, the  $u$-channel spin eigenvalues are $\{-1,1,1,1\}$, and correspond to the eigenvectors of total spin
\begin{align}
\frac{1}{\sqrt{2}}(0,1,-1,0)^\mathsmaller{T}, \quad
(0,0, 0,1)^\mathsmaller{T}, \quad
\frac{1}{\sqrt{2}}(0,1, 1,0)^\mathsmaller{T}, \quad
(1,0, 0,0)^\mathsmaller{T}.
\end{align}
In the following, we shall therefore project the asymptotic states of pairs of $S,D,\bar{D}$ fermions onto eigenstates of total spin.

\paragraph{Gauge diagonalisation.}  
Since the interactions respect the $\SUL$ symmetry, this amounts to projecting on $\SUL$ representations of the incoming or outgoing pairs. For two multiplets transforming under the representations $\mathbb{R}_1$ and $\mathbb{R}_2$ of a gauge group, the Coulomb potential generated by the gauge-boson exchange in the configuration $\mathbb{R} \subset \mathbb{R}_1 \otimes \mathbb{R}_2$ of the pair is 
$V^{\mathsmaller{\mathbb{R}}} (r) = -\alpha^{\mathsmaller{\mathbb{R}}} / r$ with~\cite{Kats:2009bv}
\begin{align}
\alpha^{\mathsmaller{\mathbb{R}}} = \frac{\alpha}{2} [C_2(\mathbb{R}_1) + C_2(\mathbb{R}_2) - C_2(\mathbb{R})] ,
\label{eq:alpha_group}
\end{align}
where $\alpha$ is the fine structure constant of the group, and $C_2 ({\mathbb{R}})$ is the quadratic Casimir operator of the representation $\mathbb{R}$. For $SU(2)$, $\mathbb{2} \otimes \mathbb{2} = \mathbb{1} + \mathbb{3}$, and $C_2(\mathbb{2})= 3/4$, $C_2(\mathbb{3}) = 2$. Thus, the $D\bar{D}$, $DD$ and $\bar{D}\bar{D}$ pairs appear in $\SUL$ singlet and triplet configurations, with $\alpha_2^{\mathsmaller{\mathbb{1}}} = 3\alpha_2/4$ and $\alpha_2^{\mathsmaller{\mathbb{3}}} = -\alpha_2/4$ respectively. For the identical-particle pairs $DD$ and $\bar{D}\bar{D}$, the (anti)symmetry of the $\SUL$ (singlet) triplet states in gauge space generates also the factor $(-1)^{I+1}$ for the $u$-channel diagrams, where here $I=0,1$ stands for the Weak isospin. Note that the $\UY$ potentials are of course not affected by the $\SUL$ diagonalisation.\footnote{All these results can be easily recovered by organising the gauge factors of  \cref{eq:K2PI_DDbartoDDbar,eq:K2PI_DDtoDD} in $4\times 4$ matrices with elements $\{ij,i'j'\}$ and diagonalising them.}

The remaining task is the $H$-mediated interaction of \cref{eq:K2PI_DDbartoSS}. This occurs only in the $\SUL$ singlet state. Projecting $D\bar{D}$ on the singlet, we obtain 
\begin{align}
\frac{\delta^{ij}}{\sqrt{2}} (y^2\delta_{ij}) = \sqrt{2} y^2 .
\label{eq:DDbarToSS_projection}
\end{align}

\paragraph{Kernel (anti)symmetrisation.} 
We close this discussion by noting the importance of considering properly both the $t$- and $u$-channel diagrams when identical particles are present in the initial and/or final states, as e.g.~in the $D\bar{D} \leftrightarrow SS$ and $DD\leftrightarrow DD$ interactions of the present model. Using the correct kernel, as derived in \cref{App:IdenticalParticles}, and taking into account the above yields the overall factor 
\begin{align}
[1-(-1)^{\ell}(-1)^{s+1} (-1)^{I+1}]/2
\label{eq:IdenticalParticles_AntiSymmetryFactor}
\end{align} 
that enforces the proper particle statistics. This result is valid for pairs of fermions as well as pairs of scalars.

\medskip

Collecting the above, in \cref{tab:potentials} we summarise the potentials generated by the one-boson-exchange diagrams of \cref{fig:2PI}.
\begin{table}[h!]
\centering
\renewcommand*{\arraystretch}{1.8}
\begin{align*}
\begin{array}{ |c|c|c|rl|c|c|} 
\hline
%%%%%%%%%%%%%%%%%%%%%%%%%%%%%%%%%%%%%%%%%%%%%%%%%%%%%%%%%%%%%%%%%%%%%%%%%%%%%%%%%%%%%%%%%%%%%%%%%%%%%%
\multirow{2}{*}{\centering Interaction}					
& \multirow{2}{*}{\centering $\UY$}	
& \multirow{2}{*}{\centering $\SUL$}		
&\multicolumn{2}{c|}{\multirow{2}{*}{\centering Potential}}
&\multicolumn{2}{c|}{\text{Sign of the potential}}
\\ \cline{6-7}
&
&
&
&
&\ell + s ={\rm even}
&\ell + s ={\rm odd}
\\ \hline \hline
%%%%%%%%%%%%%%%%%%%%%%%%%%%%%%%%%%%%%%%%%%%%%%%%%%%%%%%%%%%%%%%%%%%%%%%%%%%%%%%%%%%%%%%%%%%%%%%%%%%%%%
D\bar{D}\leftrightarrow SS			
&0		
&\mathbb{1}	
&- \[\dfrac{1+(-1)^{\ell+s}}{2}\]
&\sqrt{2} \aH \ \dfrac{e^{- \mH r}}{r}
&\text{attractive}
&0
\\ \hline
%%%%%%%%%%%%%%%%%%%%%%%%%%%%%%%%%%%%%%%%%%%%%%%%%%%%%%%%%%%%%%%%%%%%%%%%%%%%%%%%%%%%%%%%%%%%%%%%%%%%%%
 \multirow{2}{*}{\centering $D\bar{D}\leftrightarrow D\bar{D}$}  
&\multirow{2}{*}{\centering 0}
&\mathbb{1}	
&-
&\dfrac{(\alpha_1 + 3\alpha_2)/4}{r}
&\text{attractive}
&\text{attractive}
\\ \cline{3-7}
&
&\mathbb{3} 
&-
&\dfrac{(\alpha_1 - \alpha_2)/4}{r}
&\text{repulsive}
&\text{repulsive}
\\ \hline
%%%%%%%%%%%%%%%%%%%%%%%%%%%%%%%%%%%%%%%%%%%%%%%%%%%%%%%%%%%%%%%%%%%%%%%%%%%%%%%%%%%%%%%%%%%%%%%%%%%%%%
 \multirow{2}{*}{\centering $DD\leftrightarrow DD$}  
&\multirow{2}{*}{\centering 1}	
&\mathbb{1}	
&\[\dfrac{1-(-1)^{\ell+s}}{2}\]
&\dfrac{(\alpha_1 - 3\alpha_2)/4}{r}
&0
&\text{attractive}
\\ \cline{3-7}
&
&\mathbb{3}	
&\[\dfrac{1+(-1)^{\ell+s}}{2}\]
&\dfrac{(\alpha_1+\alpha_2)/4}{r}
&\text{repulsive}
&0
\\ \hline
%%%%%%%%%%%%%%%%%%%%%%%%%%%%%%%%%%%%%%%%%%%%%%%%%%%%%%%%%%%%%%%%%%%%%%%%%%%%%%%%%%%%%%%%%%%%%%%%%%%%%%
SD\leftrightarrow SD				
&1/2	
&\mathbb{2}	
&-(-1)^{\ell+s} 
&\aH \ \dfrac{e^{-\mH r}}{r} 
&\text{attractive}
&\text{repulsive}
\\ \hline
%%%%%%%%%%%%%%%%%%%%%%%%%%%%%%%%%%%%%%%%%%%%%%%%%%%%%%%%%%%%%%%%%%%%%%%%%%%%%%%%%%%%%%%%%%%%%%%%%%%%%%
\end{array}
\end{align*}
\caption{\label{tab:potentials} 
The static potentials generated by the $W$, $B$ and $H$-exchange diagrams shown in \cref{fig:2PI}. 
$\ell$ and $s$ denote the orbital angular momentum mode and the total spin respectively.}
\end{table}

\newpage
\subsection{Asymptotic scattering and bound states \label{sec:LongRangeDynamics_AsymptoticStates}}

The potentials of \cref{tab:potentials} determine the asymptotic states. For all gauge assignments except the singlet states, $\SUL \times \UY =(\mathbb{1},0)$, finding the corresponding wavefunctions is rather straightforward; it amounts to solving a single Schr\"odinger equation with the corresponding potential, and antisymmetrising the wavefunction in the case of identical particles. For the gauge-singlet states, the $D\bar{D} \leftrightarrow SS$ interaction implies that a system of coupled Schr\"odinger equations must be solved. We work out this case in detail in \cref{sec:LongRangeDynamics_AsymptoticStates_Mixed}, after we discuss the general properties of the wavefunctions and the underlying hierarchy of scales in \cref{sec:LongRangeDynamics_AsymptoticStates_Wavefunctions}.  All the results on the wavefunctions of the scattering and bound states are summarised in \cref{tab:ScatteringStates,tab:BoundStates}.

\subsubsection{Wavefunctions and hierarchy of scales \label{sec:LongRangeDynamics_AsymptoticStates_Wavefunctions}}

The non-relativistic potentials of \cref{tab:potentials} include both Coulomb and Yukawa contributions. The latter does not allow for analytical solutions. In order to obtain analytical expressions for the wavefunctions and ultimately the various cross-sections of interest, we shall neglect the Higgs mass in the potentials of \cref{tab:potentials}, 
\begin{align}
\mH \to 0, 
\label{eq:mHtoZero}
\end{align}
although we will retain it in the phase-space suppression of BSF via $H$ emission computed in \cref{Sec:BSF}, as well as in the $H$ propagator of BSF via off-shell $H$ exchange with the thermal bath, computed in \cref{Sec:BSF_OffShell}. We discuss the range of validity of this approximation in \cref{sec:LongRangeDynamics_AsymptoticStates_CoulombApprox}.

In the approximation \eqref{eq:mHtoZero}, we can express all wavefunctions in terms of those for a Coulomb potential $V(r)=-\alpha/r$,  which we shall denote as $\varphi ({\bf r};~\alpha)$ and we now summarise. For clarity, we denote by $\aS$ and $\aB$  the couplings of the scattering and bound states. The momentum of each particle in the CM frame in the scattering states is ${\bf k} \equiv \mu {\bf v}_{\rm rel}$, with ${\bf v}_{\rm rel}$ being the relative velocity. The Bohr momenta in the scattering and bound states are $\kappaS \equiv \mu \aS$ and $\kappaB \equiv \mu \aB$. For convenience, we define the parameter $\zetaS \equiv \kappaS/k = \aS / \vrel$ (in \cref{Sec:BSF} we will also use $\zetaB \equiv \kappaB/k = \aB/\vrel$), and the variables $\xS \equiv k r$ and $\xB \equiv \kappaB r$. The energy eigenvalues of the scattering and bound states are
\begin{align}
{\cal E}_{\bf k} = \frac{\bf k^2}{2\mu} = \frac{\mu \vrel^2}{2} ,
\qquad
{\cal E}_{n} =-\frac{\kappaB^2}{2\mu\, n^2} =-\frac{\mu \aB^2}{2n^2} .
\label{eq:EnergyEigenvalues}
\end{align} 
The scattering state wavefunctions can be decomposed in partial waves, 
\begin{align}
\varphi_{\bf k} ({\bf r};~\aS) 
&= \sum_{\ellS=0}^{\infty} (2\ellS +1) 
\[ \frac{\varphi_{|{\bf k}|,\ellS} (\xS;~\aS)}{\xS} \] 
P_{\ellS} ({\bf \hat{k} \cdot \hat{r}}) ,
\label{eq:varphi_k}
\end{align}
where
\begin{align}
\varphi_{|{\bf k}|,\ellS}^{} (\xS;~\aS) 
&=-
\frac{\sqrt{S_0(\zetaS)}}{(2\ellS+1)! }
\, \frac{\Gamma(1+\ellS-\im \zetaS)}{\Gamma(1-\im \zetaS)}
\nn \\
&\times 
e^{-i \xS}
\, \xS
\, (2\im \xS)^{\ellS} 
\, {}_1F_1 (1+\ellS + \im \zetaS; ~ 2\ellS + 2; ~ 2 \im \xS) ,
\label{eq:varphi_k_ell}
\end{align}
and
\begin{align}
S_0(\zetaS) \equiv \frac{2\pi \zetaS}{1-e^{-2\pi\zetaS}} 
\label{eq:S0}
\end{align}
is the well-known $s$-wave Sommerfeld factor. The bound-state wavefunctions are
\begin{align}
\varphi_{n\ell m} ({\bf r};\aB) 
&= 
\kappa^{3/2} Y_{\ell m} ({\Omega_{\bf r}})
\ \frac{2}{n^2 (2\ell+1)!} 
\[\frac{(n+\ell)!}{(n-\ell-1)!}\]^{1/2}
\nn \\
&\times 
e^{-\xB/n} 
\ (2\xB/n)^{\ell} 
\ {}_1F_1 \(-n+\ell+1;~ 2\ell+2;~ 2\xB/n \) ,
\label{eq:varphi_nlm}
\end{align}
where the normalisation of the spherical harmonics is  
$\int d\Omega \, Y_{\ell m} (\Omega) Y_{\ell' m'}^* (\Omega) = \delta_{\ell \ell'} \delta_{m m'}$.

\medskip

The emergence of the Sommerfeld effect and the existence of bound states emanate from the different scales involved in the interactions of the $D$, $\bar{D}$ and $S$ particles, 
\begin{align}
\mu \vrel^2/2 \ll \mu \vrel \ll \mu < m 
\qquad \text{and} \qquad
\mu \alpha^2/(2n^2)  \ll \mu |\alpha| / n \ll \mu < m ,
\label{eq:Hierarchies}
\end{align}
where here $\alpha = \aS$ or $\aB$. (Note that $\aS$ may be negative.) In computing the BSF cross-sections in \cref{Sec:BSF,Sec:BSF_OffShell}, we make approximations based on these hierarchies.

The hierarchies \eqref{eq:Hierarchies} imply that the couplings \eqref{eq:alphas_def} must be evaluated at the appropriate momentum transfer in every occurrence. The average momentum transfers are (cf.~\cref{Sec:BSF} for the last two)
\begin{center}
\begin{tabular}{ll}
annihilation vertices:
& $m$ \\ 
scattering-state potentials: 
& $\mu \vrel$ \\
bound-state potentials: 
& $\mu \aB$ \\
emission vertices for BSF: 
& $(\mu/2) (\aB^2/n^2 + \vrel^2)$ \\
emission vertices for bound-to-bound transitions:
& $(\mu/2) |\aB^2/n^2 - {\aB'}^2/n'^2|$
\end{tabular}
\end{center}
Here for simplicity we will neglect the running of the couplings, although it is easy to restore the scale dependence in all of our analytical results.\footnote{In computing the DM freeze-out in \cite{Oncala:2021swy}, we adopt the values of the gauge couplings $\alpha_1$ and $\alpha_2$ at the $Z$ pole, 
$\alpha_1 (\mZ) \simeq 0.00973$ and 
$\alpha_2 (\mZ) \simeq 0.0339$.  
Since $\aH$ increases with $Q$, we consider the quoted values of $\aH$ to correspond to the highest relevant scale, $Q=m$, such that the theory remains well-defined at all $Q\leqslant m$. For large values of $\aH$, a Landau pole appears at fairly low energies (but larger than $m$); however this may be cured by other new physics around those scales. For the renormalisation group equations in the present model, we refer to~\cite{Wang:2018lhk}.}

%%%%%%%%%%%%%%%%%%%%%%%%%%%%%%%%%%%%%%%%%%%%%%%%%%%%%%%%%%%%%%%%%%
\subsubsection{Mixed $SU_L(2) \times U_Y (1) = (\mathbb{1},0)$ states \label{sec:LongRangeDynamics_AsymptoticStates_Mixed}}

\begin{figure}[t]
\begin{tikzpicture}[line width=1pt, scale=0.85]
\begin{scope}[shift={(0,0)}]
\begin{scope}[shift={(-1.9,0)}]
\draw[doublefermion] 	(-1.5,1) -- (-0.5,1);\draw[doublefermionbar] 	(1.5,1) -- (0.5,1);
\draw[doublefermionbar] (-1.5,0) -- (-0.5,0);\draw[doublefermion] 		(1.5,0) -- (0.5,0);%
\draw[fill=lightgray,draw=none,shift={(0,0.5)}] (-0.7,-0.7) rectangle (0.7,0.7);
\node at (0,0.5){$G_{\mathsmaller{D\bar{D}D\bar{D}}}^{(4)}$};
\end{scope}
\node at (0,0.5){$=$};
\begin{scope}[shift={(0.4,0)}]
\draw[doublefermion] 		(0,1) -- (1,1);
\draw[doublefermionbar] 	(0,0) -- (1,0);	
\end{scope}
\node at (1.8,0.5){$+$};
\begin{scope}[shift={(5.3,0)}]
\draw[doublefermion] 	(-3,1) -- (-2,1);
\draw[doublefermionbar]	(-3,0) -- (-2,0);
\draw[doublefermion] 	(-1,1) -- (1.5,1);
\draw[doublefermionbar] (-1,0) -- (1.5,0);
\draw[doublefermionbar] (2.75,1) -- (1.7,1);
\draw[doublefermion]    (2.75,0) -- (1.7,0);
\draw[fill=white,shift={(-1.25,0.5)}] (-1,-0.7) rectangle (1,0.7);
\node at (-1.25,0.5){$\im {\cal K}_{\mathsmaller{D\bar{D} \leftrightarrow D\bar{D}}}$};
\draw[fill=lightgray,draw=none,shift={(+1.25,0.5)}] (-0.7,-0.7) rectangle (0.7,0.7);
\node at (+1.25,0.5){$G_{\mathsmaller{D\bar{D}D\bar{D}}}^{(4)}$};
\end{scope}
\node at (8.4,0.5){$+$};
\begin{scope}[shift={(11.5,0)}]
\draw[doublefermion] 	(-2.75,1) -- (-1.75,1);
\draw[doublefermionbar]	(-2.75,0) -- (-1.75,0);
\draw (-1,1) -- (1.5,1);
\draw (-1,0) -- (1.5,0);
\draw[doublefermionbar] (2.5,1) -- (1.45,1);
\draw[doublefermion]    (2.5,0) -- (1.45,0);
\draw[fill=white,shift={(-1,0.5)}] (-0.9,-0.7) rectangle (0.9,0.7);
\node at (-1,0.5){$\im {\cal K}_{\mathsmaller{D\bar{D} \leftrightarrow SS}}$};
\draw[fill=lightgray,draw=none,shift={(+1.1,0.5)}] (-0.7,-0.7) rectangle (0.7,0.7);
\node at (+1.1,0.5){$G_{\mathsmaller{SSD\bar{D}}}^{(4)}$};
\end{scope}
\end{scope}
%%%%%%%%%%%%%%%%%%%%%%%%%%%%%%%%%%%%%%%%%%%%%%%%%%%%%%%%%%%%%%%%
%%%%%%%%%%%%%%%%%%%%%%%%%%%%%%%%%%%%%%%%%%%%%%%%%%%%%%%%%%%%%%%%
%%%%%%%%%%%%%%%%%%%%%%%%%%%%%%%%%%%%%%%%%%%%%%%%%%%%%%%%%%%%%%%%
%
\begin{scope}[shift={(0,-2.2)}]
\begin{scope}[shift={(-1.9,0)}]
\draw (-1.5,1) -- (1.5,1);
\draw (-1.5,0) -- (1.5,0);%
\draw[fill=lightgray,draw=none,shift={(0,0.5)}] (-0.7,-0.7) rectangle (0.7,0.7);
\node at (0,0.5){$G_{\mathsmaller{SSSS}}^{(4)}$};
\end{scope}
\node at (0,0.5){$=$};
\begin{scope}[shift={(0.4,0)}]
\draw	(0,1) -- (1,1);
\draw 	(0,0) -- (1,0);	
\end{scope}
\node at (1.8,0.5){$+$};
\begin{scope}[shift={(5.3,0)}]
\draw 	(-3,1) -- (-2,1);
\draw	(-3,0) -- (-2,0);
\draw[doublefermion] 	(-1,1) -- (1.5,1);
\draw[doublefermionbar] (-1,0) -- (1.5,0);
\draw (2.75,1) -- (1.7,1);
\draw (2.75,0) -- (1.7,0);
\draw[fill=white,shift={(-1.25,0.5)}] (-1,-0.7) rectangle (1,0.7);
\node at (-1.25,0.5){$\im {\cal K}_{\mathsmaller{SS \leftrightarrow D\bar{D}}}$};
\draw[fill=lightgray,draw=none,shift={(+1.25,0.5)}] (-0.7,-0.7) rectangle (0.7,0.7);
\node at (+1.25,0.5){$G_{\mathsmaller{D\bar{D}SS}}^{(4)}$};
\end{scope}
\end{scope}
\end{tikzpicture}
\caption{\label{fig:ResummationCoupled}
The resummation of 2PI diagrams for the gauge-singlet states, $\SUL \times \UY =(\mathbb{1},0)$. Single and double lines correspond to $S$ and $D$ respectively.}
\end{figure}
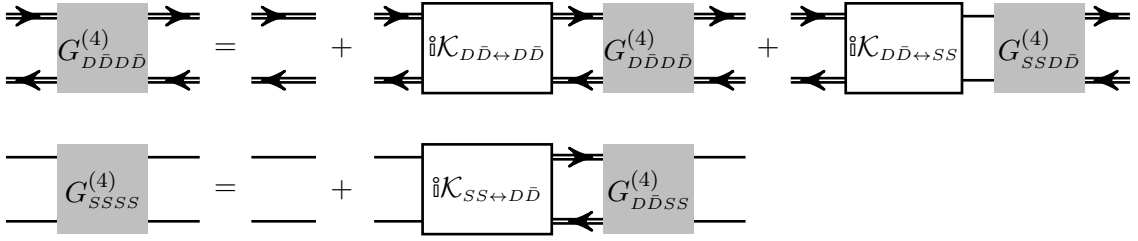

The $H$ exchange mixes $SS$ and $D\bar{D}$ gauge singlet states. The coupled resummation of the $D\bar{D} \leftrightarrow D\bar{D}$ and $D\bar{D} \leftrightarrow SS$ potentials is shown schematically in \cref{fig:ResummationCoupled}. We define the wavefunctions of the gauge-singlet states
\begin{align}
\Phi^j ({\bf r}) = \(
\begin{array}{c}
[\phi ({\bf r})]_{\SS}^j 
\\  
\[\phi ({\bf r})\]_{\DDbar}^j
\end{array}
\) ,
\label{eq:GaugeSinglet_Phi_def}
\end{align}
where the component wavefunctions are
\begin{subequations}
\label{eq:GaugeSinglet_phis_def}
\label[pluralequation]{eqs:GaugeSinglet_phis_def}	
\begin{align}
[\phi ({\bf x})]_{\SS}^j			
&\equiv \< \Omega| T \, S(x/2) S(-x/2) | {\cal S}_{(\mathbb{1},0)}^j \>_{x^0 =0} , 
\label{eq:GaugeSinglet_phiSS_def} \\
[\phi ({\bf x})]_{\DDbar}^j	
&\equiv \< \Omega| T \, D(x/2) \bar{D}(-x/2) | {\cal S}_{(\mathbb{1},0)}^j \>_{x^0 =0} .
\label{eq:GaugeSinglet_phiDDbar_def} 
\end{align}
\end{subequations}
Here, ${\cal S}_{(\mathbb{1},0)}^j$ denotes the gauge-singlet states. Along with their wavefunctions $\Phi^j$, they carry quantum numbers that define their energy, angular momentum and spin, and that we have here left implicit.  The superscript $j$ aims to differentiate between states with the same quantum numbers but different boundary conditions.
$\Omega$ here stands for the vacuum of the interacting theory, $T$ is the time ordering operator, $S$, $D$ and $\bar{D}$ are the field operators.

The resummation sketched in \cref{fig:ResummationCoupled} implies that $\Phi$ obey the Schr\"odinger equations
\begin{align}
\[
-\frac{\nabla^2}{2\mu}	
+ \mathbb{V}_{(\mathbb{1},0)} (r)
\] \Phi ({\bf r})
= {\cal E} \Phi ({\bf r}) ,
\label{eq:GaugeSinglet_Schrodinger_Coupled}
\end{align}
where $\mathbb{V}_{(\mathbb{1},0)}$ is the $2 \times 2$ potential matrix of the gauge-singlet states
\begin{align}
\mathbb{V}_{(\mathbb{1},0)} (r) = -\dfrac{1}{r}\(
\begin{array}{cc}
0 			
& \delta_{\ell+s, \rm even} \sqrt{2}\aH \, \exp (-\mH r)   
\\
\delta_{\ell+s, \rm even} \sqrt{2}\aH	 \, \exp (-\mH r)	
& (\alpha_1+3\alpha_2)/4
\end{array}
\) .
\label{eq:GaugeSinglet_Vmatrix}
\end{align}
In the limit $\mH \to 0$, the system \eqref{eq:GaugeSinglet_Schrodinger_Coupled} easily decouples. We first define the couplings
\begin{subequations}
\label{eq:GaugeSinglet_alphas_def}
\label[pluralequation]{eqs:GaugeSinglet_alphas_def}
\begin{align}
\aR &\equiv \frac{1}{2} \[ \sqrt{ [(\alpha_1+3\alpha_2)/4]^2 + 8\aH^2 \, \delta_{\ell+s, \rm even} } - (\alpha_1+3\alpha_2)/4 \] ,
\label{eq:GaugeSinglet_alphaR_def} 
\\
\aA &\equiv \frac{1}{2} \[ \sqrt{ [(\alpha_1+3\alpha_2)/4]^2 + 8\aH^2 \, \delta_{\ell+s, \rm even} } + (\alpha_1+3\alpha_2)/4 \] ,
\label{eq:GaugeSinglet_alphaA_def}
\end{align}
\end{subequations}
as well as the unitary matrix
\begin{align}
\arraycolsep=1ex\def\arraystretch{1}
P \equiv \dfrac{1}{\sqrt{\aA+\aR}}\(
\begin{array}{cc}
  \sqrt{\aA}
& \sqrt{\aR} \\
 -\sqrt{\aR}
& \sqrt{\aA} 
\end{array}
\) ,
\label{eq:GaugeSinglet_RotationMatrix}
\end{align}
whose columns are the normalised $\mathbb{V}_{(\mathbb{1},0)}$ eigenvectors with eigenvalues $-\aR$ and $\aA$. We note that $\aR, \aA \geqslant 0$. Then, the rotated wavefunctions 
$\hat{\Phi} \equiv P^{\dagger} \Phi$ obey the Schr\"odinger equations 
\begin{align}
\[-\frac{\nabla^2}{2\mu} + \hat{\mathbb{V}}_{(\mathbb{1},0)} (r)\] 
\hat{\Phi} ({\bf r})
= {\cal E} \hat{\Phi} ({\bf r}) ,
\label{eq:GaugeSinglet_Schrodinger_Rotated}
\end{align}
with the potential 
\begin{align}
\hat{\mathbb{V}}_{(\mathbb{1},0)} = P^{\dagger} \mathbb{V}_{(\mathbb{1},0)} P 
= -\frac{1}{r}  \( \begin{array}{cc}   -\aR & 0 \\ 0 & \aA  \end{array} \) .
\label{eq:GaugeSinglet_Vmatrix_Rotated}
\end{align}
We now seek scattering and bound state solutions to \cref{eq:GaugeSinglet_Schrodinger_Rotated}.

\paragraph{Scattering states.} 
The scattering state solutions of \cref{eq:GaugeSinglet_Schrodinger_Rotated} are given by \cref{eq:varphi_k}, albeit the normalisation of each component is allowed to vary and will be determined by the boundary conditions on $\Phi^j$. Analysing \cref{eq:GaugeSinglet_Schrodinger_Rotated} in partial waves, the solutions are
\begin{align}
%\VarPhi
\hat{\Phi}_{|\bf k|,\ell}^j (\xS) = \(
\begin{array}{c}
\NR^j \, \varphi_{|\bf k|,\ell} (\xS;~-\aR) \\
\NA^j \, \varphi_{|\bf k|,\ell} (\xS;~~\aA)
\end{array}
\),
\label{eq:GaugeSinglet_tildePhi_sol}
\end{align}
where at $\xS\to \infty$, the wavefunctions $\varphi_{{\bf k},\ell} (\xS;~\alpha)$ obey
\begin{align}
\frac{d \varphi_{|{\bf k}|,\ell} (\xS;~\alpha)}{d\xS}  - \im \varphi_{|{\bf k}|,\ell} (\xS;~\alpha) 
= e^{-\im (\xS-\ell \pi)} .
\label{eq:Coulomb_ScattState_BoundaryCond}
\end{align}

We seek scattering-state solutions to \cref{eq:GaugeSinglet_Schrodinger_Coupled} that asymptote at $r\to \infty$ to a pure $SS$ or $D\bar{D}$ state. In terms of partial waves, this implies that at $r\to \infty$, 
\begin{subequations}
\label{eq:GaugeSinglet_ScattState_BoundaryCond}
\label[pluralequation]{eqs:GaugeSinglet_ScattState_BoundaryCond}
\begin{align}
&SS\text{-like state:}&
\frac{d [\phi_\ell (\xS)]_i^{\SS}}{d\xS}  - \im [\phi_\ell (\xS)]_i^{\SS} 
&= \delta_i^{\SS} e^{-\im (\xS-\ell \pi)} \times \sqrt{2} \delta_{\ell+s,\rm even}, 
\label{eq:GaugeSinglet_ScattState_BoundaryCond_SS}
\\
&D\bar{D}\text{-like state:}&
\frac{d [\phi_\ell (\xS)]_i^{\DDbar}}{d\xS}  - \im [\phi_\ell (\xS)]_i^{\DDbar} 
&= \delta_i^{\DDbar} e^{-\im (\xS-\ell \pi)} , 
\label{eq:GaugeSinglet_ScattState_BoundaryCond_DDbar}
\end{align}
\end{subequations}
where $i = SS, D\bar{D}$ denotes the component. In \cref{eq:GaugeSinglet_ScattState_BoundaryCond_SS}, we included the antisymmetrisation factor $\sqrt{2} \delta_{\ell+s,\rm even}$ due to the identical particles in the $SS$-like state, with $s=$ 0 or 1 being the total spin (cf.~\cref{app:IdenticalParticles_Wavefunctions}).

Since $\hat{\Phi}_{|\bf k|,\ell} (\xS) = P^\dagger \Phi_{|\bf k|,\ell} (\xS)$, the asymptotic behaviours \eqref{eq:Coulomb_ScattState_BoundaryCond} and \eqref{eq:GaugeSinglet_ScattState_BoundaryCond} imply
\begin{subequations}
\label{eq:GaugeSinglet_tildePhi_Norm}
\label[pluralequation]{eqs:GaugeSinglet_tildePhi_Norm}
\begin{align}
\(\begin{array}{c} \NR^{\SS} \\ \NA^{\SS} \end{array} \) 
&= P^\dagger \(\begin{array}{c} \sqrt{2} \delta_{\ell+s,\rm even} \\ 0 \end{array} \) 
= \frac{\sqrt{2} \delta_{\ell+s,\rm even}}{\sqrt{\aA+\aR}}  
\(\begin{array}{c} \sqrt{\aA} \\ \sqrt{\aR} \end{array} \) ,
\label{eq:GaugeSinglet_tildePhi_Norm_SS}
\\
\(\begin{array}{c} \NR^{\DDbar} \\ \NA^{\DDbar} \end{array} \) 
&= P^\dagger \(\begin{array}{c} 0 \\ 1 \end{array} \)  
= \frac{1}{\sqrt{\aA+\aR}}  
\(\begin{array}{cc} -&\sqrt{\aR} \\ &\sqrt{\aA} \end{array} \) .
\label{eq:GaugeSinglet_tildePhi_Norm_DDbar}
\end{align}
\end{subequations}
We recall that $\aR$ and $\aA$ depend on $\ell+s$, and note that while the $D\bar{D}$-like state should \emph{not} be antisymmetrised, the $SS$ components of the symmetric $\ell+s=$~odd modes do vanish due to the antisymmetrisation of the kernel ($\aR=0$ for $\ell+s=$~odd). 
%Equivalently, we could have required that the $SS$ components of both of the $SS$-like and $D\bar{D}$-like states are antisymmetrised.  
%
%
Combining \cref{eq:GaugeSinglet_tildePhi_sol,eq:GaugeSinglet_tildePhi_Norm}, we obtain the wavefunctions $\Phi_{|\bf k|,\ell}^j ({\bf r}) = P \hat{\Phi}_{|\bf k|,\ell}^j ({\bf r})$.  The results are summarised in \cref{tab:ScatteringStates}.

\paragraph{Bound states.} It is easy to see that \cref{eq:GaugeSinglet_Schrodinger_Rotated} has only one set of bound-state solutions, 
$\hat{\Phi}_{n\ell m} ({\rm r}) = (0,~~\varphi_{n\ell m} ({\bf r};~\aA))^\mathsmaller{T}$, with ${\cal E}_n = -\mu \aA^2/(2n^2)$. The unrotated wavefunctions are $\Phi_{n\ell m} ({\rm r}) = P \hat{\Phi}_{n\ell m} ({\rm r})$. The result is summarised in \cref{tab:BoundStates}.

%%%%%%%%%%%%%%%%%%%%%%%%%%%%%%%%%%%%%%%%%%%%%%%%%%%%%%%%%%%%%%%%%%

\begin{table}[h!]
\centering
\renewcommand*{\arraystretch}{1.8}
\begin{align*}
\begin{array}{ |c|c|c|c|rl| } 
\hline
\UY	
&\SUL	
&\text{State}
&\parbox{3em}{\centering Com-ponent}	
&\multicolumn{2}{c|}{\text{Partial-wave wavefunctions}} 
\\[1ex]  \hline \hline
%%%%%%%%%%%%%%%%%%%%%%%%%%%%%%%%%%%%%%%%%%%%%%%%%%%%%%%%%%%%%%%%%%%%%%%%%%%%%%%%%%%%%%%%%%%%%%%%%%%%
 \multirow{2}{*}{\centering 0}	
&\multirow{2}{*}{\centering $\mathbb{1}$}	
&\multirow{2}{*}{\centering $SS$-like}    
&SS
& \delta_{\ell+s,{\rm even}} \ \dfrac{\sqrt{2}}{\aA+\aR} 
%\[\dfrac{1+(-1)^{\ell+s}}{\sqrt{2}}\] \! \dfrac{1}{\aA+\aR} 
&\[\aR\phikell  (\xS; \aA) + \aA\phikell (\xS; -\aR)\] 
\\[0.5ex]  \cline{4-6}
&
&
&D\bar{D} 
& \delta_{\ell+s,{\rm even}} \ \dfrac{\sqrt{2\aA\aR}}{\aA+\aR} 
%\[\dfrac{1+(-1)^{\ell+s}}{\sqrt{2}}\] \! \dfrac{\sqrt{\aA\aR}}{\aA+\aR} 
&\[\phikell (\xS; \aA) - \phikell (\xS; -\aR)\]
\\[0.5ex] \hline
%%%%%%%%%%%%%%%%%%%%%%%%%%%%%%%%%%%%%%%%%%%%%%%%%%%%%%%%%%%%%%%%%%%%%%%%%%%%%%%%%%%%%%%%%%%%%%%%%%%%
 \multirow{2}{*}{\centering 0}	
&\multirow{2}{*}{\centering $\mathbb{1}$}	
&\multirow{2}{*}{\centering $D\bar{D}$-like}      
&SS			
& \dfrac{\sqrt{\aA\aR}}{\aA+\aR}
&\[\phikell (\xS; \aA) - \phikell (\xS; -\aR)\]     
\\[0.5ex] \cline{4-6} 
&
&      
&D\bar{D}	
& \dfrac{1}{{\aA}+{\aR}} 
&\[\aA\phikell (\xS; \aA)+\aR\phikell (\xS; -\aR)\]     
\\[0.5ex]  \hline
%%%%%%%%%%%%%%%%%%%%%%%%%%%%%%%%%%%%%%%%%%%%%%%%%%%%%%%%%%%%%%%%%%%%%%%%%%%%%%%%%%%%%%%%%%%%%%%%%%%%
0	
&\mathbb{3}	
&D\bar{D}	
&D\bar{D}	
&
&\phikell \(\xS;~( \alpha_1- \alpha_2)/4 \)     
\\  \hline
%%%%%%%%%%%%%%%%%%%%%%%%%%%%%%%%%%%%%%%%%%%%%%%%%%%%%%%%%%%%%%%%%%%%%%%%%%%%%%%%%%%%%%%%%%%%%%%%%%%%
1	
&\mathbb{1}	
&DD      	
&DD			
& \delta_{\ell+s,{\rm odd}} \ \sqrt{2} 
%\(\dfrac{1- (-1)^{\ell+s}}{\sqrt{2}} \) 
&\phikell \(\xS;~(-\alpha_1+3\alpha_2)/4 \)     
\\  \hline
%%%%%%%%%%%%%%%%%%%%%%%%%%%%%%%%%%%%%%%%%%%%%%%%%%%%%%%%%%%%%%%%%%%%%%%%%%%%%%%%%%%%%%%%%%%%%%%%%%%%
1	
&\mathbb{3}	
&DD			
&DD			
& \delta_{\ell+s,{\rm even}} \ \sqrt{2} 
%\(\dfrac{1+ (-1)^{\ell+s}}{\sqrt{2}} \) 
&\phikell \(\xS;~-(\alpha_1+ \alpha_2)/4 \)     
\\  \hline
%%%%%%%%%%%%%%%%%%%%%%%%%%%%%%%%%%%%%%%%%%%%%%%%%%%%%%%%%%%%%%%%%%%%%%%%%%%%%%%%%%%%%%%%%%%%%%%%%%%%
1/2	&\mathbb{2}	&DS			&DS			
&
&\phikell \(\xS;~(-1)^{\ell+s}\aH\)      
\\ \hline
\end{array}
\end{align*}
\caption{\label{tab:ScatteringStates} 
The scattering states and their wavefunctions in the limit $\mH \to 0$. 
Here, $\phikell(\xS;~\alpha)$ denotes the $\ell$ mode of a scattering state with momentum ${\bf k}$ for a Coulomb potential $V(r) = -\alpha / r$; the position variable is $\xS \equiv kr$. The couplings $\aA$ and $\aR$, defined in \cref{eqs:GaugeSinglet_alphas_def}, depend on $\ell+s$. For $\ell+s=$~odd, the $(\mathbb{1,0})$ mixed states decouple.}

%\bigskip

\renewcommand*{\arraystretch}{2.2}
\begin{align*}
\begin{array}{ |c|c|c|c|rl| } 
\hline
%%%%%%%%%%%%%%%%%%%%%%%%%%%%%%%%%%%%%%%%%%%%%%%%%%%%%%%%%%%%%%%%%%%%%%%%%%%%%%%%%%%%%%%%%%%%%%%%%%%%
\UY	
&\SUL	
&\parbox{ 8ex}{\centering Binding \\ energy}	
&\parbox{14ex}{\centering Bound state/ \\ component}	
&\multicolumn{2}{c|}{\text{Wavefunctions}} 
\\[1ex] \hline \hline
%%%%%%%%%%%%%%%%%%%%%%%%%%%%%%%%%%%%%%%%%%%%%%%%%%%%%%%%%%%%%%%%%%%%%%%%%%%%%%%%%%%%%%%%%%%%%%%%%%%%
%%%%%%%%%%%%%%%%%%%%%%%%%%%%%%%%%%%%%%%%%%%%%%%%%%%%%%%%%%%%%%%%%%%%%%%%%%%%%%%%%%%%%%%%%%%%%%%%%%%%
 \multirow{2}{*}{0}      
&\multirow{2}{*}{$\mathbb{1}$}
&\multirow{2}{*}{$\dfrac{\mu \aA^2}{2n^2}$}
%%%%%%%%%%%%%%%%%%%%%%%%%%%%%%%
&SS			
&\sqrt{\dfrac{\aR}{\aA+\aR}} \ &\varphi_{n\ell m}({\bf r};~\aA)     
\\   \cline{4-6}
&&     
&D\bar{D}	
&\sqrt{\dfrac{\aA}{\aA+\aR}} \ &\varphi_{n\ell m}({\bf r};~\aA)     
\\ \hline
%%%%%%%%%%%%%%%%%%%%%%%%%%%%%%%%%%%%%%%%%%%%%%%%%%%%%%%%%%%%%%%%%%%%%%%%%%%%%%%%%%%%%%%%%%%%%%%%%%%%
1	
&\mathbb{1}	
&\dfrac{\mu  \[(-\alpha_1+3\alpha_2)/4\]^2   }{2n^2} 
&DD	
&\delta_{\ell+s,\rm odd} \sqrt{2} 
&\varphi_{n\ell m} \({\bf r};~(-\alpha_1+3\alpha_2)/4 \)
\\ \hline
%%%%%%%%%%%%%%%%%%%%%%%%%%%%%%%%%%%%%%%%%%%%%%%%%%%%%%%%%%%%%%%%%%%%%%%%%%%%%%%%%%%%%%%%%%%%%%%%%%%%
1/2	
&\mathbb{2}	
&\dfrac{\mu\aH^2}{2n^2}	
&DS	
&\delta_{\ell+s,\rm even} \
&\varphi_{n\ell m}({\bf r};~\aH) 
\\  \hline
%%%%%%%%%%%%%%%%%%%%%%%%%%%%%%%%%%%%%%%%%%%%%%%%%%%%%%%%%%%%%%%%%%%%%%%%%%%%%%%%%%%%%%%%%%%%%%%%%%%%
\end{array}
\end{align*}
\caption{\label{tab:BoundStates} 
The bound states and their wavefunctions in the limit $\mH \to 0$. Here $\varphi_{n\ell m}({\bf r};~\alpha)$ denotes the bound state wavefunction with quantum numbers $\{n\ell m\}$ for a Coulomb potential $V(r) = -\alpha / r$. The couplings $\aA$ and $\aR$ are defined in \cref{eqs:GaugeSinglet_alphas_def}, and depend on $\ell+s$. For $\ell+s=$~odd, the $(\mathbb{1,0})$ state becomes purely $D\bar{D}$.}
\end{table}

%\clearpage

\subsubsection{Validity of the Coulomb approximation $\mH \to 0$ \label{sec:LongRangeDynamics_AsymptoticStates_CoulombApprox}}

\paragraph{Scattering states.}
The scattering states are listed in \cref{tab:ScatteringStates}. The $DS$ states with $\ell+s=$~even are subject to an attractive Higgs-mediated potential, and the Coulomb approximation is good as long as~\cite{Petraki:2016cnz} 
\begin{align} 
\mu \vrel \gtrsim \mH .
\label{eq:CoulombCondition_Scattering}
\end{align} 
The condition becomes somewhat stronger for the $DS$ scattering states with $\ell+s=$~odd, where the Higgs-mediated potential is repulsive.
Moreover, it is relaxed or strengthened in the presence of an additional attractive or repulsive Coulomb potential due to $B$ or $W$ exchange, as is the case with the $SS$-like and $D\bar{D}$-like scattering states for $\ell+s=$~even~\cite[figs.~2,~3]{Harz:2017dlj}.

\paragraph{Bound states.}

The bound states of the present model are listed in \cref{tab:BoundStates}. The $DS$ states with $\ell+s=$~even are bound only by the Higgs-mediated potential; they exist if $(\mu \aH/n) / \mH >0.84$, and become essentially Coulombic if this condition is strengthened only by a factor of a few~\cite{Petraki:2016cnz}. (For example, the binding energy of the ground state exceeds 80\% of its Coulomb value if $\mu\aH / \mH >10$~\cite[fig.~13]{Petraki:2016cnz}.) 
The mixed $SS/D\bar{D}$ states are bound by the combined attraction of the $B,W$ and $H$ bosons, which removes the condition of existence and relaxes the condition for the Coulomb approximation~\cite[fig.~6]{Harz:2019rro}. 
Thus, the strongest condition for the Coulomb approximation to be satisfactory is
\begin{align}
\mu \aH/n > {\rm few} \times \mH.
\label{eq:CoulombCondition_Bound}
\end{align}
We note that this condition is essentially satisfied everywhere BSF via Higgs emission is phenomenologically significant (cf.~\cref{Sec:BSF}.) Indeed, the energy available to be dissipated must exceed the Higgs mass, $(\mu/2)[(\aH/n)^2+\vrel^2] > \mH$, while BSF is most significant when $\vrel \lesssim \aH/n$. Since $\aH<1$, this yields a stronger condition. 

Bound states can also form via $B$ or $W$ emission, in which case there is no phase-space suppression that ensures the validity of the Coulomb approximation. However, in~\cite{Oncala:2021swy} we shall see that these processes are less significant for the DM thermal decoupling in the present model.

\medskip

We further discuss the Coulomb approximation for the DM freeze-out in ref.~\cite{Oncala:2021swy}.

\clearpage
\subsection{Annihilation \label{sec:LongRangeDynamics_Annihilation}}

\begin{figure}[t!]
\centering
{}\hfill
\includegraphics[width=0.45\textwidth]{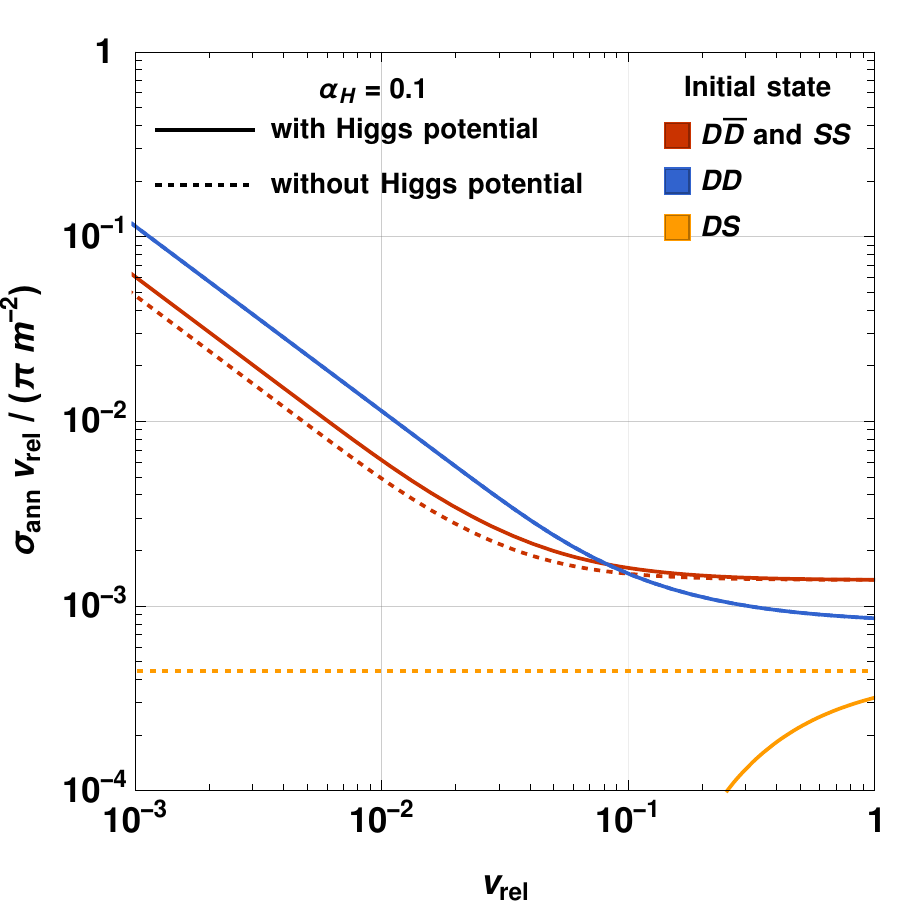}
\hfill
\includegraphics[width=0.45\textwidth]{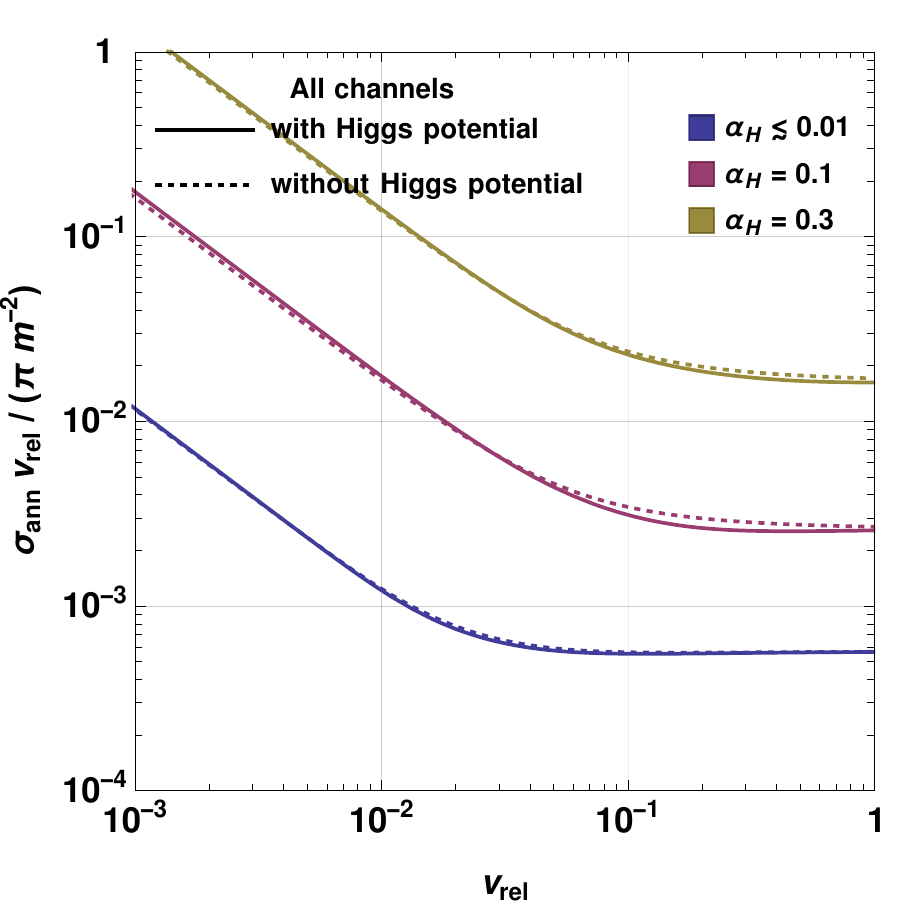}
\hfill{}
\caption{\label{fig:Annihilation} 
The $s$-wave annihilation cross-sections, by initial state (\emph{left}) and total (\emph{right}), averaged over the dof of the incoming particles, with and without the Higgs-mediated potential (cf.~\cref{tab:Annihilation}). Because the processes affected by the latter have either low multiplicity or small perturbative cross-sections, the Sommerfeld effect at low velocities arises mostly due to the $B^\mu$ and $W^\mu$ gauge bosons. 
For $\aH \lesssim 0.01$, the perturbative annihilation cross-section is also not substantially affected by the coupling to the Higgs (right panel, blue line).
Note that we have weighted the contribution of each annihilation channel with the number of DM particles eliminated in each process as estimated upon thermal averaging (cf.~ref.~\cite{Oncala:2021swy}.)}
\end{figure}

The $S$, $D$ and $\bar{D}$ fermions annihilate into SM particles via various processes that we list in \cref{tab:Annihilation} together with their tree-level cross-sections and Sommerfeld factors. We consider $s$-wave contributions only. Because the non-relativistic potentials between the annihilating particles depend in many cases on their spin and gauge representations, we project the initial states on eigenstates of total spin and Weak isospin. With the help of \cref{tab:ScatteringStates}, it is straightforward to obtain the Sommerfeld factors for all states except the spin-0 $(\mathbb{1},0)$ ones, which involve mixing of the $SS$ and $D\bar{D}$ channels and we discuss in detail below. The Sommerfeld factors are expressed in term of the $S_0(\zeta)$ function defined in \cref{eq:S0}, and the variables
\begin{align}
\zeta_1 \equiv \alpha_1/\vrel, \quad
\zeta_2 \equiv \alpha_2/\vrel, \quad
\zetaH \equiv \aH/\vrel, \quad
\zetaA \equiv \aA/\vrel,  \quad
\zetaR \equiv \aR/\vrel, 
\label{eq:zetas}
\end{align}
where the couplings $\alpha_1$, $\alpha_2$, $\aH$, $\aA$ and $\aR$ have been defined in \cref{eq:alphas_def,eq:GaugeSinglet_alphas_def}. In \cref{fig:Annihilation}, we present the total $s$-wave 2-to-2 annihilation cross-section, averaged over the dof of the incoming particles.

\subsubsection*{Mixed $(\mathbb{1},0)$ spin-0 states}

The annihilation amplitudes of the $SS$-like and $D\bar{D}$-like states (denoted by ${\cal M}$) are related to the perturbative amplitudes (denoted by ${\cal A}$) as follows
\begin{subequations}
\label{eq:M_MixedStates}
\label[pluralequation]{eqs:M_mixed}	
\begin{align}
\im {\cal M}^{\SS\mathsmaller{\text{-like} \to f}} ({\bf k}) &= \int \frac{d^3 \bf k'}{(2\pi)^3}
\[
 [\tilde{\phi}_{\bf k} ({\bf k'})]^{\SS}_{\SS}~\im {\cal A}_{\SS\mathsmaller{\to f}} ({\bf k'})
+[\tilde{\phi}_{\bf k} ({\bf k'})]^{\SS}_{\DDbar}~\im {\cal A}_{\DDbar \mathsmaller{\to f}} ({\bf k'})
\],
\label{eq:MSS_mixed} \\
\im {\cal M}^{\DDbar\mathsmaller{\text{-like} \to f}} ({\bf k}) &= \int \frac{d^3 \bf k'}{(2\pi)^3}
\[
[\tilde{\phi}_{\bf k} ({\bf k'})]^{\DDbar}_{\SS}~\im {\cal A}_{\SS\mathsmaller{\to f}} ({\bf k'})
+[\tilde{\phi}_{\bf k} ({\bf k'})]^{\DDbar}_{\DDbar}~\im {\cal A}_{\DDbar \mathsmaller{\to f}} ({\bf k'})
\],
\label{eq:MDDbar_mixed}
\end{align}
\end{subequations}
where $f$ stands for the final state, and $[\tilde{\phi}_{\bf k} ({\bf k'})]_i^j$ are the momentum-space wavefuntions, with the $j$ and $i$ indices denoting the state and the component respectively, as in \cref{sec:LongRangeDynamics_AsymptoticStates}. 
Since the perturbative $s$-wave $SS$ annihilation vanishes, in our approximation there is no interference between $SS$ and $D\bar{D}$ channels. Taking into account that the perturbative $s$-wave annihilation amplitudes are independent of the momentum to lowest order in $\vrel$, we obtain the standard result,
\begin{subequations}
\label{eq:sigmav_MixedStates}	
\label[pluralequation]{eqs:sigmav_MixedStates}	
\begin{align}
(\sigma \vrel)^{\SS\mathsmaller{\text{-like} \to f}} &= 
\left| \[\phi_{\bf k} ({\bf 0})\]_{\DDbar}^{\SS} \right|^2
(\sigma \vrel)_{\DDbar \to f} ,
\label{eq:sigmav_SS-like}
\\
(\sigma \vrel)^{\DDbar \mathsmaller{\text{-like} \to f}} &= 
\left| \[\phi_{\bf k} ({\bf 0})\]_{\DDbar}^{\DDbar} \right|^2
(\sigma \vrel)_{\DDbar \to f} , 
\label{eq:sigmav_DDbar-like}
\end{align}
\end{subequations}
where $\[\phi_{\bf k} ({\bf r})\]_i^j$ are the position-space wavefunctions computed in \cref{sec:LongRangeDynamics_AsymptoticStates}. Using the results quoted in \cref{tab:ScatteringStates}, we find 
\begin{subequations}
\label{eq:Sommerfeld_MixedStates}	
\label[pluralequation]{eqs:Sommerfeld_MixedStates}	
\begin{align}
\left| \[\phi_{\bf k} ({\bf 0})\]_{\DDbar}^{\SS} \right|^2_{\text{spin-0}} 
&=\dfrac{2\aA\aR \[\sqrt{S_0(\zetaA)} - \sqrt{S_0(-\zetaR)}\]^2} {(\aA+\aR)^2} ,
\label{eq:Sommerfeld_DDbarOfSS}
\\
\left| \[\phi_{\bf k} ({\bf 0})\]_{\DDbar}^{\DDbar} \right|^2
&=\dfrac{\[\aA \sqrt{S_0(\zetaA)} +\aR \sqrt{S_0(-\zetaR)}\]^2}{(\aA+\aR)^2} .
\label{eq:Sommerfeld_DDbarOfDDbar}
\end{align}
\end{subequations}
Note that \cref{eq:Sommerfeld_DDbarOfSS} includes the symmetry factor of the spin-0 $SS$-like state, and that $\left| \[\phi_{\bf k} ({\bf 0})\]_{\DDbar}^{\SS} \right|^2_{\text{spin-1}} =0$.

We are ultimately interested in the reduction of the DM density via the various annihilation processes. 
For the spin-0 $(\mathbb{1},0)$ states, the rate is
\begin{align}
(\nS\nS)_{(\mathbb{1},0)}^{\text{spin-0}} 
\langle \sigma \vrel \rangle^{\SS\mathsmaller{\text{-like} \to f}}
+2(\nD\nDbar)_{(\mathbb{1},0)}^{\text{spin-0}} 
\langle \sigma \vrel \rangle^{\DDbar\mathsmaller{\text{-like} \to f}} ,
\end{align}
where $n$ and $\langle\cdot\rangle$ denote densities and thermal averages, and the factor 2 in the second term appears because $D$ and $\bar{D}$ are not identical (cf.~ref.~\cite{Oncala:2021swy}.) In the limit $\mS = \mD$, the densities are $(\nS\nS)_{(\mathbb{1},0)}^{\text{spin-0}} = (\nD\nDbar)_{(\mathbb{1},0)}^{\text{spin-0}}$, thus the DM density destruction rate can be computed by regarding that the spin-0 $(\mathbb{1},0)$ $D\bar{D}$ perturbative annihilation cross-sections are enhanced by the factor
\begin{align}
%\Sigma \equiv 
\frac12 \left| \[\phi_{\bf k} ({\bf 0})\]_{\DDbar}^{\SS} \right|^2 
+ \left| \[\phi_{\bf k} ({\bf 0})\]_{\DDbar}^{\DDbar} \right|^2
= \dfrac{\aA S_0(\zetaA) + \aR S_0(-\zetaR)}{\aA+\aR} .
\label{eq:GaugeSingletSpin0_Sommerfeld}
\end{align}
We quote this result in \cref{tab:Annihilation}, but emphasise that  $(\nS\nS)_{(\mathbb{1},0)}^{\text{spin-0}}$ and $(\nD\nDbar)_{(\mathbb{1},0)}^{\text{spin-0}}$ depend exponentially on the corresponding masses, thus \cref{eq:GaugeSingletSpin0_Sommerfeld} ceases to be a good approximation already for fairly small mass differences $|\mD-\mS|$.

%%%%%%%%%%%%%%%%%%%%%%%%%%%%%%%%%%%%%%%%%%%%%%%%%%%%%%%%%%%%%%%%%%%%
\begin{table}[t!]
\centering
\definecolor{RowTitle}{rgb}{0.88,1,1}
\definecolor{RowA}{gray}{0.95}
\definecolor{RowB}{gray}{0.88}

\renewcommand*{\arraystretch}{1.1}
\begin{align*}
\begin{array}{|c|c|c|c|c|c|c|}
\hline 
\rowcolor{RowTitle}
 \text{Channel}
&\UY
&\SUL
&\text{Spin}
&\text{dof}
&(\sigma_{_0} \vrel) / (\pi m^{-2})
&\text{Sommerfeld factor}
\\ \hline \hline
%%%%%%%%%%%%%%%%%%%%%%%%%%%%%%%%%%%%%%%%%%%%%%%%%%%%%%%%%%%%%%%%%
%%%%%%%%%%%%%%%%%%%%%%%%%%%%%%%%%%%%%%%%%%%%%%%%%%%%%%%%%%%%%%%%%
\rowcolor{RowA}
SS \to HH^\dagger
&0
&\mathbb{1}
&0,1
&4
&0
&-
\\ %\hline 
%%%%%%%%%%%%%%%%%%%%%%%%%%%%%%%%%%%%%%%%%%%%%%%%%%%%%%%%%%%%%%%%%
%%%%%%%%%%%%%%%%%%%%%%%%%%%%%%%%%%%%%%%%%%%%%%%%%%%%%%%%%%%%%%%%%
\rowcolor{RowB}
%%\multirow{4}{*}{$D\bar{D} \to HH^\dagger$}
&%\multirow{4}{*}{0}
&%\multirow{2}{*}{$\mathbb{1}$}
&0
&1
&0
&-
\\ %\cline{4-7} 
\rowcolor{RowB}
&%
&\multirow{-2}{*}{$\mathbb{1}$}
&1 
&3
& (\alpha_1+2\aH)^2/12
& S_0 \[(\zeta_1+3\zeta_2)/4\]
%%%%%%%%%%%%%%%%%%%%%%%%%%%%%%%%%
\\ 
%\cline{3-7} 
\rowcolor{RowB}
&%
&%
&0
&3
&0
&-
\\ %\cline{4-7} 
 \rowcolor{RowB}
 \multirow{-4}{*}{$D\bar{D} \to HH^\dagger$}
&\multirow{-4}{*}{0}
&\multirow{-2}{*}{$\mathbb{3}$}
&1
&9
& (\alpha_2+2\aH)^2/12
& S_0 \[(\zeta_1-\zeta_2)/4\]
\\ %\hline
%%%%%%%%%%%%%%%%%%%%%%%%%%%%%%%%%%%%%%%%%%%%%%%%%%%%%%%%%%%%%%%%%
%%%%%%%%%%%%%%%%%%%%%%%%%%%%%%%%%%%%%%%%%%%%%%%%%%%%%%%%%%%%%%%%%
 \rowcolor{RowA}
%%\multirow{4}{*}{$D\bar{D} \to WW$}
&%\multirow{4}{*}{0}
&%\multirow{2}{*}{$\mathbb{1}$}
&0
&1
&3\alpha_2^2/2
&\dfrac{\aA S_0(\zetaA) + \aR S_0(-\zetaR)}{\aA+\aR}
%\Sigma~~\text{[cf.~\cref{eq:GaugeSingletSpin0_Sommerfeld}]}
\\ %\cline{4-7} 
\rowcolor{RowA}
&%
&\multirow{-2}{*}{$\mathbb{1}$}
&1
&3
&0
&-
%%%%%%%%%%%%%%%%%%%%%%%%%%%%%%%%%
\\ %\cline{3-7} 
\rowcolor{RowA}
%%%
&%%
&%
&0
&3
&0
&-
\\ %\cline{4-7} 
 \rowcolor{RowA}
 \multirow{-4}{*}{$D\bar{D} \to WW$}
&\multirow{-4}{*}{0}
&\multirow{-2}{*}{$\mathbb{3}$}
&1
&9
& \alpha_2^2/12
& S_0 \[(\zeta_1-\zeta_2)/4\]
\\ %\hline 
%%%%%%%%%%%%%%%%%%%%%%%%%%%%%%%%%%%%%%%%%%%%%%%%%%%%%%%%%%%%%%%%%
%%%%%%%%%%%%%%%%%%%%%%%%%%%%%%%%%%%%%%%%%%%%%%%%%%%%%%%%%%%%%%%%%
\rowcolor{RowB}
%\multirow{3}{*}{$D\bar{D} \to BB$}
&%
&%
&0
&1
& \alpha_1^2/2
&\dfrac{\aA S_0(\zetaA) + \aR S_0(-\zetaR)}{\aA+\aR}
\\ %\cline{4-7} 
\rowcolor{RowB}
%%%
&%%
&\multirow{-2}{*}{$\mathbb{1}$}
&1
&3
&0
&-
%%%%%%%%%%%%%%%%%%%%%%%%%%%%%%%%%
\\ %\cline{3-7} 
 \rowcolor{RowB}
 \multirow{-3}{*}{$D\bar{D} \to BB$}
&\multirow{-3}{*}{0}
&\mathbb{3}
&0,1
&12
&0
&-
\\ %\hline 
%%%%%%%%%%%%%%%%%%%%%%%%%%%%%%%%%%%%%%%%%%%%%%%%%%%%%%%%%%%%%%%%%
%%%%%%%%%%%%%%%%%%%%%%%%%%%%%%%%%%%%%%%%%%%%%%%%%%%%%%%%%%%%%%%%%
\rowcolor{RowA}
%%\multirow{3}{*}{$D\bar{D} \to WB$}
&%\multirow{3}{*}{0}
&\mathbb{1}
&0,1
&4
&0
&-
%%%%%%%%%%%%%%%%%%%%%%%%%%%%%%%%%
\\ %\cline{3-7}
\rowcolor{RowA}
&%
&%\multirow{2}{*}{$\mathbb{3}$}
&0
&3
&\alpha_1\alpha_2
& S_0 \[ (\zeta_1-\zeta_2)/4 \]
\\ %\cline{4-7}
 \rowcolor{RowA}
 \multirow{-3}{*}{$D\bar{D} \to WB$}
&\multirow{-3}{*}{0}
&\multirow{-2}{*}{$\mathbb{3}$}
&1
&9
&0
&-
\\ %\hline 
%%%%%%%%%%%%%%%%%%%%%%%%%%%%%%%%%%%%%%%%%%%%%%%%%%%%%%%%%%%%%%%%%
%%%%%%%%%%%%%%%%%%%%%%%%%%%%%%%%%%%%%%%%%%%%%%%%%%%%%%%%%%%%%%%%%
 \rowcolor{RowB}
%%\multirow{4}{*}{$D\bar{D} \to F_{\mathsmaller{L}} \bar{F}_{\mathsmaller{L}}$}
&%\multirow{4}{*}{0}
&%\multirow{2}{*}{$\mathbb{1}$}
&0
&1
&0
&-
\\ %\cline{4-7} 
\rowcolor{RowB}
%%%
&%%
&\multirow{-2}{*}{$\mathbb{1}$}
&1
&3
&\parbox{15ex}{\centering 
$2 Y_{\mathsmaller{L}}^2 \alpha_1^2/3,$ \\ 
$\sum Y_{\mathsmaller{L}}^2 = 1$}
&S_0 \[ (\zeta_1+3\zeta_2)/4 \]
%%%%%%%%%%%%%%%%%%%%%%%%%%%%%%%%%
\\ %\cline{3-7} 
\rowcolor{RowB}
%%%
&%%
&%\multirow{2}{*}{$\mathbb{3}$}
&0
&3
&0
&-
\\ %\cline{4-7} 
 \rowcolor{RowB}
 \multirow{-4}{*}{$D\bar{D} \to F_{\mathsmaller{L}} \bar{F}_{\mathsmaller{L}}$}
&\multirow{-4}{*}{0}
&\multirow{-2}{*}{$\mathbb{3}$}
&1
&9
&\parbox{15ex}{\centering $N_\mathsmaller{L} \times \alpha_2^2/6$ \\ $N_\mathsmaller{L} = 12$}
& S_0 \[ (\zeta_1-\zeta_2)/4 \]
\\ %\hline 
%%%%%%%%%%%%%%%%%%%%%%%%%%%%%%%%%%%%%%%%%%%%%%%%%%%%%%%%%%%%%%%%%
%%%%%%%%%%%%%%%%%%%%%%%%%%%%%%%%%%%%%%%%%%%%%%%%%%%%%%%%%%%%%%%%%
 \rowcolor{RowA}
%%\multirow{3}{*}{$D\bar{D} \to f_{\mathsmaller{R}} \bar{f}_{\mathsmaller{R}}$}
&%\multirow{3}{*}{0}
&%\multirow{2}{*}{$\mathbb{1}$}
&0
&1
&0
&-
\\ %\cline{4-7}
\rowcolor{RowA}
&%
&\multirow{-2}{*}{$\mathbb{1}$}
&1
&3
&\parbox{15ex}{\centering 
$Y_{\mathsmaller{R}}^2 \alpha_1^2/3,$ \\ 
$\sum Y_{\mathsmaller{R}}^2 = 8$}
& S_0 \[ (\zeta_1+3\zeta_2)/4 \]
%%%%%%%%%%%%%%%%%%%%%%%%%%%%%%%%%
\\ %\cline{3-7}
 \rowcolor{RowA}
 \multirow{-3}{*}{$D\bar{D} \to f_{\mathsmaller{R}} \bar{f}_{\mathsmaller{R}}$}
&\multirow{-3}{*}{0}
&\mathbb{3}
&0,1
&12
&0
&-
\\ %\hline
%%%%%%%%%%%%%%%%%%%%%%%%%%%%%%%%%%%%%%%%%%%%%%%%%%%%%%%%%%%%%%%%%
%%%%%%%%%%%%%%%%%%%%%%%%%%%%%%%%%%%%%%%%%%%%%%%%%%%%%%%%%%%%%%%%%
 \rowcolor{RowB}
%%\multirow{3}{*}{$DD \to H H$}
&%\multirow{3}{*}{1}
&%\multirow{2}{*}{$\mathbb{1}$}
&0
&1
&0
&-
\\ %\cline{4-7} 
\rowcolor{RowB}
%%%
&%%
&\multirow{-2}{*}{$\mathbb{1}$}
&1
&3
&4\aH^2/3 
&S_0 \[(-\zeta_1+3\zeta_2)/4\]
%%%%%%%%%%%%%%%%%%%%%%%%%%%%%%%%%
\\ %\cline{3-7} 
 \rowcolor{RowB}
 \multirow{-3}{*}{$DD \to H H$}
&\multirow{-3}{*}{1}
&\mathbb{3}
&0,1
&12
&0
&-
\\ %\hline
%%%%%%%%%%%%%%%%%%%%%%%%%%%%%%%%%%%%%%%%%%%%%%%%%%%%%%%%%%%%%%%%%
%%%%%%%%%%%%%%%%%%%%%%%%%%%%%%%%%%%%%%%%%%%%%%%%%%%%%%%%%%%%%%%%%
 \rowcolor{RowA}
%%\multirow{2}{*}{$DS \to W H$}
&%\multirow{2}{*}{1/2}
&%\multirow{2}{*}{$\mathbb{2}$}
&0
&2
&0
&-
\\ %\cline{4-7}
 \rowcolor{RowA}
 \multirow{-2}{*}{$DS \to W H$}
&\multirow{-2}{*}{1/2}
&\multirow{-2}{*}{$\mathbb{2}$}
&1
&6
&\alpha_2 \aH/2
&S_0(-\zetaH)
\\ %\hline
%%%%%%%%%%%%%%%%%%%%%%%%%%%%%%%%%%%%%%%%%%%%%%%%%%%%%%%%%%%%%%%%%
%%%%%%%%%%%%%%%%%%%%%%%%%%%%%%%%%%%%%%%%%%%%%%%%%%%%%%%%%%%%%%%%%
 \rowcolor{RowB}
%%\multirow{2}{*}{$DS \to B H$}
&%\multirow{2}{*}{1/2}
&%\multirow{2}{*}{$\mathbb{2}$}
&0
&2
&0
&-%S_0(+\zetaH)
\\ %\cline{4-7} 
 \rowcolor{RowB}
 \multirow{-2}{*}{$DS \to B H$}
&\multirow{-2}{*}{1/2}
&\multirow{-2}{*}{$\mathbb{2}$}
&1
&6
&\alpha_1 \aH/6
&S_0(-\zetaH)
\\ %\hline
%%%%%%%%%%%%%%%%%%%%%%%%%%%%%%%%%%%%%%%%%%%%%%%%%%%%%%%%%%%%%%%%%
%%%%%%%%%%%%%%%%%%%%%%%%%%%%%%%%%%%%%%%%%%%%%%%%%%%%%%%%%%%%%%%%%
\rowcolor{RowA}
DS \to 
 f_{\mathsmaller{R}} \bar{F}_{\mathsmaller{L}},
~F_{\mathsmaller{L}} \bar{f}_{\mathsmaller{R}}
&1/2
&\mathbb{2}
&0,1
&8
&0
&-
\\ \hline
%%%%%%%%%%%%%%%%%%%%%%%%%%%%%%%%%%%%%%%%%%%%%%%%%%%%%%%%%%%%%%%%%
%%%%%%%%%%%%%%%%%%%%%%%%%%%%%%%%%%%%%%%%%%%%%%%%%%%%%%%%%%%%%%%%%
%DS \to F_{\mathsmaller{L}} \bar{f}_{\mathsmaller{R}}
%&1/2
%&\mathbb{2}
%&0,1
%&8
%&0
%&-
%\\ %\hline
%%%%%%%%%%%%%%%%%%%%%%%%%%%%%%%%%%%%%%%%%%%%%%%%%%%%%%%%%%%%%%%%%
%%%%%%%%%%%%%%%%%%%%%%%%%%%%%%%%%%%%%%%%%%%%%%%%%%%%%%%%%%%%%%%%%
\end{array}
\end{align*}
\caption{\label{tab:Annihilation}
Annihilation processes, their tree-level $s$-wave velocity-weighted cross-sections $\sigma_0\vrel$ and Sommerfeld factors. All $\sigma_0 \vrel$ are averaged over the degrees of freedom of the corresponding \emph{projected} scattering state (5th column).  
%The $DD$ and $DS$ annihilations have also conjugate counterparts. 
For $DD$, $\sigma_0 \vrel$ includes the symmetry factor due to the identical initial-state particles. For the gauge-singlet spin-0 $D\bar{D}$ channels, see text for discussion. $S_0 (\zeta)$ and the various $\zeta$ parameters are defined in \cref{eq:S0,eq:zetas}. 
}
\end{table}
%%%%%%%%%%%%%%%%%%%%%%%%%%%%%%%%%%%%%%%%%%%%%%%%%%%%%%%%%%%%%%%%%%%%
%%%%%%%%%%%%%%%%%%%%%%%%%%%%%%%%%%%%%%%%%%%%%%%%%%%%%%%%%%%%%%%%%%%%
%%%%%%%%%%%%%%%%%%%%%%%%%%%%%%%%%%%%%%%%%%%%%%%%%%%%%%%%%%%%%%%%%%%%

\clearpage
\subsection{Ground-level bound states and their decay rates \label{sec:LongRangeDynamics_BoundStates}}

Besides annihilating directly into radiation, the $S$, $D$ and $\bar{D}$ fermions can form unstable bound states that decay rapidly into radiation, thereby enhancing the DM destruction rate. However, the DM annihilation via BSF is impeded by the inverse (ionisation) processes. The efficiency of BSF in reducing the DM density depends on the interplay of bound-state ionisation and decay processes and bound-to-bound transitions~\cite{vonHarling:2014kha}.

The ionisation processes become inefficient as the temperature drops around or below the binding energy. This occurs earlier for the most deeply bound states, which also have the largest decay rates. Thus, the tightest bound states have the greatest effect on the DM density, and for this reasons we shall consider the ground level of each bound state species only, $\{n\ell m\}=\{100\}$, which has the largest binding energy.\footnote{In addition to its ionisation becoming inefficient earlier, the cross-section for capture into the ground state is typically larger than those of excited states, if BSF occurs via vector emission~\cite{vonHarling:2014kha,Petraki:2015hla,Petraki:2016cnz,Harz:2018csl}. This strengthens the argument for considering only the ground states of each bound species. However, in ref.~\cite{Oncala:2019yvj} it was shown that if BSF occurs via emission of a charged scalar (here the Higgs doublet), the capture into excited states can be comparable to the capture into the ground state. It is thus possible that excited states have substantial impact in the present model. We leave this investigation for future work.} 
We list the ground-level bound states in \cref{tab:BoundStates_GroundLevel}. 

The BSF cross-sections and bound-to-bound transition rate are computed in \cref{Sec:BSF,Sec:BSF_OffShell}. Here we discuss the bound-state decay into radiation and the relevance of bound-to-bound transitions.

\subsubsection*{Decay into radiation}

The decay rate of bound states with zero angular momentum into radiation is
\begin{align}
\Gamma^\dec_{{\cal B} (X_1X_2) \to f} = (\sigma_0 \vrel)_{X_1X_2 \to f} \times |\psi_{n00} (0)|^2,
\label{eq:BoundStates_DecayRate_def}
\end{align}
where $X_1X_2$ represent the constituent fields of the bound state and $f$ stands for the final state particles. $(\sigma_0 \vrel)_{X_1X_2 \to f}$ is the $s$-wave annihilation cross-section of an $X_1X_2$ scattering state with the same quantum numbers (spin, gauge and global) as the bound state, averaged over the dof that correspond to those quantum numbers only (rather than all the dof of an $X_1X_2$ scattering state.) For an attractive Coulomb potential of strength $\alpha$, the squared ground-state wavefunction evaluated at the origin is $|\psi_{n00} (0)|^2 = \kappaB^3/\pi = \mu^3 \alpha^3 / \pi$, where $\kappaB = \mu \alpha$ is the Bohr momentum.

Taking into account the $s$-wave annihilation processes of \cref{tab:Annihilation}, we compute the total decay rates of the ground states and list them in \cref{tab:BoundStates_GroundLevel}. The decay of the mixed $SS/D\bar{D}$ bound state occurs via its $D\bar{D}$ component, and the rate is computed analogously to the annihilation of the mixed scattering states described in \cref{sec:LongRangeDynamics_Annihilation}. 
For the $DD$ bound state, a factor 2 due to the antisymmetrisation of the wavefunction has already been included in the corresponding $\sigma_0 \vrel$ in \cref{tab:Annihilation} and should not be included again when computing the decay rate of the bound state.

As seen in \cref{tab:Annihilation}, the $DS$ bound state cannot decay into two particles. It may decay instead into three bosons, however the corresponding rates are suppressed by higher powers of the couplings, 
${\cal O}(\alpha_1^2 \aH^4,~\alpha_2^2 \aH^4,~\alpha_1 \alpha_2 \aH^4,~\aH^6)$, as well as the three-body final-state phase space. The $DS$ bound state instead decays much faster into the tighter $SS/D\bar{D}$ bound state, as we shall now see.

\begin{table}[t!]
\centering
\renewcommand*{\arraystretch}{2}
\begin{align*}
\begin{array}{ |c|c|c|c|c|c|c| } 
\hline
%%%%%%%%%%%%%%%%%%%%%%%%%%%%%%%%%%%%%%%%%%%%%%%%%%%%%%%%%%%%%%%%%%%%%%%%%%%%%%%%%%%%%%%%%%%%%%%%%%%%
\parbox{9.5ex}{\centering Bound \\ state (${\cal B}$)}
&\UY	
&\SUL
&\text{Spin}	
&\parbox{4ex}{\centering dof \\ ($g_{\B}$)}	
&\parbox{17ex}{\centering Bohr \\ momentum ($\kappaB$)}	
&\text{Decay rate}~(\Gamma_{\B})
\\[1ex] \hline \hline
%%%%%%%%%%%%%%%%%%%%%%%%%%%%%%%%%%%%%%%%%%%%%%%%%%%%%%%%%%%%%%%%%%%%%%%%%%%%%%%%%%%%%%%%%%%%%%%%%%%%
%%%%%%%%%%%%%%%%%%%%%%%%%%%%%%%%%%%%%%%%%%%%%%%%%%%%%%%%%%%%%%%%%%%%%%%%%%%%%%%%%%%%%%%%%%%%%%%%%%%%
SS/D\bar{D}	
&0
&\mathbb{1}
&0
&1
& \dfrac{m\aA}{2}
& \dfrac{m \aA^3 (\alpha_1^2+3\alpha_2^2)}{16} \(\dfrac{\aA}{\aA+\aR}\)
\\ \hline
%%%%%%%%%%%%%%%%%%%%%%%%%%%%%%%%%%%%%%%%%%%%%%%%%%%%%%%%%%%%%%%%%%%%%%%%%%%%%%%%%%%%%%%%%%%%%%%%%%%%
D\bar{D}	
&0	
&\mathbb{1}	
&1
&3
& \dfrac{m (\alpha_1+3\alpha_2)}{8} 
& \dfrac{m (\alpha_1+3\alpha_2)^3 [(\alpha_1+2\aH)^2+40 \alpha_1^2]}{2^{11} \cdot 3}
\\ \hline
%%%%%%%%%%%%%%%%%%%%%%%%%%%%%%%%%%%%%%%%%%%%%%%%%%%%%%%%%%%%%%%%%%%%%%%%%%%%%%%%%%%%%%%%%%%%%%%%%%%%
DD	
&1	
&\mathbb{1}	
&1
&3
& \dfrac{m(-\alpha_1+3\alpha_2)}{8}
& \dfrac{m (-\alpha_1+3\alpha_2)^3 \aH^2}{2^7 \cdot 3}
\\ \hline
%%%%%%%%%%%%%%%%%%%%%%%%%%%%%%%%%%%%%%%%%%%%%%%%%%%%%%%%%%%%%%%%%%%%%%%%%%%%%%%%%%%%%%%%%%%%%%%%%%%%
DS	
&1/2	
&\mathbb{2}	
&0
&2
& \dfrac{m\aH}{2}
& \parbox{5.5cm}{\centering
\vphantom{a} 
Negligible. \\[1ex]	
Transition to ${\cal B}(SS/D\bar{D})$: \\	 
\cref{eq:BoundToBound_SDToSSDDbar_Hemission_Rate}, \\
\cref{eq:HoffshellBoundToBound_gamma_factorisation,eq:HoffshellBSF_Rfactors,eq:HoffshellBSF_Rpm,eq:HoffshellBSF_Rfactor_tot}}
\\[2.5em]  \hline
%%%%%%%%%%%%%%%%%%%%%%%%%%%%%%%%%%%%%%%%%%%%%%%%%%%%%%%%%%%%%%%%%%%%%%%%%%%%%%%%%%%%%%%%%%%%%%%%%%%%
\end{array}
\end{align*}
\caption{\label{tab:BoundStates_GroundLevel} 
The ground-level bound states,  $\{n\ell m\}=\{100\}$, and their rates of decay into radiation, in the limit $\mH \to 0$. The decay rate of the $DS$ bound state is suppressed, and we reference instead the formulae for its transition rate to the $SS/D\bar{D}$ bound state.  
In the first row, $\aA$ and $\aR$ are found from \cref{eqs:GaugeSinglet_alphas_def} for $\ell=s=0$.
The binding energy of each bound state is $|{\cal E}_{\B}| = \kappaB^2/m$. 
}
\end{table}

\subsubsection*{Transitions into deeper bound levels}

Besides decaying directly into radiation, bound states may transition into lower-lying bound levels via dissipation of energy. The bound-to-bound transition rates are computed in a similar fashion to BSF processes, as we shall see in \cref{Sec:BSF,Sec:BSF_OffShell}. Here we note that in radiative transitions (either scattering-to-bound or bound-to-bound), spin is conserved at leading order in the non-relativistic regime. Spin-flipping transitions can occur via emission of a vector boson, but rely on spin-orbit interaction and are suppressed by higher powers of the couplings. Consequently, they are subdominant to the direct bound-state decay rates into two relativistic species. 

Considering the bound states of \cref{tab:BoundStates_GroundLevel}, there are only two spin-conserving transitions, of which only one may occur with emission of a single boson contained in the theory. Noting that $\aA \geqslant \aH$ (cf.~\cref{eq:GaugeSinglet_alphaA_def}), this transition is
\begin{align}
{\cal B} (DS) \to {\cal B} (SS/D\bar{D}) + H. 
\label{eq:Transition_DStoSSDDbar_Hemission}
\end{align}
In fact, much like BSF, bound-to-bound transitions may occur either radiatively or via scattering on the relativistic thermal bath. In \cref{Sec:BSF,Sec:BSF_OffShell}, we compute the corresponding rates for the transition \eqref{eq:Transition_DStoSSDDbar_Hemission} and reference the results in \cref{tab:BoundStates_GroundLevel}.

%%%%%%%%%%%%%%%%%%%%%%%%%%%%%%%%%%%%%%%%%%%%%%%%%%%%%%%%%%%%%%%%%%%%
%%%%%%%%%%%%%%%%%%%%%%%%%%%%%%%%%%%%%%%%%%%%%%%%%%%%%%%%%%%%%%%%%%%%
%%%%%%%%%%%%%%%%%%%%%%%%%%%%%%%%%%%%%%%%%%%%%%%%%%%%%%%%%%%%%%%%%%%%
\clearpage
\section{Radiative bound-state formation and bound-to-bound transitions \label{Sec:BSF}}

We now compute the cross-sections for the radiative formation of the ground-level bound states of \cref{tab:BoundStates_GroundLevel}. We first outline the elements of the computation, explain the approximations involved, and define some useful quantities. Then we proceed with the computation of the amplitudes and cross-sections. The final results are listed in \cref{tab:BSF_SSDDbar,tab:BSF_DDbar,tab:BSF_DD,tab:BSF_SD} and illustrated in \cref{fig:BSF_onshell_CrossSections}.

\subsection{Preliminaries \label{sec:BSF_preliminaries}}

\begin{figure}[t!]
\centering	
\begin{tikzpicture}[line width=1pt, scale=1]
%%%%%%%%%%%%%%%%%%%%%%%%%%%%%%%%%%%%%%%%%%%%%%%%%%%%%%%%%%%%%%%%%%%%%%%%%%%%%%%%%%%%
\begin{scope}[shift={(0,0)}]
%%%%%%%% Incoming Fermions (Scalars)
\draw (-2.5,+0.5) -- (-0.50,+0.50);
\draw (-0.5,+0.5) -- (-0.05,+0.05);
\draw (-0.5,-0.5) -- (-0.05,-0.05);
\draw (-2.5,-0.5) -- (-0.50,-0.50);
%%%%%%%% Outgoing Fermions (Scalars)
\draw (2.5,+0.5) -- (0.5,+0.5);
\draw (0.5,+0.5) -- (0.1,+0.1);
\draw (0.5,-0.5) -- (0.1,-0.1);
\draw (2.5,-0.5) -- (0.5,-0.5);
%%%%%%%% Radiation and thermal bath
\draw[scalarnoarrow] (0,0.25) -- (0,1.5);
\draw[->] (-0.2,0.8) -- (-0.2,1.3);\node at (-0.6,1) {$P_{\rad}$};
%%%%%%%% G4, A functions and labels
\draw[fill=lightgray,draw=none,shift={(-1.5,0)}] (-0.6,-0.7) rectangle (0.6,0.7);
\draw[fill=lightgray,draw=none,shift={(+1.5,0)}] (-0.6,-0.7) rectangle (0.6,0.7);
\filldraw[lightgray]  (0,0) circle (12pt);
%%%
\node at (-1.5,0) {$G^{(4)}_{\rm in}$};
\node at (+1.5,0) {$G^{(4)}_{\rm out}$};
\node at ( 0,  0) {$\im {\cal A}_{\mathsmaller{T}}$};
%%%%%%%%
\end{scope}
%%%%%%%%%%%%%%%%%%%%%%%%%%%%%%%%%%%%%%%%%%%%%%%%%%%%%%%%%%%%%%%%%%%%%%%%%%%%%%%%%%%%
\end{tikzpicture}
\caption{\label{fig:BSF_general} Radiative bound-state formation and bound-to-bound transitions.}
\end{figure}
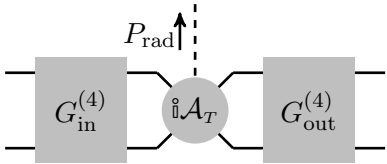

\paragraph{Full amplitudes ${\cal M}_{{\bf k}\to n\ell m}$ and ${\cal M}_{n' \ell' m' \to n\ell m}$.} 
As depicted schematically in \cref{fig:BSF_general}, the full amplitudes consist of the radiative transition part ${\cal A}_{\mathsmaller{T}}$ computed perturbatively and convoluted with the initial and final-state wavefunctions~\cite{Petraki:2015hla},
\begin{subequations}
\label{eq:FullAmplitudes_Def}
\label[pluralequation]{eqs:FullAmplitudes_Def}
\begin{align}
\im {\cal M}_{{\bf k} \to n\ell m} &= 
\int \frac{d^3 {\bf k'}}{(2\pi)^3} \frac{d^3 {\bf p}}{(2\pi)^3}
\ \tilde{\phi}_{\bf k} ({\bf k'})
\ \im {\cal A}_{\mathsmaller{T}} ({\bf k',p}) 
\ \frac{[\tilde{\psi}_{n\ell m} ({\bf p})]^\dagger}{\sqrt{2\mu}} .
\label{eq:BSF_FullAmplitude_Def}
\\
\im {\cal M}_{n'\ell'm' \to n\ell m} &= 
\int \frac{d^3 {\bf p'}}{(2\pi)^3} \frac{d^3 {\bf p}}{(2\pi)^3}
\ \frac{\tilde{\psi}_{n'\ell' m'} ({\bf p'})}{\sqrt{2\mu}}
\ \im {\cal A}_{\mathsmaller{T}} ({\bf p',p}) 
\ \frac{[\tilde{\psi}_{n\ell m} ({\bf p})]^\dagger}{\sqrt{2\mu}} .
\label{eq:BoundToBound_FullAmplitude_Def}
\end{align}
\end{subequations}
\Cref{eqs:FullAmplitudes_Def} can accommodate the possibility that the incoming and/or outgoing states are superpositions of different Fock states, as is the case with the mixed $SS/D\bar{D}$ states discussed in \cref{sec:LongRangeDynamics_AsymptoticStates}. Then, the wavefunctions are vectors and ${\cal A}_{\mathsmaller{T}}$ becomes a matrix.

\paragraph{Transition amplitudes ${\cal A}_{\mathsmaller{T}}$.}
The radiative parts of the BSF and transition diagrams are shown in \cref{fig:BSF_SSDDbar,fig:BSF_DDbar,fig:BSF_DD,fig:BSF_SD}. 
Following refs.~\cite{Petraki:2015hla,Harz:2018csl,Oncala:2019yvj}, we compute the amplitudes ${\cal A}_{\mathsmaller{T}}$, applying the standard approximations due to the hierarchy of scales and retaining only the leading order terms. Among else, we shall use the following approximate spinor identities valid for low momentum changes, $p\simeq p'$,
\begin{subequations}
\label{eq:SpinorIdentities}	
\begin{align}
\bar{u} (p,s) \, u (p',s') &\simeq +2m \ \delta^{s s'},
&\bar{u} (p,s) \, \gamma^\mu \, u (p',s') &\simeq +2 p^\mu \delta^{s s'} ,
\label{eq:spinor_ubaru}
\\
\bar{v} (p,s) \, v_2 (p',s') &\simeq -2m \ \delta^{s s'},
&\bar{v} (p,s) \, \gamma^\mu \, v (p',s') &\simeq +2 p^\mu \delta^{s s'} .
\label{eq:spinor_vbarv}
\end{align}
\end{subequations}
Moreover, we emphasise that the signs arising from the fermion permutations needed to perform the Wick contractions must be carefully taken into account, as they often differ among various diagrams contributing to the same amplitude. An example calculation is presented in detail in \cref{App:BSF_PerturbativeExample}.

\paragraph{Overlap integrals.} 
The scattering and bound state wavefunctions are listed in \cref{tab:ScatteringStates,tab:BoundStates}. To express ${\cal M}_{{\bf k}\to n\ell m}$ and  ${\cal M}_{n'\ell'm'\to n\ell m}$ compactly, we define the Coulombic overlap integrals~\cite{Petraki:2015hla,Petraki:2016cnz,Harz:2018csl,Oncala:2019yvj}
\begin{subequations}
\label{eq:OverlapIntegrals_def}
\label[pluralequation]{eqs:OverlapIntegrals_def}	
\begin{align}
{\cal R}_{{\bf k},n\ell m} (\aS,\aB) 
&\equiv 
(\mu \aB)^{3/2}
\int \frac{d^3 {\bf p}}{(2\pi)^3}
\ \tilde{\varphi}_{\bf k} ({\bf p};\aS)
\ \tilde{\varphi}_{n\ell m}^* ({\bf p};\aB) ,
\label{eq:OverlapIntegral_R_def}
\\
\boldsymbol{{\cal J}}_{{\bf k},n\ell m} (\aS,\aB) 
&\equiv 
\int \frac{d^3 {\bf p}}{(2\pi)^3}
\ {\bf p}
\ \tilde{\varphi}_{\bf k} ({\bf p};\aS)
\ \tilde{\varphi}_{n\ell m}^* ({\bf p};\aB) ,
\label{eq:OverlapIntegral_J_def}
\\
\boldsymbol{{\cal Y}}_{{\bf k},n\ell m}^\W (\aS,\aB) 
&\equiv 
8\pi \mu \alpha_2
\int \frac{d^3 {\bf k'}}{(2\pi)^3} \frac{d^3 {\bf p}}{(2\pi)^3}
\frac{{\bf k'-p}}{({\bf k'-p})^4}
\, \tilde{\varphi}_{\bf k} ({\bf k'};\aS)
\, \tilde{\varphi}_{n\ell m}^* ({\bf p};\aB),
\label{eq:OverlapIntegral_YW_def}
\\
\boldsymbol{{\cal Y}}_{{\bf k},n\ell m}^\H (\aS,\aB) 
&\equiv 
8\pi \mu \aH
\!\int\! \frac{d^3 {\bf k'}}{(2\pi)^3} \frac{d^3 {\bf p}}{(2\pi)^3}
\frac{{\bf k'-p}}{[({\bf k'-p})^2 + \mH^2]^2}
\tilde{\varphi}_{\bf k} ({\bf k'};\aS)
\tilde{\varphi}_{n\ell m}^* ({\bf p};\aB),
\label{eq:OverlapIntegral_YH_def}
\end{align}
and
\begin{align}
{\cal R}_{n'\ell'm,n\ell m} (\aB',\aB) 
&\equiv 
\int \frac{d^3 {\bf p}}{(2\pi)^3}
\ \tilde{\varphi}_{n'\ell'm'} ({\bf p};\aB')
\ \tilde{\varphi}_{n\ell m}^* ({\bf p};\aB) ,
\label{eq:OverlapIntegral_RboundTobound_def}
\end{align}
\end{subequations}
where $\tilde{\varphi}_{\bf k}$ and $\tilde{\varphi}_{n\ell m}$ are Fourier transforms of the Coulomb wavefunctions \eqref{eq:varphi_k} and \eqref{eq:varphi_nlm}.\footnote{Note that ${\cal R}_{{\bf k},n\ell m}$, ${\cal R}_{n'\ell'm,n\ell m}$ are dimensionless, while $\boldsymbol{{\cal J}}_{{\bf k},n\ell m}$ and $\boldsymbol{{\cal Y}}_{{\bf k},n\ell m}$ have mass-dimension $-1/2$.} 
In the Coulomb regime, the overlap integrals can be computed analytically, after Fourier transforming into position space. The scattering-bound integrals of \cref{eq:OverlapIntegral_R_def,eq:OverlapIntegral_J_def,eq:OverlapIntegral_YW_def,eq:OverlapIntegral_YH_def} have been computed in refs.~\cite{Petraki:2015hla,Petraki:2016cnz,Harz:2018csl,Oncala:2019yvj}. We compute the bound-bound integral \eqref{eq:OverlapIntegral_RboundTobound_def} in \cref{App:OverlapIntegral_RBoundToBound}. For the ground-level bound states, $\{n\ell m\}=\{100\}$, the results are
\begin{subequations}
\label{eq:OverlapIntegrals_100}
\label[pluralequation]{eqs:OverlapIntegrals_100}
\begin{align}
{\cal R}_{{\bf k},100} (\aS,\aB) 
&= \[ 
2^6 \pi \(1-\frac{\zetaS}{\zetaB}\)^2   
\(\frac{\zetaB^2}{1+\zetaB^2}\) 
\Ssc (\zetaS,\zetaB) \]^{1/2} , 
\label{eq:OverlapIntegral_R_100}
\\
\boldsymbol{{\cal J}}_{{\bf k},100} (\aS,\aB) 
&= \hat{\bf k} \[ 
\frac{2^6 \pi}{\mu \aB}  
\(\frac{\zetaB^2}{1+\zetaB^2}\) 
\Svec (\zetaS,\zetaB) \]^{1/2}  , 
\label{eq:OverlapIntegral_J_100}
\\
\boldsymbol{{\cal Y}}_{{\bf k},100}^\W (\aS,\aB) 
&= (\alpha_2/\aB) \boldsymbol{{\cal J}}_{{\bf k},100} (\aS,\aB),
\label{eq:OverlapIntegral_YW_100}
\\
\lim_{\mH\to0} [\boldsymbol{{\cal Y}}_{{\bf k},100}^\H (\aS,\aB)]
&= (\aH/\aB) \ \boldsymbol{{\cal J}}_{{\bf k},100} (\aS,\aB) ,
\label{eq:OverlapIntegral_YH_100}
\end{align}
and
\begin{align}
{\cal R}_{100,100} (\aB',\aB) = 8 (\aB\aB')^{3/2}/(\aB+\aB')^3,
\label{eq:OverlapIntegral_R_100to100}
\end{align}
\end{subequations}
where
\begin{align}
\zetaS \equiv \aS / \vrel \quad {\rm and} \quad \zetaB \equiv \aB/\vrel 
\label{eq:zetaSzetaB_def}
\end{align} 
entail the coupling strengths $\aS$ and $\aB$ of the potential in the scattering and bound states respectively, and we have defined the functions~\cite{vonHarling:2014kha,Petraki:2015hla,Petraki:2016cnz,Harz:2018csl,Oncala:2019yvj}
\begin{subequations}
\label{eq:SBSF}
\label[pluralequation]{eqs:SBSF}
\begin{align}
\Ssc (\zetaS,\zetaB) &\equiv 
\(\frac{2\pi \zetaS}{1- e^{-2\pi \zetaS}}\)
\[\frac{\zetaB^6 e^{-4\zetaS \, {\rm arccot} (\zetaB)} }{(1+\zetaB^2)^3}  \] .
\label{eq:Ssc}
\\
\Svec (\zetaS,\zetaB) &\equiv 
\(\frac{2\pi \zetaS}{1- e^{-2\pi \zetaS}}\) (1+\zetaS^2)
\[\frac{\zetaB^4 e^{-4\zetaS \, {\rm arccot} (\zetaB)} } {(1+\zetaB^2)^3} \] .  
\label{eq:Svec}
\end{align}
\end{subequations}
Note that $\Ssc$ and $\Svec$ include the $s$- and $p$-wave Sommerfeld factors, $S_0(\zetaS) \equiv 2\pi \zetaS / (1+e^{-2\pi \zetaS})$ and  $S_1(\zetaS) = S_0(\zetaS) (1+\zetaS^2)$, respectively. Indeed, \cref{eq:OverlapIntegral_R_100} arises from the $\ellS=0$ and
\cref{eq:OverlapIntegral_J_100,eq:OverlapIntegral_YW_100,eq:OverlapIntegral_YH_100} arise from the $\ellS=1$ 
modes of the scattering state wavefunctions.
In \cref{eq:OverlapIntegral_YH_100} we took the limit $\mH\to 0$ to be consistent with our approximations, although it is easy to obtain an analytical result for $\mH \neq 0$ (but using the Coulomb wavefunctions.)

\paragraph{BSF cross-sections.}
The cross-sections for BSF via emission of a massless vector ($B$ or $W$) or scalar ($H$ or $H^\dagger$) boson are, respectively~\cite{Petraki:2015hla,Petraki:2016cnz}
\begin{subequations}
\label{eq:BSF_dsigmadOmega}
\label[pluralequation]{eqs:BSF_dsigmadOmega}
\begin{align}
\vrel \, \frac{d\sigma_{{\bf k} \to n\ell m}^{\V}}{d\Omega} &= 
\frac{|\PVvec|}{2^7\pi^2 m^3}
\ |\boldsymbol{{\cal M}}_{{\bf k} \to n\ell m}|^2 
\[1- \(\hat{\bf P}_{\mathsmaller{V}} \cdot \boldsymbol{\hat{{\cal M}}}_{{\bf k} \to n\ell m} \)^2\],
\label{eq:BSF_dsigmadOmega_Vector}
\\
\vrel \, \frac{d\sigma_{{\bf k} \to n\ell m}^{\H}}{d\Omega} &= 
\frac{|\PHvec|}{2^7\pi^2 m^3}
|{\cal M}_{{\bf k} \to n\ell m}|^2 ,
\label{eq:BSF_dsigmadOmega_Scalar}
\end{align}
\end{subequations}
where $\PVvec$ and $\PHvec$ are the momenta of the emitted bosons, which dissipate the kinetic energy of the relative motion and the binding energy, 
$|\PVvec|$ or $\sqrt{\PHvec^2+\mH^2} = \omega_{{\bf k}\to n\ell m}$, with
\begin{align}
\omega_{{\bf k}\to n\ell m} \simeq 
{\cal E}_{\bf k} - {\cal E}_{n\ell m} 
\simeq (m/4) \(\aB^2/n^2 + \vrel^2\) .
\label{eq:omega_BSF}
\end{align}
In \cref{eq:BSF_dsigmadOmega_Vector}, we have summed over polarisations of the emitted vector. As we shall see in the following, to working order, the amplitudes for BSF via vector emission are $\boldsymbol{{\cal M}}_{{\bf k}\to 100} \propto {\bf k}$, while the amplitudes for BSF via scalar emission are independent of the solid angle $\Omega$. Then, \cref{eqs:BSF_dsigmadOmega} simplify to~\cite{Petraki:2015hla,Petraki:2016cnz}
\begin{subequations}
\label{eq:BSF_sigmas}
\label[pluralequation]{eqs:BSF_sigmas}
\begin{align}
\sigma_{{\bf k} \to n\ell m}^{\V} \vrel &= 
\frac{\aB^2}{2^6 \cdot 3 \pi m^2} \(\frac{1+\zetaB^2}{\zetaB^2}\)
\ |\boldsymbol{{\cal M}}_{{\bf k} \to n\ell m}|^2 ,
\label{eq:BSF_sigma_Vector}
\\
\sigma_{{\bf k} \to n\ell m}^{\H} \vrel &= 
\frac{\aB^2}{2^7 \pi m^2} \(\frac{1+\zetaB^2}{\zetaB^2}\)
\ |{\cal M}_{{\bf k} \to n\ell m}|^2  
\ \hH (\omega_{{\bf k}\to n\ell m}),
\label{eq:BSF_sigma_Scalar}
\end{align}
\end{subequations}	
where $\hH \equiv |\PHvec| / \EH$ is the phase-space suppression due to the Higgs mass,
\begin{align}
\hH (\omega) \equiv 
\(1-\mH^2 / \omega^2\)^{1/2}.
\label{eq:hH}
\end{align}

\paragraph{Bound-to-bound transition rates.} 
Similarly to the above, the rates for the radiative de-excitation of bound states are
\begin{subequations}
\label{eq:BoundToBound_dGammadOmega}
\label[pluralequation]{eqs:BoundToBound_dGammadOmega}
\begin{align}
\dfrac{d\Gamma_{n'\ell'm' \to n\ell m}^{\V}}{d\Omega} &=
\frac{|\PVvec|}{2^7\pi^2 m^2}
\ |\boldsymbol{{\cal M}}_{n'\ell'm' \to n\ell m}|^2 
\[1- \(\hat{\bf P}_{\mathsmaller{V}} \cdot \boldsymbol{\hat{{\cal M}}}_{n'\ell'm' \to n\ell m} \)^2\],
\label{eq:BoundToBound_dGammadOmega_Vector}
\\
\dfrac{d\Gamma_{n'\ell'm' \to n\ell m}^{\H}}{d\Omega} &=
\frac{|\PHvec|}{2^7\pi^2 m^2}
|{\cal M}_{n'\ell'm' \to n\ell m}|^2 ,
\label{eq:BoundToBound_dGammadOmega_Scalar}
\end{align}
\end{subequations}	
where $|\PVvec|$ and $|\PHvec|$ are determined again by the amount of dissipated energy, which is now the difference between the binding energies of the two states, 
\begin{align}
\omega_{n'\ell'm'\to n\ell m} \simeq 
{\cal E}_{n'\ell' m'} -{\cal E}_{n\ell m} \simeq (m/4) \(\aB^2/n^2 - \aB'^2/n'^2\) .
\label{eq:omega_BoundToBound}
\end{align}
For monopole transitions via $H$ or $H^\dagger$ emission, the amplitude is independent of the $\PHvec$ direction, and \cref{eq:BoundToBound_dGammadOmega_Scalar} simplifies to
\begin{align}
\Gamma_{n'\ell'm' \to n\ell m}^{\H} &=
\frac{\aB'^2/n'^2 - \aB^2/n^2}{2^7 \pi m}
\ |{\cal M}_{n'\ell'm' \to n\ell m}|^2 
\ \hH (\omega_{n'\ell'm'\to n\ell m}) ,
\label{eq:BoundToBound_Gamma_Scalar}
\end{align}
where the phase-space suppression $\hH$ is defined in \cref{eq:hH}.

\clearpage
\subsection{$SS/D\bar{D}$ bound states: $(\mathbb{1},0)$, spin 0, $n\ell m =\{100\}$ \label{sec:BSF_SSDDbar}}
%%%%%%%%%%%%%%%%%%%%%%%%%%%%%%%%%%%%%%%%%%%%%%%%%%%%%%%%%%%%%%%%%%%%
\begin{table}[p]
\centering
\renewcommand*{\arraystretch}{1}
\begin{align*}
\begin{array}{|c|c|c|c|c|c|rl|}
%%%%%%%%%%%%%%%%%%%%%%%%%%%%%%%%%%%%%%%%%%%%%%%%%%%%%%%%%%%%%%%%%%%%%%%%%%%%%%%%%%%%%%%%%%%%%%%%%%%%%%%%
\multicolumn{8}{c}{\boldsymbol{
\textbf{Bound state}~SS/D\bar{D}:
~~~(\mathbb{1},0),
~~\textbf{spin 0}, 
~~\{n\ell m\} = \{100\}
}}
\\ \hline
%%%%%%%%%%%%%%%%%%%%%%%%%%%%%%%%%%%%%%%%%%%%%%%%%%%%%%%%%%%%%%%%%%%%%%%%%%%%%%%%%%%%%%%%%%%%%%%%%%%%%%%%
\multicolumn{5}{|c|}{\text{Scattering state (spin 0)}}
&\multirow{2}{*}{\text{\parbox{6ex}{\centering Rad boson}}}
&\multicolumn{2}{c|}{\text{Cross-section}}
\\ \cline{1-5}
\text{State}		
& U_{\mathsmaller{Y}}(1) 
&SU_{\mathsmaller{L}}(2) 
& \text{dof} 
& \ellS 
&
&\multicolumn{2}{c|}{(\sigma_\BSF \vrel) / (\pi m^{-2})}
\\ \hline\hline
%%%%%%%%%%%%%%%%%%%%%%%%%%%%%%%%%%%%%%%%%%%%%%%%%%%%%%%%%%%%%%%%%%%%%%%%%%%%%%%%%%%%%%%%%%%%%%%%%%%%%%%%
D\bar{D}
& 0   		
& \mathbb{1} 
& 1  
& 1 
& B 			
&\dfrac{2^7}{3} 
\dfrac{\alpha_1 \aA^2}{\aA+\aR} 
\[1 + \dfrac{\aH}{\aA}\sqrt{\dfrac{8\aR}{\aA}} \, \]^2
&\Svec \(\dfrac{\zeta_1+3\zeta_2}{4},\zetaA\) 
\\ \hline
%%%%%%%%%%%%%%%%%%%%%%%%%%%%%%%%%%%%%%%%%%%%%%%%%%%%%%%%%%%%%%%%%%%%%%%%%%%%%%%%%%%%%%%%%%%%%%%%%%%%%%%%
SS 					
& 0   
& \mathbb{1} 
& 1  
& 1 
& B 			
& \multicolumn{2}{c|}{0\text{ (due to antisymmetry of $SS$ scattering state)}}
\\ \hline
%%%%%%%%%%%%%%%%%%%%%%%%%%%%%%%%%%%%%%%%%%%%%%%%%%%%%%%%%%%%%%%%%%%%%%%%%%%%%%%%%%%%%%%%%%%%%%%%%%%%%%%%
D\bar{D} 			
& 0   
& \mathbb{3} 
& 3 
& 1 
& W 			
&\dfrac{2^5}{3}\dfrac{\alpha_2 \aA^2}{\aA+\aR} \!
\[1\!+\!\dfrac{\alpha_2}{\aA}\!+\!\dfrac{\aH}{\aA}\sqrt{\dfrac{8\aR}{\aA}}\,\]^2
&\Svec \(\dfrac{\zeta_1-\zeta_2}{4},\zetaA\)
\\ \hline
%%%%%%%%%%%%%%%%%%%%%%%%%%%%%%%%%%%%%%%%%%%%%%%%%%%%%%%%%%%%%%%%%%%%%%%%%%%%%%%%%%%%%%%%%%%%%%%%%%%%%%%%
DS 					
&+1/2 
& \mathbb{2} 
& 2  
& 0 
& H			
&\dfrac{2^7 \aH}{\aA} \dfrac{\(\sqrt{\aA} + \sqrt{8\aR}\)^2}{\aA+\aR} 
\(1-\dfrac{\aH}{\aA}\)^2
&\Ssc \(\zetaH,\zetaA\) \, \hH (\omega)
\\ \hline
%%%%%%%%%%%%%%%%%%%%%%%%%%%%%%%%%%%%%%%%%%%%%%%%%%%%%%%%%%%%%%%%%%%%%%%%%%%%%%%%%%%%%%%%%%%%%%%%%%%%%%%%
\bar{D}S 			
&-1/2 
& \mathbb{2} 
& 2  
& 0 
& H^\dagger 	
&\multicolumn{2}{c|}{\text{same as above}}
\\ \hline
%%%%%%%%%%%%%%%%%%%%%%%%%%%%%%%%%%%%%%%%%%%%%%%%%%%%%%%%%%%%%%%%%%%%%%%%%%%%%%%%%%%%%%%%%%%%%%%%%%%%%%%%
\end{array}
\end{align*}
\vspace{-15pt}
\captionof{table}{\label{tab:BSF_SSDDbar}	
Radiative processes and cross-sections for capture into the ground level of the $SS/D\bar{D}$ bound states. Here, $\aA$, $\aR$ are obtained from \cref{eqs:GaugeSinglet_alphas_def} for $\ell=s=0$. 
Each cross-section is averaged over the dof of the corresponding scattering state (4th column.) $\Svec$ and $\Ssc$ are defined in \cref{eqs:SBSF}, and $\hH$ in \cref{eq:hH}. Here, $\omega=m(\aA^2+\vrel^2)/4$.}
%%%%%%%%%%%%%%%%%%%%%%%%%%%%%%%%%%%%%%%%%%%%%%%%%%%%%%%%%%%%%%%%%%%%
\medskip
%%%%%%%%%%%%%%%%%%%%%%%%%%%%%%%%%%%%%%%%%%%%%%%%%%%%%%%%%%%%%%%%%%%%
\begin{tikzpicture}[line width=1pt, scale=1]
%%%%%%%%%%%%%%%%%%%%%%%%%%%%%%%%%%%%%%%%%%%%%%%%%%%%%%%%%%%%%%%%%%%%%%%%%%%%%%%%%%%%%%%%%%%%%
%%%%%%%%%%%%%%%%%%%%%%%%%%%%%%%%%%%%%%%%%%%%%%%%%%%%%%%%%%%%%%%%%%%%%%%%%%%%%%%%%%%%%%%%%%%%%
%%%%%%%%%%%%%%%%%%%%%%%%%%%%%%%%%%%%%%%%%%%%%%%%%%%%%%%%%%%%%%%%%%%%%%%%%%%%%%%%%%%%%%%%%%%%%
\begin{scope}[shift={(0,0)}]
%%%%%%%%%%%%%%%%%%%%%%%%%%%%%
\begin{scope}[shift={(0,0)}]
%%%%% field lines
\draw[doublefermion] 	(-1,+0.5)--(0,+0.5);\draw[doublefermion] 	(0,+0.5)--(1,+0.5);
\draw[doublefermionbar] (-1,-0.5)--(0,-0.5);\draw[doublefermionbar] (0,-0.5)--(1,-0.5);
\draw[vector] (0.3,1.5)--(0,+0.5);%\node at (0.6,1.7) {$W$};
%%%%% spins
\node at (-1.25,+0.45) {$s_1$};\node at (+1.25,+0.45) {$r_1$};
\node at (-1.25,-0.45) {$s_2$};\node at (+1.25,-0.45) {$r_2$};
%%%%% momenta
\draw[->] (-0.9,+0.75) -- (-0.4,+0.75);\node at (-1.2,+1) {$K/2+k^{(\prime)}$};
\draw[->] (-0.9,-0.75) -- (-0.4,-0.75);\node at (-1.2,-1.1) {$K/2-k^{(\prime)}$};
\draw[->] (+0.4,+0.75) -- (+0.9,+0.75);\node at (+1.1,+1) {$P/2+p$};
\draw[->] (+0.4,-0.75) -- (+0.9,-0.75);\node at (+1.1,-1.1) {$P/2-p$};
\draw[->] (-0.1,1.1)--(0.05,1.55);\node at (-0.25,1.6) {$\PB$};
%%%%% SU2L
\node at (-0.7,+0.25) {$i$};\node at (+0.4,+0.25) {$i'$};
\node at (-0.7,-0.25) {$j$};\node at (+0.4,-0.25) {$j'$};
\end{scope}
%%%%%%%%%%%%%%%%%%%%%%%%%%%%%
\begin{scope}[shift={(3,0)}]
\draw[doublefermion] 	(-1,+0.5)--(0,+0.5);\draw[doublefermion] 	(0,+0.5)--(1,+0.5);
\draw[doublefermionbar] (-1,-0.5)--(0,-0.5);\draw[doublefermionbar] (0,-0.5)--(1,-0.5);
\draw[vector] (0.3,-1.5)--(0,-0.5);
\end{scope}
%%%%%%%%%%%%%%%%%%%%%%%%%%%%%
%\begin{scope}[shift={(5.5,0)}]
%\draw[doublefermion] 	(-1,+0.5)--(0,+0.5);\draw[doublefermion] 	(0,+0.5)--(1,+0.5);
%\draw[doublefermionbar] (-1,-0.5)--(0,-0.5);\draw[doublefermionbar] (0,-0.5)--(1,-0.5);
%\draw[vector] (0,+0.5)--(0,0);\draw[gluon] (0,-0.5)--(0,0);
%\draw[vector] (0.65,0)--(0,0);
%\end{scope}
%%%%%%%%%%%%%%%%%%%%%%%%%%%%%
\begin{scope}[shift={(8.5,0)}]
\draw[doublefermion] 	(-1,+0.5)--(0,+0.5);\draw 	(0,+0.5)--(1,+0.5);
\draw[doublefermionbar] (-1,-0.5)--(0,-0.5);\draw (0,-0.5)--(1,-0.5);
\draw[scalar] (0,+0.5)--(0,-0.2);\draw[scalar] (0,-0.2)--(0,-0.5);
\draw[vector] (0.65,0)--(0,0);
%%%%%
\draw[->] (+0.4,+0.75) -- (+0.9,+0.75);\node at (+0.6,+1) {$P/2+p$};
\draw[->] (+0.4,-0.75) -- (+0.9,-0.75);\node at (+0.6,-1) {$P/2-p$};
\node at (+1.25,+0.45) {$r_1$};\node at (+1.25,-0.45) {$r_2$};
\end{scope}
%%%%%%%%%%%%%%%%%%%%%%%%%%%%%
\begin{scope}[shift={(11.3,0)}]
\draw[doublefermion] 	(-1,+0.5)--(0,+0.5);\draw 	(0,+0.5)--(1,+0.5);
\draw[doublefermionbar] (-1,-0.5)--(0,-0.5);\draw (0,-0.5)--(1,-0.5);
\draw[scalar] (0,+0.5)--(0,-0.2);\draw[scalar] (0,-0.2)--(0,-0.5);
\draw[vector] (0.65,0)--(0,0);
%%%%%
\draw[->] (+0.4,+0.75) -- (+0.9,+0.75);\node at (+0.6,+1) {$P/2-p$};
\draw[->] (+0.4,-0.75) -- (+0.9,-0.75);\node at (+0.6,-1) {$P/2+p$};
\node at (+1.25,+0.45) {$r_2$};\node at (+1.25,-0.45) {$r_1$};
\end{scope}
%%%%%%%%%%%%%%%%%%%%%%%%%%%%%
\end{scope}
%%%%%%%%%%%%%%%%%%%%%%%%%%%%%%%%%%%%%%%%%%%%%%%%%%%%%%%%%%%%%%%%%%%%%%%%%%%%%%%%%%%%%%%%%%%%%
%%%%%%%%%%%%%%%%%%%%%%%%%%%%%%%%%%%%%%%%%%%%%%%%%%%%%%%%%%%%%%%%%%%%%%%%%%%%%%%%%%%%%%%%%%%%%
%%%%%%%%%%%%%%%%%%%%%%%%%%%%%%%%%%%%%%%%%%%%%%%%%%%%%%%%%%%%%%%%%%%%%%%%%%%%%%%%%%%%%%%%%%%%%
\begin{scope}[shift={(0,-3.5)}]
%%%%%%%%%%%%%%%%%%%%%%%%%%%%%
\begin{scope}[shift={(0,0)}]
%%%%% field lines
\draw[doublefermion] 	(-1,+0.5)--(0,+0.5);\draw[doublefermion] 	(0,+0.5)--(1,+0.5);
\draw[doublefermionbar] (-1,-0.5)--(0,-0.5);\draw[doublefermionbar] (0,-0.5)--(1,-0.5);
\draw[gluon] (0.3,1.5)--(0,+0.5);%\node at (0.6,1.7) {$W$};
%%%%% spins
\node at (-1.25,+0.45) {$s_1$};\node at (+1.25,+0.45) {$r_1$};
\node at (-1.25,-0.45) {$s_2$};\node at (+1.25,-0.45) {$r_2$};
%%%%% momenta
\draw[->] (-0.9,+0.75) -- (-0.4,+0.75);\node at (-1.2,+1) {$K/2+k^{(\prime)}$};
\draw[->] (-0.9,-0.75) -- (-0.4,-0.75);\node at (-1.2,-1.1) {$K/2-k^{(\prime)}$};
\draw[->] (+0.4,+0.75) -- (+0.9,+0.75);\node at (+1.1,+1) {$P/2+p$};
\draw[->] (+0.4,-0.75) -- (+0.9,-0.75);\node at (+1.1,-1.1) {$P/2-p$};
\draw[->] (-0.1,1.1)--(0.05,1.55);\node at (-0.25,1.6) {$\PW$};
%%%%% SU2L
\node at (-0.7,+0.25) {$i$};\node at (+0.4,+0.25) {$i'$};
\node at (-0.7,-0.25) {$j$};\node at (+0.4,-0.25) {$j'$};
\node at (0.3,1.7) {$a$};
\end{scope}
%%%%%%%%%%%%%%%%%%%%%%%%%%%%%
\begin{scope}[shift={(3,0)}]
\draw[doublefermion] 	(-1,+0.5)--(0,+0.5);\draw[doublefermion] 	(0,+0.5)--(1,+0.5);
\draw[doublefermionbar] (-1,-0.5)--(0,-0.5);\draw[doublefermionbar] (0,-0.5)--(1,-0.5);
\draw[gluon] (0.3,-1.5)--(0,-0.5);
\end{scope}
%%%%%%%%%%%%%%%%%%%%%%%%%%%%%
\begin{scope}[shift={(5.5,0)}]
\draw[doublefermion] 	(-1,+0.5)--(0,+0.5);\draw[doublefermion] 	(0,+0.5)--(1,+0.5);
\draw[doublefermionbar] (-1,-0.5)--(0,-0.5);\draw[doublefermionbar] (0,-0.5)--(1,-0.5);
\draw[gluon] (0,+0.5)--(0,0);\draw[gluon] (0,-0.5)--(0,0);
\draw[gluon] (0.65,0)--(0,0);
\end{scope}
%%%%%%%%%%%%%%%%%%%%%%%%%%%%%
\begin{scope}[shift={(8.5,0)}]
\draw[doublefermion] 	(-1,+0.5)--(0,+0.5);\draw 	(0,+0.5)--(1,+0.5);
\draw[doublefermionbar] (-1,-0.5)--(0,-0.5);\draw (0,-0.5)--(1,-0.5);
\draw[scalar] (0,+0.5)--(0,-0.2);\draw[scalar] (0,-0.2)--(0,-0.5);
\draw[gluon] (0.65,0)--(0,0);
%%%%%
\draw[->] (+0.4,+0.75) -- (+0.9,+0.75);\node at (+0.6,+1) {$P/2+p$};
\draw[->] (+0.4,-0.75) -- (+0.9,-0.75);\node at (+0.6,-1) {$P/2-p$};
\node at (+1.25,+0.45) {$r_1$};\node at (+1.25,-0.45) {$r_2$};
\end{scope}
%%%%%%%%%%%%%%%%%%%%%%%%%%%%%
\begin{scope}[shift={(11.3,0)}]
\draw[doublefermion] 	(-1,+0.5)--(0,+0.5);\draw 	(0,+0.5)--(1,+0.5);
\draw[doublefermionbar] (-1,-0.5)--(0,-0.5);\draw (0,-0.5)--(1,-0.5);
\draw[scalar] (0,+0.5)--(0,-0.2);\draw[scalar] (0,-0.2)--(0,-0.5);
\draw[gluon] (0.65,0)--(0,0);
%%%%%
\draw[->] (+0.4,+0.75) -- (+0.9,+0.75);\node at (+0.6,+1) {$P/2-p$};
\draw[->] (+0.4,-0.75) -- (+0.9,-0.75);\node at (+0.6,-1) {$P/2+p$};
\node at (+1.25,+0.45) {$r_2$};\node at (+1.25,-0.45) {$r_1$};
\end{scope}
%%%%%%%%%%%%%%%%%%%%%%%%%%%%%
\end{scope}
%%%%%%%%%%%%%%%%%%%%%%%%%%%%%%%%%%%%%%%%%%%%%%%%%%%%%%%%%%%%%%%%%%%%%%%%%%%%%%%%%%%%%%%%%%%%%
%%%%%%%%%%%%%%%%%%%%%%%%%%%%%%%%%%%%%%%%%%%%%%%%%%%%%%%%%%%%%%%%%%%%%%%%%%%%%%%%%%%%%%%%%%%%%
%%%%%%%%%%%%%%%%%%%%%%%%%%%%%%%%%%%%%%%%%%%%%%%%%%%%%%%%%%%%%%%%%%%%%%%%%%%%%%%%%%%%%%%%%%%%%
\begin{scope}[shift={(0,-6.75)}]
%%%%%%%%%%%%%%%%%%%%%%%%%%%%%
\begin{scope}[shift={(0,0)}]
%%%%% field lines
\draw[doublefermion] 	(-1,+0.5)--(0,+0.5);\draw[doublefermion] 	(0,+0.5)--(1,+0.5);
\draw[fermionnoarrow]   (-1,-0.5)--(0,-0.5);\draw[doublefermionbar] (0,-0.5)--(1,-0.5);
\draw[scalar] (0,-0.5)--(0.3,-1.5);%\node at (0.6,1.7) {$W$};
%%%%% spins
\node at (-1.25,+0.45) {$s_1$};\node at (+1.25,+0.45) {$r_1$};
\node at (-1.25,-0.45) {$s_2$};\node at (+1.25,-0.45) {$r_2$};
%%%%% momenta
\draw[->] (-0.9,+0.75) -- (-0.4,+0.75);\node at (-1.2,+1) {$K/2+k^{(\prime)}$};
\draw[->] (-0.9,-0.75) -- (-0.4,-0.75);\node at (-1.2,-1.1) {$K/2-k^{(\prime)}$};
\draw[->] (+0.4,+0.75) -- (+0.9,+0.75);\node at (+1.1,+1) {$P/2+p$};
\draw[->] (+0.4,-0.75) -- (+0.9,-0.75);\node at (+1.1,-1.1) {$P/2-p$};
\draw[->] (-0.1,-1.1)--(0.05,-1.55);\node at (-0.25,-1.6) {$\PH$};
%%%%% SU2L
\node at (-0.7,+0.25) {$i$};\node at (+0.4,+0.25) {$i'$};
							\node at (+0.4,-0.25) {$j'$};
							\node at (+0.3,-1.7) {$h$};
\end{scope}
%%%%%%%%%%%%%%%%%%%%%%%%%%%%%%
%\begin{scope}[shift={(3,0)}]
%\draw[doublefermion] 	(-1,+0.5)--(0,+0.5);\draw[doublefermion] 	(0,+0.5)--(1,+0.5);
%\draw (-1,-0.5)--(0,-0.5);\draw[doublefermionbar] (0,-0.5)--(1,-0.5);
%\draw[gluon] (0,+0.5)--(0,0);\draw[scalar] (0,-0.5)--(0,0.12);
%\draw[scalar] (0,0)--(0.7,0);
%\end{scope}
%%%%%%%%%%%%%%%%%%%%%%%%%%%%%%
%\begin{scope}[shift={(5.5,0)}]
%\draw[doublefermion] 	(-1,+0.5)--(0,+0.5);\draw[doublefermion] 	(0,+0.5)--(1,+0.5);
%\draw (-1,-0.5)--(0,-0.5);\draw[doublefermionbar] (0,-0.5)--(1,-0.5);
%\draw[vector] (0,+0.5)--(0,0);\draw[scalar] (0,-0.5)--(0,0.12);
%\draw[scalar] (0,0)--(0.7,0);
%\end{scope}
%%%%%%%%%%%%%%%%%%%%%%%%%%%%%%
\begin{scope}[shift={(8.5,0)}]
%%%%% field lines
\draw[doublefermion] 	(-1,+0.5)--(0,+0.5);\draw 	(0,+0.5)--(1,+0.5);
\draw   				(-1,-0.5)--(1,-0.5);
\draw[scalar] (0,+0.5)--(0.3,+1.5);
%%%%% momenta
\draw[->] (+0.4,+0.75) -- (+0.9,+0.75);\node at (+0.9,+1) {$P/2+p$};
\draw[->] (+0.4,-0.75) -- (+0.9,-0.75);\node at (+0.9,-1.1) {$P/2-p$};
\node at (+1.25,+0.45) {$r_1$};\node at (+1.25,-0.45) {$r_2$};
\end{scope}
%%%%%%%%%%%%%%%%%%%%%%%%%%%%%
\begin{scope}[shift={(11.3,0)}]
%%%%% field lines
\draw[doublefermion] 	(-1,+0.5)--(0,+0.5);\draw 	(0,+0.5)--(1,+0.5);
\draw   				(-1,-0.5)--(1,-0.5);
\draw[scalar] (0,+0.5)--(0.3,+1.5);
%%%%% momenta
\draw[->] (+0.4,+0.75) -- (+0.9,+0.75);\node at (+0.9,+1.0) {$P/2-p$};
\draw[->] (+0.4,-0.75) -- (+0.9,-0.75);\node at (+0.9,-1.1) {$P/2+p$};
\node at (+1.25,+0.45) {$r_2$};\node at (+1.25,-0.45) {$r_1$};
\end{scope}
%%%%%%%%%%%%%%%%%%%%%%%%%%%%%
\end{scope}
%%%%%%%%%%%%%%%%%%%%%%%%%%%%%%%%%%%%%%%%%%%%%%%%%%%%%%%%%%%%%%%%%%%%%%%%%%%%%%%%%%%%%%%%%%%%%
%%%%%%%%%%%%%%%%%%%%%%%%%%%%%%%%%%%%%%%%%%%%%%%%%%%%%%%%%%%%%%%%%%%%%%%%%%%%%%%%%%%%%%%%%%%%%
%%%%%%%%%%%%%%%%%%%%%%%%%%%%%%%%%%%%%%%%%%%%%%%%%%%%%%%%%%%%%%%%%%%%%%%%%%%%%%%%%%%%%%%%%%%%%
\end{tikzpicture}
\vspace{-18pt}

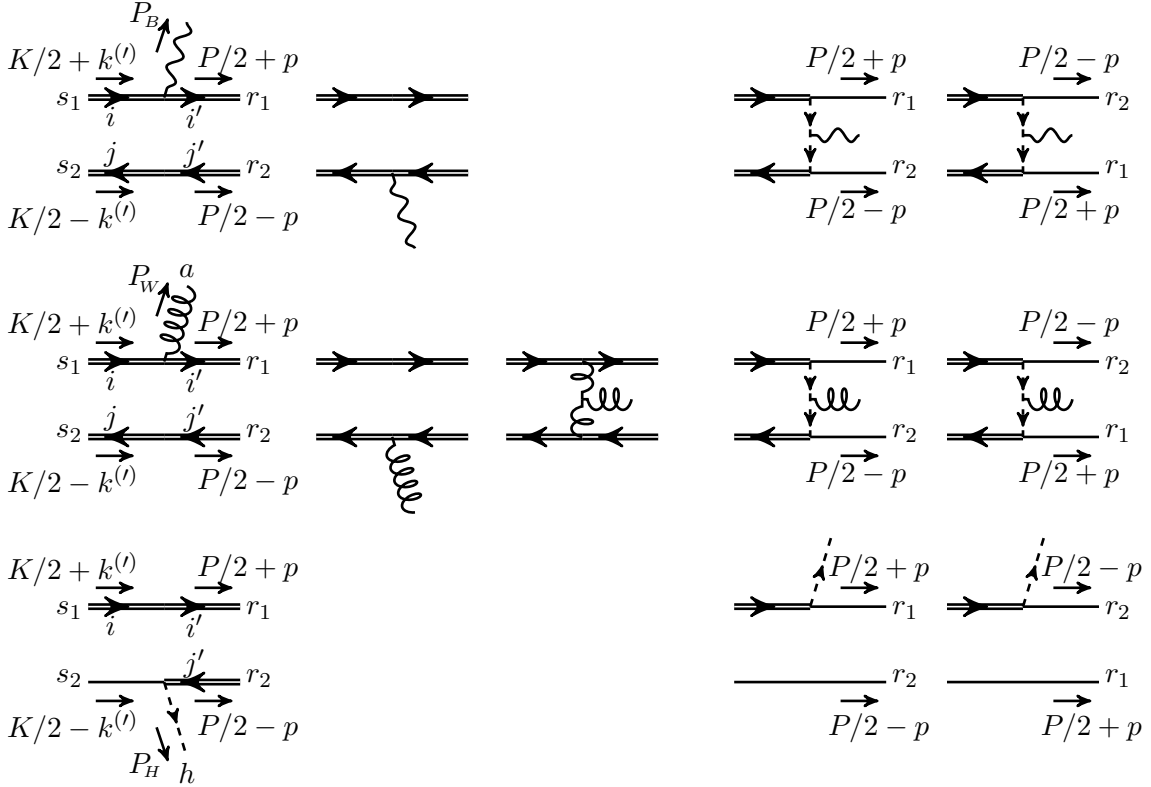
\captionof{figure}{\label{fig:BSF_SSDDbar}
The radiative parts of the diagrams contributing to the formation of $SS/D\bar{D}$ bound states. 
\emph{Top row:} $D\bar{D} \to {\cal B}(SS/D\bar{D}) + B$. 
\emph{Middle row:} $D\bar{D} \to {\cal B}(SS/D\bar{D}) + W$. 
\emph{Bottom row:} $DS \to {\cal B}(SS/D\bar{D}) + H$, which has also a conjugate counterpart (not shown.) 
Single and double lines correspond to $S$ and $D$ fermions, while vector, gluon and dashed lines to $B$, $W$ and $H$ bosons. The arrows on the field lines denote the flow of Hypercharge. Wherever not shown, the momenta, spins and gauge indices of the external particles can be deduced from the other graphs.
}
\end{table}
%%%%%%%%%%%%%%%%%%%%%%%%%%%%%%%%%%%%%%%%%%%%%%%%%%%%%%%%%%%%%%%%%%%%

The BSF processes are listed in \cref{tab:BSF_SSDDbar}, and the radiative part of the diagrams contributing to these processes are shown in \cref{fig:BSF_SSDDbar}. We project the bound-state fields on the spin-0 state via 
$[U_{\text{spin-0}}^{-1}]^{r_2 r_1} = \epsilon^{r_1r_2}/\sqrt{2}$,  
and the $D\bar{D}$ component on the $\SUL$ singlet via $\delta_{i'j'}/\sqrt{2}$.

\subsubsection{$D\bar{D} \to {\cal B} (SS/D\bar{D}) + B$}
The perturbative parts of the amplitude are
\begin{subequations}
\label{eq:BSF_A_DDbarToSSDDbar_Bemission}
\label[pluralequation]{eqs:BSF_A_DDbarToSSDDbar_Bemission}
\begin{align}
\im \boldsymbol{{\cal A}}_{ij}^{s_1s_2}
&\[D\bar{D} \to (SS)_{(\mathbb{1},0)}^{\text{spin-0}} + B\]
\simeq 
\nn \\
&\simeq \im 
\, \delta_{ij} 
\, \frac{\epsilon^{s_1 s_2}}{\sqrt{2}} 
\, y^2  \, g_1 \YH  \, 4 m^2 
\[\frac{2(\bf k' - p)}{[({\bf k'} - {\bf p})^2 +\mH^2]^2} 
+ \frac{2(\bf k' + p)}{[({\bf k'} + {\bf p})^2 +\mH^2]^2} \] ,
\label{eq:BSF_A_DDbarToSSSpin0_Bemission} 
\\
\im \boldsymbol{{\cal A}}_{ij}^{s_1s_2}
&\[D\bar{D}\to (D\bar{D})_{(\mathbb{1},0)}^{\text{spin-0}} +  B \]
\simeq 
\nn \\
&\simeq \im
\, \frac{\delta_{ij}}{\sqrt{2}} 
\, \frac{\epsilon^{s_1 s_2}}{\sqrt{2}} 
\, g_1 \YD \, 2m  \, 2{\bf p} 
\, (2\pi)^3
\[\delta^3 ({\bf k'} - {\bf p} - \PBvec/2) 
+ \delta^3 ({\bf k'} - {\bf p} + \PBvec/2)\] ,
\label{eq:BSF_A_DDbarToDDbarSingletSpin0_Bemission}
\end{align}
\end{subequations}
In \cref{eq:BSF_A_DDbarToSSSpin0_Bemission}, the fermion permutations introduced factors $(-1)$ and $(+1)$  for the $t$- and $u$-channel diagrams. The projection on the antisymmetric spin-0 eigenstate alloted another factor $(-1)$ to the $u$-channel. 
% diagram with respect to its $t$-channel counterpart. 
The full amplitude \eqref{eq:BSF_FullAmplitude_Def} is
\begin{align}
\im \boldsymbol{{\cal M}}_{ij}^{s_1s_2} 
&= \dfrac{1}{\sqrt{2\mu}}
\int \frac{d^3 {\bf k'}}{(2\pi)^3} \frac{d^3 {\bf p}}{(2\pi)^3}
\ \tilde{\varphi}_{\bf k}^{} \({\bf k'};\dfrac{\alpha_1+3\alpha_2}{4}\) 
\nn \\
&\times \[
\im \boldsymbol{{\cal A}}_{ij}^{s_1s_2}
\[D\bar{D} \to (SS)_{(\mathbb{1},0)}^{\text{spin-0}} + B\]
\sqrt{\dfrac{\aR}{\aA+\aR}}
\ \tilde{\varphi}_{100}^\dagger ({\bf p}; \aA)
\right. 
\nn \\
&\left. +~
\im \boldsymbol{{\cal A}}_{ij}^{s_1s_2}
\[D\bar{D} \to (D\bar{D})_{(\mathbb{1},0)}^{\text{spin-0}} + B\]
\sqrt{\dfrac{\aA}{\aA+\aR}}
\ \tilde{\varphi}_{100}^\dagger ({\bf p}; \aA)
\],
\end{align}
where only the $\ellS=1$ component of the scattering state wavefunction is meant to be kept, and here $\aA$ and $\aR$ should be evaluated from \cref{eqs:GaugeSinglet_alphas_def} for $\ell=s=0$. This becomes
\begin{align}
\im \boldsymbol{{\cal M}}_{ij}^{s_1s_2}
&\simeq \im \delta_{ij} \epsilon^{s_1 s_2} \( \frac{4\pi\alpha_1}{\aA+\aR} \ 4m \)^{1/2}
\times
\nn \\
&\times \left\{
\sqrt{8\aR} \, \boldsymbol{{\cal Y}}_{{\bf k},100}^\H \(\frac{\alpha_1+3\alpha_2}{4}, \aA\)
+\sqrt{\aA} 
\, \boldsymbol{{\cal J}}_{{\bf k},100} \(\frac{\alpha_1+3\alpha_2}{4}, \aA\) 
\right\}.
\label{eq:BSF_M_DDbarToSSDDbar_Bemission}
\end{align}
Note that we have neglected the $\pm \PBvec/2$ terms inside the $\delta$-functions of \cref{eq:BSF_A_DDbarToDDbarSingletSpin0_Bemission} that give rise to higher order corrections~\cite{Petraki:2015hla,Petraki:2016cnz}. Squaring and summing over the initial-state gauge indices and spins selects the $(\mathbb{1},0)$ spin-0 $D\bar{D}$ scattering state, which has one dof. Using the overlap integrals \eqref{eq:OverlapIntegrals_100}, we find
\begin{align}
\sum_{s_1,s_2}\sum_{i,j}
\left|
\boldsymbol{{\cal M}}_{ij}^{s_1s_2}
\right|^2
\simeq 2^{13} \pi^2 \(\frac{\alpha_1}{\aA+\aR}\)
\(1+\frac{\aH}{\aA}\sqrt{\frac{8\aR}{\aA}}\)^2 
\(\frac{\zetaA^2}{1+\zetaA^2}\)
\Svec \(\frac{\zeta_1 + 3\zeta_2}{4}, \zetaA \).
\label{eq:BSF_Msquared_DDbarToSSDDbar_Bemission}
\end{align}
The cross-section is obtained from \cref{eq:BSF_sigma_Vector} setting $\aB \to \aA$, and is shown in \cref{tab:BSF_SSDDbar}.

\subsubsection{$D\bar{D} \to {\cal B} (SS/D\bar{D}) + W$}
The perturbative parts of the amplitude are
\begin{subequations}
\label{eq:BSF_A_DDbarToSSDDbar_Wemission}
\begin{align}
\im (\boldsymbol{{\cal A}}^{s_1s_2})_{ij}^a 
\[D\bar{D}\to (SS)_{(\mathbb{1},0)}^{\text{spin-0}} + W\]
&\simeq \im 
\, t^a_{ji} 
\, \frac{\epsilon^{s_1 s_2}}{\sqrt{2}} 
\, y^2 \, g_2 \, 4 m^2 \times
\nn \\
&\times 
\[\frac{2(\bf k' - p)}{[({\bf k'} - {\bf p})^2 +\mH^2]^2} 
+ \frac{2(\bf k' + p)}{[({\bf k'} + {\bf p})^2 +\mH^2]^2} \] ,
\label{eq:BSF_A_DDbarToSS_Wemission}
\\
\im (\boldsymbol{{\cal A}}^{s_1s_2})_{ij}^a 
\[D\bar{D}\to (D\bar{D})_{(\mathbb{1},0)}^{\text{spin-0}} +  W \]
&\simeq \im
\, \frac{t^a_{ji}}{\sqrt{2}} 
\, \frac{\epsilon^{s_1 s_2}}{\sqrt{2}} 
\, g_2 \, 2m \times 
\left\{
2m \ 4\pi\alpha_2 \ \frac{2({\bf k'-p})}{({\bf k'-p})^4}
\right.
\label{eq:BSF_A_DDbarToDDbar_Wemission}
\\
&\left.
+2{\bf p} \, (2\pi)^3 
\[\delta^3 ({\bf k'} - {\bf p} - \PWvec/2) 
\!+\! \delta^3 ({\bf k'} - {\bf p} + \PWvec/2) \] 
\right\},
\nn 
\end{align}
\end{subequations}
where the signs of the $t$- and $u$-channel diagrams in \cref{eq:BSF_A_DDbarToSS_Wemission} are as in $D\bar{D} \to SS + B$ above, and the first factor 2 in the first term of \cref{eq:BSF_A_DDbarToDDbar_Wemission} is the quadratic Casimir of $\SUL$ (see ref.~\cite{Harz:2018csl} for details of this computation.) The full amplitude \eqref{eq:BSF_FullAmplitude_Def} is
\begin{align}
\im (\boldsymbol{{\cal M}}^{s_1s_2})_{ij}^a 
&\simeq \im (t^a)_{ij} \epsilon^{s_1 s_2} \( \frac{4\pi\alpha_2}{\aA+\aR} \ 4m \)^{1/2}
\times \left\{
\sqrt{8\aR} \, \boldsymbol{{\cal Y}}_{{\bf k},100}^\H \(\frac{\alpha_1-\alpha_2}{4}, \aA\)
\right. \nn \\ 
&\left.
+\sqrt{\aA} \[
 \boldsymbol{{\cal J}}_{{\bf k},100}    \(\frac{\alpha_1-\alpha_2}{4}, \aA\) 
+\boldsymbol{{\cal Y}}_{{\bf k},100}^\W \(\frac{\alpha_1-\alpha_2}{4}, \aA\) 
\] \right\},
\label{eq:BSF_M_DDbarToSSDDbar_Wemission}
\end{align}
where again $\aA$ and $\aR$ should be evaluated from \cref{eqs:GaugeSinglet_alphas_def} for $\ell=s=0$, and we have neglected the $\pm \PWvec/2$ terms inside the $\delta$-functions in \cref{eq:BSF_A_DDbarToDDbar_Wemission}. 
Squaring, summing over the initial and final state gauge indices and spins selects the $(\mathbb{3},0)$ spin-0 $D\bar{D}$ state, which has three dof. Using \cref{eq:OverlapIntegrals_100,eq:SBSF}, we find 
\begin{align}
\frac{1}{3} \sum_{s_1, s_2} \sum_{i,j,a}
|(\boldsymbol{{\cal M}}^{s_1s_2})_{ij}^a |^2
&\simeq 2^{11} \pi^2 \(\frac{\alpha_2}{\aA+\aR}\)
\(1+\frac{\alpha_2}{\aA}+\frac{\aH}{\aA}\sqrt{\frac{8\aR}{\aA}}\)^2 
\nn \\ 
&\times
\(\frac{\zetaA^2}{1+\zetaA^2}\) 
\Svec \(\frac{\zeta_1 - \zeta_2}{4}, \zetaA \).
\label{eq:BSF_Msquared_DDbarToSSDDbar_Wemission}
\end{align}
The cross-section is obtained from \cref{eq:BSF_sigma_Vector} setting $\aB \to \aA$, and is shown in \cref{tab:BSF_SSDDbar}.

\subsubsection{$DS \to {\cal B} (SS/D\bar{D}) + H$}
The perturbative parts of the amplitude are
\begin{subequations}
\label{eq:BSF_A_SDToSSDDbar_Hemission}
\label[pluralequation]{eqs:BSF_A_SDToSSDDbar_Hemission}
\begin{align}
\im {\cal A}^{s_1s_2}_{i,h} 
\[DS \to (SS)_{(\mathbb{1},0)}^{\text{spin-0}} + H\]
&\simeq-\im
\, \delta_{ih}
\, \frac{\epsilon^{s_1 s_2}}{\sqrt{2}} 
\, y \, 4m^2 \times 
\label{eq:BSF_A_SDToSS_Hemission}
\\
&\times (2\pi)^3 
\[\delta^3 ({\bf k'} - {\bf p} - \PHvec/2)  
+ \delta^3 ({\bf k'} - {\bf p} + \PHvec/2)\] , 
\nn 
\\
\im {\cal A}^{s_1s_2}_{i,h} 
\[ DS \to (D\bar{D})_{(\mathbb{1},0)}^{\text{spin-0}} + H \]
&\simeq-\im
\, \frac{\delta_{ih}}{\sqrt{2}}
\, \frac{\epsilon^{s_1 s_2}}{\sqrt{2}} 
\, y \, 4m^2 
\, (2\pi)^3 \delta^3 ({\bf k'} - {\bf p} + \PHvec/2) .
\label{eq:BSF_A_SDToDDbar_Hemission}
\end{align}
\end{subequations}
where now the fermion permutations introduced factors $(+1)$ and $(-1)$ for the $t$- and $u$-channel $DS \to SS + H$ diagrams respectively, and $(-1)$ for the $DS \to D\bar{D} + H$ diagram. We note that there are two diagrams where an off-shell vector boson ($B$ or $W$) emitted from one leg and an off-shell Higgs emitted from the other leg fuse to produce the final-state Higgs. In \cref{App:VectorScalarFusion}, we show that these diagrams are of higher order, thus we do not consider them here. The full amplitude \eqref{eq:BSF_FullAmplitude_Def} is
\begin{align}
\im {\cal M}^{s_1s_2}_{i,h} 
\simeq-\im \, \delta_{ih} \, \epsilon^{s_1 s_2}
\, \sqrt{\frac{2^{7} \pi \aH}{\aA^{3}}} 
\, \frac{\sqrt{\aA}+\sqrt{8 \aR}}{\sqrt{\aA+\aR}}
\, {\cal R}_{{\bf k},100} (\aH,\aA) , 
\label{eq:BSF_M_SDToSSDDbar_Hemission}
\end{align}
where we have neglected the $\pm \PHvec/2$ terms inside the $\delta$-functions in \cref{eqs:BSF_A_SDToSSDDbar_Hemission}. 
Squaring and summing over the initial and final state gauge indices and spins selects the $(\mathbb{2},1/2)$ spin-0 $DS$ state, which has two dof. Using the overlap integrals \eqref{eq:OverlapIntegrals_100}, we find 
\begin{align}
\frac{1}{2} \sum_{s_1,s_2} \sum_{i,h}
|{\cal M}^{s_1s_2}_{i,h} |^2
&\simeq 2^{14} \pi^2 
\ \frac{\aH}{\aA^3}
\ \frac{\(\sqrt{\aA}+\sqrt{8\aR}\)^2}{\aA+\aR} 
\(1-\frac{\zetaH}{\zetaA}\)^2
\(\frac{\zetaA^2}{1+\zetaA^2}\) 
\Ssc \(\zetaH, \zetaA \).
\label{eq:BSF_Msquared_SDToSSDDbar_Hemission}
\end{align}
The cross-section is obtained from \cref{eq:BSF_sigma_Scalar} setting $\aB \to \aA$, and is shown in \cref{tab:BSF_SSDDbar}.

\subsubsection{Bound-to-bound transition ${\cal B} (DS) \to {\cal B} (SS/D\bar{D}) + H$}

Using the perturbative amplitudes~\eqref{eq:BSF_A_SDToSSDDbar_Hemission}, we may now compute the rate of bound-to-bound transition \eqref{eq:Transition_DStoSSDDbar_Hemission}. Projecting on the spin-0 $DS$ state, and taking into account the bound-state wavefuntions of \cref{tab:BoundStates}, we find that the full amplitude is, analogously to \cref{eq:BSF_M_SDToSSDDbar_Hemission}, given by
\begin{align}
\im {\cal M}_{i,h} 
\simeq-\im \, \delta_{ih} \, m
\, \sqrt{2^{5} \pi \aH}
\, \frac{\sqrt{\aA}+\sqrt{8 \aR}}{\sqrt{\aA+\aR}}
\, {\cal R}_{100,100} (\aH,\aA) , 
\label{eq:BoundToBound_M_SDToSSDDbar_Hemission}
\end{align}
Squaring and averaging over the initial and final state gauge indices, and using the overlap integral \eqref{eq:OverlapIntegral_R_100to100}, we find 
\begin{align}
\frac{1}{2} \sum_{i,h}
|{\cal M}_{i,h} |^2
&\simeq 2^{11} \pi m^2
\ \aH
\ \frac{\(\sqrt{\aA}+\sqrt{8\aR}\)^2}{\aA+\aR} 
\ \dfrac{(\aH\aA)^3}{(\aH+\aA)^6} .
\label{eq:BoundToBound_Msquared_SDToSSDDbar_Hemission}
\end{align}
The transition rate is found from \cref{eq:BoundToBound_Gamma_Scalar} to be
\begin{align}
\Gamma_{\DS \to \SSDDbar} = 2^4 m  
\, \aH
\, (\aA^2 - \aH^2)
\, \frac{\(\sqrt{\aA}+\sqrt{8\aR}\)^2}{\aA+\aR} 
\, \dfrac{(\aA\aH)^3}{(\aA+\aH)^6} 
\[1- \dfrac{16\mH^2}{m^2 (\aA^2-\aH^2)^2}\]^{1/2} .
\label{eq:BoundToBound_SDToSSDDbar_Hemission_Rate}
\end{align}

\clearpage
\subsection{$D\bar{D}$ bound states: $(\mathbb{1},0)$, spin 1, $n\ell m =\{100\}$ \label{sec:BSF_DDbar}}
%%%%%%%%%%%%%%%%%%%%%%%%%%%%%%%%%%%%%%%%%%%%%%%%%%%%%%%%%%%%%%%%%%%%
\begin{table}[p]
\centering
\begin{align*}
\begin{array}{|c|c|c|c|c|c|rl|}
%%%%%%%%%%%%%%%%%%%%%%%%%%%%%%%%%%%%%%%%%%%%%%%%%%%%%%%%%%%%%%%%%%%%%%%%%%%%%%%%%%%%%%%%%%%%%%%%%%%%%%%%
\multicolumn{8}{c}{\boldsymbol{
\textbf{Bound state}~D\bar{D}:
~~~(\mathbb{1},0),
~~\textbf{spin 1}, 
~~\{n\ell m\} = \{100\}
}}
\\ \hline
%%%%%%%%%%%%%%%%%%%%%%%%%%%%%%%%%%%%%%%%%%%%%%%%%%%%%%%%%%%%%%%%%%%%%%%%%%%%%%%%%%%%%%%%%%%%%%%%%%%%%%%%
\multicolumn{5}{|c|}{\text{Scattering state (spin 1)}}
&\multirow{2}{*}{\text{\parbox{6ex}{\centering Rad boson}}}
&\multicolumn{2}{c|}{\text{Cross-section}}
\\ \cline{1-5}
\text{State}		
& U_{\mathsmaller{Y}}(1) 
&SU_{\mathsmaller{L}}(2) 
& \text{dof} 
& \ellS 
&
&\multicolumn{2}{c|}{(\sigma_\BSF \vrel) / (\pi m^{-2})}
\\ \hline\hline
%%%%%%%%%%%%%%%%%%%%%%%%%%%%%%%%%%%%%%%%%%%%%%%%%%%%%%%%%%%%%%%%%%%%%%%%%%%%%%%%%%%%%%%%%%%%%%%%%%%%%%
\multirow{2}{*}{$D\bar{D}$-like}
& \multirow{2}{*}{0}   
& \multirow{2}{*}{$\mathbb{1}$} 
& \multirow{2}{*}{3}  
& \multirow{2}{*}{1} 
& \multirow{2}{*}{$B$} 			
&\dfrac{2^7 \alpha_1 \aB}{3}  \dfrac{\aA^2}{(\aA+\aR)^2}
& \[\(1-\dfrac{2\aH}{\aB} \sqrt{\dfrac{\aR}{\aA}}\) \Svec^{1/2} \(\zetaA,\zetaB \) \right.
\\[1em]
&
&
&
&
&			
& \multicolumn{2}{r|}{\left.
+ \(\dfrac{\aR}{\aA}+\dfrac{2\aH}{\aB} \sqrt{\dfrac{\aR}{\aA}}\) \Svec^{1/2} \(-\zetaR,\zetaB \)
\]^2}
\\[2ex] \hline
%%%%%%%%%%%%%%%%%%%%%%%%%%%%%%%%%%%%%%%%%%%%%%%%%%%%%%%%%%%%%%%%%%%%%%%%%%%%%%%%%%%%%%%%%%%%%%%%%%%%%%
\multirow{2}{*}{$SS$-like} 		
& \multirow{2}{*}{0}   
& \multirow{2}{*}{$\mathbb{1}$} 
& \multirow{2}{*}{3}  
& \multirow{2}{*}{1} 
& \multirow{2}{*}{$B$} 			
&\dfrac{2^8 \alpha_1 \aB}{3}  \dfrac{\aA\aR}{(\aA+\aR)^2}
& \[\(1-\dfrac{\aH}{\aB} \sqrt{\dfrac{8\aR}{\aA}}\) \!\Svec^{1/2} \(\zetaA,\zetaB \) \right.
\\[1em]
&
&
&
&
&			
& \multicolumn{2}{r|}{\left.
- \(1+\dfrac{\aH}{\aB} \sqrt{\dfrac{8\aA}{\aR}}\) \Svec^{1/2} \(-\zetaR,\zetaB \)
\]^2}
\\[2ex] \hline
%%%%%%%%%%%%%%%%%%%%%%%%%%%%%%%%%%%%%%%%%%%%%%%%%%%%%%%%%%%%%%%%%%%%%%%%%%%%%%%%%%%%%%%%%%%%%%%%%%%%%%
D\bar{D} 			
& 0   
& \mathbb{3} 
& 9 
& 1 
& W 			
& \dfrac{2^7 \alpha_2\aB}{3} \(1+\dfrac{\alpha_2}{\aB}\)^2
& \Svec \(\dfrac{\zeta_1-\zeta_2}{4},\zetaB\)
\\ \hline
%%%%%%%%%%%%%%%%%%%%%%%%%%%%%%%%%%%%%%%%%%%%%%%%%%%%%%%%%%%%%%%%%%%%%%%%%%%%%%%%%%%%%%%%%%%%%%%%%%%%%%%%
DS 					
&+1/2 
& \mathbb{2} 
& 6  
& 0 
& H			
& \dfrac{2^{7} \aH}{\aB} \(1+\dfrac{\aH}{\aB}\)^2
& \Ssc \(-\zetaH,\zetaB\) \hH(\omega)
\\ \hline
%%%%%%%%%%%%%%%%%%%%%%%%%%%%%%%%%%%%%%%%%%%%%%%%%%%%%%%%%%%%%%%%%%%%%%%%%%%%%%%%%%%%%%%%%%%%%%%%%%%%%%%%
\bar{D}S 			
&-1/2 
& \mathbb{2} 
& 6  
& 0 
& H^\dagger 	
&\multicolumn{2}{c|}{\text{same as above}}
\\ \hline
%%%%%%%%%%%%%%%%%%%%%%%%%%%%%%%%%%%%%%%%%%%%%%%%%%%%%%%%%%%%%%%%%%%%%%%%%%%%%%%%%%%%%%%%%%%%%%%%%%%%%%%%
\end{array}
\end{align*}
\captionof{table}{\label{tab:BSF_DDbar}	
Same as \cref{tab:BSF_SSDDbar} for the $D\bar{D}$ bound states. 
Here, the bound-state coupling is $\aB = (\alpha_1+3\alpha_2)/4$, and correspondingly 
$\zetaB = (\zeta_1+3\zeta_2)/4$. 
In the first two processes, $\aA$ and $\aR$ should be evaluated from  \cref{eq:GaugeSinglet_alphas_def} with $\ellS=s=1$ for the scattering state.
For the phase-space suppression $\hH$, here $\omega = m\[(\alpha_1+3\alpha_2)^2/16+\vrel^2\]/4$.}
%%%%%%%%%%%%%%%%%%%%%%%%%%%%%%%%%%%%%%%%%%%%%%%%%%%%%%%%%%%%%%%%%%%%
\medskip
%%%%%%%%%%%%%%%%%%%%%%%%%%%%%%%%%%%%%%%%%%%%%%%%%%%%%%%%%%%%%%%%%%%%
\begin{tikzpicture}[line width=1pt, scale=1]
%%%%%%%%%%%%%%%%%%%%%%%%%%%%%%%%%%%%%%%%%%%%%%%%%%%%%%%%%%%%%%%%%%%%%%%%%%%%%%%%%%%%%%%%%%%%%
%%%%%%%%%%%%%%%%%%%%%%%%%%%%%%%%%%%%%%%%%%%%%%%%%%%%%%%%%%%%%%%%%%%%%%%%%%%%%%%%%%%%%%%%%%%%%
%%%%%%%%%%%%%%%%%%%%%%%%%%%%%%%%%%%%%%%%%%%%%%%%%%%%%%%%%%%%%%%%%%%%%%%%%%%%%%%%%%%%%%%%%%%%%
\begin{scope}[shift={(0,0)}]
%%%%%%%%%%%%%%%%%%%%%%%%%%%%%
\begin{scope}[shift={(0,0)}]
%%%%% field lines
\draw[doublefermion] 	(-1,+0.5)--(0,+0.5);\draw[doublefermion] 	(0,+0.5)--(1,+0.5);
\draw[doublefermionbar] (-1,-0.5)--(0,-0.5);\draw[doublefermionbar] (0,-0.5)--(1,-0.5);
\draw[vector] (0.3,1.5)--(0,+0.5);%\node at (0.6,1.7) {$W$};
%%%%% spins
\node at (-1.25,+0.45) {$s_1$};\node at (+1.25,+0.45) {$r_1$};
\node at (-1.25,-0.45) {$s_2$};\node at (+1.25,-0.45) {$r_2$};
%%%%% momenta
\draw[->] (-0.9,+0.75) -- (-0.4,+0.75);\node at (-1.2,+1) {$K/2+k^{(\prime)}$};
\draw[->] (-0.9,-0.75) -- (-0.4,-0.75);\node at (-1.2,-1.1) {$K/2-k^{(\prime)}$};
\draw[->] (+0.4,+0.75) -- (+0.9,+0.75);\node at (+1.1,+1) {$P/2+p$};
\draw[->] (+0.4,-0.75) -- (+0.9,-0.75);\node at (+1.1,-1.1) {$P/2-p$};
\draw[->] (-0.1,1.1)--(0.05,1.55);\node at (-0.25,1.6) {$\PB$};
%%%%% SU2L
\node at (-0.7,+0.25) {$i$};\node at (+0.4,+0.25) {$i'$};
\node at (-0.7,-0.25) {$j$};\node at (+0.4,-0.25) {$j'$};
\end{scope}
%%%%%%%%%%%%%%%%%%%%%%%%%%%%%
\begin{scope}[shift={(3,0)}]
\draw[doublefermion] 	(-1,+0.5)--(0,+0.5);\draw[doublefermion] 	(0,+0.5)--(1,+0.5);
\draw[doublefermionbar] (-1,-0.5)--(0,-0.5);\draw[doublefermionbar] (0,-0.5)--(1,-0.5);
\draw[vector] (0.3,-1.5)--(0,-0.5);
\end{scope}
%%%%%%%%%%%%%%%%%%%%%%%%%%%%%
%\begin{scope}[shift={(5.5,0)}]
%\draw[doublefermion] 	(-1,+0.5)--(0,+0.5);\draw[doublefermion] 	(0,+0.5)--(1,+0.5);
%\draw[doublefermionbar] (-1,-0.5)--(0,-0.5);\draw[doublefermionbar] (0,-0.5)--(1,-0.5);
%\draw[vector] (0,+0.5)--(0,0);\draw[gluon] (0,-0.5)--(0,0);
%\draw[vector] (0.65,0)--(0,0);
%\end{scope}
%%%%%%%%%%%%%%%%%%%%%%%%%%%%%
\begin{scope}[shift={(8.5,0)}]
\draw (-1,+0.5)--(0,+0.5);\draw[doublefermion] 		(0,+0.5)--(1,+0.5);
\draw (-1,-0.5)--(0,-0.5);\draw[doublefermionbar] 	(0,-0.5)--(1,-0.5);
\draw[scalar] (0,-0.5)--(0,0.2);\draw[scalar] (0,0)--(0,+0.5);
\draw[vector] (0.65,0)--(0,0);
%%%%%
\draw[->] (-0.9,+0.75) -- (-0.4,+0.75);\node at (-0.75,+1) {$K/2+k^{(\prime)}$};
\draw[->] (-0.9,-0.75) -- (-0.4,-0.75);\node at (-0.75,-1) {$K/2-k^{(\prime)}$};
\node at (-1.25,+0.45) {$s_1$};\node at (-1.25,-0.45) {$s_2$};
%%%%% SU2L
\node at (+0.4,+0.8) {$i'$};
\node at (+0.4,-0.8) {$j'$};
\end{scope}
%%%%%%%%%%%%%%%%%%%%%%%%%%%%%
\begin{scope}[shift={(11.3,0)}]
\draw (-1,+0.5)--(0,+0.5);\draw[doublefermion] 		(0,+0.5)--(1,+0.5);
\draw (-1,-0.5)--(0,-0.5);\draw[doublefermionbar] 	(0,-0.5)--(1,-0.5);
\draw[scalar] (0,-0.5)--(0,0.2);\draw[scalar] (0,0)--(0,+0.5);
\draw[vector] (0.65,0)--(0,0);
%%%%%
\draw[->] (-0.9,+0.75) -- (-0.4,+0.75);\node at (-0.6,+1) {$K/2-k^{(\prime)}$};
\draw[->] (-0.9,-0.75) -- (-0.4,-0.75);\node at (-0.6,-1) {$K/2+k^{(\prime)}$};
\node at (-1.25,+0.45) {$s_2$};\node at (-1.25,-0.45) {$s_1$};
\end{scope}
%%%%%%%%%%%%%%%%%%%%%%%%%%%%%
\end{scope}
%%%%%%%%%%%%%%%%%%%%%%%%%%%%%%%%%%%%%%%%%%%%%%%%%%%%%%%%%%%%%%%%%%%%%%%%%%%%%%%%%%%%%%%%%%%%%
%%%%%%%%%%%%%%%%%%%%%%%%%%%%%%%%%%%%%%%%%%%%%%%%%%%%%%%%%%%%%%%%%%%%%%%%%%%%%%%%%%%%%%%%%%%%%
%%%%%%%%%%%%%%%%%%%%%%%%%%%%%%%%%%%%%%%%%%%%%%%%%%%%%%%%%%%%%%%%%%%%%%%%%%%%%%%%%%%%%%%%%%%%%
\begin{scope}[shift={(0,-3.75)}]
%%%%%%%%%%%%%%%%%%%%%%%%%%%%%
\begin{scope}[shift={(0,0)}]
%%%%% field lines
\draw[doublefermion] 	(-1,+0.5)--(0,+0.5);\draw[doublefermion] 	(0,+0.5)--(1,+0.5);
\draw[doublefermionbar] (-1,-0.5)--(0,-0.5);\draw[doublefermionbar] (0,-0.5)--(1,-0.5);
\draw[gluon] (0.3,1.5)--(0,+0.5);%\node at (0.6,1.7) {$W$};
%%%%% spins
\node at (-1.25,+0.45) {$s_1$};\node at (+1.25,+0.45) {$r_1$};
\node at (-1.25,-0.45) {$s_2$};\node at (+1.25,-0.45) {$r_2$};
%%%%% momenta
\draw[->] (-0.9,+0.75) -- (-0.4,+0.75);\node at (-1.2,+1) {$K/2+k^{(\prime)}$};
\draw[->] (-0.9,-0.75) -- (-0.4,-0.75);\node at (-1.2,-1.1) {$K/2-k^{(\prime)}$};
\draw[->] (+0.4,+0.75) -- (+0.9,+0.75);\node at (+1.1,+1) {$P/2+p$};
\draw[->] (+0.4,-0.75) -- (+0.9,-0.75);\node at (+1.1,-1.1) {$P/2-p$};
\draw[->] (-0.1,1.1)--(0.05,1.55);\node at (-0.25,1.6) {$\PW$};
%%%%% SU2L
\node at (-0.7,+0.25) {$i$};\node at (+0.4,+0.25) {$i'$};
\node at (-0.7,-0.25) {$j$};\node at (+0.4,-0.25) {$j'$};
\node at (0.3,1.7) {$a$};
\end{scope}
%%%%%%%%%%%%%%%%%%%%%%%%%%%%%
\begin{scope}[shift={(3,0)}]
\draw[doublefermion] 	(-1,+0.5)--(0,+0.5);\draw[doublefermion] 	(0,+0.5)--(1,+0.5);
\draw[doublefermionbar] (-1,-0.5)--(0,-0.5);\draw[doublefermionbar] (0,-0.5)--(1,-0.5);
\draw[gluon] (0.3,-1.5)--(0,-0.5);
\end{scope}
%%%%%%%%%%%%%%%%%%%%%%%%%%%%%
\begin{scope}[shift={(5.5,0)}]
\draw[doublefermion] 	(-1,+0.5)--(0,+0.5);\draw[doublefermion] 	(0,+0.5)--(1,+0.5);
\draw[doublefermionbar] (-1,-0.5)--(0,-0.5);\draw[doublefermionbar] (0,-0.5)--(1,-0.5);
\draw[gluon] (0,+0.5)--(0,0);\draw[gluon] (0,-0.5)--(0,0);
\draw[gluon] (0.65,0)--(0,0);
\end{scope}
%%%%%%%%%%%%%%%%%%%%%%%%%%%%%
\end{scope}
%%%%%%%%%%%%%%%%%%%%%%%%%%%%%%%%%%%%%%%%%%%%%%%%%%%%%%%%%%%%%%%%%%%%%%%%%%%%%%%%%%%%%%%%%%%%%
%%%%%%%%%%%%%%%%%%%%%%%%%%%%%%%%%%%%%%%%%%%%%%%%%%%%%%%%%%%%%%%%%%%%%%%%%%%%%%%%%%%%%%%%%%%%%
%%%%%%%%%%%%%%%%%%%%%%%%%%%%%%%%%%%%%%%%%%%%%%%%%%%%%%%%%%%%%%%%%%%%%%%%%%%%%%%%%%%%%%%%%%%%%
\begin{scope}[shift={(10,-3.75)}]%[shift={(0,-7.5)}]
%%%%%%%%%%%%%%%%%%%%%%%%%%%%%
\begin{scope}[shift={(0,0)}]
%%%%% field lines
\draw[doublefermion] 	(-1,+0.5)--(0,+0.5);\draw[doublefermion] 	(0,+0.5)--(1,+0.5);
\draw[fermionnoarrow]   (-1,-0.5)--(0,-0.5);\draw[doublefermionbar] (0,-0.5)--(1,-0.5);
\draw[scalar] (0,-0.5)--(0.3,-1.5);%\node at (0.6,1.7) {$W$};
%%%%% spins
\node at (-1.25,+0.45) {$s_1$};\node at (+1.25,+0.45) {$r_1$};
\node at (-1.25,-0.45) {$s_2$};\node at (+1.25,-0.45) {$r_2$};
%%%%% momenta
\draw[->] (-0.9,+0.75) -- (-0.4,+0.75);\node at (-1.2,+1) {$K/2+k^{(\prime)}$};
\draw[->] (-0.9,-0.75) -- (-0.4,-0.75);\node at (-1.2,-1.1) {$K/2-k^{(\prime)}$};
\draw[->] (+0.4,+0.75) -- (+0.9,+0.75);\node at (+1.1,+1) {$P/2+p$};
\draw[->] (+0.4,-0.75) -- (+0.9,-0.75);\node at (+1.1,-1.1) {$P/2-p$};
\draw[->] (-0.1,-1.1)--(0.05,-1.55);\node at (-0.25,-1.6) {$\PH$};
%%%%% SU2L
\node at (-0.7,+0.25) {$i$};\node at (+0.4,+0.25) {$i'$};
\node at (+0.4,-0.25) {$j'$};
\node at (0.3,-1.7) {$h$};
\end{scope}
%%%%%%%%%%%%%%%%%%%%%%%%%%%%%
%\begin{scope}[shift={(3,0)}]
%\draw[doublefermion] 	(-1,+0.5)--(0,+0.5);\draw[doublefermion] 	(0,+0.5)--(1,+0.5);
%\draw (-1,-0.5)--(0,-0.5);\draw[doublefermionbar] (0,-0.5)--(1,-0.5);
%\draw[gluon] (0,+0.5)--(0,0);\draw[scalar] (0,-0.5)--(0,0.12);
%\draw[scalar] (0,0)--(0.7,0);
%\end{scope}
%%%%%%%%%%%%%%%%%%%%%%%%%%%%%
%\begin{scope}[shift={(5.5,0)}]
%\draw[doublefermion] 	(-1,+0.5)--(0,+0.5);\draw[doublefermion] 	(0,+0.5)--(1,+0.5);
%\draw (-1,-0.5)--(0,-0.5);\draw[doublefermionbar] (0,-0.5)--(1,-0.5);
%\draw[vector] (0,+0.5)--(0,0);\draw[scalar] (0,-0.5)--(0,0.12);
%\draw[scalar] (0,0)--(0.7,0);
%\end{scope}
%%%%%%%%%%%%%%%%%%%%%%%%%%%%%
\end{scope}
%%%%%%%%%%%%%%%%%%%%%%%%%%%%%%%%%%%%%%%%%%%%%%%%%%%%%%%%%%%%%%%%%%%%%%%%%%%%%%%%%%%%%%%%%%%%%
%%%%%%%%%%%%%%%%%%%%%%%%%%%%%%%%%%%%%%%%%%%%%%%%%%%%%%%%%%%%%%%%%%%%%%%%%%%%%%%%%%%%%%%%%%%%%
%%%%%%%%%%%%%%%%%%%%%%%%%%%%%%%%%%%%%%%%%%%%%%%%%%%%%%%%%%%%%%%%%%%%%%%%%%%%%%%%%%%%%%%%%%%%%
\end{tikzpicture}

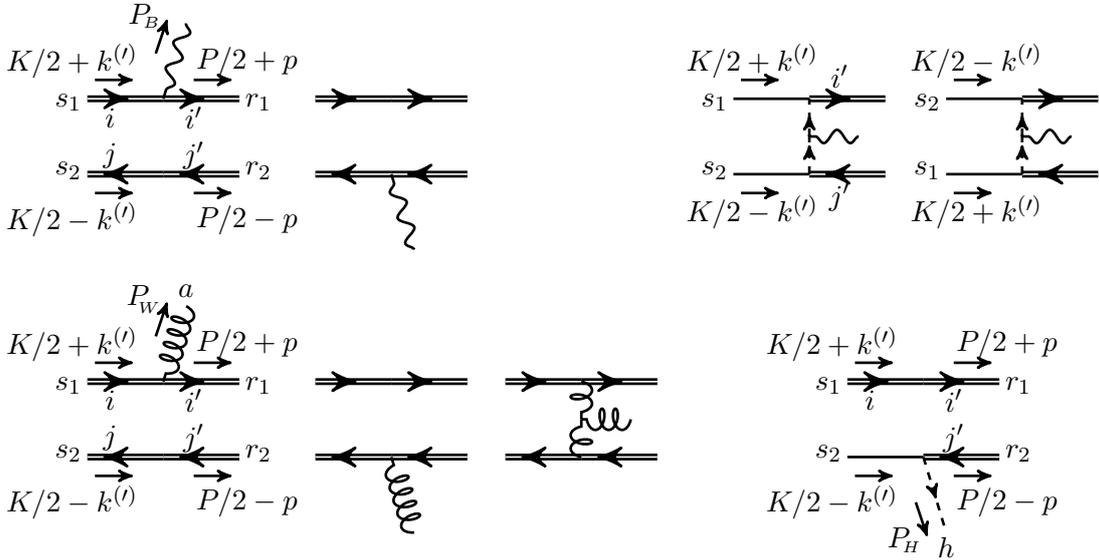
\captionof{figure}{\label{fig:BSF_DDbar}
Same as in \cref{fig:BSF_SSDDbar}, for the formation of $D\bar{D}$ bound states. 
\emph{Top:} $D\bar{D}$-like $\to {\cal B}(D\bar{D}) + B$ and  $SS$-like $\to {\cal B}(D\bar{D}) + B$. 
\emph{Bottom left:} $D\bar{D} \to {\cal B}(D\bar{D}) + W$. 
\emph{Bottom right:} $DS \to {\cal B}(SS/D\bar{D}) + H$.
}
\end{table}
%%%%%%%%%%%%%%%%%%%%%%%%%%%%%%%%%%%%%%%%%%%%%%%%%%%%%%%%%%%%%%%%%%%%

The BSF processes are listed in \cref{tab:BSF_DDbar}, and the radiative parts of the diagrams contributing to these processes are shown in \cref{fig:BSF_DDbar}. For all processes below, we project the bound-state fields on the $\SUL$ singlet via $\delta_{i'j'}/\sqrt{2}$. Moreover, since spin is conserved at working order, as already seen in \cref{sec:BSF_SSDDbar}, we project both the scattering and the bound states on the spin-1 configuration  by contracting the spin indices with
$[U_{\text{spin-1}}^{\sigma}]^{s_1s_2} [U_{\text{spin-1}}^{\rho \dagger}]^{r_2r_1}$, where the indices $\sigma, \rho=-1,0,1$ run through the three states of the spin-1 multiplets. While the Clebsh-Gordan coefficients contained in $U_{\text{spin-1}}$ are well-known, we shall not need them explicitly. We will instead only invoke that the operators 
$[U_{\text{spin-1}}^{\sigma \dagger}]^{s_1s_2}$ are symmetric in $s_1$, $s_2$, and $[U_{\text{spin-1}}^{\sigma}]^{s_1s_2} [U_{\text{spin-1}}^{\rho \dagger}]^{s_2s_1} = \delta^{\sigma \rho}$.

\subsubsection{$D\bar{D}$-like $\to {\cal B} (D\bar{D}) + B$}

Besides the projections mentioned above, here we also project the $D\bar{D}$ component of the scattering state on the $\SUL$ singlet configuration via $\delta_{ij}/\sqrt{2}$. The perturbative parts of the amplitude are
\begin{subequations}
\label{eq:BSF_A_MixedToDDbarSingletSpin1_Bemission} 
\label[pluralequation]{eqs:BSF_A_MixedToDDbarSingletSpin1_Bemission} 
\begin{align}
\im \boldsymbol{{\cal A}}^{\sigma \rho} 
&\[(SS)^{\text{spin-1}} \to (D\bar{D})_{(\mathbb{1},0)}^{\text{spin-1}} + B\] \simeq
\nn \\
&\simeq -\im \sqrt{2} 
\, \delta^{\sigma \rho}
\, y^2  \, g_1 \YH  \, 4 m^2 
\[\frac{2(\bf k' - p)}{[({\bf k'} - {\bf p})^2 +\mH^2]^2} 
+ \frac{2(\bf k' + p)}{[({\bf k'} + {\bf p})^2 +\mH^2]^2} \] ,
\label{eq:BSF_A_SSToDDbarSingletSpin1_Bemission}
\\
\im \boldsymbol{{\cal A}}^{\sigma \rho}  
&\[(D\bar{D})_{(\mathbb{1},0)}^{\text{spin-1}}\to (D\bar{D})_{(\mathbb{1},0)}^{\text{spin-1}} +  B \] 
\simeq 
\nn \\
&\simeq +\im \, \delta^{\sigma \rho}
\, g_1 \YD \, 2m  \, 2{\bf p} 
\, (2\pi)^3
\[\delta^3 ({\bf k'} - {\bf p} - \PBvec/2) 
+ \delta^3 ({\bf k'} - {\bf p} + \PBvec/2)\] .
\label{eq:BSF_A_DDbarToDDbarSingletSpin1_Bemission}
\end{align}
\end{subequations}
In \cref{eq:BSF_A_SSToDDbarSingletSpin1_Bemission}, the fermion permutations introduced factors $(-1)$ and $(+1)$  for the $t$- and $u$-channel diagrams.  Upon projection on the symmetric spin-1 eigenstate, their relative sign does not change. 
The factor $\sqrt{2}$ appearing in the beginning of \cref{eq:BSF_A_SSToDDbarSingletSpin1_Bemission} arises from the projection onto the $\SUL$ singlet final $D\bar{D}$ state. 
Using the wavefunctions listed in \cref{tab:ScatteringStates,tab:BoundStates}, we find the full amplitude \eqref{eq:BSF_FullAmplitude_Def} to be
\begin{align}
\im \boldsymbol{{\cal M}}^{\sigma \rho} 
&\simeq \im \delta^{\sigma \rho}
\, \frac{2}{\aA+\aR} \sqrt{4\pi\alpha_1 \, 4m}
\, \times
\nn \\
&\times \left\{
2\sqrt{2\aA\aR} \[
- \boldsymbol{{\cal Y}}_{{\bf k},100}^\H \(\aA,\frac{\alpha_1+3\alpha_2}{4}\)
+ \boldsymbol{{\cal Y}}_{{\bf k},100}^\H \(-\aR,\frac{\alpha_1+3\alpha_2}{4}\)
\] 
\right. \nn \\
&\left.
+\aA \boldsymbol{{\cal J}}_{{\bf k},100} \(\aA,\frac{\alpha_1+3\alpha_2}{4}\) 
+\aR \boldsymbol{{\cal J}}_{{\bf k},100} \(-\aR,\frac{\alpha_1+3\alpha_2}{4}\) 
\right\},
\label{eq:BSF_M_DDbarlikeToDDbarSingletSpin1_Bemission}
\end{align}
where here $\aA$ and $\aR$ should be evaluated from \cref{eqs:GaugeSinglet_alphas_def} for $\ellS=s=1$ (scattering state). As before, we have neglected the $\pm \PBvec/2$ terms inside the $\delta$-functions in \cref{eq:BSF_A_DDbarToDDbarSingletSpin1_Bemission}. Next, we square, sum over the spins, and average over the three dof of the incoming spin-1 state. Using the overlap integrals \cref{eq:OverlapIntegrals_100}, we find
\begin{align}
\frac{1}{3}  
&\sum_{\sigma,\rho=1}^3 \left|\boldsymbol{{\cal M}}^{\sigma\rho}\right|^2 
\simeq 
\frac{2^{13}\pi^2 \alpha_1}{\aB}
\frac{\aA^2}{(\aA+\aR)^2}
\(\frac{\zetaB^2}{1+\zetaB^2}\) \times
\nn \\
&\times 
\[
  \(1-\frac{2\aH}{\aB} \sqrt{\frac{\aR}{\aA}}\) \Svec^{1/2} \(\zetaA,\zetaB \)
+ \(\frac{\aR}{\aA}+\frac{2\aH}{\aB} \sqrt{\frac{\aR}{\aA}}\) \Svec^{1/2} \(-\zetaR,\zetaB \)
\]^2 ,
\label{eq:BSF_Msquared_DDbarlikeToDDbarSingletSpin1_Bemission}
\end{align}
where here $\aB =(\alpha_1+3\alpha_2)/4$ and correspondingly $\zetaB =(\zeta_1+3\zeta_2)/4$. The cross-section is obtained from \cref{eq:BSF_sigma_Vector}, and is shown in \cref{tab:BSF_DDbar}.

\subsubsection{$SS$-like $\to {\cal B} (D\bar{D}) + B$}
	
Using the perturbative parts \eqref{eq:BSF_A_MixedToDDbarSingletSpin1_Bemission}, and the wavefunctions listed in \cref{tab:ScatteringStates,tab:BoundStates}, we find the full amplitude \eqref{eq:BSF_FullAmplitude_Def}
\begin{align}
\im \boldsymbol{{\cal M}}^{\sigma \rho} 
&\simeq \im \delta^{\sigma \rho}
\, \frac{2}{\aA+\aR} \sqrt{4\pi\alpha_1 \, 4m}
\, \times
\nn \\
&\times \left\{
-4 \[
 \aR \boldsymbol{{\cal Y}}_{{\bf k},100}^\H \(\aA,\frac{\alpha_1+3\alpha_2}{4}\)
+\aA \boldsymbol{{\cal Y}}_{{\bf k},100}^\H \(-\aR,\frac{\alpha_1+3\alpha_2}{4}\)
\] 
\right. \nn \\ &\left. 
+\sqrt{2\aA\aR} \[
+\boldsymbol{{\cal J}}_{{\bf k},100} \(\aA,\frac{\alpha_1+3\alpha_2}{4}\) 
-\boldsymbol{{\cal J}}_{{\bf k},100} \(-\aR,\frac{\alpha_1+3\alpha_2}{4}\) \]
\right\},
\label{eq:BSF_M_SSlikeToDDbarSingletSpin1_Bemission}
\end{align}
where again $\aA$ and $\aR$ should be evaluated from \cref{eqs:GaugeSinglet_alphas_def} for $\ellS=s=1$ (scattering state), and with the help of the overlap integrals \eqref{eq:OverlapIntegrals_100},
\begin{align}
\frac{1}{3} 
&\sum_{\sigma,\rho=1}^3 \left|\boldsymbol{{\cal M}}^{\sigma\rho}\right|^2 
\simeq 
\frac{2^{14}\pi^2 \alpha_1}{\aB}
\frac{\aA\aR}{(\aA+\aR)^2}
\(\frac{\zetaB^2}{1+\zetaB^2}\) \times
\nn \\
&\times 
\[
\(1-\frac{2\aH}{\aB} \sqrt{\frac{2\aR}{\aA}}\) \Svec^{1/2} \(\zetaA,\zetaB \)
- \(1+\frac{2\aH}{\aB} \sqrt{\frac{2\aA}{\aR}}\) \Svec^{1/2} \(-\zetaR,\zetaB \)
\]^2 ,
\label{eq:BSF_Msquared_SSlikeToDDbarSingletSpin1_Bemission}
\end{align}
with $\aB =(\alpha_1+3\alpha_2)/4$ and $\zetaB =(\zeta_1+3\zeta_2)/4$. The cross-section is obtained from \cref{eq:BSF_sigma_Vector}, and is shown in \cref{tab:BSF_DDbar}.

\subsubsection{$D\bar{D} \to {\cal B} (D\bar{D}) + W$}

The perturbative part of the amplitude is (cf.~\cref{eq:BSF_A_DDbarToDDbar_Wemission})
\begin{multline}
\im (\boldsymbol{{\cal A}}^{\sigma\rho})_{ij}^a
\[ (D\bar{D})^{\text{spin-1}} \to (D\bar{D})_{(\mathbb{1},0)}^{\text{spin-1}} +  W \]
\simeq \im \delta^{\sigma\rho} \, \frac{t^a_{ji}}{\sqrt{2}}  \, g_2 \, 2m \times 
\\
\times \left\{
2m \ 4\pi\alpha_2 \ \frac{2({\bf k'-p})}{({\bf k'-p})^4}
+2{\bf p} \, (2\pi)^3 
\[\delta^3 ({\bf k'} - {\bf p} - \PWvec/2) + \delta^3 ({\bf k'} - {\bf p} + \PWvec/2) \] 
\right\}. 
\label{eq:BSF_A_DDbarToDDbarSingletSpin1_Wemission}
\end{multline}
Using the wavefunctions listed in \cref{tab:ScatteringStates,tab:BoundStates}, and anticipating that the scattering state will be an $\SUL$ triplet, we find the full amplitude \eqref{eq:BSF_FullAmplitude_Def},
\begin{align}
\im (\boldsymbol{{\cal M}}^{\sigma\rho})_{ij}^a
&\simeq \im \delta^{\sigma\rho} t^a_{ji}  
\sqrt{4 \pi \alpha_2  \  2^5 m} \times 
\nn \\
&\times \left\{
  {\cal Y}_{{\bf k},100}^\W \(\frac{\alpha_1-\alpha_2}{4}, \frac{\alpha_1+3\alpha_2}{4}\)
+ {\cal J}_{{\bf k},100} \(\frac{\alpha_1-\alpha_2}{4}, \frac{\alpha_1+3\alpha_2}{4}\)
\right\}. 
\label{eq:BSF_M_DDbarToDDbarSingletSpin1_Wemission}
\end{align}
Squaring and summing over gauge and spin indices, projects the scattering state on the spin-1 $\SUL$ triplet that has 9 dof. Using the overlap integrals \eqref{eq:OverlapIntegrals_100}, we find
\begin{align}
\frac{1}{9} 
\sum_{\sigma,\rho}\sum_{i,j,a}
\left| \[\boldsymbol{{\cal M}}^{\sigma\rho}\]_{ij}^a \right|^2 
&\simeq 
2^{13} \pi^2 \ \frac{\alpha_2}{\aB} 
\(1+\frac{\alpha_2}{\aB} \)^2 
\(\frac{\zetaB^2}{1+\zetaB^2}\)
\, \Svec\(\frac{\zeta_1-\zeta_2}{4}, \zetaB\) .
\label{eq:BSF_Msquared_DDbarToDDbarSingletSpin1_Wemission}
\end{align}
The cross-section is obtained from \cref{eq:BSF_sigma_Vector}, and is shown in \cref{tab:BSF_DDbar}. It agrees with the results of refs.~\cite{Harz:2018csl,Harz:2019rro} appropriately adjusted.

\subsubsection{$DS \to {\cal B} (D\bar{D}) + H$}

The perturbative part of the amplitude is
\begin{align}
\im {\cal A}_{i,h}^{\sigma\rho} 
\[ (DS)^{\text{spin-1}} \to (D\bar{D})_{(\mathbb{1},0)}^{\text{spin-1}} +  H \]
\simeq
\im \delta^{\sigma \rho} \, \frac{\delta_{ih}}{\sqrt{2}}\, y \, 4m^2 \,
(2\pi)^3 \delta^3({\bf k'-p}+\PHvec/2) .
\label{eq:BSF_A_DSToDDbarSingletSpin1_Hemission}
\end{align}
Using the wavefunctions listed in \cref{tab:ScatteringStates,tab:BoundStates}, we find the full amplitude \eqref{eq:BSF_FullAmplitude_Def},
\begin{align}
\im {\cal M}^{\sigma\rho}_{i,h}
&\simeq \im \delta^{\sigma\rho} \delta_{ih}
\sqrt{\frac{2^8\pi \aH}{[(\alpha_1+3\alpha_2)/4]^3}}
\ {\cal R}_{{\bf k},100} \(-\aH, \frac{\alpha_1+3\alpha_2}{4} \) .
\label{eq:BSF_M_DSToDDbarSingletSpin1_Hemission}
\end{align}
and taking into account the overlap integral \eqref{eq:OverlapIntegral_R_100},
\begin{align}
\frac{1}{6} \sum_{\sigma,\rho} \sum_{i,h}
\left| {\cal M}^{\sigma\rho}_{i,h} \right|^2
&\simeq
\frac{2^{14} \pi^2 \aH}{\aB^3}
\(1+\frac{\aH}{\aB}\)^2
\(\frac{\zetaB^2}{1+\zetaB^2}\)
\Ssc \(-\zetaH, \zetaB \) ,
\label{eq:BSF_Msquared_DSToDDbarSingletSpin1_Hemission}
\end{align}
with $\aB =(\alpha_1+3\alpha_2)/4$ and $\zetaB =(\zeta_1+3\zeta_2)/4$. 
The cross-section is obtained from \cref{eq:BSF_sigma_Scalar}, and is shown in \cref{tab:BSF_DDbar}.

\clearpage
\subsection{$DD$ bound states: $(\mathbb{1},1)$, spin 1, $n\ell m =\{100\}$ \label{sec:BSF_DD}}
%%%%%%%%%%%%%%%%%%%%%%%%%%%%%%%%%%%%%%%%%%%%%%%%%%%%%%%%%%%%%%%%%%%%
\begin{table}[p]
\centering
%%%%%%%%%%%%%%%%%%%%%%%%%%%%%%%%%%%%%%%%%%%%%%%%%%%%%%%%%%%%%%%%%%%%%%%%%%%%%%%%%%%%%%%%%%%%%%%%%%%%%%%%
%%%%%%%%%%%%%%%%%%%%%%%%%%%%%%%%%%%%%%%%%%%%%%%%%%%%%%%%%%%%%%%%%%%%%%%%%%%%%%%%%%%%%%%%%%%%%%%%%%%%%%%%
%%%%%%%%%%%%%%%%%%%%%%%%%%%%%%%%%%%%%%%%%%%%%%%%%%%%%%%%%%%%%%%%%%%%%%%%%%%%%%%%%%%%%%%%%%%%%%%%%%%%%%%%
\begin{align*}
&\begin{array}{|c|c|c|c|c|c|rl|}
%%%%%%%%%%%%%%%%%%%%%%%%%%%%%%%%%%%%%%%%%%%%%%%%%%%%%%%%%%%%%%%%%%%%%%%%%%%%%%%%%%%%%%%%%%%%%%%%%%%%%%%%
\multicolumn{8}{c}{\boldsymbol{
\textbf{Bound state}~DD:
~~~(\mathbb{1},1),
~~\textbf{spin 1}, 
~~\{n\ell m\} = \{100\}
}}
\\ \hline
%%%%%%%%%%%%%%%%%%%%%%%%%%%%%%%%%%%%%%%%%%%%%%%%%%%%%%%%%%%%%%%%%%%%%%%%%%%%%%%%%%%%%%%%%%%%%%%%%%%%%%%%
\multicolumn{5}{|c|}{\text{Scattering state (spin 1)}}
&\multirow{2}{*}{\text{\parbox{6ex}{\centering Rad boson}}}
&\multicolumn{2}{c|}{\text{Cross-section}}
\\ \cline{1-5}
\text{State}		
& U_{\mathsmaller{Y}}(1) 
&SU_{\mathsmaller{L}}(2) 
& \text{dof} 
& \ellS 
&
&\multicolumn{2}{c|}{(\sigma_\BSF \vrel) / (\pi m^{-2})}
\\ \hline\hline
%%%%%%%%%%%%%%%%%%%%%%%%%%%%%%%%%%%%%%%%%%%%%%%%%%%%%%%%%%%%%%%%%%%%%%%%%%%%%%%%%%%%%%%%%%%%%%%%%%%%%%%%
DD 
& 1   
& \mathbb{1} 
& 3  
& 1 
& B 			
& \multicolumn{2}{c|}{\text{0 (due to antisymmetry of $DD$ scattering state)}}
\\ \hline
%%%%%%%%%%%%%%%%%%%%%%%%%%%%%%%%%%%%%%%%%%%%%%%%%%%%%%%%%%%%%%%%%%%%%%%%%%%%%%%%%%%%%%%%%%%%%%%%%%%%%%%%
DD 
& 1   
& \mathbb{3} 
& 9 
& 1 
& W 			
& \dfrac{2^{11} \aB \alpha_2}{3} \(1+\dfrac{\alpha_2}{\aB}\)^2
& \Svec \(-\dfrac{\zeta_1+\zeta_2}{4} , \zetaB \) 
\\ \hline
%%%%%%%%%%%%%%%%%%%%%%%%%%%%%%%%%%%%%%%%%%%%%%%%%%%%%%%%%%%%%%%%%%%%%%%%%%%%%%%%%%%%%%%%%%%%%%%%%%%%%%%%
DS 
& +1/2
& \mathbb{2} 
& 6  
& 0 
& H^\dagger	
& \dfrac{2^{11} \aH}{\aB} \(1+\dfrac{\aH}{\aB}\)^2
& \Ssc \(-\zetaH , \zetaB \) \, \hH(\omega)
\\ \hline
%%%%%%%%%%%%%%%%%%%%%%%%%%%%%%%%%%%%%%%%%%%%%%%%%%%%%%%%%%%%%%%%%%%%%%%%%%%%%%%%%%%%%%%%%%%%%%%%%%%%%%%%
\end{array}
\end{align*}
\captionof{table}{\label{tab:BSF_DD}
Same as \cref{tab:BSF_SSDDbar} for the $DD$ bound states. 
Here, $\aB = (-\alpha_1+3\alpha_2)/4$, $\zetaB=(-\zeta_1+3\zeta_2)/4$, and 
$\omega=m\[(-\alpha_1+3\alpha_2)^2 / 16 + \vrel^2\]/4$.
All processes have conjugate counterparts.}
%%%%%%%%%%%%%%%%%%%%%%%%%%%%%%%%%%%%%%%%%%%%%%%%%%%%%%%%%%%%%%%%%%%%
\medskip
%%%%%%%%%%%%%%%%%%%%%%%%%%%%%%%%%%%%%%%%%%%%%%%%%%%%%%%%%%%%%%%%%%%%
\begin{tikzpicture}[line width=1pt, scale=1]
%%%%%%%%%%%%%%%%%%%%%%%%%%%%%%%%%%%%%%%%%%%%%%%%%%%%%%%%%%%%%%%%%%%%%%%%%%%%%%%%%%%%%%%%%%%%%
%%%%%%%%%%%%%%%%%%%%%%%%%%%%%%%%%%%%%%%%%%%%%%%%%%%%%%%%%%%%%%%%%%%%%%%%%%%%%%%%%%%%%%%%%%%%%
%%%%%%%%%%%%%%%%%%%%%%%%%%%%%%%%%%%%%%%%%%%%%%%%%%%%%%%%%%%%%%%%%%%%%%%%%%%%%%%%%%%%%%%%%%%%%
\begin{scope}[shift={(0,0)}]
%%%%%%%%%%%%%%%%%%%%%%%%%%%%%
\begin{scope}[shift={(0,0)}]
%%%%% field lines
\draw[doublefermion] (-1,+0.5)--(0,+0.5);\draw[doublefermion] 	(0,+0.5)--(1,+0.5);
\draw[doublefermion] (-1,-0.5)--(0,-0.5);\draw[doublefermion] (0,-0.5)--(1,-0.5);
\draw[gluon] (0.3,1.5)--(0,+0.5);%\node at (0.6,1.7) {$W$};
%%%%% spins
\node at (-1.25,+0.45) {$s_1$};\node at (+1.25,+0.45) {$r_1$};
\node at (-1.25,-0.45) {$s_2$};\node at (+1.25,-0.45) {$r_2$};
%%%%% momenta
\draw[->] (-0.9,+0.75) -- (-0.4,+0.75);\node at (-1.2,+1) {$K/2+k^{(\prime)}$};
\draw[->] (-0.9,-0.75) -- (-0.4,-0.75);\node at (-1.2,-1.1) {$K/2-k^{(\prime)}$};
\draw[->] (+0.4,+0.75) -- (+0.9,+0.75);\node at (+1.1,+1) {$P/2+p$};
\draw[->] (+0.4,-0.75) -- (+0.9,-0.75);\node at (+1.1,-1.1) {$P/2-p$};
\draw[->] (-0.1,1.1)--(0.05,1.55);\node at (-0.35,1.6) {$\PW$};
%%%%% SU2L
\node at (-0.6,+0.25) {$i$};\node at (+0.5,+0.25) {$i'$};
\node at (-0.6,-0.23) {$j$};\node at (+0.5,-0.23) {$j'$};
\node at (0.3,1.7) {$a$};
\end{scope}
%%%%%%%%%%%%%%%%%%%%%%%%%%%%%
\begin{scope}[shift={(5.5,0)}]
%%%%% field lines
\draw[doublefermion] (-1,+0.5)--(0,+0.5);\draw[doublefermion] 	(0,+0.5)--(1,+0.5);
\draw[doublefermion] (-1,-0.5)--(0,-0.5);\draw[doublefermion] (0,-0.5)--(1,-0.5);
\draw[gluon] (0.3,-1.5)--(0,-0.5);%\node at (0.6,1.7) {$W$};
%%%%% spins
\node at (-1.25,+0.45) {$s_1$};\node at (+1.25,+0.45) {$r_1$};
\node at (-1.25,-0.45) {$s_2$};\node at (+1.25,-0.45) {$r_2$};
%%%%% momenta
\draw[->] (-0.9,+0.75) -- (-0.4,+0.75);\node at (-1.2,+1) {$K/2+k^{(\prime)}$};
\draw[->] (-0.9,-0.75) -- (-0.4,-0.75);\node at (-1.2,-1.1) {$K/2-k^{(\prime)}$};
\draw[->] (+0.4,+0.75) -- (+0.9,+0.75);\node at (+1.1,+1) {$P/2+p$};
\draw[->] (+0.4,-0.75) -- (+0.9,-0.75);\node at (+1.1,-1.1) {$P/2-p$};
\draw[->] (-0.1,-1.1)--(0.05,-1.55);\node at (-0.35,-1.6) {$\PW$};
%%%%% SU2L
\node at (-0.6,+0.25) {$i$};\node at (+0.5,+0.25) {$i'$};
\node at (-0.6,-0.23) {$j$};\node at (+0.5,-0.23) {$j'$};
\node at (0.3,-1.7) {$a$};
\end{scope}
%%%%%%%%%%%%%%%%%%%%%%%%%%%%%
\begin{scope}[shift={(11,0)}]
\draw[doublefermion] (-1,+0.5)--(0,+0.5);\draw[doublefermion] (0,+0.5)--(1,+0.5);
\draw[doublefermion] (-1,-0.5)--(0,-0.5);\draw[doublefermion] (0,-0.5)--(1,-0.5);
\draw[gluon] (0,+0.5)--(0,0);\draw[gluon] (0,-0.5)--(0,0);
\draw[gluon] (0.65,0)--(0,0);
%%%%% spins
\node at (-1.25,+0.45) {$s_1$};\node at (+1.25,+0.45) {$r_1$};
\node at (-1.25,-0.45) {$s_2$};\node at (+1.25,-0.45) {$r_2$};
%%%%% momenta
\draw[->] (-0.9,+0.75) -- (-0.4,+0.75);\node at (-1.2,+1) {$K/2+k^{(\prime)}$};
\draw[->] (-0.9,-0.75) -- (-0.4,-0.75);\node at (-1.2,-1.1) {$K/2-k^{(\prime)}$};
\draw[->] (+0.4,+0.75) -- (+0.9,+0.75);\node at (+1.1,+1) {$P/2+p$};
\draw[->] (+0.4,-0.75) -- (+0.9,-0.75);\node at (+1.1,-1.1) {$P/2-p$};
\draw[->] (+0.4, 0.25) -- (+0.8, 0.25);\node at (+1.05, 0.1) {$\PW$};
\end{scope}
%%%%%%%%%%%%%%%%%%%%%%%%%%%%%
\end{scope}
%%%%%%%%%%%%%%%%%%%%%%%%%%%%%%%%%%%%%%%%%%%%%%%%%%%%%%%%%%%%%%%%%%%%%%%%%%%%%%%%%%%%%%%%%%%%%
%%%%%%%%%%%%%%%%%%%%%%%%%%%%%%%%%%%%%%%%%%%%%%%%%%%%%%%%%%%%%%%%%%%%%%%%%%%%%%%%%%%%%%%%%%%%%
%%%%%%%%%%%%%%%%%%%%%%%%%%%%%%%%%%%%%%%%%%%%%%%%%%%%%%%%%%%%%%%%%%%%%%%%%%%%%%%%%%%%%%%%%%%%%
\begin{scope}[shift={(0,-4)}]
%%%%%%%%%%%%%%%%%%%%%%%%%%%%%
\begin{scope}[shift={(0,0)}]
%%%%% field lines
\draw[doublefermion] (-1,+0.5)--(0,+0.5);\draw[doublefermion] (0,+0.5)--(1,+0.5);
\draw[doublefermion] (-1,-0.5)--(0,-0.5);\draw[doublefermion] (0,-0.5)--(1,-0.5);
\draw[gluon] (0.3,1.5)--(0,+0.5);%\node at (0.6,1.7) {$W$};
%%%%% spins
\node at (-1.25,+0.45) {$s_1$};\node at (+1.25,+0.45) {$r_2$};
\node at (-1.25,-0.45) {$s_2$};\node at (+1.25,-0.45) {$r_1$};
%%%%% momenta
\draw[->] (-0.9,+0.75) -- (-0.4,+0.75);\node at (-1.2,+1) {$K/2+k^{(\prime)}$};
\draw[->] (-0.9,-0.75) -- (-0.4,-0.75);\node at (-1.2,-1.1) {$K/2-k^{(\prime)}$};
\draw[->] (+0.4,+0.75) -- (+0.9,+0.75);\node at (+1.1,+1) {$P/2-p$};
\draw[->] (+0.4,-0.75) -- (+0.9,-0.75);\node at (+1.1,-1.1) {$P/2+p$};
\draw[->] (-0.1,1.1)--(0.05,1.55);\node at (-0.35,1.6) {$\PW$};
%%%%% SU2L
\node at (-0.6,+0.25) {$i$};\node at (+0.5,+0.25) {$j'$};
\node at (-0.6,-0.23) {$j$};\node at (+0.5,-0.23) {$i'$};
\node at (0.3,1.7) {$a$};
\end{scope}
%%%%%%%%%%%%%%%%%%%%%%%%%%%%%
\begin{scope}[shift={(5.5,0)}]
%%%%% field lines
\draw[doublefermion] (-1,+0.5)--(0,+0.5);\draw[doublefermion] 	(0,+0.5)--(1,+0.5);
\draw[doublefermion] (-1,-0.5)--(0,-0.5);\draw[doublefermion] (0,-0.5)--(1,-0.5);
\draw[gluon] (0.3,-1.5)--(0,-0.5);%\node at (0.6,1.7) {$W$};
%%%%% spins
\node at (-1.25,+0.45) {$s_1$};\node at (+1.25,+0.45) {$r_2$};
\node at (-1.25,-0.45) {$s_2$};\node at (+1.25,-0.45) {$r_1$};
%%%%% momenta
\draw[->] (-0.9,+0.75) -- (-0.4,+0.75);\node at (-1.2,+1) 	{$K/2+k^{(\prime)}$};
\draw[->] (-0.9,-0.75) -- (-0.4,-0.75);\node at (-1.2,-1.1) {$K/2-k^{(\prime)}$};
\draw[->] (+0.4,+0.75) -- (+0.9,+0.75);\node at (+1.1,+1) 	{$P/2-p$};
\draw[->] (+0.4,-0.75) -- (+0.9,-0.75);\node at (+1.1,-1.1) {$P/2+p$};
\draw[->] (-0.1,-1.1)--(0.05,-1.55);\node at (-0.35,-1.6) {$\PW$};
%%%%% SU2L
\node at (-0.6,+0.25) {$i$};\node at (+0.5,+0.25) {$j'$};
\node at (-0.6,-0.23) {$j$};\node at (+0.5,-0.23) {$i'$};
\node at (0.3,-1.7) {$a$};
\end{scope}
%%%%%%%%%%%%%%%%%%%%%%%%%%%%%
\begin{scope}[shift={(11,0)}]
\draw[doublefermion] (-1,+0.5)--(0,+0.5);\draw[doublefermion] (0,+0.5)--(1,+0.5);
\draw[doublefermion] (-1,-0.5)--(0,-0.5);\draw[doublefermion] (0,-0.5)--(1,-0.5);
\draw[gluon] (0,+0.5)--(0,0);\draw[gluon] (0,-0.5)--(0,0);
\draw[gluon] (0.65,0)--(0,0);
%%%%% spins
\node at (-1.25,+0.45) {$s_1$};\node at (+1.25,+0.45) {$r_2$};
\node at (-1.25,-0.45) {$s_2$};\node at (+1.25,-0.45) {$r_1$};
%%%%% momenta
\draw[->] (-0.9,+0.75) -- (-0.4,+0.75);\node at (-1.2,+1) 	{$K/2+k^{(\prime)}$};
\draw[->] (-0.9,-0.75) -- (-0.4,-0.75);\node at (-1.2,-1.1) {$K/2-k^{(\prime)}$};
\draw[->] (+0.4,+0.75) -- (+0.9,+0.75);\node at (+1.1,+1) 	{$P/2-p$};
\draw[->] (+0.4,-0.75) -- (+0.9,-0.75);\node at (+1.1,-1.1) {$P/2+p$};
\draw[->] (+0.4, 0.25) -- (+0.8, 0.25);\node at (+1.05, 0.1){$\PW$};
\end{scope}
%%%%%%%%%%%%%%%%%%%%%%%%%%%%%
\end{scope}
%%%%%%%%%%%%%%%%%%%%%%%%%%%%%%%%%%%%%%%%%%%%%%%%%%%%%%%%%%%%%%%%%%%%%%%%%%%%%%%%%%%%%%%%%%%%%
%%%%%%%%%%%%%%%%%%%%%%%%%%%%%%%%%%%%%%%%%%%%%%%%%%%%%%%%%%%%%%%%%%%%%%%%%%%%%%%%%%%%%%%%%%%%%
%%%%%%%%%%%%%%%%%%%%%%%%%%%%%%%%%%%%%%%%%%%%%%%%%%%%%%%%%%%%%%%%%%%%%%%%%%%%%%%%%%%%%%%%%%%%%
\begin{scope}[shift={(0,-8)}]%[shift={(0,-9)}]
%%%%%%%%%%%%%%%%%%%%%%%%%%%%%
\begin{scope}[shift={(2.75,0)}]%[shift={(0,0)}]
%%%%% field lines
\draw[doublefermion] 	(-1,+0.5)--(0,+0.5);\draw[doublefermion] (0,+0.5)--(1,+0.5);
\draw[fermionnoarrow]   (-1,-0.5)--(0,-0.5);\draw[doublefermion] (0,-0.5)--(1,-0.5);
\draw[scalarbar] (0,-0.5)--(0.3,-1.5);%\node at (0.6,1.7) {$W$};
%%%%% spins
\node at (-1.25,+0.45) {$s_1$};\node at (+1.25,+0.45) {$r_1$};
\node at (-1.25,-0.45) {$s_2$};\node at (+1.25,-0.45) {$r_2$};
%%%%% momenta
\draw[->] (-0.9,+0.75) -- (-0.4,+0.75);\node at (-1.2,+1) {$K/2+k^{(\prime)}$};
\draw[->] (-0.9,-0.75) -- (-0.4,-0.75);\node at (-1.2,-1.1) {$K/2-k^{(\prime)}$};
\draw[->] (+0.4,+0.75) -- (+0.9,+0.75);\node at (+1.1,+1) {$P/2+p$};
\draw[->] (+0.4,-0.75) -- (+0.9,-0.75);\node at (+1.1,-1.1) {$P/2-p$};
\draw[->] (-0.1,-1.1)--(0.05,-1.55);\node at (-0.25,-1.6) {$\PHdagger$};
%%%%% SU2L
\node at (-0.6,+0.25) {$i$};\node at (+0.5,+0.25) {$i'$};
\node at (+0.5,-0.23) {$j'$};
\node at (0.3,-1.7) {$h$};
\end{scope}
\begin{scope}[shift={(8.25,0)}]%[shift={(0,0)}]
%%%%% field lines
\draw[doublefermion] 	(-1,+0.5)--(0,+0.5);\draw[doublefermion] (0,+0.5)--(1,+0.5);
\draw[fermionnoarrow]   (-1,-0.5)--(0,-0.5);\draw[doublefermion] (0,-0.5)--(1,-0.5);
\draw[scalarbar] (0,-0.5)--(0.3,-1.5);%\node at (0.6,1.7) {$W$};
%%%%% spins
\node at (-1.25,+0.45) {$s_1$};\node at (+1.25,+0.45) {$r_2$};
\node at (-1.25,-0.45) {$s_2$};\node at (+1.25,-0.45) {$r_1$};
%%%%% momenta
\draw[->] (-0.9,+0.75) -- (-0.4,+0.75);\node at (-1.2,+1) 	{$K/2+k^{(\prime)}$};
\draw[->] (-0.9,-0.75) -- (-0.4,-0.75);\node at (-1.2,-1.1) {$K/2-k^{(\prime)}$};
\draw[->] (+0.4,+0.75) -- (+0.9,+0.75);\node at (+1.1,+1) 	{$P/2-p$};
\draw[->] (+0.4,-0.75) -- (+0.9,-0.75);\node at (+1.1,-1.1) {$P/2+p$};
\draw[->] (-0.1,-1.1)--(0.05,-1.55);   \node at (-0.25,-1.6){$\PHdagger$};
%%%%% SU2L
\node at (-0.6,+0.25) {$i$};\node at (+0.5,+0.25) {$j'$};
\node at (+0.5,-0.23) {$i'$};
\node at (0.3,-1.7) {$h$};
\end{scope}
\end{scope}
%%%%%%%%%%%%%%%%%%%%%%%%%%%%%%%%%%%%%%%%%%%%%%%%%%%%%%%%%%%%%%%%%%%%%%%%%%%%%%%%%%%%%%%%%%%%%
%%%%%%%%%%%%%%%%%%%%%%%%%%%%%%%%%%%%%%%%%%%%%%%%%%%%%%%%%%%%%%%%%%%%%%%%%%%%%%%%%%%%%%%%%%%%%
%%%%%%%%%%%%%%%%%%%%%%%%%%%%%%%%%%%%%%%%%%%%%%%%%%%%%%%%%%%%%%%%%%%%%%%%%%%%%%%%%%%%%%%%%%%%%
\end{tikzpicture}

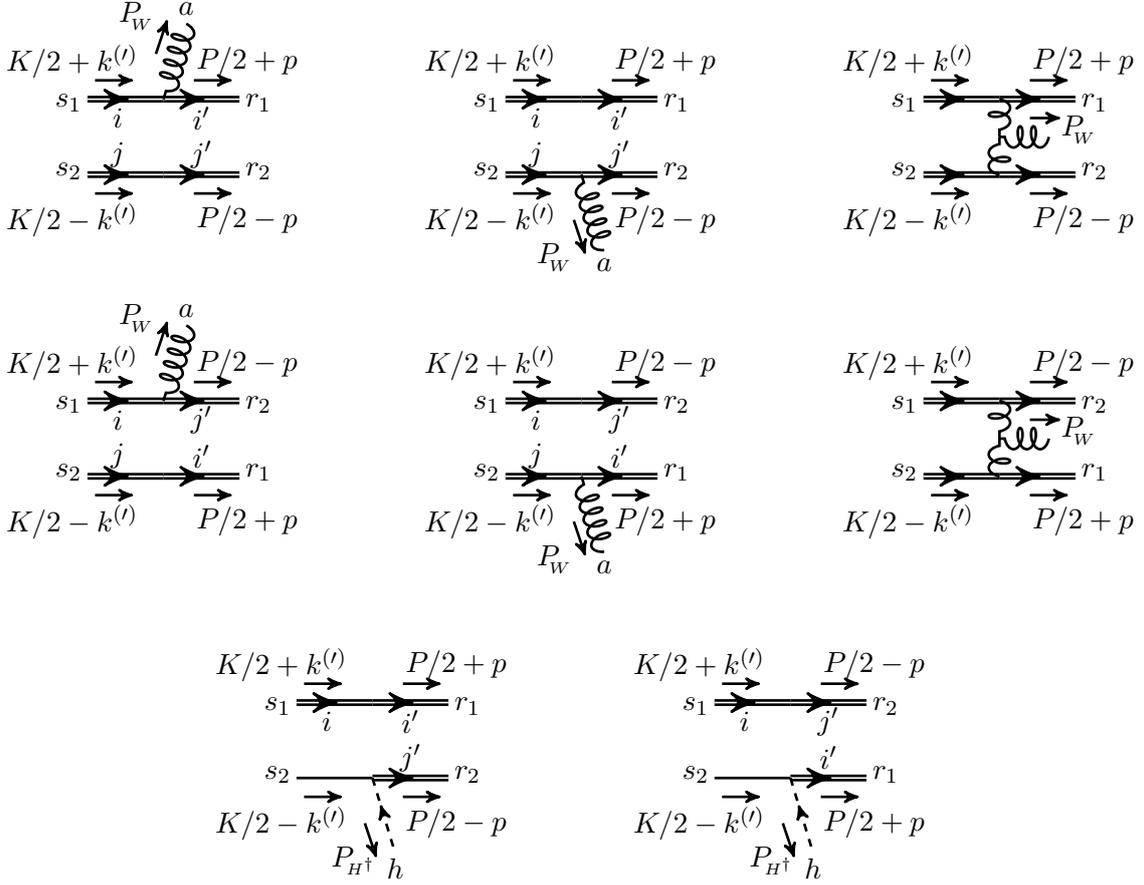
\captionof{figure}{\label{fig:BSF_DD}
Same as \cref{fig:BSF_SSDDbar}, for the formation of $DD$ bound states. 
\emph{Top two rows:} $DD \to {\cal B}(DD) + W$.  
%The left four diagrams give also $DD \to {\cal B}(DD) + B$.
\emph{Bottom row:} $DS \to {\cal B}(DD) + H^\dagger$.
}
\end{table}
%%%%%%%%%%%%%%%%%%%%%%%%%%%%%%%%%%%%%%%%%%%%%%%%%%%%%%%%%%%%%%%%%%%%

The BSF processes are listed in \cref{tab:BSF_DD}, and the radiative part of the diagrams contributing to these processes are shown in \cref{fig:BSF_DD}. In all processes below, we project the bound-state fields on the $\SUL$ singlet via $\epsilon_{i'j'}/\sqrt{2}$. Moreover, as in \cref{sec:BSF_DDbar}, we project both the scattering and the bound states on the spin-1 configuration  via
$[U_{\text{spin-1}}^{\sigma}]^{s_1s_2} [U_{\text{spin-1}}^{\rho \dagger}]^{r_2r_1}$, and invoke that 
$[U_{\text{spin-1}}^{\sigma \dagger}]^{s_1s_2}$ are symmetric in $s_1$, $s_2$, and $[U_{\text{spin-1}}^{\sigma}]^{s_1s_2} [U_{\text{spin-1}}^{\rho \dagger}]^{s_1s_2} = \delta^{\sigma \rho}$.

\subsubsection{$DD \to {\cal B} (DD) + W$}
The perturbative part of the amplitude is 
\begin{align}
&\im (\boldsymbol{{\cal A}}^{\sigma\rho})_{ij}^a
\[ (DD)^{\text{spin-1}} \to (DD)_{(\mathbb{1},0)}^{\text{spin-1}} +  W \]
\simeq \im \delta^{\sigma\rho}  \, g_2 \, 2m \times 
\nn \\
&\times \left\{ 
\[2m \ 4\pi\alpha_2 \ \frac{2({\bf k'-p})}{({\bf k'-p})^4}
\(-\im f^{abc} t^b_{i'i} t^c_{j'j} \frac{\epsilon_{i'j'}}{\sqrt{2}}\)
+2{\bf p} \, (2\pi)^3 \delta^3 ({\bf k'} - {\bf p})
\( t^a_{i'i} \frac{\epsilon_{i'j}}{\sqrt{2}} - t^a_{j'j} \frac{\epsilon_{ij'}}{\sqrt{2}}\)\] 
\right. \nn \\ &\left. 
- (i \leftrightarrow j,~{\bf k \to - k}) \right\}, 
\label{eq:BSF_A_DDToDDSingletSpin1_Wemission}
\end{align}
where the last line accounts for the $u$-channel diagrams. The different number of fermion permutations in the $t$- and $u$-channel diagrams introduces a relative $(-1)$ factor between the two, while the projection on the symmetric spin-1 state does not.  Here we have neglected the $\pm \PWvec/2$ terms inside the $\delta$-functions already at the level of the perturbative amplitude. 

It is easy to check that the gauge factors in \cref{eq:BSF_A_DDToDDSingletSpin1_Wemission} are symmetric in $i \leftrightarrow j$, as expected, since the scattering state must be an $\SUL$ triplet. Convoluting with the scattering state wavefunction and setting ${\bf k \to -k}$ for the $u$ channel renders the latter identical to the $t$-channel up to the extra factor $-(-1)^{\ellS} = +1$ since $\ellS=1$. Thus, the $t$ and $u$ channels add up, and we find the full amplitude \eqref{eq:BSF_FullAmplitude_Def} to be
\begin{align}
\im (\boldsymbol{{\cal M}}^{\sigma\rho})_{ij}^a
&\simeq \im \delta^{\sigma\rho}  \sqrt{4\pi \alpha_2 \, 2^{8} m } 
\times 
\[2\boldsymbol{{\cal Y}}^{\W}_{{\bf k},100} \(-\frac{\alpha_1+\alpha_2}{4}, \frac{-\alpha_1+3\alpha_2}{4}\)
\(-\im f^{abc} t^b_{i'i} t^c_{j'j} \frac{\epsilon_{i'j'}}{\sqrt{2}}\)
\right. \nn \\  &\left.
+ \boldsymbol{{\cal J}}_{{\bf k},100} \(-\frac{\alpha_1+\alpha_2}{4}, \frac{-\alpha_1+3\alpha_2}{4}\)
\( t^a_{i'i} \frac{\epsilon_{i'j}}{\sqrt{2}} - t^a_{j'j} \frac{\epsilon_{ij'}}{\sqrt{2}}\)\] ,
\label{eq:BSF_M_DDToDDSingletSpin1_Wemission}
\end{align}
where we also took into account the symmetry factors of the scattering and bound state wavefunctions, as stated in \cref{tab:ScatteringStates,tab:BoundStates}.
Considering the relation \eqref{eq:OverlapIntegral_YW_100} between the overlap integrals, the above simplifies to
\begin{align}
\im (\boldsymbol{{\cal M}}^{\sigma\rho})_{ij}^a
&\simeq \im \delta^{\sigma\rho}  \sqrt{4\pi \alpha_2 \, 2^{8} m } 
\times 
\boldsymbol{{\cal J}}_{{\bf k},100} 
\(-\frac{\alpha_1+\alpha_2}{4}, \aB \)
G_{ij}^a ,
\label{eq:BSF_M_DDToDDSingletSpin1_Wemission_mod}
\end{align}
where here $\aB = (-\alpha_1+3\alpha_2)/4$, and $G_{ij}^a$ is the gauge factor
\begin{align}
G_{ij}^a \equiv 
t^a_{i'i} \frac{\epsilon_{i'j}}{\sqrt{2}} - t^a_{j'j} \frac{\epsilon_{ij'}}{\sqrt{2}}
+ \frac{\alpha_2}{\aB} \(-\im 2f^{abc} t^b_{i'i} t^c_{j'j} \frac{\epsilon_{i'j'}}{\sqrt{2}}\) ,
\label{eq:BSF_GaugeFactor_DDToDDSingletSpin1_Wemission}
\end{align}
with
\begin{align}
G_{ij}^a {G_{ij}^a}^* =  3 \(1+ \frac{\alpha_2}{\aB}\)^2 .
\label{eq:BSF_GaugeFactorSquared_DDToDDSingletSpin1_Wemission}
\end{align}
Squaring \cref{eq:BSF_M_DDToDDSingletSpin1_Wemission_mod} and summing over gauge indices and spins projects the scattering state on the spin-1 $\SUL$ triplet configuration that has nine dof. Considering the overlap integral \eqref{eq:OverlapIntegral_J_100}, we find
\begin{align}
\frac{1}{9} \sum_{\sigma, \rho} \sum_{i,j,a}
|(\boldsymbol{{\cal M}}^{\sigma\rho})_{ij}^a|^2
&\simeq 2^{17} \pi^2 \frac{\alpha_2}{\aB}
\(1+ \frac{\alpha_2}{\aB}\)^2
\(\frac{\zetaB^2}{1+\zetaB^2}\)
\Svec\( -\frac{\zeta_1+\zeta_2}{4},\zetaB \) .
\label{eq:BSF_Msquared_DDToDDSingletSpin1_Wemission_mod}
\end{align}
The cross-section is obtained from \cref{eq:BSF_sigma_Vector} and is shown in \cref{tab:BSF_DD}.

\subsubsection{$DS \to {\cal B} (DD) + H^\dagger$}
The perturbative part of the amplitude is
\begin{align}
\im {\cal A}^{\sigma\rho}_{i,h}
\[ (DS)^{\text{spin-1}} \to (DD)_{(\mathbb{1},0)}^{\text{spin-1}} +  H \]
\simeq
&-\im \delta^{\sigma\rho} \, \frac{\epsilon_{ih}}{\sqrt{2}}  
\, y \, 4m^2 \, (2\pi)^3 \delta^3 ({\bf k'-p}+\PHvec/2)
\nn \\
&+\im \delta^{\sigma\rho} \, \frac{\epsilon_{hi}}{\sqrt{2}}  
\, y \, 4m^2 \, (2\pi)^3 \delta^3 ({\bf k'+p}+\PHvec/2) ,
\label{eq:BSF_A_DSToDDSingletSpin1_Hemission}
\end{align}
where the fermion permutations alloted factors $(+1)$ and $(-1)$ to the $t$ and $u$ channels respectively. As seen from \cref{eq:BSF_A_DSToDDSingletSpin1_Hemission}, the resulting relative sign is canceled upon contraction of the bound-state fields on the $\SUL$ singlet state. Convoluting \cref{eq:BSF_A_DSToDDSingletSpin1_Hemission} with the scattering and bound state wavefunctions found in \cref{tab:ScatteringStates,tab:BoundStates}, and setting ${\bf p \to -p}$ renders the $u$-channel contribution the same as $t$-channel, with the extra factor $(-1)^\ell=+1$ for $\ell=0$. Thus the $t$ and $u$ channels add up, and we find
\begin{align}
\im {\cal M}^{\sigma\rho}_{i,h}
\simeq
&-\im \delta^{\sigma\rho} \, \epsilon_{ih}   
\sqrt{\frac{2^{12}\pi \aH}{\aB^3}} \, {\cal R}_{{\bf k},100} (-\aH,\aB) .
\label{eq:BSF_M_DSToDDSingletSpin1_Hemission}
\end{align}
Here $\aB = (-\alpha_1+3\alpha_2)/4$, and we have included the symmetry factors of the scattering and bound state wavefunctions. Squaring and summing over the gauge indices and spins, and using the overlap integral \cref{eq:OverlapIntegral_R_100}, we find
\begin{align}
\frac{1}{6}\sum_{\sigma, \rho} \sum_{i,h}
|{\cal M}^{\sigma\rho}_{i,h}|^2
\simeq
\frac{2^{18}\pi^2 \aH}{\aB^3} 
\(1+\frac{\aH}{\aB}\)^2
\( \frac{\zetaB^2}{1+\zetaB^2} \)
\Ssc (-\zetaH,\zetaB) ,
\label{eq:BSF_Msquared_DSToDDSingletSpin1_Hemission}
\end{align}
where we averaged over the six dof of the spin-1 $\SUL$ doublet scattering state. The cross-section is obtained from \cref{eq:BSF_sigma_Scalar} and is shown in \cref{tab:BSF_DD}.

\clearpage
\subsection{$DS$ bound states: $(\mathbb{2},1/2)$, spin 0, $n\ell m =\{100\}$ \label{sec:BSF_SD}}

%%%%%%%%%%%%%%%%%%%%%%%%%%%%%%%%%%%%%%%%%%%%%%%%%%%%%%%%%%%%%%%%%%%%
\begin{table}[t!]
\centering
%%%%%%%%%%%%%%%%%%%%%%%%%%%%%%%%%%%%%%%%%%%%%%%%%%%%%%%%%%%%%%%%%%%%%%%%%%%%%%%%%%%%%%%%%%%%%%%%%%%%%%%%
%%%%%%%%%%%%%%%%%%%%%%%%%%%%%%%%%%%%%%%%%%%%%%%%%%%%%%%%%%%%%%%%%%%%%%%%%%%%%%%%%%%%%%%%%%%%%%%%%%%%%%%%
%%%%%%%%%%%%%%%%%%%%%%%%%%%%%%%%%%%%%%%%%%%%%%%%%%%%%%%%%%%%%%%%%%%%%%%%%%%%%%%%%%%%%%%%%%%%%%%%%%%%%%%%
\begin{align*}
\begin{array}{|c|c|c|c|c|c|c|}
%%%%%%%%%%%%%%%%%%%%%%%%%%%%%%%%%%%%%%%%%%%%%%%%%%%%%%%%%%%%%%%%%%%%%%%%%%%%%%%%%%%%%%%%%%%%%%%%%%%%%%%%
\multicolumn{7}{c}{\boldsymbol{
\textbf{Bound state}~DS:
~~~(\mathbb{2},1/2),
~~\textbf{spin 0}, 
~~\{n\ell m\} = \{100\}
}}
\\ \hline
%%%%%%%%%%%%%%%%%%%%%%%%%%%%%%%%%%%%%%%%%%%%%%%%%%%%%%%%%%%%%%%%%%%%%%%%%%%%%%%%%%%%%%%%%%%%%%%%%%%%%%%%
\multicolumn{5}{|c|}{\text{Scattering state (spin 0)}}
&\multirow{2}{*}{\text{\parbox{6.5ex}{\centering Rad Boson}}}
&\text{Cross-section}
\\ \cline{1-5}
  \text{State}		
& U_{\mathsmaller{Y}}(1) 
&SU_{\mathsmaller{L}}(2) 
& \text{dof} 
& \ellS 
&
&(\sigma_\BSF \vrel) / (\pi m^{-2})
\\ \hline\hline
%%%%%%%%%%%%%%%%%%%%%%%%%%%%%%%%%%%%%%%%%%%%%%%%%%%%%%%%%%%%%%%%%%%%%%%%%%%%%%%%%%%%%%%%%%%%%%%%%%%%%%%%
DS 					
&+1/2 
& \mathbb{2} 
& 2  
& 1 
& B 		
& 2^{5} 3\: \aH \alpha_1 \, \Svec (-\zetaH,\zetaH)
\\ \hline
%%%%%%%%%%%%%%%%%%%%%%%%%%%%%%%%%%%%%%%%%%%%%%%%%%%%%%%%%%%%%%%%%%%%%%%%%%%%%%%%%%%%%%%%%%%%%%%%%%%%%%%%
DS 					
&+1/2 
& \mathbb{2} 
& 2  
& 1 
& W 		
& 2^{5} 3^2 \aH \alpha_2 \, \Svec (-\zetaH,\zetaH)
\\ \hline
%%%%%%%%%%%%%%%%%%%%%%%%%%%%%%%%%%%%%%%%%%%%%%%%%%%%%%%%%%%%%%%%%%%%%%%%%%%%%%%%%%%%%%%%%%%%%%%%%%%%%%%%
\multirow{2}{*}{$SS$-like}		
& \multirow{2}{*}{0}
& \multirow{2}{*}{$\mathbb{1}$} 
& \multirow{2}{*}{1}  
& \multirow{2}{*}{0} 
& \multirow{2}{*}{$H^\dagger$} 
& \dfrac{2^{12} \aA^2}{(\aA+\aR)^2} 
\[\(1+\sqrt{\dfrac{\aR}{8\aA}}\) \(1+\dfrac{\aR}{\aH}\) \Ssc^{1/2} (-\zetaR,\zetaH) \right.
\\[1em] 
&
&
&
&
&
& \left. 
-\(\dfrac{\aR}{\aA}-\sqrt{\dfrac{\aR}{8\aA}}\) 
\(1-\dfrac{\aA}{\aH}\) \Ssc^{1/2} (\zetaA,\zetaH) 
\]^2 \hH(\omega)
\\ \hline
%%%%%%%%%%%%%%%%%%%%%%%%%%%%%%%%%%%%%%%%%%%%%%%%%%%%%%%%%%%%%%%%%%%%%%%%%%%%%%%%%%%%%%%%%%%%%%%%%%%%%%%%
\multirow{2}{*}{$D\bar{D}$-like}		
& \multirow{2}{*}{0}
& \multirow{2}{*}{$\mathbb{1}$} 
& \multirow{2}{*}{1}  
& \multirow{2}{*}{0} 
& \multirow{2}{*}{$H^\dagger$}
& \dfrac{2^{8} \aA^2}{(\aA+\aR)^2} \[
\!\(\dfrac{\aR}{\aA} - \sqrt{\dfrac{8\aR}{\aA}}\) 
\!\(1+\dfrac{\aR}{\aH}\) \Ssc^{1/2} (-\zetaR,\zetaH) \right.
\\[1em] 
&
&
&
&
&
& \left. 
+\(1 + \sqrt{\dfrac{8\aR}{\aA}}\) 
\(1-\dfrac{\aA}{\aH}\) \Ssc^{1/2} (\zetaA,\zetaH) 
\]^2 \hH(\omega)
\\\hline
%%%%%%%%%%%%%%%%%%%%%%%%%%%%%%%%%%%%%%%%%%%%%%%%%%%%%%%%%%%%%%%%%%%%%%%%%%%%%%%%%%%%%%%%%%%%%%%%%%%%%%%%
D\bar{D} 			
&  0  
& \mathbb{3} 
& 3  
& 0 
& H^\dagger 
& 2^8 \(1-\dfrac{\alpha_1-\alpha_2}{\aH}\)^2
\Ssc \( \dfrac{\zeta_1-\zeta_2}{4},\zetaH\) \, \hH(\omega) 
\\ \hline
%%%%%%%%%%%%%%%%%%%%%%%%%%%%%%%%%%%%%%%%%%%%%%%%%%%%%%%%%%%%%%%%%%%%%%%%%%%%%%%%%%%%%%%%%%%%%%%%%%%%%%%%
DD 					
&  1  
& \mathbb{1} 
& 1  
& 0 
& H 		
& \text{0 (due to antisymmetry of $DD$ scattering state)}
\\ \hline
%%%%%%%%%%%%%%%%%%%%%%%%%%%%%%%%%%%%%%%%%%%%%%%%%%%%%%%%%%%%%%%%%%%%%%%%%%%%%%%%%%%%%%%%%%%%%%%%%%%%%%%%
DD 					
&  1  
& \mathbb{3} 
& 3  
& 0 
& H 		
& 2^{11} \(1+\dfrac{\alpha_1+\alpha_2}{\aH}\)^2  
\Ssc \(-\dfrac{\zeta_1+\zeta_2}{4},\zetaH\) \, \hH(\omega)
\\ \hline
%%%%%%%%%%%%%%%%%%%%%%%%%%%%%%%%%%%%%%%%%%%%%%%%%%%%%%%%%%%%%%%%%%%%%%%%%%%%%%%%%%%%%%%%%%%%%%%%%%%%%%%%
\end{array}
\end{align*}
\captionof{table}{\label{tab:BSF_SD}
Same as \cref{tab:BSF_SD} for the $DS$ bound states. For the $SS$-like and $D\bar{D}$-like states, $\aA$ and $\aR$ should be evaluated from \cref{eq:GaugeSinglet_alphas_def} for $\ellS=s=0$. Here, $\omega = m(\aH^2+\vrel^2)/4$.  All processes have conjugate counterparts.}
\end{table}
%%%%%%%%%%%%%%%%%%%%%%%%%%%%%%%%%%%%%%%%%%%%%%%%%%%%%%%%%%%%%%%%%%%%
%\medskip
%%%%%%%%%%%%%%%%%%%%%%%%%%%%%%%%%%%%%%%%%%%%%%%%%%%%%%%%%%%%%%%%%%%%
\begin{figure}[t!]
\input{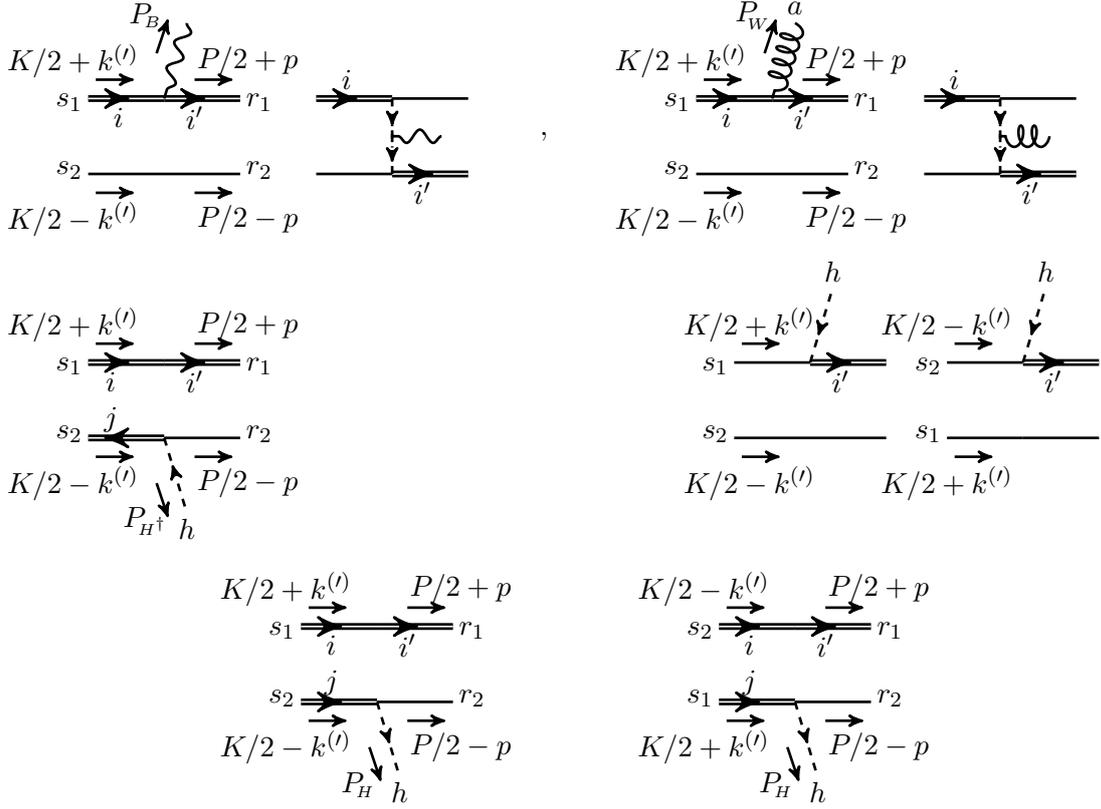}
\captionof{figure}{\label{fig:BSF_SD}
Same as \cref{fig:BSF_SSDDbar}, for the formation of $DS$ bound states. 
\emph{Top row:} $DS \to {\cal B}(DS) + B$ (left) and $DS \to {\cal B}(DS) + W$ (right). 
\emph{Middle row:} $D\bar{D}$-like $\to {\cal B}(DS) + H^\dagger$ and $SS$-like $\to {\cal B}(DS) + H^\dagger$.   The diagram on the left gives also $D\bar{D} \to {\cal B}(DS) + H^\dagger$. 
\emph{Bottom row:} $DD \to {\cal B}(DS) + H$.
}
\end{figure}
%%%%%%%%%%%%%%%%%%%%%%%%%%%%%%%%%%%%%%%%%%%%%%%%%%%%%%%%%%%%%%%%%%%%

The BSF processes are listed in \cref{tab:BSF_SD}, and the radiative part of the diagrams contributing to these processes are shown in \cref{fig:BSF_SD}. 
We project the bound-state fields on the spin-0 state via 
$U_{\text{spin-0}}^{s_1 s_2} [U_{\text{spin-0}}^{-1}]^{r_2 r_1} 
= (\epsilon^{s_1s_2}/\sqrt{2}) (\epsilon^{r_1r_2}/\sqrt{2})$.

\subsubsection{$DS \to {\cal B} (DS) + B$}
The perturbative part of the amplitude is
\begin{align}
\im \boldsymbol{{\cal A}}_{i,i'}
\[ (DS)^{\text{spin-0}} \to (DS)^{\text{spin-0}} +  B \]
&\simeq \im \delta_{ii'}\,  g_1 \YD \, 2m \, 2 {\bf p} 
\, (2\pi)^3 \delta^3 ({\bf k'-p}-\PBvec/2)
\nn \\
&+ \im \delta_{ii'}\,  g_1 \YH \, y^2 \, 4m^2 \, \frac{2 ({\bf k'+p})}{[({\bf k+p})^2+\mH^2]^2} ,
\label{eq:BSF_A_DSToDS_Bemission}
\end{align}
where the fermion permutations alloted factors $(+1)$ and $(-1)$ to the $t$ and $u$ channels.  Upon projection on the spin-0 states, the $u$ channel acquired another factor $(-1)$. Using the wavefunctions of \cref{tab:ScatteringStates,tab:BoundStates}, we find the full amplitude \eqref{eq:BSF_FullAmplitude_Def} to be
\begin{align}
\im \boldsymbol{{\cal M}}_{i,i'}
&\simeq \im \delta_{ii'}\,  \sqrt{4\pi \alpha_1 \, 4m} \, 
\[ \boldsymbol{{\cal J}}_{{\bf k},100} (-\aH,\aH)
+2\boldsymbol{{\cal Y}}_{{\bf k},100}^\H (-\aH,\aH) \],
\label{eq:BSF_M_DSToDS_Bemission}
\end{align}
Next we square, sum over the gauge indices and average over the two dof of the spin-0 scattering state. Using the overlap integrals \eqref{eq:OverlapIntegrals_100}, we obtain
\begin{align}
\frac{1}{2} \sum_{i,i'} |\boldsymbol{{\cal M}}_{i,i'}|^2
&\simeq \dfrac{2^{11} 3^2 \pi^2 \alpha_1}{\aH} 
\(\dfrac{\zetaH^2}{1+\zetaH^2}\) \Svec(-\zetaH,\zetaH) .
\label{eq:BSF_Msquared_DSToDS_Bemission}
\end{align}
The cross-section is found from \cref{eq:BSF_sigma_Vector} for $\aB = \aH$ and is shown in \cref{tab:BSF_SD}.

\subsubsection{$DS \to {\cal B} (DS) + W$}
The perturbative part of the amplitude is
\begin{align}
\im \boldsymbol{{\cal A}}_{i,i'}^a
\[ (DS)^{\text{spin-0}} \to (DS)^{\text{spin-0}} +  W \]
&\simeq \im t^a_{i'i}\,  g_2 \, 2m \, 2 {\bf p} 
\, (2\pi)^3 \delta^3 ({\bf k'-p}-\PBvec/2)
\nn \\
&+ \im t^a_{i'i}\,  g_2 \, y^2 \, 4m^2 \, \frac{2 ({\bf k'+p})}{[({\bf k+p})^2+\mH^2]^2} ,
\label{eq:BSF_A_DSToDS_Wemission}
\end{align}
where the signs are determined as in \cref{eq:BSF_A_DSToDS_Bemission}. The full amplitude is
\begin{align}
\im \boldsymbol{{\cal M}}_{i,i'}^a
&\simeq \im t^a_{i'i}\,  \sqrt{4\pi \alpha_2 \, 2^4 m} \, 
\[ \boldsymbol{{\cal J}}_{{\bf k},100} (-\aH,\aH)
+2\boldsymbol{{\cal Y}}_{{\bf k},100}^\H (-\aH,\aH) \],
\label{eq:BSF_M_DSToDS_Wemission}
\end{align}
and
\begin{align}
\frac{1}{2} \sum_{i,i',a} |\boldsymbol{{\cal M}}_{i,i'}^a|^2
&\simeq \dfrac{2^{11} 3^3 \pi^2 \alpha_2}{\aH}
\(\dfrac{\zetaH^2}{1+\zetaH^2}\) \Svec(-\zetaH,\zetaH) ,
\label{eq:BSF_Msquared_DSToDS_Wemission}
\end{align}
where we used ${\rm Tr}(t^a t^a) = 3/2$. The cross-section is found from \cref{eq:BSF_sigma_Vector} for $\aB = \aH$ and is shown in \cref{tab:BSF_SD}.

\subsubsection{$SS$-like $\to {\cal B} (DS) + H^\dagger$}
The perturbative parts of the amplitude are
\begin{subequations}
\begin{align}
\im {\cal A}_{i'h}
\[ (SS)^{\text{spin-0}} \to (DS)^{\text{spin-0}} +  H^\dagger \]
\simeq
&-\im \delta_{ih} \, y \, 4m^2 \, (2\pi)^3 \delta^3 ({\bf k'-p}-\PHdaggervec/2)
\nn \\
&-\im \delta_{ih} \, y \, 4m^2 \, (2\pi)^3 \delta^3 ({\bf k'+p}+\PHdaggervec/2)
\label{eq:BSF_A_SSToDS_Hdaggeremission}
\\
\im {\cal A}_{ij,i'h}
\[ (D\bar{D})^{\text{spin-0}} \to (DS)^{\text{spin-0}} +  H^\dagger \]
\simeq 
&\im \delta_{ii'} \delta_{jh} \, y \, 4m^2 
(2\pi)^3 \delta^3 ({\bf k'-p}+\PHdaggervec/2) .
\label{eq:BSF_A_DDbarToDS_Hdaggeremission}
\end{align}
In \cref{eq:BSF_A_SSToDS_Hdaggeremission}, the fermion permutations alloted signs $(+1)$ and $(-1)$ factors to the $t$ and $u$ channels. Upon projection on the spin-0 state, the $u$ channel acquired another factor $(-1)$. We now project the $D\bar{D}$ component of the scattering state in \cref{eq:BSF_A_DDbarToDS_Hdaggeremission}, on the $\SUL$ singlet configuration via $\delta_{ij}/\sqrt{2}$,
\begin{align}
\im {\cal A}_{i'h}
\[ (D\bar{D})^{\text{spin-0}}_{(\mathbb{1},0)} \to (DS)^{\text{spin-0}} +  H^\dagger \]
\simeq 
&\im \: \frac{\delta_{i'h}}{\sqrt{2}} \, y \, 4m^2 
(2\pi)^3 \delta^3 ({\bf k'-p}+\PHdaggervec/2) .
\label{eq:BSF_A_DDbarSingletToDS_Hdaggeremission}
\end{align}
\end{subequations}

Considering the wavefunctions of \cref{tab:ScatteringStates,tab:BoundStates}, the full amplitude \eqref{eq:BSF_FullAmplitude_Def} is
\begin{align}
\im {\cal M}_{i'h} 
&\simeq-\im \delta_{i'h}
\sqrt{\dfrac{2^{12} \pi}{\aH^2}}  \(\dfrac{\aA}{\aA+\aR}\) \times
\nn \\
&\times 
\[\(1+\sqrt{\dfrac{\aR}{8\aA}}\)  {\cal R}_{{\bf k},100} (-\aR,\aH)
-\(\dfrac{\aR}{\aA} - \sqrt{\dfrac{\aR}{8\aA}}\) {\cal R}_{{\bf k},100} (\aA,\aH) \] ,
\label{eq:BSF_M_SSlikeToDS_Hdaggeremission}
\end{align}
where $\aA$ and $\aR$ should be evaluated from \cref{eq:GaugeSinglet_alphas_def} for $\ellS=s=0$. Using the overlap integrals \eqref{eq:OverlapIntegrals_100}, 
\begin{align}
&\sum_{i',h} | {\cal M}_{i'h} |^2
\simeq
\dfrac{2^{19} \pi^2}{\aH^2}  
\(\dfrac{\aA}{\aA+\aR}\)^2 
\(\dfrac{\zetaH^2}{1+\zetaH^2}\)
\times
\label{eq:BSF_Msquared_SSlikeToDS_Hdaggeremission}
\\
&\times 
\[\(1+\sqrt{\dfrac{\aR}{8\aA}}\) \(1+\dfrac{\aR}{\aH}\)  \Ssc^{1/2} (-\zetaR,\zetaH)
-\(\dfrac{\aR}{\aA} - \sqrt{\dfrac{\aR}{8\aA}}\)  \(1-\dfrac{\aA}{\aH}\) \Ssc^{1/2} (\zetaA,\zetaH) \]^2 .
\nn 
\end{align}
The cross-section is found from \cref{eq:BSF_sigma_Scalar} for $\aB = \aH$ and is shown in \cref{tab:BSF_SD}.

\subsubsection{$D\bar{D}$-like $\to {\cal B} (DS) + H^\dagger$}
Starting from the perturbative parts \eqref{eq:BSF_A_SSToDS_Hdaggeremission} and \eqref{eq:BSF_A_DDbarSingletToDS_Hdaggeremission}, and considering the wavefunctions of \cref{tab:ScatteringStates,tab:BoundStates}, the full amplitude \eqref{eq:BSF_FullAmplitude_Def} is
\begin{align}
\im {\cal M}_{i'h} 
&\simeq \im \delta_{i'h}
\sqrt{\dfrac{2^{8} \pi}{\aH^2}}  \(\dfrac{\aA}{\aA+\aR}\) \times
\nn \\
&\times 
\[\(\dfrac{\aR}{\aA} - \sqrt{\dfrac{8\aR}{\aA}}\)  {\cal R}_{{\bf k},100} (-\aR,\aH)
 +\(1 + \sqrt{\dfrac{8\aR}{\aA}}\) {\cal R}_{{\bf k},100} (\aA,\aH) \] ,
\label{eq:BSF_M_DDbarlikeToDS_Hdaggeremission}
\end{align}
where again $\aA$ and $\aR$ should be evaluated from \cref{eq:GaugeSinglet_alphas_def} for $\ellS=s=0$. Using the overlap integrals \eqref{eq:OverlapIntegrals_100}, 
\begin{align}
&\sum_{i',h} | {\cal M}_{i'h} |^2
\simeq
\dfrac{2^{15} \pi^2}{\aH^2}  
\(\dfrac{\aA}{\aA+\aR}\)^2 
\(\dfrac{\zetaH^2}{1+\zetaH^2}\)
\times
\label{eq:BSF_Msquared_DDbarlikeToDS_Hdaggeremission}
\\
&\times 
\[\(\dfrac{\aR}{\aA} - \sqrt{\dfrac{8\aR}{\aA}}\) \(1+\dfrac{\aR}{\aH}\)  \Ssc^{1/2} (-\zetaR,\zetaH)
 +\(1 + \sqrt{\dfrac{8\aR}{\aA}}\) \(1-\dfrac{\aA}{\aH}\) \Ssc^{1/2} (\zetaA,\zetaH) \]^2 .
\nn 
\end{align}
The cross-section is found from \cref{eq:BSF_sigma_Scalar} for $\aB = \aH$ and is shown in \cref{tab:BSF_SD}.

\subsubsection{$D\bar{D} \to {\cal B} (DS) + H^\dagger$}
The perturbative part of the amplitude is given in \cref{eq:BSF_A_DDbarToDS_Hdaggeremission}. We project it the $D\bar{D}$ scattering state on the $\SUL$ triplet configuration via 
$t^a_{ji} /\sqrt{C(\mathbb{2})} = \sqrt{2}t^a_{ji}$, where $C(\mathbb{2})=1/2$ is the Casimir of the $\SUL$ doublet representation, and obtain
\begin{align}
\im {\cal A}_{i'h}^a
\[ (D\bar{D})^{\text{spin-0}}_{(\mathbb{3,0})} \to (DS)^{\text{spin-0}}_{(\mathbb{2,1/2})} +  H^\dagger \]
&\simeq \im \sqrt{2} t^a_{hi'} y \, 4m^2 
\, (2\pi)^3 \delta^3 ({\bf k'-p}+\PHdaggervec/2) .
\label{eq:BSF_A_DDbarTripletToDS_Hdaggeremission}
\end{align}
Considering the wavefunctions of \cref{tab:ScatteringStates,tab:BoundStates}, the full amplitude \eqref{eq:BSF_FullAmplitude_Def} is
\begin{align}
\im {\cal M}_{i'h}^a
&\simeq \im (\sqrt{2}t^a_{hi'}) \sqrt{\dfrac{2^{9}\pi}{\aH^2}}  
\ {\cal R}_{{\bf k},100} \(\frac{\alpha_1-\alpha_2}{4},\aH\) .
\label{eq:BSF_M_DDbarToDS_Hdaggeremission}
\end{align}
Squaring, summing over the final state gauge indices, and averaging over the three dof of the scattering state, we obtain
\begin{align}
\frac{1}{3} \sum_{i',h,a}
|{\cal M}_{i'h}^a|^2
&\simeq 
\dfrac{2^{15}\pi^2}{\aH^2}
\(1-\dfrac{\alpha_1-\alpha_2}{\aH}\)^2
\(\dfrac{\zetaH^2}{1+\zetaH^2}\)
\Ssc \(\frac{\zeta_1-\zeta_2}{4},\zetaH \) ,
\label{eq:BSF_Msquared_DDbarToDS_Hdaggeremission}
\end{align}
where we used ${\rm Tr}(\sqrt{2}t^a \sqrt{2}t^a) = 3$. The cross-section is found from \cref{eq:BSF_sigma_Scalar} for $\aB = \aH$ and is shown in \cref{tab:BSF_SD}.

\subsubsection{$DD \to {\cal B} (DS) + H$}

The perturbative part of the amplitude is
\begin{multline}
\im {\cal A}_{ij,i'h}
\[ (DD)^{\text{spin-0}} \to (DS)^{\text{spin-0}} +  H \] \simeq
\\
\simeq-\im y \, 4m^2 \[ 
 \delta_{ii'} \delta_{jh} (2\pi)^3 \delta^3 ({\bf k'-p}+\PHvec/2)
+\delta_{ji'} \delta_{ih} (2\pi)^3 \delta^3 ({\bf k'+p}-\PHvec/2)
\] ,
\label{eq:BSF_A_DDToDS_Hemission}
\end{multline}
where the fermion permutations alloted signs $(+1)$ and $(-1)$ factors to the $t$ and $u$ channels. Upon projection on the spin-0 state, the $u$ channel acquired another factor $(-1)$. 
We now project the $DD$ scattering state on the $\SUL$ triplet configuration via the symbolic operator $(U_{\mathbb{3}})_{ij}^a = (U_{\mathbb{3}})_{ji}^a$, which satisfies 
$(U_{\mathbb{3}})_{ij}^a (U_{\mathbb{3}}^\dagger)_{ji}^b = \delta^{ab}$,
\begin{multline}
\im {\cal A}_{i'h}^a
\[ (DD)^{\text{spin-0}}_{(\mathbb{3},1)} \to (DS)^{\text{spin-0}}_{(\mathbb{2},1/2)} +  H \] \simeq
\\
\simeq-\im (U_{\mathbb{3}})_{i'h} \: y \, 4m^2 \[ 
(2\pi)^3 \delta^3 ({\bf k'-p}+\PHvec/2) + (2\pi)^3 \delta^3 ({\bf k'+p}-\PHvec/2)
\] .
\label{eq:BSF_A_DDToDS_Hemission_proj}
\end{multline}
Considering the wavefunctions of \cref{tab:ScatteringStates,tab:BoundStates}, the full amplitude \eqref{eq:BSF_FullAmplitude_Def} is
\begin{align}
\im {\cal M}_{i'h}^a \simeq
-\im (U_{\mathbb{3}})_{i'h} \sqrt{\frac{2^{12} \pi}{\aH^2}} \,
{\cal R}_{{\bf k},100} \(-\dfrac{\alpha_1+\alpha_2}{4},\aH\) ,
\label{eq:BSF_M_DDToDS_Hemission_proj}
\end{align}
where we included the symmetry factor of the $DD$ wavefunction. Squaring, summing over the final state gauge indices, and averaging over the three dof of the scattering state, and using the overlap integral \eqref{eq:OverlapIntegral_R_100}, we obtain
\begin{align}
\frac{1}{3} \sum_{i',h,a}
|{\cal M}_{i'h}^a|^2
&\simeq 
\dfrac{2^{18}\pi^2}{\aH^2}
\(1+\dfrac{\alpha_1+\alpha_2}{\aH}\)^2
\(\dfrac{\zetaH^2}{1+\zetaH^2}\)
\Ssc \(-\dfrac{\zeta_1+\zeta_2}{4},\zetaH \) .
\label{eq:BSF_Msquared_DDToDS_Hemission}
\end{align}
The cross-section is found from \cref{eq:BSF_sigma_Scalar} for $\aB = \aH$ and is shown in \cref{tab:BSF_SD}.

%%%%%%%%%%%%%%%%%%%%%%%%%%%%%%%%%%%%%%%%%%%%%%%%%%%%%%%%%%%%%%%%%%%%
\begin{figure}[t!]
\centering
\includegraphics[height=0.71\textheight]{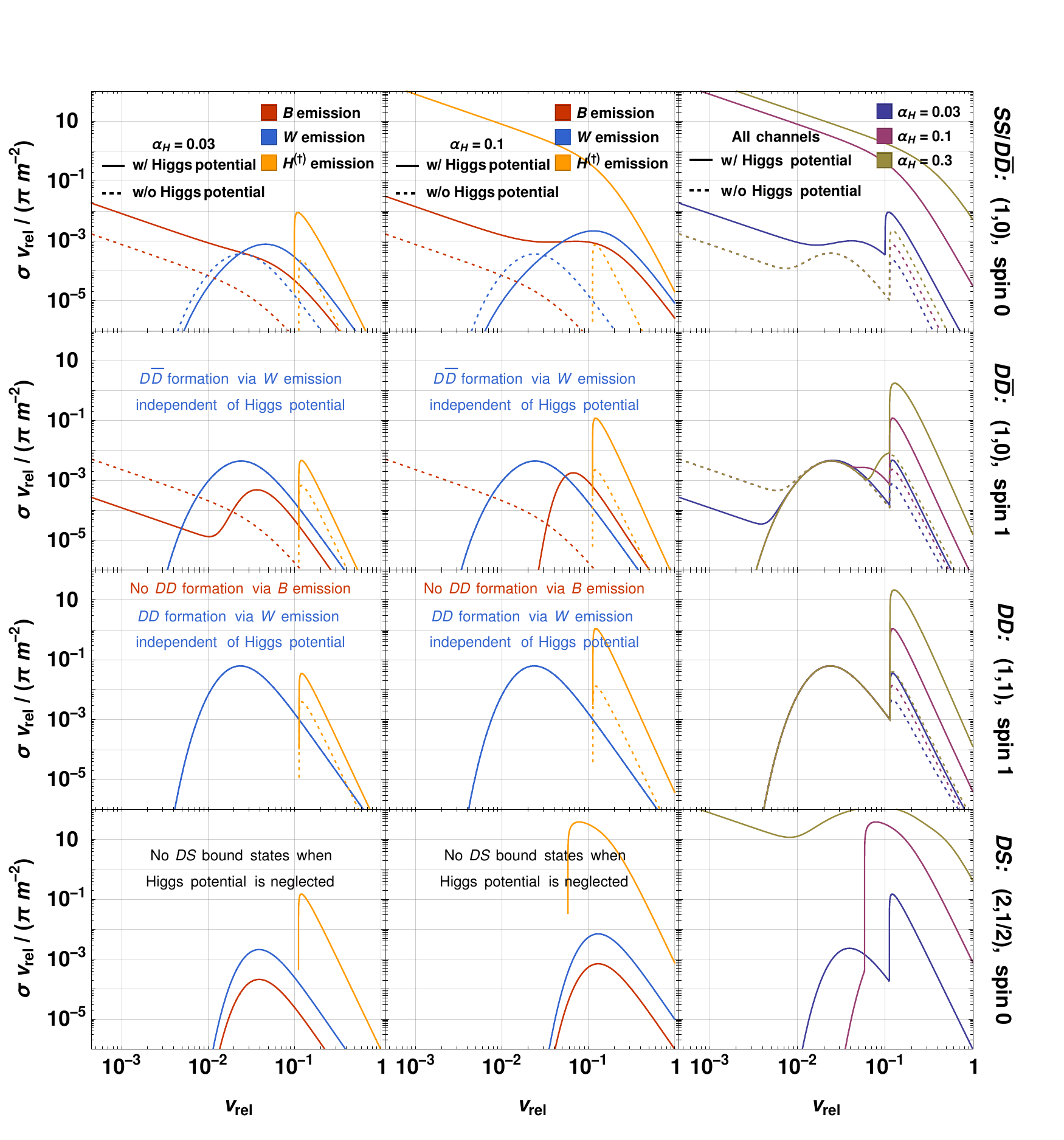}
\caption{\label{fig:BSF_onshell_CrossSections} 
The radiative BSF cross-sections vs relative velocity. The four rows correspond to capture into the ground levels of the bound states marked on the right. 
In the $DD$ and $DS$ panels, we have included the capture into the conjugate bound states, and all BSF channels have been weighted with the number of DM particles eliminated in each process as estimated upon thermal averaging (cf.~ref.~\cite{Oncala:2021swy}.)
%%%
\emph{Left and middle columns}: The contributions of $B$, $W$ and $H^{(\dagger)}$ emission, for $\aH = 0.03$ and $0.1$. Comparing with \cref{fig:Annihilation} at $\vrel \sim 0.1$ relevant for DM freeze-out, BSF can be comparable to or significantly faster (in the case of $H^{(\dagger)}$ emission) than annihilation, $\sigma_{\ann} \vrel /(\pi m^{-2}) \sim 10^{-3}$.
%%%
\emph{Right column}: The sum of the $B$-, $W$- and $H^{(\dagger)}$-emission contributions, for different values of $\aH$. 
%%% 
In both columns, we show the cross-sections considering and neglecting the Higgs-mediated potential. 
%%%
The various $\sigma \vrel$ normalised to $\pi m^{-2}$ are independent of the DM mass, except for the cutoff on BSF via $H^{(\dagger)}$ emission due the Higgs mass; for this purpose, we have used $m=50$~TeV and temperature $T=300$~GeV which sets $\mH \simeq 168$~GeV. 
Note that we take the Higgs potential to be Coulombic (see text for discussion.)
}
\end{figure}
%%%%%%%%%%%%%%%%%%%%%%%%%%%%%%%%%%%%%%%%%%%%%%%%%%%%%%%%%%%%%%%%%%%%

\clearpage
\subsection{Unitarity and BSF via Higgs emission \label{sec:BSF_Unitarity}}

The unitarity of the $S$ matrix implies an upper limit on the partial-wave inelastic cross-sections, $\sigma_{\ell}^{\rm inel} \leqslant \sigma_{\ell}^{\rm uni} = (2\ell +1)\pi/k^2$, where $\ell$ is the partial wave and $k$ is the momentum of either of the interacting particles in the CM frame. In the non-relativistic regime, $k=\mu \vrel$ with $\mu$ being the reduced mass, thus
\begin{align}
\sigma^{\rm uni}_{\ell} \vrel \simeq \dfrac{(2\ell+1) \pi}{\mu^2 \vrel}.
\label{eq:UnitarityLimit_sigmav}
\end{align}

As already discussed in ref.~\cite{Oncala:2019yvj}, the high efficiency of BSF via  charged scalar emission (here, the Higgs doublet) implies that the unitarity limit \eqref{eq:UnitarityLimit_sigmav} may be saturated already for rather small values of $\aH$. If the incoming particles interact via an attractive long-range force,  then this occurs for the continuum of velocities $\vrel \lesssim \aB$, otherwise only for a finite range or discrete values of $\vrel$ (cf.~\cref{fig:BSF_onshell_Unitarity}.) The apparent violation of unitarity at larger $\aH$ by the computations of \cref{sec:BSF_SSDDbar,sec:BSF_DDbar,sec:BSF_DD,sec:BSF_SD}, whether it occurs for an infinite or finite range of $\vrel$, indicates that these computations must be amended. At small values of $\aH$, higher order corrections to the perturbative transition amplitudes ${\cal A}_{\mathsmaller{T}}$ are expected to be insignificant. Restoring unitarity necessitates instead that the two-particle interactions at infinity are resummed~\cite{Baldes:2017gzw}.

The results of \cref{sec:BSF_SSDDbar,sec:BSF_DDbar,sec:BSF_DD,sec:BSF_SD} already include the resummation of the (leading-order) long-range interaction between the incoming particles, computed in \cref{sec:LongRangeDynamics_Potential}.  However, according to the optical theorem, all elastic and inelastic processes to which the incoming state may participate contribute to its self-energy. Typically, contact-type interactions can be neglected, as they do not distort significantly the wavefuctions of the interacting particles. Nevertheless, if a contact-type interaction is very strong --  as is the case when the corresponding cross-section approaches (or even appears to exceed) the unitarity limit \eqref{eq:UnitarityLimit_sigmav} -- then its contribution to the two-particle self-energy may be significant. 

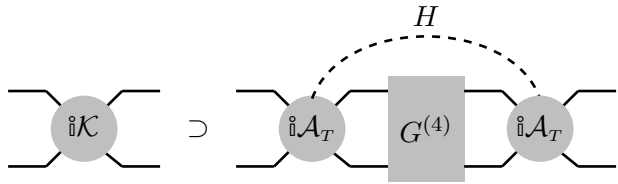
\begin{figure}[t!]
\centering	
\begin{tikzpicture}[line width=1pt, scale=1]
%%%%%%%%%%%%%%%%%%%%%%%%%%%%%%%%%%%%%%%%%%%%%%%%%%%%%%%%%%%%%%%%%%%%%%%%%%%%%%%%%%%%
\begin{scope}[shift={(-4.5,0)}]
%%%%%%%% Incoming Fermions (Scalars)
\draw (-1.0,+0.5) -- (-0.50,+0.50);\draw (-0.5,+0.5) -- (-0.05,+0.05);
\draw (-1.0,-0.5) -- (-0.50,-0.50);\draw (-0.5,-0.5) -- (-0.05,-0.05);
%%%%%%%% Intermediate Fermions (Scalars)
\draw (+0.05,+0.05) -- (0.5,+0.50);\draw (+0.50,+0.50) -- (1.0,+0.50);
\draw (-0.05,-0.05) -- (0.5,-0.50);\draw (+0.50,-0.50) -- (1.0,-0.50);
%%%%%%%% A node
\filldraw[lightgray]  (0,0) circle (12pt);
\node at ( 0.0,0) {$\im {\cal K}_{}$};
\node at ( 1.5,0) {$\supset$};
\end{scope}
%%%%%%%%%%%%%%%%%%%%%%%%%%%%%%%%%%%%%%%%%%%%%%%%%%%%%%%%%%%%%%%%%%%%%%%%%%%%%%%%%%%%
%%%%%%%%%%%%%%%%%%%%%%%%%%%%%%%%%%%%%%%%%%%%%%%%%%%%%%%%%%%%%%%%%%%%%%%%%%%%%%%%%%%%
\begin{scope}[shift={(0,0)}]
%%%%%%%%%%%%%%%%%%%%%%%%%%%%%%%%%%%%%%%%%
\begin{scope}[shift={(-1.5,0)}]
%%%%%%%% Incoming Fermions (Scalars)
\draw (-1.0,+0.5) -- (-0.50,+0.50);\draw (-0.5,+0.5) -- (-0.05,+0.05);
\draw (-1.0,-0.5) -- (-0.50,-0.50);\draw (-0.5,-0.5) -- (-0.05,-0.05);
%%%%%%%% Intermediate Fermions (Scalars)
\draw (+0.05,+0.05) -- (0.5,+0.50);\draw (+0.50,+0.50) -- (1.0,+0.50);
\draw (-0.05,-0.05) -- (0.5,-0.50);\draw (+0.50,-0.50) -- (1.0,-0.50);
%%%%%%%% A node
\filldraw[lightgray]  (0,0) circle (12pt);
\node at ( 0,0) {$\im {\cal A}_{\mathsmaller{T}}$};
\end{scope}
%%%%%%%%%%%%%%%%%%%%%%%%%%%%%%%%%%%%%%%%%
%%%%%%%% G4
\draw[fill=lightgray,draw=none,shift={(0,0)}] (-0.5,-0.7) rectangle (0.5,0.7);
\node at (0,0) {$G^{(4)}$};
%%%%%%%% H exchange
\draw[scalarnoarrow] (-1.5,0.4) to[out=70,in=110] (1.5,0.4);
\node at (0,1.5) {$H$};
%%%%%%%%%%%%%%%%%%%%%%%%%%%%%%%%%%%%%%%%%
\begin{scope}[shift={(+1.5,0)}]
%%%%%%%% Incoming Fermions (Scalars)
\draw (-1.0,+0.5) -- (-0.50,+0.50);\draw (-0.5,+0.5) -- (-0.05,+0.05);
\draw (-1.0,-0.5) -- (-0.50,-0.50);\draw (-0.5,-0.5) -- (-0.05,-0.05);
%%%%%%%% Intermediate Fermions (Scalars)
\draw (+0.05,+0.05) -- (0.5,+0.50);\draw (+0.50,+0.50) -- (1.0,+0.50);
\draw (-0.05,-0.05) -- (0.5,-0.50);\draw (+0.50,-0.50) -- (1.0,-0.50);
%%%%%%%% A node
\filldraw[lightgray]  (0,0) circle (12pt);
\node at ( 0,0) {$\im {\cal A}_{\mathsmaller{T}}$};
\end{scope}
%%%%%%%%%%%%%%%%%%%%%%%%%%%%%%%%%%%%%%%%%
\end{scope}
%%%%%%%%%%%%%%%%%%%%%%%%%%%%%%%%%%%%%%%%%%%%%%%%%%%%%%%%%%%%%%%%%%%%%%%%%%%%%%%%%%%%
\end{tikzpicture}
\caption{\label{fig:BSF_onshell_BSFresummation} The contributions to the 2PI kernels arising from inelastic processes that involve  Higgs emission. The solid lines stand for any of the $S$, $D$ or $\bar{D}$ particles, and $G^{(4)}$ includes their scattering and bound states. The dashed line represents the Higgs doublet. ${\cal A}_{\mathsmaller{T}}$ are the perturbative transition amplitudes with $H^{(\dagger)}$ emission, computed in \cref{sec:BSF_SSDDbar,sec:BSF_DDbar,sec:BSF_DD,sec:BSF_SD} }
\end{figure}

\begin{figure}[t!]
\centering
\includegraphics[height=0.71\textheight]{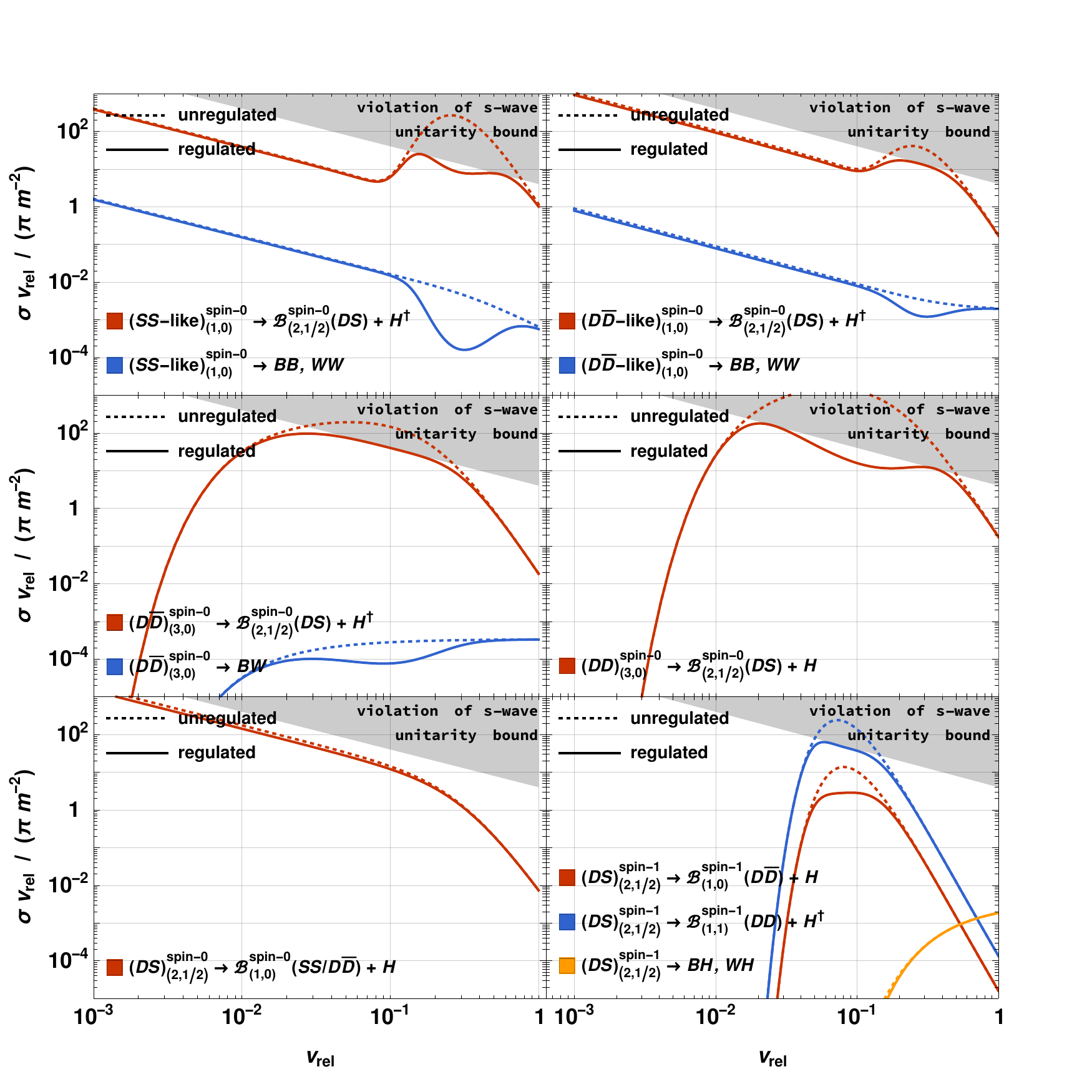}
\caption{\label{fig:BSF_onshell_Unitarity} 
The effect of the perscription \eqref{eq:BSF_UnitarityRegularisation} on the inelastic cross-sections. While only BSF via Higgs emission approaches or appears to violate the unitarity limit, the regurarisation affects all cross-sections with the same intial state.   
We have used $\aH = 0.2$, $m=50~\TeV$ and $T=300~\GeV$ which corresponds to $\mH\simeq 168~\GeV$.}
\end{figure}

The contributions to the 2PI kernels arising from inelastic processes that involve  Higgs emission are shown in \cref{fig:BSF_onshell_BSFresummation}. These diagrams include both scattering and bound intermediate states of the $S$, $D$ and $\bar{D}$ particles, and therefore include bremsstrahlung, BSF and bound-to-bound transitions. Note that in \cref{fig:BSF_onshell_BSFresummation} we have not included the resummation of the long-range kernels of \cref{sec:LongRangeDynamics_Potential} in the incoming and outgoing pairs, as this would result in double-counting; only 2PI diagrams must be included in the kernels that determine the potential.

The proper resummation of the diagrams of  \cref{fig:BSF_onshell_BSFresummation}  requires developing suitable formalism, and is beyond the scope of the present work. To ensure that our cross-sections are consistent with partial-wave unitarity, we shall instead adapt the result of ref.~\cite{Blum:2016nrz} that resummed the box diagrams arising from the perturbative part of $s$-wave annihilation into radiation (hard scattering), to compute the effect on the scattering-state wavefunctions and ultimately on the full cross-sections for $s$-wave annihilation into radiation. This procedure regulates the annihilation cross-sections as follows~\cite[eq.~(40)]{Blum:2016nrz},
\begin{align}
\sigma_{s\text{-wave}}^{\rm reg} (i \to f) =
\dfrac{\sigma_{s\text{-wave}} (i \to f)}
{\(1  +  \dfrac
{\sum_{f'} \sigma_{s\text{-wave}} (i \to f')}
{4\sigma_{s\text{-wave}}^{\rm uni}} 
\)^2} ,
\label{eq:BSF_UnitarityRegularisation}
\end{align}
where we have generalised the result of ref.~\cite{Blum:2016nrz} to multiple annihilation channels. \Cref{eq:BSF_UnitarityRegularisation} ensures that the unitarity limit is respected by the total $s$-wave inelastic cross-section, since it implies 
$r^{\rm reg} = r /(1+r/4)^2 \leqslant 1$, with 
$r^{\rm (reg)} \equiv \[\sum_f  \sigma_{s\text{-wave}}^{\rm (reg)} (i\to f)\] / \sigma_{s\text{-wave}}^{\rm uni}$.

We emphasise that the assumptions made in deriving \cref{eq:BSF_UnitarityRegularisation} are not strictly satisfied in our case, for at least two reasons: (i) Reference~\cite{Blum:2016nrz} assumed that for the resummed inelastic processes (hard scattering), $\sigma \vrel$ is independent of $\vrel$. For BSF via Higgs emission, the corresponding cross-sections can be found from \cref{tab:BSF_SSDDbar,tab:BSF_DDbar,tab:BSF_DD,tab:BSF_SD} by setting $\zetaS \to 0$; \cref{eq:Ssc} then shows that they depend on $\vrel$ for $\vrel \gtrsim \aB$. (ii) In the present case, the new contributions to the kernel may affect both the initial (scattering) and final (bound) state wavefunctions, while only the former is relevant for annihilation into radiation in the analysis of ref.~\cite{Blum:2016nrz}.  
Nevertheless, we shall adopt \cref{eq:BSF_UnitarityRegularisation} as a perscription that regulates the inelastic cross-sections in the velocity range where the base calculation violates unitarity, while leaving them essentially unaffected outside that range. 
We leave a more precise treatment for future work.

In \cref{fig:BSF_onshell_Unitarity}, we show how the prescription \eqref{eq:BSF_UnitarityRegularisation} adjusts the inelastic cross-sections for the scattering states that may participate in BSF via Higgs emission. We include the total inelastic cross-section (BSF plus annihilation) in the resummation, although only BSF is significant. However, both BSF and annihilation are regulated by the same factor, which implies that the annihilation cross-sections are affected significantly even while themselves being well below the unitarity limit.

\clearpage
\section{Bound-state formation and bound-to-bound transitions via scattering\label{Sec:BSF_OffShell}}

The dissipation of energy necessary for the capture into bound states or transitions between bound levels, may occur via exchange of an \emph{off-shell} mediator with particles of the thermal bath~\cite{Kim:2016zyy,Biondini:2017ufr,Biondini:2018pwp,Biondini:2018ovz,Biondini:2019int,Binder:2019erp,Binder:2020efn}.\footnote{Other rearragement processes have been considered in ref.~\cite{Geller:2018biy}.}  
References~\cite{Binder:2019erp,Binder:2020efn} showed, in the context of a $U(1)$ model, that the cross-sections for BSF via scattering factorise into the radiative ones (with any phase-space suppression due to the mass of the emitted vector removed), and a factor that depends on the thermal bath and the interaction that mediates the scattering.

In the following, we consider BSF and bound-to-bound transitions via off-shell Higgs exchange. We derive a similar factorisation and then compute the BSF cross-sections and transition rates. We also adapt the results of refs.~\cite{Binder:2019erp,Binder:2020efn} to our model, for BSF and transitions via off-shell $B$ and $W$ exchange.

\subsection{$H$ exchange \label{sec:BSF_OffShell_Higgs}}

\begin{figure}[t!]
\centering	
\begin{tikzpicture}[line width=1pt, scale=1]
%%%%%%%%%%%%%%%%%%%%%%%%%%%%%%%%%%%%%%%%%%%%%%%%%%%%%%%%%%%%%%%%%%%%%%%%%%%%%%%%%%%%
\begin{scope}[shift={(0,0)}]
%%%%%%%% Incoming Fermions (Scalars)
\draw (-2.5,+0.5) -- (-0.50,+0.50);
\draw (-0.5,+0.5) -- (-0.05,+0.05);
\draw (-0.5,-0.5) -- (-0.05,-0.05);
\draw (-2.5,-0.5) -- (-0.50,-0.50);
%%%%%%%% Outgoing Fermions (Scalars)
\draw (2.5,+0.5) -- (0.5,+0.5);
\draw (0.5,+0.5) -- (0.1,+0.1);
\draw (0.5,-0.5) -- (0.1,-0.1);
\draw (2.5,-0.5) -- (0.5,-0.5);
%%%%%%%% Radiation and thermal bath
\draw[scalar] (0,0.25) -- (0,1.5);
\draw (-2.5,1.5) -- (2.5,1.5);
\node at (-1.25,2) {$
\begin{array}{ccc}	
 e_{\mathsmaller{R}},
&d_{\mathsmaller{R}}, 
&\overline{u}_{\mathsmaller{R}}
\\
 \overline{L}_{\mathsmaller{L}},
&\overline{Q}_{\mathsmaller{L}},
&Q_{\mathsmaller{L}}
\end{array}
$};
\node at (+1.25,2) {$
\begin{array}{ccc}		
 L_{\mathsmaller{L}},
&Q_{\mathsmaller{L}},
&\overline{Q}_{\mathsmaller{L}}
\\
 \overline{e}_{\mathsmaller{R}},
&\overline{d}_{\mathsmaller{R}},
&u_{\mathsmaller{R}}
\end{array}
$};
\draw[->] (-0.2,0.8) -- (-0.2,1.3);\node at (-0.50,1) {$\PH$};
\draw[->] (-2.5,1.3) -- (-2.0,1.3);\node at (-2.25,1) {$q_i$};
\draw[->] (+2.0,1.3) -- (+2.5,1.3);\node at (+2.25,1) {$q_f$};
%%%%%%%% G4, A functions and labels
\draw[fill=lightgray,draw=none,shift={(-1.5,0)}] (-0.6,-0.7) rectangle (0.6,0.7);
\draw[fill=lightgray,draw=none,shift={(+1.5,0)}] (-0.6,-0.7) rectangle (0.6,0.7);
\filldraw[lightgray]  (0,0) circle (12pt);
%%%
\node at (-1.5,0) {$G^{(4)}_{\rm in}$};
\node at (+1.5,0) {$G^{(4)}_{\rm out}$};
\node at ( 0,  0) {$\im {\cal A}_{\mathsmaller{T}}$};
%%%%%%%%
\end{scope}
%%%%%%%%%%%%%%%%%%%%%%%%%%%%%%%%%%%%%%%%%%%%%%%%%%%%%%%%%%%%%%%%%%%%%%%%%%%%%%%%%%%%
%%%%%%%%%%%%%%%%%%%%%%%%%%%%%%%%%%%%%%%%%%%%%%%%%%%%%%%%%%%%%%%%%%%%%%%%%%%%%%%%%%%%
\begin{scope}[shift={(-4,-3.5)}]
%%%%%%%% Incoming Fermions / Scalars
\draw (-2.5,+0.5) -- (-0.50,+0.50);
\draw (-0.5,+0.5) -- (-0.05,+0.05);
\draw (-0.5,-0.5) -- (-0.05,-0.05);
\draw (-2.5,-0.5) -- (-0.50,-0.50);
%%%%%%%% Outgoing Fermions / Scalars
\draw (2.5,+0.5) -- (0.5,+0.5);
\draw (0.5,+0.5) -- (0.1,+0.1);
\draw (0.5,-0.5) -- (0.1,-0.1);
\draw (2.5,-0.5) -- (0.5,-0.5);
%%%%%%%% Radiation and thermal bath
\draw[scalar] (0,0.25) -- (0,1.5);
\draw[vector] (-2.5,1.5) -- (0,1.5);
\draw[scalar] (0,1.5) -- (2.5,1.5);
\node at (-1.25,1.9) {$B^{\mu},~W^{a,\mu}$};
\node at (+1.25,1.9) {$H$};
\draw[->] (-0.2,0.8) -- (-0.2,1.3);\node at (-0.50,1) {$\PH$};
\draw[->] (-2.5,1.3) -- (-2.0,1.3);\node at (-2.25,1) {$q_i$};
\draw[->] (+2.0,1.3) -- (+2.5,1.3);\node at (+2.25,1) {$q_f$};
%%%%%%%% G4, A functions and labels
\draw[fill=lightgray,draw=none,shift={(-1.5,0)}] (-0.6,-0.7) rectangle (0.6,0.7);
\draw[fill=lightgray,draw=none,shift={(+1.5,0)}] (-0.6,-0.7) rectangle (0.6,0.7);
\filldraw[lightgray]  (0,0) circle (12pt);
%%%
\node at (-1.5,0) {$G^{(4)}_{\rm in}$};
\node at (+1.5,0) {$G^{(4)}_{\rm out}$};
\node at ( 0,  0) {$\im {\cal A}_{\mathsmaller{T}}$};
%%%%%%%%
\end{scope}
%%%%%%%%%%%%%%%%%%%%%%%%%%%%%%%%%%%%%%%%%%%%%%%%%%%%%%%%%%%%%%%%%%%%%%%%%%%%%%%%%%%%
%%%%%%%%%%%%%%%%%%%%%%%%%%%%%%%%%%%%%%%%%%%%%%%%%%%%%%%%%%%%%%%%%%%%%%%%%%%%%%%%%%%%
\begin{scope}[shift={(+4,-3.5)}]
%%%%%%%% Incoming Fermions / Scalars
\draw (-2.5,+0.5) -- (-0.50,+0.50);
\draw (-0.5,+0.5) -- (-0.05,+0.05);
\draw (-0.5,-0.5) -- (-0.05,-0.05);
\draw (-2.5,-0.5) -- (-0.50,-0.50);
%%%%%%%% Outgoing Fermions / Scalars
\draw (2.5,+0.5) -- (0.5,+0.5);
\draw (0.5,+0.5) -- (0.1,+0.1);
\draw (0.5,-0.5) -- (0.1,-0.1);
\draw (2.5,-0.5) -- (0.5,-0.5);
%%%%%%%% Radiation and thermal bath
\draw[scalar] (0,0.25) -- (0,1.5);
\draw[scalarbar] (-2.5,1.5) -- (0,1.5);
\draw[vector] (0,1.5) -- (2.5,1.5);
\node at (-1.25,1.9) {$H^\dagger$};
\node at (+1.25,1.9) {$B^{\mu},~W^{a,\mu}$};
\draw[->] (-0.2,0.8) -- (-0.2,1.3);\node at (-0.50,1) {$\PH$};
\draw[->] (-2.5,1.3) -- (-2.0,1.3);\node at (-2.25,1) {$q_i$};
\draw[->] (+2.0,1.3) -- (+2.5,1.3);\node at (+2.25,1) {$q_f$};
%%%%%%%% G4, A functions and labels
\draw[fill=lightgray,draw=none,shift={(-1.5,0)}] (-0.6,-0.7) rectangle (0.6,0.7);
\draw[fill=lightgray,draw=none,shift={(+1.5,0)}] (-0.6,-0.7) rectangle (0.6,0.7);
\filldraw[lightgray]  (0,0) circle (12pt);
%%%
\node at (-1.5,0) {$G^{(4)}_{\rm in}$};
\node at (+1.5,0) {$G^{(4)}_{\rm out}$};
\node at ( 0,  0) {$\im {\cal A}_{\mathsmaller{T}}$};
%%%%%%%%
\end{scope}
%%%%%%%%%%%%%%%%%%%%%%%%%%%%%%%%%%%%%%%%%%%%%%%%%%%%%%%%%%%%%%%%%%%%%%%%%%%%%%%%%%%%
\end{tikzpicture}
\caption{\label{fig:BSF_OffShellHiggs} Bound-state formation or bound-to-bound transitions via exchange of an off-shell Higgs doublet with the SM particles. The arrows on the field lines denote the flow of Hypercharge. All processes have their conjugate counterparts that occur via $H^{\dagger}$ exchange.}
\end{figure}
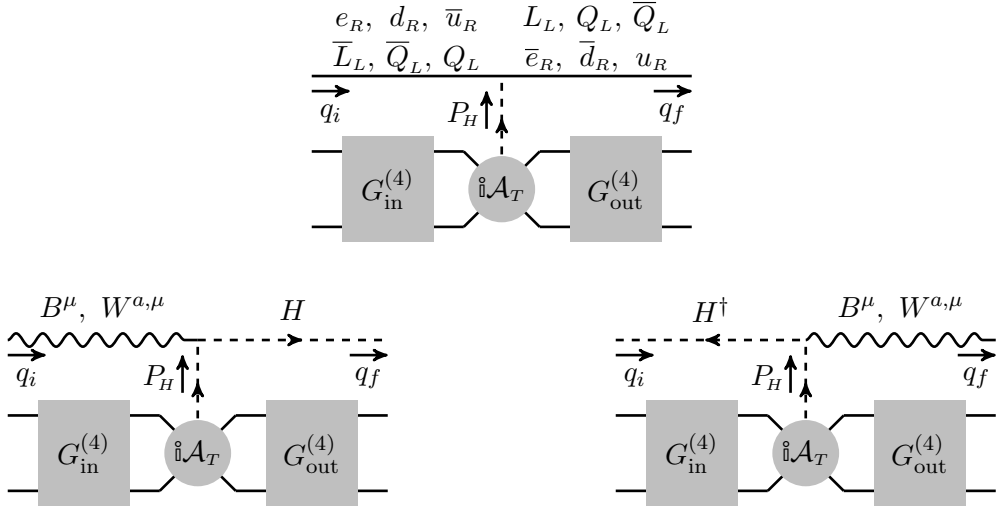

\subsubsection{Factorisation of the effective BSF cross-sections and transition rates \label{sec:BSF_OffShellHiggs_CrossSectionFactor}}

For simplicity, we lay out the discussion in terms of the BSF cross-sections only. The derivation for bound-to-bound transition rates is analogous. 

The thermally averaged rate per unit volume for BSF via off-shell Higgs exchange is
\begin{align}
\dfrac{d\<\Gamma_{n\ell m}^{\HoffBSF}\>}{dV}=
g_1 g_2 \int \dfrac{d^3k_1}{(2\pi)^3} \dfrac{d^3k_2}{(2\pi)^3}
f_+(k_1^0) f_+(k_2^0) [1+f_-(\omega_{{\bf k} \to n\ell m})]
\ \sigma_{{\bf k} \to n\ell m}^{\HoffBSF} \vrel ,
\label{eq:HoffshellBSF_dGammadV}
\end{align}
where we defined 
\begin{align}
\sigma_{{\bf k} \to n\ell m}^{\HoffBSF} \vrel 
&\equiv \dfrac{[1+f_-(\omega_{{\bf k} \to n\ell m})]^{-1}}{2k_1^0 2k_2^0} \int 
\dfrac{d^3 P}{(2\pi)^3 2P^0} 
\dfrac{d^3 q_i}{(2\pi)^3 2q_i^0} 
\dfrac{d^3 q_f}{(2\pi)^3 2q_f^0}
f_\pm(q_i^0)[1 \mp f_\pm(q_f^0)] \times
\nn \\
&\times
(2\pi)^4 \delta^4 (k_1 + k_2 + q_i - P - q_f)
\dfrac{1}{g_1 g_2} |{\cal M}_{{\bf k} \to n\ell m}^{\HoffBSF}|^2 .
\label{eq:HoffshellBSF_sigmav}
\end{align}
Here, the indices 1,2 correspond to the two incoming DM fields, while $q_i$ and $q_f$ denote the initial and final bath particle momenta. We consider scattering only on relativistic species, whose density in a thermal environment is large, and set $q_i^0 = |{\bf q}_i|$ and $q_f^0 = |{\bf q}_f|$. 
$f_\pm (E) = (e^{E/T} \pm 1)^{-1}$ are the phase-space occupation numbers for fermions ($+$) and bosons ($-$), and in \cref{eq:HoffshellBSF_sigmav} the upper and lower signs correspond to scattering on fermionic or bosonic dof respectively.  

$\omega_{{\bf k} \to n\ell m}$ is the energy dissipated by the DM fields in the capture process, given by \cref{eq:omega_BSF}. When the DM particles are non-relativistic, it depends only on the relative momentum of the incoming DM particles and the binding energy of the bound state, but is independent of the momenta of the bath particles. (We note that this does not hold for the 3-momentum exchange $|{\bf q}_f - {\bf q}_i|$ along the off-shell mediator.)  The factor $[1+f_-(\omega_{{\bf k} \to n\ell m})]$ in \cref{eq:HoffshellBSF_dGammadV} is compensated by the inverse factor in \cref{eq:HoffshellBSF_sigmav}; this definition ensures that the thermal averaging of the cross-section \eqref{eq:HoffshellBSF_sigmav} is the same as that of its radiative counterpart, which includes a Bose-enhancement factor for the radiated boson~\cite{vonHarling:2014kha} (cf.~ref.\cite{Oncala:2021swy}.)

Next, we conjecture that the amplitude for off-shell $H$ exchange, ${\cal M}_{{\bf k} \to n\ell m}^{\HoffBSF}$, can be factorised into the corresponding  amplitude with on-shell $H$ emission, ${\cal M}_{{\bf k} \to n\ell m}^{\HBSF}$, and a function of the momenta of the bath particles, as follows\footnote{For BSF via vector exchange/emission, the amplitude-squared does not factorise as in \cref{eq:HoffshellBSF_Mfactorization}. However, it is still possible to obtain a factorisation formula similar to \cref{eq:HoffshellBSF_sigmav_factorisation} below, for an effective cross-section, at leading order in the non-relativistic approximation~\cite{Binder:2019erp,Binder:2020efn}. (See also \cref{foot:HoffshellBSF_sigmav_factorisation}.)}
\begin{align}
\sum_{i,f~\rm dof}
|{\cal M}_{{\bf k} \to n\ell m}^{\HoffBSF}|^2 \simeq 
|{\cal M}_{{\bf k} \to n\ell m}^{\HBSF}|^2 
\times R_0(q_i \cdot q_f) ,
\label{eq:HoffshellBSF_Mfactorization}
\end{align}
where the sum on the left side runs over the dof (spin and gauge) of the initial and final bath particles. ${\cal M}_{{\bf k} \to n\ell m}^{\HBSF}$ may depend only on the momentum exchange ${\bf q} \equiv {\bf q}_f - {\bf q}_i$ rather than on ${\bf q}_i$ and ${\bf q}_f$ separately. $R_0$ must be Lorentz invariant since the amplitudes are. It may thus depend only on the 4-vector products $q_i^2=q_f^2=0$ and $q_i \cdot q_f$; in \cref{eq:HoffshellBSF_Mfactorization} we have denoted its dependence on the latter. The $R_0$ factors will be specified in \cref{sec:BSF_OffShellHiggs_Amplitudes} for the processes shown in \cref{fig:BSF_OffShellHiggs}. 
Switching the integration from ${\bf q}_f$ to ${\bf q}$, \cref{eq:HoffshellBSF_sigmav} gives 
\begin{align}
\sigma_{{\bf k} \to n\ell m}^{\HoffBSF} \vrel 
&= [1+f_-(\omega_{{\bf k} \to n\ell m})]^{-1}
\int \dfrac{d^3 q_i}{(2\pi)^3  2|{\bf q}_i|} 
\nn \\
&\times \dfrac{1}{2k_1^0 2k_2^0} \int 
\dfrac{d^3 P}{(2\pi)^3 2P^0} 
\dfrac{d^3 q}{(2\pi)^3 2q^0} 
(2\pi)^4 \delta^4 (k_1 + k_2 - P - q)
\dfrac{1}{g_1 g_2} |{\cal M}_{{\bf k} \to n\ell m}^{\HBSF}|^2  
\nn \\
&\times 
\dfrac{2 q^0}{2|{\bf q}+{\bf q}_i|} 
f_\pm (|{\bf q}_i|) \[1 \mp f_\pm(|{\bf q}+{\bf q}_i|)\]
R_0(q_i\cdot q_f) ,
\label{eq:HoffshellBSF_sigmav_mod1}
\end{align}
where
\begin{align}
q^0 &\equiv q_f^0 - q_i^0 \simeq 
|{\bf q} + {\bf q}_i| - |{\bf q}_i|.
\label{eq:HoffshellBSF_q0_def}
\end{align}
The second line of \cref{eq:HoffshellBSF_sigmav_mod1} would form the BSF cross-section via on-shell emission, except for two complications: (i) The dispersion relation of the radiated momentum is given by \cref{eq:HoffshellBSF_q0_def} rather than the on-shell condition of the radiated boson, and depends on the variable ${\bf q}_i$. (ii) The last line of \cref{eq:HoffshellBSF_sigmav_mod1} depends on ${\bf q}$. These complications are resolved within the non-relativistic approximation, where the cross-section of BSF via scattering can be shown to be proportional to that of radiative BSF, as we shall now see.

\subsubsection*{Non-relativistic approximation}

Similarly to radiative BSF, if the incoming particles are non-relativistic and the final state particles are weakly bound, we may neglect the recoil of the bound state. Then, the energy-momentum conservation implies $q^0 \simeq \omega$ with $\omega$ given by \cref{eq:omega_BSF,eq:omega_BoundToBound} for BSF and bound-to-bound transitions. (Here we drop the $\omega$ indices for simplicity.) The dispersion relation \eqref{eq:HoffshellBSF_q0_def} yields
\begin{align}
|{\bf q}| \simeq \omega \[
\sqrt{1 + \dfrac{2|{\bf q}_i|}{\omega} + \dfrac{|{\bf q}_i|^2 \tau^2}{\omega^2}} 
- \dfrac{|{\bf q}_i| \tau}{\omega} \],
\label{eq:HoffshellBSF_q}
\end{align}
with $\tau \equiv \hat{\bf q}_i \cdot \hat{\bf q}$. It is also easy to show that $q_i \cdot q_f = ({\bf q}^2-\omega^2)/2$. Using the $\delta^4$-function to perform the integration over $d^3P \, d|{\bf q}|$ (as is standard in the computation of 2-to-2 cross-sections), \cref{eq:HoffshellBSF_sigmav_mod1} gives
\begin{align}
\sigma_{{\bf k} \to n\ell m}^{\HoffBSF} \vrel 
&\simeq 
[1+f_-(\omega)]^{-1} 
\int \dfrac{d^3 q_i}{(2\pi)^3  2|{\bf q}_i|} 
\dfrac{\omega}{\omega+|{\bf q}_i|} 
f_\pm (|{\bf q}_i|) \[1 \mp f_\pm(\omega+|{\bf q}_i|)\]
\nn \\
&\times \dfrac{1}{2k_1^0 2k_2^0} \dfrac{1}{2P^0 2\omega} 
\int \dfrac{d\Omega_{\bf q}}{4\pi^2} 
\ {\bf q}^2 
\( \dfrac{\omega + |{\bf q}_i|}{|{\bf q}| + |{\bf q}_i| \tau} \)
\ \dfrac{1}{g_1 g_2} |{\cal M}_{{\bf k} \to n\ell m}^{\HBSF}|^2  
\nn \\
&\times 
R_0 \( \frac{{\bf q}^2-\omega^2}{2} \) ,
\label{eq:HoffshellBSF_sigmav_mod2}
\end{align}
with $P^0 \simeq 2m$ and $|{\bf q}|$ given by \cref{eq:HoffshellBSF_q}. Note that the second line of \cref{eq:HoffshellBSF_sigmav_mod2} differs from 
$(\sigma_{{\bf k} \to n\ell m}^{\HBSF} \vrel)$ by the factor 
$\left. 
\dfrac{\omega + |{\bf q}_i|}{|{\bf q}| + |{\bf q}_i| \tau} 
\right/  \dfrac{\omega}{|{\bf q}|}$  
due to the different dispersion relation of the radiated momentum $q$, here given by \cref{eq:HoffshellBSF_q0_def}.

The radiative BSF amplitudes are typically computed by expanding in powers of the radiated momentum ${\bf q}$~\cite{Petraki:2016cnz}.\footnote{The expansion is in effect in the dimensionless combination 
$|{\bf q}|/ \sqrt{\kappa^2/n^2+{\bf k}^2} = |{\bf q}| /\sqrt{2\mu \omega}$. For BSF via on-shell emission, the radiated momentum is limited by the available energy, $|{\bf q}| \leqslant \omega$, thus the expansion parameter is always $\sqrt{\omega/(2\mu)} \ll 1$. However, for BSF via scattering, $|{\bf q}|$ can be comparable to or larger than $\sqrt{2\mu \omega}$, particularly at $T \gtrsim \kappa/n$, which puts in question the validity of the expansion. Nevertheless, for the purposes of DM freeze-out, BSF typically reaches ionisation equilibrium at high $T$, where the DM destruction rate via BSF is independent of the BSF cross-sections~\cite{Binder:2018znk} (cf.~ref.~\cite{Oncala:2021swy}.) At lower $T$, where the magnitude of the BSF cross-sections matters, the $|{\bf q}|$ expansion is a valid approximation.}
%%%%%%%%%%%%%%%%%
As seen in \cref{Sec:BSF}, the dominant contribution to the various ${\cal M}_{{\bf k} \to n\ell m}^{\HBSF}$ amplitudes arises from the zeroth order term~\cite{Oncala:2019yvj}, i.e.~${\cal M}_{{\bf k} \to n\ell m}^{\HBSF}$ are independent of ${\bf q}$ (and therefore $\tau$) at leading order. Reshuffling the various factors,   \cref{eq:HoffshellBSF_sigmav_mod2} becomes
\begin{align}
&\sigma_{{\bf k} \to n\ell m}^{\HoffBSF} \vrel 
\simeq
\dfrac{1}{8\pi^2} 
\dfrac{1}{2k_1^0 2k_2^0} \dfrac{\omega}{2P^0} 
\int d\Omega_{\bf q} 
\ \dfrac{1}{g_1 g_2} |{\cal M}_{{\bf k} \to n\ell m}^{\HBSF}|^2  \times
\label{eq:HoffshellBSF_sigmav_mod3}
\\
&\times [1+f_-(\omega)]^{-1} 
\int \dfrac{d^3 q_i}{(2\pi)^3  2|{\bf q}_i|} 
f_\pm (|{\bf q}_i|) \[1 \mp f_\pm(\omega+|{\bf q}_i|)\]
\dfrac{|{\bf q}|^2}{\omega(|{\bf q}| + |{\bf q}_i| \tau)} 
R_0 \( \frac{{\bf q}^2-\omega^2}{2} \) .
\nn
\end{align}
The integration over $d^3 q_i$ in the second line of \cref{eq:HoffshellBSF_sigmav_mod3} eliminates any dependence of the integrand on the orientation of the vector ${\bf q}$, allowing us to identify the first line as the cross-section for on-shell Higgs emission with the phase-space suppression removed (cf.~\cref{eq:BSF_dsigmadOmega_Scalar,eq:BSF_sigma_Scalar}.) Thus, the effective cross-section for off-shell Higgs exchange can be factorised at leading order as follows\footnote{Note that \cref{eq:HoffshellBSF_sigmav_factorisation} holds for the effective cross-section of BSF via off-shell Higgs exchange as defined in \cref{eq:HoffshellBSF_sigmav}. Since ${\cal M}_{{\bf k}\to n\ell m}^{\HBSF}$ is presumed to be independent of ${\bf q}$ and therefore ${\bf q \cdot k}$, it has not been necessary to integrate over the angular variables of ${\bf k}$ in order to obtain a factorised expression for the cross-section. This is in contrast to the case of vector exchange, where the amplitude does not factorise as in \cref{eq:HoffshellBSF_Mfactorization} and in fact depends on ${\bf q}_i \cdot {\bf k}$ and ${\bf q}_f \cdot {\bf k}$. A factorisation similar to \cref{eq:HoffshellBSF_sigmav_factorisation} is obtained only for the cross-section averaged over the ${\bf k}$ solid angle~\cite{Binder:2019erp,Binder:2020efn} (cf.~\cref{sec:BSF_OffShell_BW}.) 
\label{foot:HoffshellBSF_sigmav_factorisation}}
\begin{empheq}[box=\widefbox]{align}
\sigma_{{\bf k} \to n\ell m}^{\HoffBSF} \vrel \simeq 
\dfrac{\sigma_{{\bf k} \to n\ell m}^{\HBSF} \vrel}{\hH(\omega_{{\bf k}\to n\ell m})}
\times \RH (\omega_{{\bf k}\to n\ell m}),
\label{eq:HoffshellBSF_sigmav_factorisation}
\end{empheq}
where $\hH(\omega)$ is the phase-space suppression \eqref{eq:hH} of the on-shell emission due to the Higgs mass, with $\omega_{{\bf k}\to n\ell m}$ being the dissipated energy \eqref{eq:omega_BSF}, and we restored the indices for concreteness. The dimensionless factor $\RH(\omega)$ is
\begin{align}
\RH (\omega) &\equiv
\int \dfrac{d^3 q_i}{(2\pi)^3 2|{\bf q}_i|}
\dfrac{{\bf q}^2}
{\omega (|{\bf q}| + |{\bf q}_i| \tau)} 
\dfrac{ f_\pm ({\bf q}_i) \[1 \mp f_\pm(\omega+ |{\bf q}_i|)\] }{1+f_-^{}(\omega)}
\, R_0 \( \frac{{\bf q}^2-\omega^2}{2} \) .
\label{eq:HoffshellBSF_R_def}
\end{align}
We recall that $|{\bf q}|$ is given by \cref{eq:HoffshellBSF_q}. Since the entire integrand is rotationally invariant (recall that $R_0$ is Lorentz invariant), we perform the $d^3q_i$ integration by setting ${\bf q}$ on the $z$ axis. Then the azimuthal angle is parametrised by $\tau$ defined above. 
Changing integration variables from $|{\bf q}_i|$ and $\tau$ to $u\equiv|{\bf q}_i| / \omega$ and $z \equiv {\bf q}^2/\omega^2-1$, \cref{eq:HoffshellBSF_R_def} simplifies to
\begin{align}
\RH (\omega) &= \dfrac{\omega^2}{16\pi^2}
\int_0^\infty du \ 
\dfrac{f_\pm (\omega u) \[1 \mp f_\pm \(\omega (1+ u)\)\] }{1+f_-^{}(\omega)}
\int_0^{4u(1+u)} dz
\, R_0 \( z \omega^2/2 \) .
\label{eq:HoffshellBSF_R_NonRel}
\end{align}
We compute $\RH$ next. The final result can be found in \cref{eq:HoffshellBSF_Rfactors,eq:HoffshellBSF_Rpm,eq:HoffshellBSF_Rfactor_tot}.

\medskip
Following the same steps, we find that the bound-to-bound transition rate via off-shell Higgs exchange is related to the radiative one via
\begin{empheq}[box=\widefbox]{align}
\Gamma_{n'\ell'm \to n\ell m}^{\HoffBSF} \simeq 
\dfrac{\Gamma_{n'\ell'm' \to n\ell m}^{\HBSF} }
{ \hH (\omega_{n'\ell'm'\to n\ell m}) }
\times \RH (\omega_{n'\ell'm'\to n\ell m}),
\label{eq:HoffshellBoundToBound_gamma_factorisation}
\end{empheq}
where $\omega_{n'\ell'm'\to n\ell m}$ is the dissipated energy \eqref{eq:omega_BoundToBound}.

\subsubsection{Amplitudes \label{sec:BSF_OffShellHiggs_Amplitudes}}

Similarly to their radiative analogues, the amplitudes for BSF and bound-to-bound transitions via scattering consist of the perturbative transition amplitudes that encode the scattering on the bath particles, convoluted with the initial and final state wavefunctions. Focusing again on BSF (cf.~\cref{eq:BSF_FullAmplitude_Def}),
\begin{align}
\im {\cal M}_{{\bf k} \to n\ell m}^{\HoffBSF} = 
\int \frac{d^3 {\bf k'}}{(2\pi)^3} \frac{d^3 {\bf p}}{(2\pi)^3}
\ \frac{[\psi_{n\ell m} ({\bf p})]^\dagger}{\sqrt{2\mu}}
\ \im {\cal A}_{\mathsmaller{T}}^{\HoffBSF} ({\bf k',p})
\ \phi_{\bf k} ({\bf k'}) .
\label{eq:HoffshellBSF_FullAmplitude_Def}
\end{align}
We now compute ${\cal A}_{\mathsmaller{T}}^{\HoffBSF} ({\bf k',p})$ for the scattering processes shown in \cref{fig:BSF_OffShellHiggs}, and deduce from \cref{eq:HoffshellBSF_R_NonRel} the corresponding $R$ factors. 

\subsubsection*{Scattering on fermions}

The Higgs couples to the SM fermions via the operators
\begin{align}
\delta {\cal L} = 
-y_e (\delta_{ab}   \bar{L}_{\mathsmaller{L} \, a}  H_b)  e_{\mathsmaller{R}}  
-y_d (\delta_{ab}   \bar{Q}_{\mathsmaller{L} \, a}  H_b)  d_{\mathsmaller{R}}  
-y_u (\epsilon_{ab} \bar{Q}_{\mathsmaller{L} \, a}  H^\dagger_b)  u_{\mathsmaller{R}}  
+\hc,
\label{eq:SMfermionsMassTerms}
\end{align}
where the $a,b$ superscripts indicate the $\SUL$ contractions, while the family indices are suppressed. These couplings give rise to the scattering processes shown in \cref{fig:BSF_OffShellHiggs} (top). The corresponding BSF perturbative transition amplitudes (projected on the desired spin and gauge, scattering and bound states) are\footnote{In \cref{eq:HoffshellBSF_Afermions}, the sign of the $\gamma_5$ term and whether the Yukawa coupling should be $\yF$ or $\yF^*$ depend on the exact process we are considering. For scattering on antifermions, the spinors $u_i$, $u_f$ become $v_i$, $v_f$. In addition, for a scattering involving an up-type right-handed (anti)quark, $\delta_{hh'}$ should be replaced by $\epsilon_{hh'}$. However, all these differences do not affect the $R_0$ factors.} 
\begin{align}
\im \[ {\cal A}_{\mathsmaller{T}}^{\HoffBSF} ({\bf k',p}) \]_h=
\im \[ {\cal A}_{\mathsmaller{T}}^{\mathsmaller{H}  \text{-}\BSF} ({\bf k',p})\]_{h'} 
\times \dfrac{\im}{q^2 - \mH^2} 
\bar{u}_f (-\im \yF^*) \delta_{hh'} \(\dfrac{1-\gamma_5}{2}\) u_i ,
\label{eq:HoffshellBSF_Afermions}
\end{align}
where $h,h'$ are the $\SUL$ indices of the left-handed SM fermion field and the exchanged Higgs, respectively. The scattering and bound states gauge indices, if any, are left implicit. We use $\yF$ to denote collectively the SM Yukawa couplings of \cref{eq:SMfermionsMassTerms}. Inserting \cref{eq:HoffshellBSF_Afermions} into \eqref{eq:HoffshellBSF_FullAmplitude_Def}, squaring and summing over the bath particle spin and gauge dof, we arrive at the $R_0$ factors (cf.~\cref{eq:HoffshellBSF_Mfactorization}),
\begin{align}
R_0 = 2 \times |\yF|^2 
\, \dfrac{2q_i \cdot q_f}{(2q_i \cdot q_f + \mH^2)^2}  ,
\label{eq:HoffshellBSF_R0_Fermions}
\end{align}
where we introduced a factor 2 to account for the partner process controlled by the same coupling, where the initial (final) fermion becomes the final (initial) antifermion.

\subsubsection*{Scattering on bosons}

The perturbative transition amplitude for scattering on gauge bosons (\cref{fig:BSF_OffShellHiggs}, bottom left), projected on the desired spin and gauge, scattering and bound states, is
\begin{align}
\im \[{\cal A}_{\mathsmaller{T}}^{\HoffBSF} ({\bf k',p})\]^{a,\mu}_h =
\im [{\cal A}_{\mathsmaller{T}}^{\HBSF} ({\bf k',p})]_{h'} 
\times \dfrac{\im}{q^2 - \mH^2} 
\times \im g \, T^a_{hh'} \, (q_f+q)^\mu,
\label{eq:HoffshellBSF_Abosons}
\end{align}
where $h,h'$ and $a$ are the $\SUL$ indices of the outgoing and exchanged Higgs bosons and the incoming gauge boson respectively. $T^a$ and $g$ stand for the generators and the gauge coupling of the gauge group under consideration. Inserting \cref{eq:HoffshellBSF_Abosons} into \eqref{eq:HoffshellBSF_FullAmplitude_Def}, squaring and summing over the bath particle polarisations and gauge dof, we find the $R_0$ factors
\begin{align}
R_0 = 2 \times 4\pi \alpha \, C_2 (\mathbb{R}_{\H})
\, \dfrac{4 q_i \cdot q_f}{(2q_i \cdot q_f + \mH^2)^2}   ,
\label{eq:HoffshellBSF_R0_Bosons}
\end{align}
where, as before, we introduced a factor 2 to account for the partner process where $H^\dagger$ is the incoming bath particle (\cref{fig:BSF_OffShellHiggs}, bottom right). 
$C_2 (\mathbb{R}_{\H})$ is the quadratic Casimir of the Higgs representation under the gauge group considered; here, 
$C_2 (\mathbb{R}_{\H}) = \YH^2 = 1/4$ for Hypercharge and $C_2 (\mathbb{R}_{\H}) = 3/4$ for $\SUL$.

\subsubsection{BSF cross-sections and transition rates \label{sec:BSF_OffShellHiggs_CrossSections}}

Both \cref{eq:HoffshellBSF_R0_Fermions,eq:HoffshellBSF_R0_Bosons} depend only on $q_i \cdot q_f$, as presumed in \cref{eq:HoffshellBSF_Mfactorization}, and in fact in the same fashion. Inserting them into \cref{eq:HoffshellBSF_R_NonRel}, and carrying out the integration over $z$, we find the contributions of scattering on fermions and bosons to BSF,
\begin{subequations}
\label{eq:HoffshellBSF_Rfactors}
\label[pluralequation]{eqs:HoffshellBSF_Rfactors}
\begin{empheq}[box=\widefbox]{align}
\RHF
&= 2 \times \frac{|\yF|^2}{4\pi} \frac{1}{2}\times R_{+} \,,
\label{eq:HoffshellBSF_RF}
\\
\RHBH
&= 2 \times (1/4) \alpha_1 \times R_{-} \,,
\label{eq:HoffshellBSF_RB}
\\
\RHWH
&= 2 \times (3/4) \alpha_2 \times R_{-} \,,
\label{eq:HoffshellBSF_RW}
\end{empheq}
\end{subequations}
where $R_{\pm}$ are dimensionless functions of two parameters, $\omega/T$ and $\mH/\omega$, 
\begin{empheq}[box=\widefbox]{align}
R_{\pm} 
&\equiv  \dfrac{1}{2\pi}  \int_0^\infty  du
\dfrac{e^{u \, \omega/T}}{e^{u (\omega/T)}\pm 1}
\dfrac{e^{\omega/T}-1}{e^{(1+u) \omega/T}\pm1}
%\nn \\ &\times 
\left\{
\ln \[1+\dfrac{4u (1+u)}{\mH^2/\omega^2}\] 
- \dfrac{4u(1+u)}{4u(1+u) + \mH^2/\omega^2}
\right\} .
\label{eq:HoffshellBSF_Rpm}
\end{empheq}
The factor $1/2$ in \cref{eq:HoffshellBSF_RF} is due to the SM fermions being chiral. The $\RH$ factor that determines the BSF via scattering cross-section \eqref{eq:HoffshellBSF_sigmav_factorisation} is
\begin{empheq}[box=\widefbox]{align}
\RH = \sum_{\mathsmaller{F}} \RHF + \RHBH + \RHWH .  
\label{eq:HoffshellBSF_Rfactor_tot}
\end{empheq}
Among the SM fermions, the top quark yields the largest contribution as long as it remains relativistic.

The $R_{\pm}$ factors \eqref{eq:HoffshellBSF_Rpm} diverge at $\mH \to 0$, which during the DM thermal decoupling around the EWPT (cf.~ref.~\cite{Oncala:2021swy}.) This divergence can be removed by a full next-to-leading-order calculation, as done in ref.~\cite{Binder:2020efn} in the context of a $U(1)$ gauge theory. Performing such a computation for the model considered here is beyond the scope of this work. However, comparing the results of refs.~\cite{Binder:2019erp} and \cite{Binder:2020efn} for a massive and massless vector mediator respectively, we find that, upon thermal averaging, the former approximates well the latter at temperatures higher than the binding energy if the screening scale (i.e. the mediator mass) is set to $0.74 \, \omega$.\footnote{Reference~\cite{Binder:2020efn} found that using the binding energy as the minimum screening scale provides a good approximation. While this is indeed so, the above prescription ensures that the $R$ factor depends only on $\omega/T$ as predicted by the full computation, besides being a somewhat better approximation.}
Considering this, in \cref{eq:HoffshellBSF_Rpm} we shall do the replacement 
\begin{align}
\mH \to \max (\mH, \omega) .
\label{eq:HoffshellBSF_Rfactor_mHMin}
\end{align}

\medskip

\begin{figure}[t]
\centering
\includegraphics[width=0.425\textwidth]{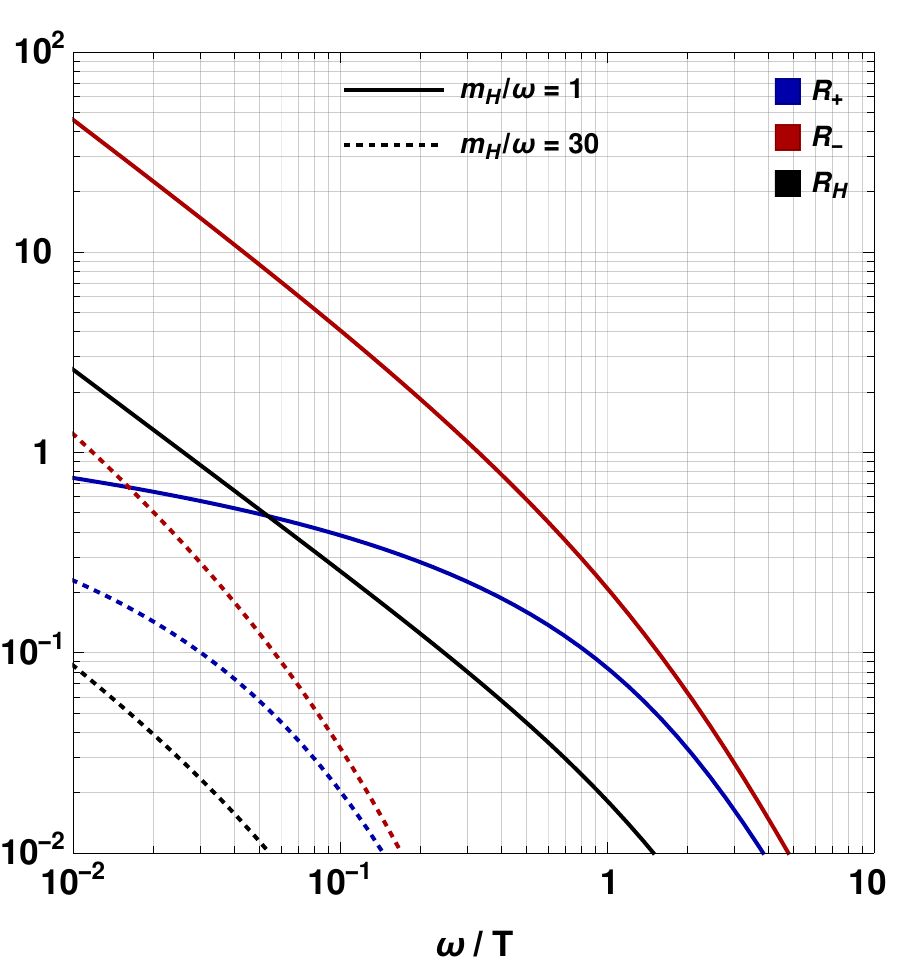}~~~~
\includegraphics[width=0.425\textwidth]{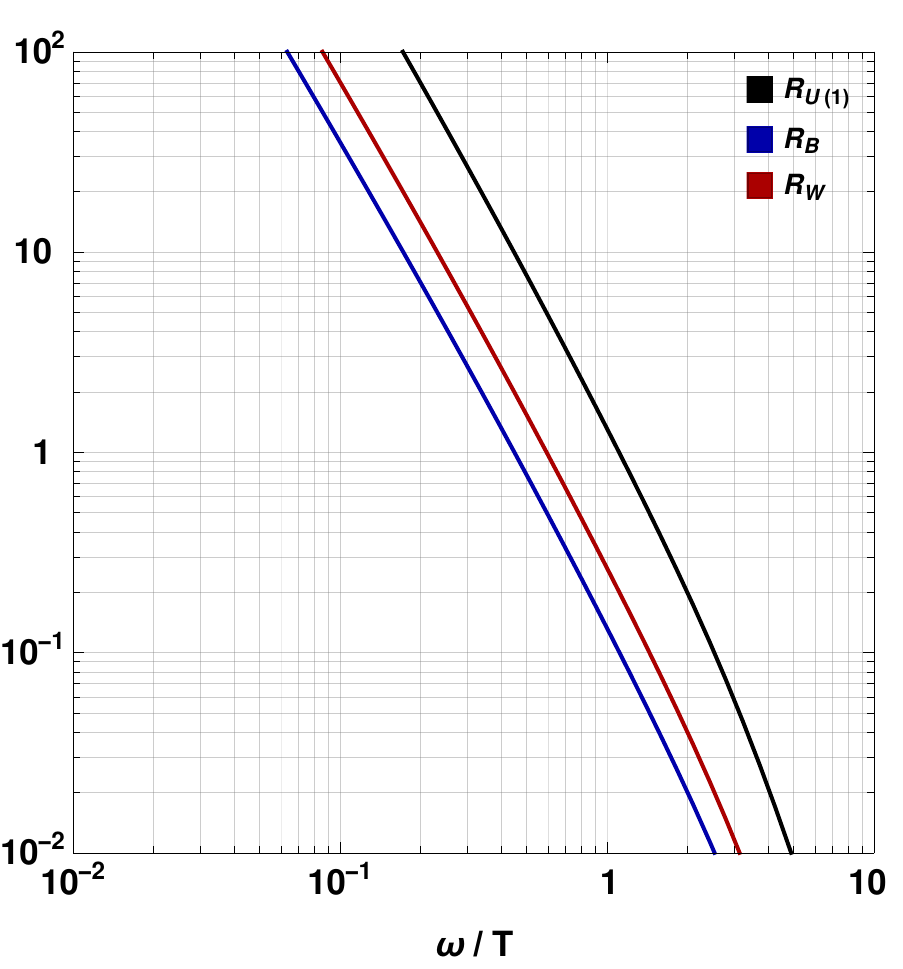}
\caption{\label{fig:offshellBSF_Rfactors} 
\emph{Left}: The $R_{\pm}$ and $\RH$ factors of \cref{eq:HoffshellBSF_Rpm,eq:HoffshellBSF_Rfactor_tot} that determine the ratio of BSF via off-shell $H^{(\dagger)}$ scattering over on-shell $H^{(\dagger)}$ emission.
\emph{Right}: The $\RB$ and $\RW$ factors of \cref{eqs:BWoffshellBSF_R}  that determine the ratio of BSF via off-shell $B$ or $W$ scattering on fermions over on-shell $B$ or $W$ emission. Also shown, the $\Rabelian$ factor from \cite{Binder:2020efn} (cf.~\cref{foot:RU1}.)
}
\end{figure}

We present $R_\pm$ and $\RH$ in \cref{fig:offshellBSF_Rfactors}. It is clear that they are more significant for $\omega/T \ll 1$. This implies that for bound-to-bound transitions, they enhance the rates at $T \gg \omega_{n'\to n} =|{\cal E}_{n'}-{\cal E}_{n}|$.  
For BSF via scattering, the $R_{\pm}$ factors weigh preferentially the contribution of DM pairs with low relative velocity. We note that even though in a thermal bath $\<\omega_{{\bf k}\to n\ell m}\> =|{\cal E}_n| + (3/2)T > T$ (cf.~\cref{eq:omega_BSF}), lower values of $\omega_{{\bf k}\to n\ell m}$ may still incur in a sizeable portion of the DM collisions while $T \gtrsim |{\cal E}_n|$.

Even when the $\RH$ factor \eqref{eq:HoffshellBSF_Rfactor_tot} is less than 1, BSF via scattering may potentially be (i) faster than radiative BSF, which is suppressed by the $\hH (\omega_{{\bf k} \to n\ell m})$ phase-space factor \eqref{eq:hH}, becoming entirely inaccessible for $\mH/\omega_{{\bf k} \to n\ell m} >1$, and (ii) significant with respect to direct annihilation and BSF via vector emission, since these processes are suppressed by one (two) extra power(s) of couplings compared to BSF via $H^{(\dagger)}$ off-shell exchange (on-shell emission.)   Analogously, bound-to-bound transitions via scattering may dominate over their radiative counterparts and/or the direct bound-state decay into radiation.

To assess realistically the impact of BSF and bound-to-bound transitions via scattering we must thermally average the cross-sections and rates of \cref{eq:HoffshellBSF_sigmav_factorisation,eq:HoffshellBoundToBound_gamma_factorisation}, and consider the interplay of bound-state formation, decay, ionisation and transition processes in the thermal bath. This is done in ref.~\cite{Oncala:2021swy}. Here we only note that considering BSF via scattering does not increase the DM destruction rate proportionally, since at early times a state of ionisation equilibrium is typically reached where the DM destruction due to BSF is independent of the actual BSF rate provided that the latter is sufficiently large~\cite{Binder:2018znk}.

\subsection{$B$ and $W$ exchange \label{sec:BSF_OffShell_BW}}

References~\cite{Binder:2019erp,Binder:2020efn} showed in the context of a $U(1)$ gauge theory that the effective cross-section for BSF via off-shell vector exchange, defined via the thermally averaged rate per volume (cf.~\cref{foot:HoffshellBSF_sigmav_factorisation})
\begin{align}
\dfrac{d\<\Gamma_{n\ell m}^{\VoffBSF}\>}{dV} = n_1 n_2 
\( \dfrac{2\mu^3}{\pi T^3} \)^{1/2}
\int d\vrel  \, \vrel^2 \, e^{-\mu \vrel^2/(2 T)}
\( 1+ \dfrac{1}{e^{\omega_{{\bf k} \to n\ell m}/T} -1 } \)
(\sigma_{{\bf k} \to n\ell m}^{\VoffBSF} \vrel) ,
\label{eq:VoffshellBSF_dGammadV}
\end{align}
is, at leading order in the non-relativistic regime, proportional to the cross-section for on-shell emission, 
$\sigma_{{\bf k} \to n\ell m}^{\VoffBSF} \vrel = 
(\sigma_{{\bf k} \to n\ell m}^{\VBSF} \vrel) \times R_{\V}$, where 
$R_{\V} = 2 \times \alpha \times \Rabelian$. 
As in \cref{sec:BSF_OffShell_Higgs}, the factor 2 accounts for the partner processes related via exchanging the initial (final) bath particle with the final (initial) bath antiparticle, and $\alpha$ is the fine structure constant of the group.  The factor $\Rabelian$ depends only on $\omega/T$ provided that $V$ is massless, and has been derived in \cite{Binder:2020efn} via a next-to-leading order calculation where the colinear and infrared the divergences are cancelled. For a massive $V$, a simple analytical formula that depends on $\omega/T$ and $m_{\V}/\omega$ has been computed in \cite{Binder:2019erp}. 
We define $\Rabelian$ to correspond to scattering on one species of relativistic Dirac fermions with charge unity, and use the results of \cite{Binder:2020efn}.\footnote{\label{foot:RU1}
$\Rabelian$ is related to $R_{2002.07145}$ defined in ref.~\cite[eq.~(4.13) and fig.~14]{Binder:2017lkj} as
$\Rabelian \equiv R_{2002.07145} / (2\pi)$. 
We thank Tobias Binder for providing the numerical values of $R_{2002.07145}$.} 

Adapting the result to the present model, the cross-sections for BSF via off-shell $B$ and $W$ exchange are related to those of on-shell emission  as follows
\begin{subequations}
\label{eq:BWoffshellBSF_sigmav_factorisation}
\label[pluralequation]{eqs:BWoffshellBSF_sigmav_factorisation}
\begin{align}
\sigma_{{\bf k} \to n\ell m}^{\BoffBSF} \vrel &\simeq 
(\sigma_{{\bf k} \to n\ell m}^{\BBSF} \vrel) 
\times \RB (\omega_{{\bf k} \to n\ell m} / T),
\label{eq:BoffshellBSF_sigmav_factorisation}
\\
\sigma_{{\bf k} \to n\ell m}^{\WoffBSF} \vrel &\simeq 
(\sigma_{{\bf k} \to n\ell m}^{\WBSF} \vrel) 
\times \RW (\omega_{{\bf k} \to n\ell m} / T),
\label{eq:WoffshellBSF_sigmav_factorisation}
\end{align}
\end{subequations}
where
\begin{subequations}
\label{eq:BWoffshellBSF_R} 
\label[pluralequation]{eqs:BWoffshellBSF_R} 
\begin{align}
\RB (\omega/T) &= 2 \times c_{\B} \alpha_1 \times \Rabelian (\omega/T),
\label{eq:BoffshellBSF_R} 
\\
\RW (\omega/T) &= 2 \times c_{\W} \alpha_2 \times \Rabelian (\omega/T) .
\label{eq:WoffshellBSF_R} 
\end{align}
\end{subequations}
The factors $c_{\B}$ and $c_{\W}$ account for scattering on the relativistic SM fermions. The contribution of a chiral fermion $F$ transforming under the representation $\mathbb{R}_{\F}$ of a gauge group is 
$c^{\F} = C(\mathbb{R}_{\F}) / 2$, where $C$ is the Casimir operator.
When all the SM fermions are relativistic, 
$c_{\B} = (1/2)\sum_{\F} Y_{\F}^2 = 5$ and 
$c_{\W} = (1/2) C(\mathbb{2}) \times 12 = 3$,  
where $C(\mathbb{2}) = 1/2$ for $\SUL$.

Note that \cref{eq:WoffshellBSF_R} includes only scattering on SM fermions via off-shell $W$ exchange. However, non-Abelian gauge bosons may also scatter on themselves due to the trilinear gauge coupling. Estimating this effect necessitates a dedicated next-to-leading order computation that is beyond the scope of this work. We shall thus neglect this contribution. 

Formulae analogous to \cref{eqs:BWoffshellBSF_R} hold for bound-to-bound transitions via off-shell $B$ and $W$ exchange, however no such transition is of interest here. 

We present the $\RB$ and $\RW$ factors in \cref{fig:offshellBSF_Rfactors}.

%%%%%%%%%%%%%%%%%%%%%%%%%%%%%%%%%%%%%%%%%%%%%%%%%%%%%%%%%%%%%%%%%%%%
%%%%%%%%%%%%%%%%%%%%%%%%%%%%%%%%%%%%%%%%%%%%%%%%%%%%%%%%%%%%%%%%%%%%
%%%%%%%%%%%%%%%%%%%%%%%%%%%%%%%%%%%%%%%%%%%%%%%%%%%%%%%%%%%%%%%%%%%%
\clearpage
\section{Conclusion \label{Sec:Conclusion}}

Renormalisable scenarios in which DM is the lightest mass eigenstate arising from the mixing of two electroweak multiplets that couple to the Higgs, are among the archetypical WIMP DM models. Here, we have considered the role of the Higgs doublet in the non-perturbative phenomena --- the Sommerfeld effect and the formation of bound states --- that take place during the thermal decoupling of multi-TeV DM from the primordial plasma.

We have shown that the effect of the Higgs doublet is two-fold: (i) it mediates a long-range interaction that affects the wavefunctions of both scattering and bound states, and (ii) its emission precipitates extremely fast monopole transitions, including capture into bound states and transitions between bound levels. 
In a companion paper~\cite{Oncala:2021swy}, we show that the above effects can reduce the relic density of stable species very significantly, thereby altering experimental constraints.

These results build on the work of ref.~\cite{Oncala:2019yvj} that showed the importance of bound-state formation via emission of a scalar charged under a symmetry (see also~\cite{Ko:2019wxq}), as well as the work of refs.~\cite{Harz:2017dlj,Harz:2019rro} that demonstrated the long-range effect of the Higgs boson between TeV-scale particles. In the present first computation of such effects involving the Higgs doublet, we have focused on the simplest model, comprised by two (nearly) mass degenerate fermionic $\SUL$ multiplets, a singlet and a doublet. Our calculations can of course be extended to other models of WIMP DM coupled to the Higgs doublet.

We have considered transitions -- both BSF and bound-to-bound transitions -- via radiative emission of an on-shell Higgs doublet, as well as via scattering on the thermal bath through off-shell Higgs-doublet exchange. We showed that the rates for the latter factorise into the former times a temperature-dependent function. This parallels the results of refs.~\cite{Binder:2019erp,Binder:2020efn} that considered capture via exchange of an off-shell gauge boson and found a similar factorisation. However, the temperature-dependent function depends on the multiple mode of the transition, and is thus different for the monopole transitions occurring via Higgs-doublet (or generally charged-scalar) exchange and the dipole transitions occuring via gauge-boson exchange.

We finish by summarising the main assumptions and approximations involved in our computations.  
(a) We have assumed that the coannihilating multiplets have equal masses, in order to obtain analytical results even for the processes that involve mixing of different states (here $SS/D\bar{D}$; for processes that do not involve mixing, this is not necessary, as the wavefunctions can be expressed in terms of the reduced mass.) 
%This is a suitable approximation since the masses of the multiplets must be quite similar in order for them to co-annihilate significantly in the early universe. 
%
(b) We have neglected the Higgs mass in the Higgs-mediated potential in order again to obtain analytical results; this approximation is discussed in \cref{sec:LongRangeDynamics_AsymptoticStates_CoulombApprox}. 
(c) The calculations are done in the symmetric electroweak phase. 
(d)  While not considered here, the capture into excited states can be significant (even if not dominant) due to the monopole transitions occurring in this class of models~\cite{Oncala:2019yvj}. 
(e) Monopole transitions can lead to the apparent violation of unitarity even for small couplings; here we have adopted an prescription to treat this issue, however a dedicated study is required, as discussed in \cref{sec:BSF_Unitarity}. 
Among the above, we believe that (d) and (e), in particular, merit further work in the future. The suitability of these approximations for the computation of the DM relic density is discussed in detail in \cite{Oncala:2021swy}.

\clearpage
\appendix
\section*{Appendices}
\section{Kernel and wavefunction (anti)symmetrisation for identical particles \label{App:IdenticalParticles}}

\begin{figure}[b!]
\centering	
\begin{tikzpicture}[line width=1pt, scale=1]
\begin{scope}[shift={(0,+1.5)}]
\node at (-1.8,0.5){$\sA$};\node at (+1.8,0.5){$\sA'$};
\node at (-1,0.9){$P/2+p$};\node at (+1,0.9){$P/2+p'$};
\draw[fermion] (-1.5, 0.5) -- (0, 0.5);\draw[fermion] (0, 0.5) -- (1.5, 0.5);
\draw[fermion] (-1.5,-0.5) -- (0,-0.5);\draw[fermion] (0,-0.5) -- (1.5,-0.5);
\node at (-1,-0.9){$P/2-p$};\node at (+1,-0.9){$P/2-p'$};
\node at (-1.8,-0.5){$\sB$};\node at (+1.8,-0.5){$\sB'$};
\draw[fill=lightgray,draw=none,shift={(0,0)}] (-0.35,-0.6) rectangle (0.35,0.6);
\node at (0,0){$G$};
\end{scope}
\node at (0,0){$=$};
\begin{scope}[shift={(-4,-4.8)}]
\begin{scope}[shift={(0,+3)}]
\node at (-1.8,0.5){$\sA$};\node at (+1.8,0.5){$\sA'$};
\node at (0,0.9){$P/2+p=P/2+p'$};
\draw[fermion] (-1.5, 0.5) -- (1.5, 0.5);
\draw[fermion] (-1.5,-0.5) -- (1.5,-0.5);
\node at (0,-0.9){$P/2-p=P/2-p'$};
\node at (-1.8,-0.5){$\sB$};\node at (+1.8,-0.5){$\sB'$};
%%%%%
\node at(4,0){$+$};
\end{scope}
\node at (0,1.5){$+$};
\begin{scope}[shift={(0,0)}]
\node at (-1.8,0.5){$\sA$};\node at (+1.8,0.5){$\sA'$};
\node at (-1,0.9){$P/2+p$};\node at (+1,0.9){$P/2+p'$};
\draw[fermion] (-1.5, 0.5) -- (0, 0.5);\draw[fermion] (0, 0.5) -- (1.5, 0.5);
\draw[fermion] (-1.5,-0.5) -- (0,-0.5);\draw[fermion] (0,-0.5) -- (1.5,-0.5);
\node at (-1,-0.9){$P/2-p$};\node at (+1,-0.9){$P/2-p'$};
\node at (-1.8,-0.5){$\sB$};\node at (+1.8,-0.5){$\sB'$};
\draw[fill=white,draw=black,shift={(0,0)}] (-0.33,-0.6) rectangle (0.33,0.6);
\node at (0,0){$\im{\cal A}$};
%%%%%
\node at(4,0){$+$};
\end{scope}
\node at (0,-1.5){$+$};
\begin{scope}[shift={(0,-3)}]
\node at (-2.55,0.5){$\sA$};\node at (0,0.25){$\rA$};\node at (+2.55,0.5){$\sA'$};
\node at (-1.75,0.9){$P/2+p$};\node at (0,0.9){$P/2+q$};\node at (+1.75,0.9){$P/2+p'$};
\draw[fermion] (-2.25, 0.5) -- (-0.75, 0.5);
\draw[fermion] (-0.75, 0.5) -- (+0.75, 0.5);
\draw[fermion] (+0.75, 0.5) -- (+2.25, 0.5);
\draw[fermion] (-2.25,-0.5) -- (-0.75,-0.5);
\draw[fermion] (-0.75,-0.5) -- (+0.75,-0.5);
\draw[fermion] (+0.75,-0.5) -- (+2.25,-0.5);
\node at (-1.75,-0.9){$P/2-p$};\node at (0,-0.9){$P/2-q$};\node at (+1.75,-0.9){$P/2-p'$};
\node at (-2.55,-0.5){$\sB$};\node at (0,-0.25){$\rB$};\node at (+2.55,-0.5){$\sB'$};
\draw[fill=white,draw=black,shift={(-0.75,0)}] (-0.33,-0.6) rectangle (0.33,0.6);
\node at (-0.75,0){$\im{\cal A}$};
\draw[fill=white,draw=black,shift={(+0.75,0)}] (-0.33,-0.6) rectangle (0.33,0.6);
\node at (+0.75,0){$\im{\cal A}$};
%%%%%
\node at(4,0){$+$};
\end{scope}
\node at (0,-4.5) {$+$};
\node at (0,-5) {$\vdots$};
\end{scope}
\begin{scope}[shift={(+4,-4.8)}]
\begin{scope}[shift={(0,+3)}]
\node at (-1.8,0.5){$\sB$};\node at (+1.8,0.5){$\sA'$};
\node at (0,0.9){$P/2-p=P/2+p'$};
\draw[fermion] (-1.5, 0.5) -- (1.5, 0.5);
\draw[fermion] (-1.5,-0.5) -- (1.5,-0.5);
\node at (0,-0.9){$P/2+p=P/2-p'$};
\node at (-1.8,-0.5){$\sA$};\node at (+1.8,-0.5){$\sB'$};
\node at (0,0) {$(-1)^f$};
\end{scope}
\node at (0,1.5){$+$};
\begin{scope}[shift={(0,0)}]
\node at (-1.8,0.5){$\sB$};\node at (+1.8,0.5){$\sA'$};
\node at (-1,0.9){$P/2-p$};\node at (+1,0.9){$P/2+p'$};
\draw[fermion] (-1.5, 0.5) -- (0, 0.5);\draw[fermion] (0, 0.5) -- (1.5, 0.5);
\draw[fermion] (-1.5,-0.5) -- (0,-0.5);\draw[fermion] (0,-0.5) -- (1.5,-0.5);
\node at (-1,-0.9){$P/2+p$};\node at (+1,-0.9){$P/2-p'$};
\node at (-1.8,-0.5){$\sA$};\node at (+1.8,-0.5){$\sB'$};
\draw[fill=white,draw=black,shift={(0,0)}] (-0.33,-0.6) rectangle (0.33,0.6);
\node at (-0.9,0){$(-1)^f$};
\node at ( 0.0,0){$\im{\cal A}$};
\end{scope}
\node at (0,-1.5){$+$};
\begin{scope}[shift={(0,-3)}]
\node at (-2.55,0.5){$\sB$};\node at (0,0.25){$\rA$};\node at (+2.55,0.5){$\sA'$};
\node at (-1.75,0.9){$P/2-p$};\node at (0,0.9){$P/2+q$};\node at (+1.75,0.9){$P/2+p'$};
\draw[fermion] (-2.25, 0.5) -- (-0.75, 0.5);
\draw[fermion] (-0.75, 0.5) -- (+0.75, 0.5);
\draw[fermion] (+0.75, 0.5) -- (+2.25, 0.5);
\draw[fermion] (-2.25,-0.5) -- (-0.75,-0.5);
\draw[fermion] (-0.75,-0.5) -- (+0.75,-0.5);
\draw[fermion] (+0.75,-0.5) -- (+2.25,-0.5);
\node at (-1.75,-0.9){$P/2+p$};\node at (0,-0.9){$P/2-q$};\node at (+1.75,-0.9){$P/2-p'$};
\node at (-2.55,-0.5){$\sA$};\node at (0,-0.25){$\rB$};\node at (+2.55,-0.5){$\sB'$};
\draw[fill=white,draw=black,shift={(-0.75,0)}] (-0.33,-0.6) rectangle (0.33,0.6);
\node at (-1.6,0){$(-1)^f$};
\node at (-0.75,0){$\im{\cal A}$};
\draw[fill=white,draw=black,shift={(+0.75,0)}] (-0.33,-0.6) rectangle (0.33,0.6);
\node at (+0.75,0){$\im{\cal A}$};
\end{scope}
\node at (0,-4.5) {$+$};
\node at (0,-5) {$\vdots$};
\end{scope}

\end{tikzpicture}
\caption{\label{fig:Resummation_TandUchannels}
Resummation of $t$-type (\emph{left}) and $u$-type (\emph{right}) 2PI diagrams for pairs of identical particles. Summation over $\rA$, $\rB$ and integration over $q$ is implied. The $u$-type diagrams carry extra factors $(-1)^f$ with respect to their $t$-type counterparts, where $f=0$ or 1 if the interacting particles are bosons or fermions respectively, due to the different number of fermion permutations needed to perform the Wick contractions.}
\end{figure}
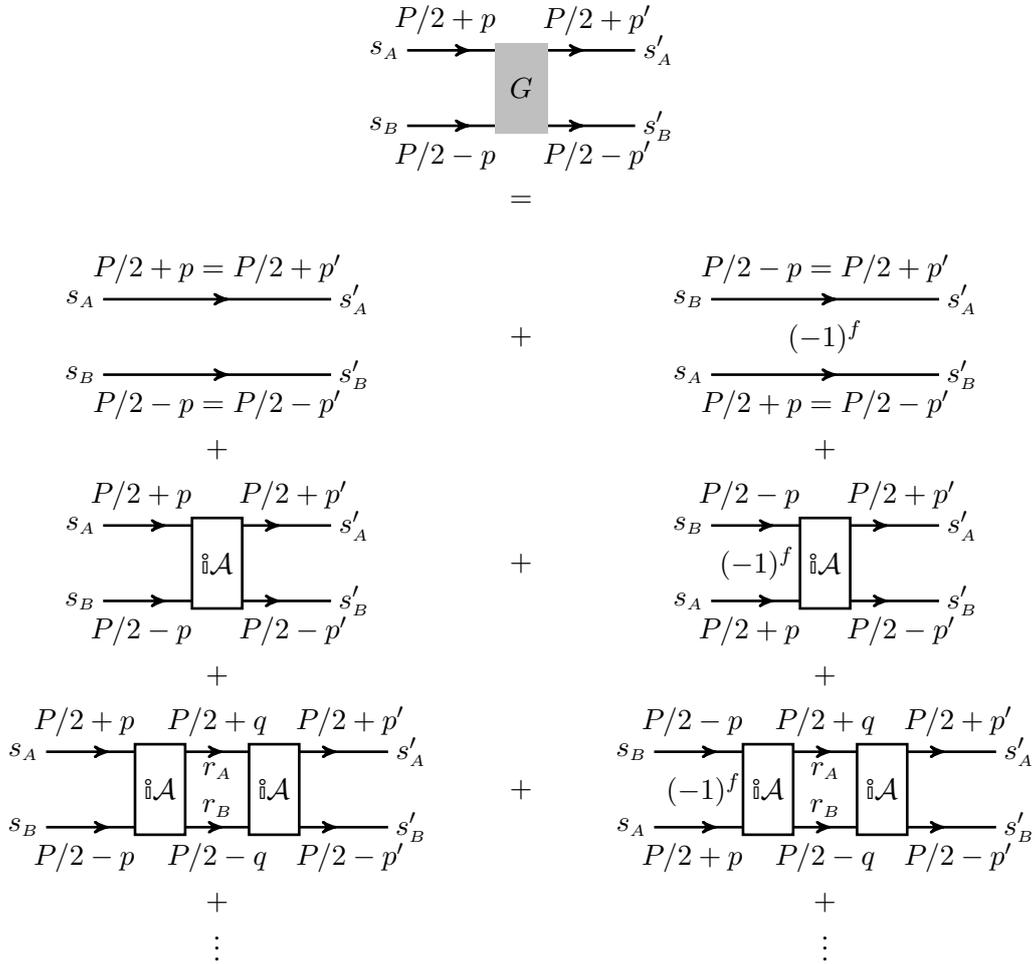

\subsection{The 2PI kernel \label{app:IdenticalParticles_Kernel}}

For identical particle (IP) pairs, $t$-channel diagrams have $u$-channel counterparts.  
However, adding them up and resumming them double-counts the loop diagrams because it corresponds to exchanging identical particles in the loops. The proper resummation necessitates using an (anti-)symmetrised kernel, as we will now show. Note that this holds not only for tree-level 2PI diagrams, but more generally for 2PI diagrams involving loops in $t$- and $u$-type configurations.

We consider the 4-point function of a pair of identical particles, $G^{\IP} \var{p}{\sA}{\sB}{p'}{\sA'}{\sB'}$, where the momentum and spin assignments are shown in \cref{fig:Resummation_TandUchannels}. To ease the notation, the dependence of $G^\IP$ on the total momentum $P$ is left implicit. Let ${\cal A}$ be a function of the same variables that stands for the sum of either the $t$- or $u$-type 2PI diagrams. Clearly, 
\begin{subequations}
\label{eq:A0}
\label[pluralequation]{eqs:A0}
\begin{align}
{\cal A} \var{p}{\sA}{\sB}{p'}{\sA'}{\sB'} = 
{\cal A} \var{-p}{\sB}{\sA}{-p'}{\sB'}{\sA'} .
\label{eq:A0_tchannel}
\end{align}
Then, the sum of complementary 2PI diagrams ($u$- or $t$-type respectively) is
\begin{align}
(-1)^f {\cal A} \var{p}{\sA}{\sB}{-p'}{\sB'}{\sA'} = 
(-1)^f {\cal A} \var{-p}{\sB}{\sA}{p'}{\sA'}{\sB'} ,
\label{eq:A0_uchannel}
\end{align}
\end{subequations}
where $f=0$~or~1 if the interacting particles are bosons or fermions. This factor arises from the different number of fermion permutations needed in the $t$- and $u$-type cases, in order to perform the Wick contractions.

The 4-point function $G^{\IP}$ includes the two ladders shown in the two columns of \cref{fig:Resummation_TandUchannels}.  These ladders are related by exchanging the momenta and spins of the initial (or final) state particles, thus we may write 
\begin{align}
G^\IP \var{p}{\sA}{\sB}{p'}{\sA'}{\sB'} 
= G_0^\IP \var{p}{\sA}{\sB}{p'}{\sA'}{\sB'} + (-1)^f G_0^\IP \var{-p}{\sB}{\sA}{p'}{\sA'}{\sB'} .
\label{eq:G}
\end{align}
Let $S(P)\equiv\im/(P^2-m^2)$.  The unamputated function $G_0$ 
is
\begin{align}
G_0^\IP  
&\var{p}{\sA}{\sB}{p'}{\sA'}{\sB'} 
= S(P/2+p') S (P/2-p') 
\ (2\pi)^4 \delta^{(4)} (p-p')  \: \delta_{\sA^{},\sA'} \delta_{\sB^{},\sB'} 
\nn \\
&+ S(P/2+p') S (P/2-p') 
\, \im {\cal A} \var{p}{\sA}{\sB}{p'}{\sA'}{\sB'} 
\, S(P/2+p) S (P/2-p)
\nn \\
&+ S(P/2+p') S (P/2-p') 
\sum_{\rA,\rB} \!\int\! \frac{d^4q}{(2\pi)^4} 
\im {\cal A} \var{q}{\rA}{\rB}{p'}{\sA'}{\sB'}
S (P/2 + q) S (P/2 - q)
\nn \\
&\times 
\im {\cal A} \var{p}{\sA}{\sB}{q}{\rA}{\rB} 
\: S (P/2 + p) \, S (P/2 - p)
+ \cdots .
\label{eq:G0}
\end{align}
\Cref{eq:G0} can be re-expressed as a Dyson-Schwinger equation with ${\cal A}$ being the kernel,
\begin{align}
&G_0^\IP \var{p}{\sA}{\sB}{p'}{\sA'}{\sB'} 
= S(P/2+p') \, S (P/2-p') \times 
\nn \\
&\times 
\[(2\pi)^4 \delta^{(4)} (p-p') \: \delta_{\sA^{},\sA'} \delta_{\sB^{},\sB'}
+ \sum_{\rA,\rB} \int \frac{d^4q}{(2\pi)^4} 
\: \im {\cal A} \var{p}{\sA}{\sB}{q}{\rA}{\rB}
\: G_0^\IP \var{q}{\rA}{\rB}{p'}{\sA'}{\sB'} \].
\label{eq:G0_DysonSchwinger}
\end{align}
In \cref{eq:G0_DysonSchwinger}, we can change the integration variable $q \to -q$ and switch $\rA \leftrightarrow \rB$. Adding up the resulting equation with \cref{eq:G0_DysonSchwinger}, we obtain
\begin{align}
&G_0^\IP \var{p}{\sA}{\sB}{p'}{\sA'}{\sB'} 
= S(P/2+p') \, S (P/2-p') \times 
\left\{ (2\pi)^4 \delta^{(4)} (p-p') \: \delta_{\sA^{},\sA'} \delta_{\sB^{},\sB'}
\right. \nn \\
&+ \frac{1}{2} \sum_{\rA,\rB} \!\int\! \frac{d^4q}{(2\pi)^4} 
\[\im {\cal A} \var{p}{\sA}{\sB}{q}{\rA}{\rB}  \: G_0^\IP \var{ q}{\rA}{\rB}{p'}{\sA'}{\sB'} 
 +\im {\cal A} \var{p}{\sA}{\sB}{-q}{\rB}{\rA} \: G_0^\IP \var{-q}{\rB}{\rA}{p'}{\sA'}{\sB'} \].
\label{eq:G0_DysonSchwinger_sym}
\end{align}
Combining \cref{eq:G,eq:G0_DysonSchwinger_sym,eq:A0}, we obtain the Dyson-Schwinger equation for $G$,
\begin{align}
G^\IP &\var{p}{\sA}{\sB}{p'}{\sA'}{\sB'} 
= S(P/2+p') \, S (P/2-p') \times 
\nn \\
&\times 
\[(2\pi)^4 \delta^{(4)} (p-p') \: \delta_{\sA^{},\sA'} \delta_{\sB^{},\sB'}
+(-1)^f (2\pi)^4 \delta^{(4)} (p+p') \: \delta_{\sA^{},\sB'} \delta_{\sA^{},\sB'}
 \right.
\nn \\
&\left.
+  \sum_{\rA,\rB} \int \frac{d^4q}{(2\pi)^4} 
\: \im {\cal K} \var{p}{\sA}{\sB}{q}{\rA}{\rB}
\: G^\IP \var{q}{\rA}{\rB}{p'}{\sA'}{\sB'} 
\] ,
\label{eq:G_DysonSchwinger}
\end{align}
where we defined 
\begin{align}
\im {\cal K} \var{p}{\sA}{\sB}{q}{\rA}{\rB} 
\equiv \frac{1}{2} \[
\im {\cal A} \var{ p}{\sA}{\sB}{q}{\rA}{\rB}
+ (-1)^f\im {\cal A} \var{-p}{\sB}{\sA}{q}{\rA}{\rB}
\] .
\label{eq:IdenticalParticles_K}
\end{align}
Evidently, \cref{eq:IdenticalParticles_K} is the average of the $t$- and $u$-type 2PI diagrams,
\begin{align}
\im {\cal K}  = \frac12 \( \im {\cal A}_t +\im {\cal A}_u \) .
\label{eq:IdenticalParticles_K_TplusU}
\end{align}
The factor $1/2$ ensures that the loop diagrams are not double-counted. From \cref{eq:IdenticalParticles_K}, we can also deduce the following relation
\begin{align}
\im {\cal K} \var{p}{\sA}{\sB}{q}{\rA}{\rB} 
= (-1)^f \im {\cal K} \var{p}{\sA}{\sB}{-q}{\rB}{\rA} ,
\label{eq:IdenticakParicles_K_symmetry}
\end{align}
which we use below in the discussion on the (anti-)symmetrisation of the wavefunctions.

\bigskip 
Finally, we note that if the interacting particles carry additional conserved numbers, e.g.~non-Abelian (gauge) charges, then appropriate factors ensuring their conservation may appear in the 0th order terms of \cref{eq:G_DysonSchwinger}, as well as inside ${\cal A}$ and consequently ${\cal K}$. However, \cref{eq:IdenticalParticles_K_TplusU} remains generally valid as is.

\subsection{Wavefunctions \label{app:IdenticalParticles_Wavefunctions}}

The 0th order terms of the Dyson-Schwinger equations determine the normalisation of the wavefunctions (see e.g.~\cite{Petraki:2015hla,Silagadze:1998ri}.) The two contributions appearing in the second line of \cref{eq:G_DysonSchwinger} ensure that the wavefunctions of identical particles are properly (anti)symmetrised, as we will now show. Instead of deriving the normalisation conditions from \cref{eq:G_DysonSchwinger}, we shall deduce them by comparing to the case of distinguishable particles (DP), whose wavefunctions are normalised as standard~\cite{Petraki:2015hla,Silagadze:1998ri}. 

For DP with equal masses, and incoming and outgoing momenta and spins as in \cref{fig:Resummation_TandUchannels},  the Dyson-Schwinger eq.~for the four-point function $G^{\DP}$ is (compare with \cref{eq:G_DysonSchwinger})
\begin{align}
&G^{\DP} \var{p}{\sA}{\sB}{p'}{\sA'}{\sB'} 
= S (P/2+p') \, S (P/2-p') \times 
\nn \\
&
\[(2\pi)^4 \delta^{(4)} (p-p') \: \delta_{\sA^{},\sA'} \delta_{\sB^{},\sB'}
+  \sum_{\rA,\rB} \int \frac{d^4q}{(2\pi)^4} 
\im {\cal K} \var{p}{\sA}{\sB}{q}{\rA}{\rB}
G^{\DP} \var{q}{\rA}{\rB}{p'}{\sA'}{\sB'} 
\] .
\label{eq:G_DP_DysonSchwinger}
\end{align}
We diagonalise \cref{eq:G_DP_DysonSchwinger} in spin space.  The factor $\delta_{\sA^{},\sA'} \delta_{\sB^{},\sB'}$ is simply the unity operator, with all its eigenvalues being 1. Thus, the contribution from the  spin-$s$ state is
\begin{align}
&G_s^{\DP} (p,p')
= S (P/2+p')  S (P/2-p') 
\[(2\pi)^4 \delta^{(4)} (p-p') 
+  \int \frac{d^4q}{(2\pi)^4} \im {\cal K}_s (p,q) G_s^{\DP} (q,p')  \] ,
\label{eq:G_DP_DysonSchwinger_s}
\end{align}
where ${\cal K}_s$ is the projected kernel. $G_s^{\DP}$ receives contributions from all energy eigenstates that schematically read~\cite{Petraki:2015hla,Silagadze:1998ri}
\begin{align}
G_{n,s}^{\DP} (p,p')
\simeq \frac
{\im \tilde{\Psi}_{n,s}^\DP (p) \ [\tilde{\Psi}_{n,s}^\DP (p')]^\star }
{2P^0 (P^0 - \omega_{n,s}+\im \epsilon)} ,
\label{eq:G_DP_contributions}
\end{align}
where here $n$ denotes collectively all quantum numbers characterising an eigenstate of energy $\omega_{n,s}$. For scattering states, these include a continuous variable that corresponds to the relative momentum of the two interacting particles, while bound states are characterised by a set of discrete quantum numbers. The (momentum space) wavefunctions $\tilde{\Psi}_{n,s}^\DP (p)$ obey the Schr\"odinger equation, and have the standard normalisation conditions that emanate from the first term in \cref{eq:G_DP_DysonSchwinger_s}.

\smallskip

Now we return to IP and the Dyson-Schwinger \cref{eq:G_DysonSchwinger}. For fermions, the operators 
$\delta_{\sA^{},\sA'} \delta_{\sB^{},\sB'}$ and 
$\delta_{\sA^{},\sB'} \delta_{\sB^{},\sA'}$
have eigenvalues $1$ and $(-1)^{s+1}$ respectively, while for bosons the eigenvalues are 1. Collectively, this is $1$ and $(-1)^{s+f}$. Thus, \cref{eq:G_DysonSchwinger} yields 
\begin{align}
&G_s^{\IP} (p,p')
= S (P/2+p')  S (P/2-p') 
\nn \\
&\times 
\[(2\pi)^4 \delta^{(4)} (p-p') + (-1)^s \, (2\pi)^4 \delta^{(4)} (p+p') 
+  \int \frac{d^4q}{(2\pi)^4} \im {\cal K}_s (p,q) G_s^{\IP} (q,p')  \] .
\label{eq:G_IP_DysonSchwinger_s}
\end{align}
The projected kernel is 
\begin{align}
\im {\cal K}_s (p,q) = 
[U^s]_{\sA \sB}
\ \im {\cal K}_s\var{p}{\sA}{\sB}{q}{\rA}{\rB}
[U^s]^{-1}_{\rA \rB} ,
\label{eq:K_IP_projection}
\end{align}
where $U^s$ is the projection operator on the spin-$s$ state, with the symmetry property  $[U^s]_{\sA \sB} = (-1)^{s+f} [U^s]_{\sB \sA}$. \Cref{eq:K_IP_projection} combined with \cref{eq:IdenticakParicles_K_symmetry} imply
\begin{align}
\im {\cal K}_s (p,q) = (-1)^{s} \ \im {\cal K}_s (p,-q) .
\label{eq:Ks_IP_symmetry}
\end{align}
The contribution to the four-point function from the $n$th energy eigenstate is
\begin{align}
G_{n,s}^{\IP} (p,p')
\simeq \frac
{\im \tilde{\Psi}_{n,s}^\IP (p) \ [\tilde{\Psi}_{n,s}^\IP (p')]^\star }
{2P^0 (P^0 - \omega_{n,s}+\im \epsilon)} ,
\label{eq:G_IP_contributions}
\end{align}
where $\tilde{\Psi}_{n,s}^\IP (p)$ are the IP wavefunctions. We now make the conjecture
\begin{align}
\tilde{\Psi}_{n,s}^\IP (p) &= \frac{1}{\sqrt{2}} \[
\tilde{\Psi}_{n,s}^\DP (p) + (-1)^{s} \tilde{\Psi}_{n,s}^\DP (-p) \] ,
\label{eq:IdenticalParticles_WF_symmetrisation_s}
\end{align}
where $\tilde{\Psi}_{n,s}^\DP (p)$ are the solutions to the DP Dyson-Schwinger \cref{eq:G_DP_DysonSchwinger_s}, assuming the kernel is the same as that of \cref{eq:G_IP_DysonSchwinger_s}. Plugging \cref{eq:IdenticalParticles_WF_symmetrisation_s} into \eqref{eq:G_IP_contributions}, and considering \eqref{eq:G_DP_contributions}, we re-express $G_{n,s}^{\IP}$ as
\begin{align}
G_{n,s}^{\IP} (p,p') 
= \frac{1}{2} \[G_{n,s}^{\DP} (p,p') + G_{n,s}^{\DP} (-p,-p')\]
+ \frac{(-1)^s}{2} \[G_{n,s}^{\DP} (p,-p') + G_{n,s}^{\DP} (-p,p')\] .
\label{eq:G_ID_vs_DP}
\end{align}
It is now easy to see that, by virtue of the DP Dyson-Schwinger \cref{eq:G_DP_DysonSchwinger_s} and the property of the IP kernel \eqref{eq:Ks_IP_symmetry}, the four-point function $G_{n,s}^{\IP} (p,p')$ of \cref{eq:G_ID_vs_DP} satisfies the IP Dyson-Schwinger \cref{eq:G_IP_DysonSchwinger_s}. Therefore, the wavefunctions \eqref{eq:IdenticalParticles_WF_symmetrisation_s} are indeed the desired solutions. Expanding in modes of definite orbital angular momentum $\ell$, for which 
\begin{align}
\tilde{\Psi}_{n,\ell s}^\DP (-p) = (-1)^\ell \tilde{\Psi}_{n,\ell s}^\DP (-p),
\label{eq:WF_IP_Parity}
\end{align} 
\cref{eq:IdenticalParticles_WF_symmetrisation_s} becomes
\begin{align}
\tilde{\Psi}_{n,\ell s}^\IP (p) &= \frac{1+ (-1)^{\ell +s}}{\sqrt{2}} \tilde{\Psi}_{n,\ell s}^\DP (p).
\label{eq:WF_IP_symmetrisation}
\end{align}
Note though that, as mentioned in \cref{app:IdenticalParticles_Kernel}, if the interacting particles carry additional conserved numbers, then appropriate (anti-)symmetrisation factors may appear in \cref{eq:WF_IP_Parity,eq:WF_IP_symmetrisation} (cf.~e.g.~$DD$ potential in~\cref{sec:LongRangeDynamics_Potential}.)

\clearpage
%%%%%%%%%%%%%%%%%%%%%%%%%%%%%%%%%%%%%%%%%%%%%%%%%%%%%%%%%%%%%%%%%%%%%%%%%%%%%%%%%%%%%%%%%%%%%%%%%
%%%%%%%%%%%%%%%%%%%%%%%%%%%%%%%%%%%%%%%%%%%%%%%%%%%%%%%%%%%%%%%%%%%%%%%%%%%%%%%%%%%%%%%%%%%%%%%%%
%%%%%%%%%%%%%%%%%%%%%%%%%%%%%%%%%%%%%%%%%%%%%%%%%%%%%%%%%%%%%%%%%%%%%%%%%%%%%%%%%%%%%%%%%%%%%%%%%
\section{Perturbative transition amplitudes: an example \label{App:BSF_PerturbativeExample}}

\begin{figure}[h!]
\centering	
\begin{tikzpicture}[line width=1pt, scale=1]
\begin{scope}[shift={(0,0)}]
%%%%% field lines
\draw[doublefermion] 	(-1,+0.5)--(0,+0.5);\draw[doublefermion] 	(0,+0.5)--(1,+0.5);
\draw[fermionnoarrow]   (-1,-0.5)--(0,-0.5);\draw[doublefermionbar] (0,-0.5)--(1,-0.5);
\draw[scalar] (0,-0.5)--(0.3,-1.5);%\node at (0.6,1.7) {$W$};
%%%%% spins
\node at (-1.25,+0.45) {$s_1$};\node at (+1.25,+0.45) {$r_1$};
\node at (-1.25,-0.45) {$s_2$};\node at (+1.25,-0.45) {$r_2$};
%%%%% momenta
\draw[->] (-0.9,+0.75) -- (-0.4,+0.75);\node at (-0.7,+1) {$k_1$};
\draw[->] (-0.9,-0.75) -- (-0.4,-0.75);\node at (-0.7,-1.1) {$k_2$};
\draw[->] (+0.4,+0.75) -- (+0.9,+0.75);\node at (+0.7,+1) {$p_1$};
\draw[->] (+0.4,-0.75) -- (+0.9,-0.75);\node at (+0.7,-1.1) {$p_2$};
\draw[->] (-0.1,-1.1)--(0.05,-1.55);\node at (-0.25,-1.6) {$\PH$};
%%%%% SU2L
\node at (-0.7,+0.25) {$i$};\node at (+0.4,+0.25) {$i'$};
							\node at (+0.4,-0.25) {$j'$};
							\node at (+0.3,-1.7) {$h$};
\end{scope}
\end{tikzpicture}
\caption{\label{fig:BSF_PerturbativeExample} 
Example of diagram contributing to the perturbative parts of the amplitudes of various transition processes considered in \cref{Sec:BSF}.}
\end{figure}
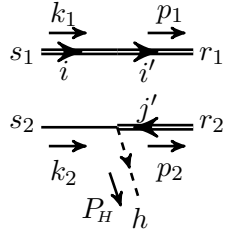

We demonstrate the calculation of diagrams contributing to the perturbative part of the transition amplitudes of \cref{Sec:BSF}. We will work out in detail the diagram shown in \cref{fig:BSF_PerturbativeExample}.

%%%%%%%%%%%%%%%%%%%%%%%%%%%%%%%%%%%%%%%%%%%%%%%%%%%%%%%%%%%%%%%%%%%%
%%%%%%%%%%%%%%%%%%%%%%%%%%%%%%%%%%%%%%%%%%%%%%%%%%%%%%%%%%%%%%%%%%%%

We first express the fields in canonical form, following~\cite{PeskinSchroeder}, 
\begin{subequations}
\label{eq:CanonicalQuantisation}
\label[pluralequation]{eqs:CanonicalQuantisation}
\begin{align}
H_j(x) 
&= \int \dfrac{d^3q}{(2\pi)^3} \dfrac{1}{\sqrt{2 E_{\H}({\bf q})}} 
\[a_j 			({\bf q}) \, e^{-\im {q \cdot x}} 
 +b_j^{\dagger} 	({\bf q}) \, e^{+\im {q \cdot x}} \] ,
\label{eq:CanonicalQuantisation_H}
\\
H^\dagger_j(x) 
&= \int \dfrac{d^3q}{(2\pi)^3} \dfrac{1}{\sqrt{2 E_{\H}({\bf q})}} 
\[b_j			({\bf q}) \, e^{-\im {q \cdot x}} 
 +a_j^{\dagger}	({\bf q}) \, e^{+\im {q \cdot x}}\] ,
\label{eq:CanonicalQuantisation_Hbar}
\\
S(x) 
&= \int \dfrac{d^3q}{(2\pi)^3} \dfrac{1}{\sqrt{2 E_{\S}({\bf q})}} 
\sum_s 
\[c			 ({\bf q},s) 	\, u ({\bf q},s) e^{-\im {q \cdot x}} 
 +c^{\dagger}({\bf q},s)	\, v ({\bf q},s) e^{+\im {q \cdot x}}\] ,
\label{eq:CanonicalQuantisation_S}
\\
\bar{S}(x) 
&= \int \dfrac{d^3q}{(2\pi)^3} \dfrac{1}{\sqrt{2 E_{\S}({\bf q})}} 
\sum_s 
\[c			 ({\bf q},s) \, \bar{v} ({\bf q},s) e^{-\im {q \cdot x}} 
 +c^{\dagger}({\bf q},s) \, \bar{u} ({\bf q},s) e^{+\im {q \cdot x}}\] ,
\label{eq:CanonicalQuantisation_Sbar}
\\
D_j(x) 
&= \int \dfrac{d^3q}{(2\pi)^3} \dfrac{1}{\sqrt{2 E_{\D}({\bf q})}} 
\sum_s 
\[d_j 			({\bf q},s) \, u ({\bf q},s) e^{-\im {q \cdot x}} 
 +f_j^{\dagger} 	({\bf q},s)	\, v ({\bf q},s) e^{+\im {q \cdot x}}\] ,
\label{eq:CanonicalQuantisation_D}
\\
\bar{D}_j(x) 
&= \int \dfrac{d^3q}{(2\pi)^3} \dfrac{1}{\sqrt{2 E_{\D}({\bf q})}} 
\sum_s 
\[f_j			({\bf q},s) \, \bar{v} ({\bf q},s) e^{-\im {q \cdot x}} 
+ d_j^{\dagger}	({\bf q},s)	\, \bar{u} ({\bf q},s) e^{+\im {q \cdot x}}\] ,
\label{eq:CanonicalQuantisation_Dbar}
\end{align}
\end{subequations}
where $j$ is the $\SUL$ index, and the various $q^0$ inside the integrals are equal to the corresponding on-shell energies,
\begin{align}
E_{\H}({\bf q}) = \sqrt{\mH^2+{\bf q}^2}, 
\qquad
E_{\S}({\bf q}) = E_{\D}({\bf q}) = \sqrt{m^2+{\bf q}^2}.
\end{align}
The creation and annihilation operators obey the (anti)commutation relations
\begin{subequations}
\label{eq:CommutationRelations}
\label[pluralequation]{eqs:CommutationRelations}
\begin{align}
[a_i({\bf p}), a_j^\dagger({\bf q}) ] = 
[b_i({\bf p}), b_j^\dagger({\bf q}) ] &= 
(2\pi)^3 \delta^3 ({\bf p-q}) \, \delta_{ij},
\label{eq:CommutationRelations_H}
\\
\{c({\bf p},r), c^\dagger({\bf q},s) \} &= 
(2\pi)^3 \delta^3 ({\bf p-q}) \, \delta_{rs},
\label{eq:CommutationRelations_S}
\\
\{d_i({\bf p},r), d_j^\dagger({\bf q},s) \} = 
\{f_i({\bf p},r), f_j^\dagger({\bf q},s) \} &= 
(2\pi)^3 \delta^3 ({\bf p-q}) \, \delta_{rs} \, \delta_{ij},
\label{eq:CommutationRelations_D}
\end{align}
\end{subequations}
with all other combinations being zero.
The one-particle states are 
\begin{subequations}
\label{eq:OneParticleStates}
\label[pluralequation]{eqs:OneParticleStates}
\begin{align}
|H_j({\bf q}) \> 
&= \sqrt{2E_{\H}({\bf q})} \, a_j^\dagger ({\bf q},s) |0\> ,
&
|H_j^\dagger ({\bf q}) \> 
&= \sqrt{2E_{\H}({\bf q})} \, b_j^\dagger ({\bf q},s) |0\> ,
\label{eq:OneParticleState_HorHdagger}
\\
|S({\bf q},s) \> 
&= \sqrt{2E_{\S}({\bf q})} \, c^\dagger ({\bf q},s) |0\> ,
\label{eq:OneParticleState_S}
\\
|D_j({\bf q},s) \> 
&= \sqrt{2E_{\D}({\bf q})} \, d_j^\dagger ({\bf q},s) |0\> ,
&
|\bar{D}_j({\bf q},s) \> 
&= \sqrt{2E_{\D}({\bf q})} \, f_j^\dagger ({\bf q},s) |0\> .
\label{eq:OneParticleState_DorDbar}
\end{align}
\end{subequations}

We now return to the diagram of \cref{fig:BSF_PerturbativeExample}. Its contribution to an amplitude is
\begin{align}
&(2\pi)^4 \delta^4(k_1+k_2-p_1-p_2-\PH) \, \im {\cal A} [DS \to D\bar{D}H] \simeq 
\nn \\
&\simeq
\<  D_{i'} ({\bf p}_1, r_1)  \, \bar{D}_{j'} ({\bf p}_2, r_2) \, H_h (\PHvec) |
~ (-\im y) \int d^4 x \, \bar{S}(x) H^\dagger(x) D(x) ~
|D_i({\bf k}_1, s_1) \, S({\bf k}_2, s_2) \>
\nn \\[1ex] 
%%%%%%%%%%%%%%%%%%%%%%%%%%%%%%%%%%%%%%%%%%%%%%%%%%%%%%%%%%%%%%%%%%%%%
&= 
(-\im y)
\sum_{t_{\S}^{},t_{\D}^{}} \sum_n
\int d^4 x \int 
%%%%%%%%%%%%%%%%%%%%%%%%%%%% 
\dfrac{d^3 q_{\S}}{(2\pi)^3} 
\dfrac{d^3 q_{\H}}{(2\pi)^3} 
\dfrac{d^3 q_{\Dbar}}{(2\pi)^3}
%%%%%%%%%%%%%%%%%%%%%%%%%%%%
\dfrac
{\sqrt{
   2E_{\D}({\bf k}_1)
\, 2E_{\S}({\bf k}_2)
\, 2E_{\D}({\bf p}_1)
\, 2E_{\D}({\bf p}_2)
\, 2E_{\H}(\PHvec)
}}
{\sqrt{
   2E_{\S}({\bf q}_{\S}^{})
\, 2E_{\H}({\bf q}_{\H}^{})
\, 2E_{\D}({\bf q}_{\Dbar}^{})
}}
%%%%%%%%%%%%%%%%%%%%%%%%%%%% 
\nn \\
&\times
\<0|
a_h (\PHvec)
f_{j'} ({\bf p}_2, r_2)
d_{i'} ({\bf p}_1, r_1)
%%%%
\[c ({\bf q}_{\S}^{}, t_{\S}) \, 
\bar{v} ({\bf q}_{\S}^{}, t_{\S}^{}) \, 
e^{-\im {q_{\S}^{} \cdot x}} 
+ c^{\dagger} ({\bf q}_{\S}^{}, t_{\S}^{}) \, 
\bar{u} ({\bf q}_{\S}^{}, t_{\S}^{}) \, 
e^{+\im {q_{\S}^{} \cdot x}}\]
\nn \\
&\times
\[b_n ({\bf q}_{\H}^{}) \, e^{-\im {q_{\H}^{} \cdot x}} 
+ a_n^{\dagger}	({\bf q}_{\H}^{}) \, e^{+\im {q_{\H}^{} \cdot x}}\]
\nn \\
&\times
\[
d_n ({\bf q}_{\Dbar}^{}, t_{\Dbar}^{}) \, 
u ({\bf q}_{\Dbar}^{}, t_{\Dbar}^{}) \,
e^{-\im {q_{\Dbar}^{} \cdot x}} 
+ f_n^{\dagger} ({\bf q}_{\Dbar}^{}, t_{\Dbar}^{}) \, 
v ({\bf q}_{\Dbar}^{}, t_{\Dbar}^{}) \, 
e^{+\im {q_{\Dbar}^{} \cdot x}}\]
%%%%
d_i^\dagger ({\bf k}_1, s_1)
c^\dagger   ({\bf k}_2, s_2)
|0\>
\nn \\[1ex] 
%%%%%%%%%%%%%%%%%%%%%%%%%%%%%%%%%%%%%%%%%%%%%%%%%%%%%%%%%%%%%%%%%%%%%
&= 
(-\im y) 
\sum_{t_{\S}^{},t_{\Dbar}^{}} \sum_n
\int d^4 x 
\int 
\dfrac{d^3 q_{\S}}{(2\pi)^3} 
\dfrac{d^3 q_{\H}}{(2\pi)^3} 
\dfrac{d^3 q_{\Dbar}}{(2\pi)^3}
%%%%%%%%%%%%%%%%%%%%%%%%%%%%
\dfrac
{\sqrt{
   2E_{\D}({\bf k}_1)
\, 2E_{\S}({\bf k}_2)
\, 2E_{\D}({\bf p}_1)
\, 2E_{\D}({\bf p}_2)
\, 2E_{\H}(\PHvec)
}}
{\sqrt{
   2E_{\S}({\bf q}_{\S}^{})
\, 2E_{\H}({\bf q}_{\H}^{})
\, 2E_{\D}({\bf q}_{\Dbar}^{})
}}
\nn \\
%%%%%%%%%%%%%%%%%%%%%%%%%%%% 
&\times \ e^{-\im (q_{\S}^{} - q_{\H}^{} - q_{\Dbar}^{}) x}
%%%%%%%%%%%%%%%%%%%%%%%%%%%% 
\times
(-1)^2 (2\pi)^3 \delta^{3} ({\bf k}_2 - {\bf q}_{\S}) \delta_{t_{\S}^{} s_2^{}}
\, \bar{v} ({\bf q}_{\S}^{}, t_{\S}^{})
\times 
(2\pi)^3 \delta^{3} (\PHvec - {\bf q}_{\H}^{}) \delta_{hn}
\nn \\
&
\times
(-1) (2\pi)^3 \delta^{3} ({\bf p}_2^{} - {\bf q}_{\Dbar}^{}) 
\delta_{t_{\Dbar}^{} r_2^{}} \delta_{j'n}
\, v ({\bf q}_{\Dbar}^{}, t_{\Dbar}^{})
\times
(2\pi)^3 \delta^{3} ({\bf k}_1^{} - {\bf p}_1^{}) \delta_{s_1^{} r_1^{}} \delta_{ii'}
\nn \\[1ex] 
%%%%%%%%%%%%%%%%%%%%%%%%%%%%%%%%%%%%%%%%%%%%%%%%%%%%%%%%%%%%%%%%%%%%%
&=
(-1)^3(-\im y)  \delta_{s_1^{} r_1^{}} \delta_{ii'} \delta_{hj'}
\, \bar{v} ({\bf k}_2^{}, s_2^{}) \, v ({\bf p}_2^{}, r_2^{})
\sqrt{2E_{\D}({\bf k}_1) \, 2E_{\D}({\bf p}_1)}
(2\pi)^3 \delta^{3} ({\bf k}_1^{} - {\bf p}_1^{})
\nn \\
&\times 
\int d^4 x \ e^{-\im (k_2^{} - \PH - p_2^{}) x},
\label{eq:PerturbativeAmplitudeExample_0}
\end{align}
where in the third step, we did the following contractions, in order,
\begin{itemize}
\item 
$c ({\bf q}_{\S}^{}, t_{\S})$ with $c^\dagger ({\bf k}_2, s_2)$, 
\item 
$a_n^\dagger ({\bf q}_{\H}^{})$ with $a_h(\PHvec)$, 
\item 
$f_n^{\dagger} ({\bf q}_{\Dbar}^{}, t_{\Dbar}^{})$ with $f_{j'} ({\bf p}_2, r_2)$,
\item 
$d_i^{\dagger} ({\bf k}_1^{}, s_1^{})$ with $d_{i'} ({\bf p}_1, r_1)$,
\end{itemize} 
and accounted for the signs arising from the permutations of the fermion operators. 
Considering that $|{\bf k}_2 - {\bf p}_2| = |\PHvec| \ll |{\bf k}_2|, |{\bf p}_2|$, we may set $\bar{v} ({\bf k}_2^{}, s_2^{}) \, v ({\bf p}_2^{}, r_2^{}) \simeq 
-2m\delta_{s_2 r_2}$. Moreover, we do the following manipulation\footnote{Because the diagram of \cref{fig:BSF_PerturbativeExample} is disconnected when considered alone, the energy-momentum conservation on the vertex suggests that it vanishes. This is an artifact of having set the incoming and outgoing $D$, $\bar{D}$ and $S$ particles on shell. In reality, because of their interactions along the ladders (cf.~\cref{fig:BSF_general}), the propagating fields are not exactly on-shell. A method for integrating out the virtuality of these particles has been suggested in \cite{Petraki:2015hla} and employed e.g.~in \cite{Harz:2018csl,Oncala:2018bvl}.} 
\begin{multline}
(2\pi)^3\delta^3 ({\bf k}_1^{} - {\bf p}_1^{})
\int d^4 x \ e^{-\im (k_2^{} - \PH - p_2^{}) x} =
(2\pi)^3\delta^3 ({\bf k}_1^{} - {\bf p}_1^{})
\int d^4 x \ e^{-\im (k_1^{} + k_2^{} - \PH - p_1^{} - p_2^{}) x}
\\
=
(2\pi)^4\delta^4 (k_1 + k_2 - \PH - p_1 - p_2) 
(2\pi)^3\delta^3 ({\bf k}_1^{} - {\bf p}_1^{}) .
\end{multline}
Finally, setting $E_{\D}({\bf k}_1)=E_{\D}({\bf p}_1)\simeq m$, \cref{eq:PerturbativeAmplitudeExample_0} yields
\begin{align}
\im {\cal A} [DS \to D\bar{D}H] \simeq
(-\im y)  \delta_{s_1^{} r_1^{}}  \delta_{s_2^{} r_2^{}} \delta_{ii'} \delta_{hj'}
\, 4m^2
(2\pi)^3 \delta^{3} ({\bf k}_1^{} - {\bf p}_1^{}) .
\label{eq:PerturbativeAmplitudeExample}
\end{align}

\clearpage
%%%%%%%%%%%%%%%%%%%%%%%%%%%%%%%%%%%%%%%%%%%%%%%%%%%%%%%%%%%%%%%%%%%%%%%%%%%%%%%%%%%%%%%%%%%%%%%%%
%%%%%%%%%%%%%%%%%%%%%%%%%%%%%%%%%%%%%%%%%%%%%%%%%%%%%%%%%%%%%%%%%%%%%%%%%%%%%%%%%%%%%%%%%%%%%%%%%
%%%%%%%%%%%%%%%%%%%%%%%%%%%%%%%%%%%%%%%%%%%%%%%%%%%%%%%%%%%%%%%%%%%%%%%%%%%%%%%%%%%%%%%%%%%%%%%%%
\section{Overlap integral for monopole bound-to-bound transitions \label{App:OverlapIntegral_RBoundToBound}}

We want to compute the overlap integrals defined in \cref{eq:OverlapIntegral_RboundTobound_def},
\begin{align}
{\cal R}_{n'\ell'm',n\ell m} (\aB',\aB) 
&\equiv 
\int \frac{d^3 {\bf p}}{(2\pi)^3}
\ \tilde{\varphi}_{n'\ell'm'} ({\bf p};\aB')
\ \tilde{\varphi}_{n\ell m}^* ({\bf p};\aB)
\nn \\ 
&= 
\int d^3 {\bf r}
\ \varphi_{n'\ell'm'} ({\bf r};\aB')
\ \varphi_{n\ell m}^* ({\bf r};\aB) ,
\label{eq:OverlapIntegral_RboundTobound_def_re}
\end{align}
where the position-space bound-level wavefunctions for the potential $V=-\aB/r$ are
\begin{align}
\varphi_{n\ell m} ({\bf r};\aB) &= 
\(\dfrac{2\kappaB}{n}\)^\frac{3}{2}  
\[\frac{(n-\ell-1)!}{2n(n+\ell)!} \]^\frac{1}{2}
e^{-\xB^{}/n} \ \(\frac{2\xB}{n}\)^{\ell} 
L_{n-\ell-1}^{2\ell+1} \(\frac{2\xB}{n}\) 
Y_{\ell m} ({\Omega_{\bf r}}) ,
\label{eq:Wavefunctions_BoundStates}
\end{align}
with $\kappaB \equiv \mu \aB$ being the Bohr momentum and $\xB \equiv \kappaB r$.
Inserting \cref{eq:Wavefunctions_BoundStates} into \eqref{eq:OverlapIntegral_RboundTobound_def_re}, 
\begin{multline}
{\cal R}_{n'\ell'm',n\ell m} (\aB',\aB)  
= \delta_{\ell \ell'} \delta_{m m'}
\[\frac{(n-\ell-1)!}{2n(n+\ell)!} \, \frac{(n'-\ell-1)!}{2n'(n'+\ell)!} \]^{1/2}
\(\dfrac{4\kappaB\kappaB'}{n n'}\)^{\ell+3/2} \times
\\
\times \int_0^\infty dr 
\, e^{-(\kappaB^{}/n+\kappaB'/n') \, r}
\, r^{2\ell+2}
\, L_{n-\ell-1}^{2\ell+1} \(\frac{2\kappaB r}{n}\) 
\, L_{n'-\ell-1}^{2\ell+1} \(\frac{2\kappaB' r}{n'}\) .
\label{eq:OverlapIntegral_RboundTobound_step1}
\end{multline}
To compute the integral, ee will use the identity~\cite[section~7.414, item 4]{Integrals_GradshteynRyzhik}
\begin{align}
&\int_0^\infty dr \, e^{-\rho r} r^{a} L_{q}^{a} (\lambda r) L_{q'}^{a} (\lambda' r) =
\label{eq:Identity} 
\\ 
&=\dfrac{\Gamma(q+q'+a+1)}{q!q'!}
\dfrac{(b-\lambda)^q \, (b-\lambda')^{q'}}{b^{q+q'+a+1}}
{}_2F_1 \(-q,~ -q';~ -q-q'-a;~ \dfrac{\rho(\rho-\lambda-\lambda')}{(\rho-\lambda)(\rho-\lambda')}\)
\nn \\
&\equiv h(a,q,q',\lambda,\lambda',\rho) , \nn 
\end{align}
where ${}_2F_1$ is the ordinary hypergeometric function. \Cref{eq:Identity} holds for ${\rm Re}(a) > -1$ and ${\rm Re}(\rho) > 0$. The overlap integral \eqref{eq:OverlapIntegral_RboundTobound_step1} can be expressed in terms of the $h$ function as
\begin{align}
{\cal R}_{n'\ell'm',n\ell m} (\aB',\aB) &=
\delta_{\ell \ell'} \delta_{m m'}
\[\frac{(n-\ell-1)!}{2n(n+\ell)!} \, \frac{(n'-\ell-1)!}{2n'(n'+\ell)!} \]^{1/2}
\(\dfrac{4\kappaB\kappaB'}{n n'}\)^{\ell+3/2}
\nn \\ 
&\times 
\[-\dfrac{d}{d\rho} h \(a,q,q',\lambda,\lambda', \rho \) \],
\label{eq:OverlapIntegral_RboundTobound_step2}
\end{align}
with
\begin{align}
a =2\ell+1, \quad
q^{(\prime)} =n^{(\prime)}-\ell-1, \quad
\lambda^{(\prime)} = 2\kappaB^{(\prime)}/n^{(\prime)}, \quad
\rho = \kappaB/n+\kappaB'/n'.
\label{eq:OverlapIntegral_parameters}
\end{align}
We find
\begin{empheq}[box=\widefbox]{align}
&{\cal R}_{n'\ell'm',n\ell m} (\aB',\aB) =
\delta_{\ell \ell'} \delta_{m m'}
\dfrac{(-1)^{n-\ell-1} \, (n+n'-1)!}{\sqrt{n n' (n+\ell)!(n'+\ell)! (n-\ell-1)! (n'-\ell-1)!}}
\nn \\
&\times
\(\dfrac{4\aB \aB'}{n n'}\)^{\ell+3/2}
(\aB-\aB') 
\(\dfrac{\aB}{n} - \dfrac{\aB'}{n'}\)^{n+n'-2\ell-3}
\(\dfrac{\aB}{n} + \dfrac{\aB'}{n'}\)^{-(n+n'+1)}
\label{eq:OverlapIntegral_RboundTobound_result}
\\
&\times {}_2F_1 \[ 1+\ell-n,~ 1+\ell-n',~ 1-n-n',~ \(\dfrac{\aB/n+\aB'/n'}{\aB/n-\aB'/n'} \)^2 \].
\nn 
\end{empheq}
Note that \cref{eq:OverlapIntegral_RboundTobound_result} vanishes if $\aB=\aB'$ and $n\neq n'$, but is equal to 1 if $\aB=\aB'$ and $n = n'$, due to the orthonormality of the wavefunctions. \Cref{eq:OverlapIntegral_RboundTobound_result} is useful for calculating monopole transitions between bound states of different potentials, i.e.~for $\aB \neq \aB'$. This result complements the computation of ref.~\cite{Oncala:2019yvj} of scattering-to-bound monopole transitions.

For $n=n'=1$ and $\ell=0$, we obtain
\begin{align}
{\cal R}_{100,100} (\aB',\aB) = 
\dfrac{8 (\aB\aB')^{3/2}}{ (\aB+\aB')^3} .
\label{eq:OverlapIntegral_R100to100}
\end{align}

%%%%%%%%%%%%%%%%%%%%%%%%%%%%%%%%%%%%%%%%%%%%%%%%%%%%%%%%%%%%%%%%%%%%%%%%%%%%%%%%%%%%%%%%%%%%%%%%%
%%%%%%%%%%%%%%%%%%%%%%%%%%%%%%%%%%%%%%%%%%%%%%%%%%%%%%%%%%%%%%%%%%%%%%%%%%%%%%%%%%%%%%%%%%%%%%%%%
%%%%%%%%%%%%%%%%%%%%%%%%%%%%%%%%%%%%%%%%%%%%%%%%%%%%%%%%%%%%%%%%%%%%%%%%%%%%%%%%%%%%%%%%%%%%%%%%%
%\clearpage
\section{Scalar emission via vector-scalar fusion \label{App:VectorScalarFusion}}

\begin{figure}[h!]
\centering	
\begin{tikzpicture}[line width=1pt, scale=1.6]
%%%%%%%%%%%%%%%%%%%%%%%%%%%%%%
\begin{scope}[shift={(3,0)}]
\draw[doublefermion] 	(-1,+0.5)--(0,+0.5);\draw[doublefermion] 	(0,+0.5)--(1,+0.5);
\draw (-1,-0.5)--(0,-0.5);\draw[doublefermionbar] (0,-0.5)--(1,-0.5);
\draw[gluon] (0,+0.5)--(0,0);\draw[scalar] (0,-0.5)--(0,0);
\draw[scalar] (0,0)--(0.7,0);
%%%%% spins
\node at (-1.25,+0.45) {$s_1$};\node at (+1.25,+0.45) {$r_1$};
\node at (-1.25,-0.45) {$s_2$};\node at (+1.25,-0.45) {$r_2$};
%%%%% momenta
\draw[->] (-0.9,+0.7) -- (-0.4,+0.7);\node at (-0.8,+0.9) {$K/2+k^{(\prime)}$};
\draw[->] (-0.9,-0.7) -- (-0.4,-0.7);\node at (-0.8,-1.0) {$K/2-k^{(\prime)}$};
\draw[->] (+0.4,+0.7) -- (+0.9,+0.7);\node at (+0.8,+0.9) {$P/2+p$};
\draw[->] (+0.4,-0.7) -- (+0.9,-0.7);\node at (+0.8,-1.0) {$P/2-p$};
\draw[->] (+0.4, 0.10) -- (+0.8, 0.10);\node at (+0.8, 0.25){$\PH$};
%%%%% SU2L
\node at (-0.7,+0.3) {$i$};\node at (+0.4,+0.3) {$i'$};
%\node at (-0.7,-0.25) {$j$};
\node at (+0.4,-0.3) {$j'$};
\node at (+0.85,-0.05) {$h$};
\end{scope}
%%%%%%%%%%%%%%%%%%%%%%%%%%%%%%
\end{tikzpicture}
\caption{\label{fig:BSF_VectorScalarFusion}
Scalar emission via vector-scalar fusion.}
\end{figure}
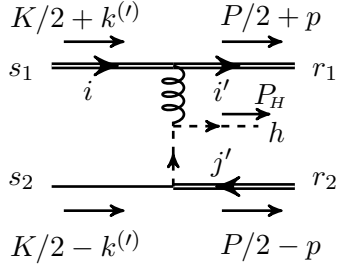

In many of the BSF processes considered in this work, the radiative parts of the amplitudes receive contributions from diagrams where off-shell vector and Higgs bosons fuse to produce the on-shell radiated Higgs boson; one such diagram is pictured in \cref{fig:BSF_VectorScalarFusion}. These diagrams resemble the ones where an on-shell vector is emitted from an off-shell vector or scalar mediator exchanged between the interacting particles (cf.~\cref{fig:BSF_SSDDbar,fig:BSF_DDbar,fig:BSF_DD,fig:BSF_SD}.) This suggests that they may be significant. Here we show that the diagrams of the type of \cref{fig:BSF_VectorScalarFusion} are of higher order than those featuring emission of a vector from an off-shell mediator. 
Moreover, BSF via vector emission is of higher order than BSF via emission of a charged scalar~\cite{Oncala:2019yvj}. Thus, the Higgs emission diagrams of the type of \cref{fig:BSF_VectorScalarFusion} are very subdominant.

%%%%%%%%%%%%%%%%%%%%%%%%%%%%%%%%%%%%%%%%%%%%%%%%%%%%%%%%%%%%%%%%%%%%
%%%%%%%%%%%%%%%%%%%%%%%%%%%%%%%%%%%%%%%%%%%%%%%%%%%%%%%%%%%%%%%%%%%%

The contribution from the diagram of \cref{fig:BSF_VectorScalarFusion} is
\begin{align}
\im {\cal A} &= 
\bar{u} (P/2+p, r_1) \, \im g_2 \gamma^\mu t^a_{i'i} \, u (K/2+k',s_1)
\[\frac{-\im g_{\mu \nu}}{(k'-p+\PH/2)^2}\]
\label{eq:BSF_A_VectorScalarFusion}
\\
&\times \im g_2 t^a_{j'j} (3\PH/2 -k'+p)^\nu
\ \frac{\im}{(k'-p-\PH/2)^2-\mH^2}
\ \bar{v} (K/2-k,s_2) (-\im y) v(P/2-p) .
\nn 
\end{align}
Applying the standard approximations due to the scale hierarchies, the above becomes
\begin{align}
\im {\cal A} &\simeq
\im g_2^2 y \, t^a_{i'i} t^a_{j'j} \, 2m 
\frac{(k'-p-\PH/2) \cdot [(K+P)/2 +k'+p]}
{({\bf k'-p})^2 [({\bf k'-p})^2 + \mH^2]} .
\label{eq:BSF_Aapprox_VectorScalarFusion}
\end{align}
The 4-vector product in the numerator is of order $\sim m^2 (\aB^2 + \vrel^2)$, which renders \cref{eq:BSF_Aapprox_VectorScalarFusion} of higher order than \cref{eq:BSF_A_DDbarToSSDDbar_Bemission,eq:BSF_A_DDbarToSSDDbar_Wemission}, and even  more so that \cref{eqs:BSF_A_SDToSSDDbar_Hemission}. 

%%%%%%%%%%%%%%%%%%%%%%%%%%%%%%%%%%%%%%%%%%%%%%%%%%%%%%%%%%%%%%%%%%%%
%%%%%%%%%%%%%%%%%%%%%%%%%%%%%%%%%%%%%%%%%%%%%%%%%%%%%%%%%%%%%%%%%%%%
%%%%%%%%%%%%%%%%%%%%%%%%%%%%%%%%%%%%%%%%%%%%%%%%%%%%%%%%%%%%%%%%%%%%
\clearpage
\section*{Acknowledgements}
We thank Karl Nordstr\"om for collaboration at the early stages of this work. This work was supported by the ANR ACHN 2015 grant (``TheIntricateDark" project), and by the NWO Vidi grant ``Self-interacting asymmetric dark matter".

%%%%%%%%%%%%%%%%%%%%%%%%%%%%%%%%%%%%%%%%%%%%%%%%%%%%%%%%%%%%%%%%%%%%
%%%%%%%%%%%%%%%%%%%%%%%%%%%%%%%%%%%%%%%%%%%%%%%%%%%%%%%%%%%%%%%%%%%%
%%%%%%%%%%%%%%%%%%%%%%%%%%%%%%%%%%%%%%%%%%%%%%%%%%%%%%%%%%%%%%%%%%%%
%\clearpage
\bibliography{Bibliography.bib}

\end{document}